\documentclass{aa}  

\usepackage{graphicx}
\usepackage{txfonts}
\usepackage{caption}
\usepackage{multirow}
\usepackage{color}
\usepackage{placeins} 

\begin{document}

   \title{New inclination changing eclipsing binaries\\in the Magellanic Clouds\thanks{Full Table4 is only available in electronic form at the CDS via anonymous ftp to \texttt{cdsarc.u-strasbg.fr} (130.79.128.5) or via \texttt{http://cdsweb.u-strasbg.fr/cgi-bin/qcat?J/A+A/XX}}}

   \author{J.~Jury\v{s}ek\inst{1,2}\fnmsep\thanks{\email{jurysek@fzu.cz}}
          , P.~Zasche\inst{2},
          M.~Wolf\inst{2},
          J.~Vra\v{s}til\inst{2},
          D.~Vokrouhlick\'y\inst{2},
          M.~Skarka\inst{3,4},
          J.~Li\v{s}ka\inst{3,5},
          J.~Jan\'ik\inst{3},\\
          M.~Zejda\inst{3},
          P.~Kurf\"{u}rst\inst{3}
          \and
          E.~Paunzen\inst{3}
          }

   \institute{
   Institute of Physics, The Czech Academy of Sciences, 
   Na Slovance 1999/2, CZ-181 21 Praha 8, Czech Republic\\
   \and
   Astronomical Institute of the Charles University, 
   Faculty of Mathematics and Physics, V Hole\v{s}ovi\v{c}k\'ach 2, 
   CZ-180 00 Praha 8, Czech Republic\\
   \and
   Department of Theoretical Physics and Astrophysics, Masaryk University, Kotl\'a\v{r}sk\'a 2,
   CZ-611 37 Brno, Czech Republic \\
   \and
   Konkoly Observatory, Research Centre for Astronomy and Earth Sciences, 
   Hungarian Academy of Sciences, Konkoly Thege Mikl\'os \'ut 15-17, 
   H-1121 Budapest, Hungary \\
   \and
   Central European Institute of Technology, Brno University of Technology,
   Purky\v{n}ova 656/123, CZ-612 00 Brno, Czech Republic \\
   }

   \date{}

 
  \abstract
   {
   Multiple stellar systems are unique laboratories for astrophysics. Analysis of their
 orbital dynamics, if well characterized from their observations, may reveal invaluable
 information about the physical properties of the participating stars. Unfortunately,
 there are only a few known and well described multiple systems, this is even more so for systems located outside the Milky Way galaxy. 
 A particularly interesting situation occurs when the inner binary in a compact triple
 system is eclipsing. This is because the stellar interaction, typically resulting in precession
 of orbital planes, may be observable as a variation of depth of the eclipses on a long timescale.
   }
   {
   We aim to present a novel method to determine compact triples using publicly available photometric data
 from large surveys. Here we apply it to eclipsing binaries (EBs) in Magellanic Clouds from OGLE III database.
 Our tool consists of identifying the cases where the orbital plane of EB evolves in accord with 
 expectations from the interaction with a third star.
   }
   {
   We analyzed light curves (LCs) of 26121 LMC and 6138 SMC EBs with the goal to identify those
 for which the orbital inclination varies in time. Archival LCs of the selected systems, when
 complemented by our own observations with Danish 1.54-m telescope, were thoroughly analyzed using
 the PHOEBE program. This provided physical parameters of components of each system. Time dependence
 of the EB’s inclination was described using the theory of orbital-plane precession. By observing
 the parameter-dependence of the precession rate, we were able to constrain the third companion mass
 and its orbital period around EB.
     }
   {
    We identified 58 candidates of new compact triples in Magellanic Clouds. This is the largest published
 sample of such systems so far. Eight of them were analyzed thoroughly and physical parameters of inner
 binary were determined together with an estimation of basic characteristics of the third star. Prior
 to our work, only one such system was well characterized outside the Milky Way galaxy. Therefore, we increased
 this sample in a significant way. These data may provide important clues about stellar formation
 mechanisms for objects with different metalicity than found in our galactic neighborhood.
   }
   {}

        
   \authorrunning{J. Jury\v{s}ek et al.}        
   \maketitle
%

\section{Introduction}

It is well known that substantial part of all types of star in the solar 
neighborhood form binary or multiple stellar systems 
\citep[e.g.,][]{abt1983, guinan}. These systems include stars of all spectral types and all stages of their evolution. Proper
description of their dynamical characteristics may result in determination of their physical parameters, 
and thus provide clues to their formation paths \citep[e.g.,][]{goodwin_kroupa2005}.

Despite the huge effort of astrophysicists in the field of stellar multiplicity in recent years, 
some mysteries about multiple systems still remain. For instance, recent results 
of thorough analysis of data from the \textit{Kepler} mission have shown that there 
is a significant drop in population of triple systems with the period of a third 
component $P_2 < 200$~days
\citep{borkovits2016}. That is in accord with earlier results of \citet{tokovinin2014} who pointed out a 
similar drop of systems with the $P_2 < 1000$~days. The lull of systems at these intermediate $P_2$ values 
appears to be real and subject to severe selection effects. The theoretical explanation, however, still remains unknown.

\begin{figure*}[!t]
\centering
\begin{tabular}{cc}
\includegraphics[width=88mm]{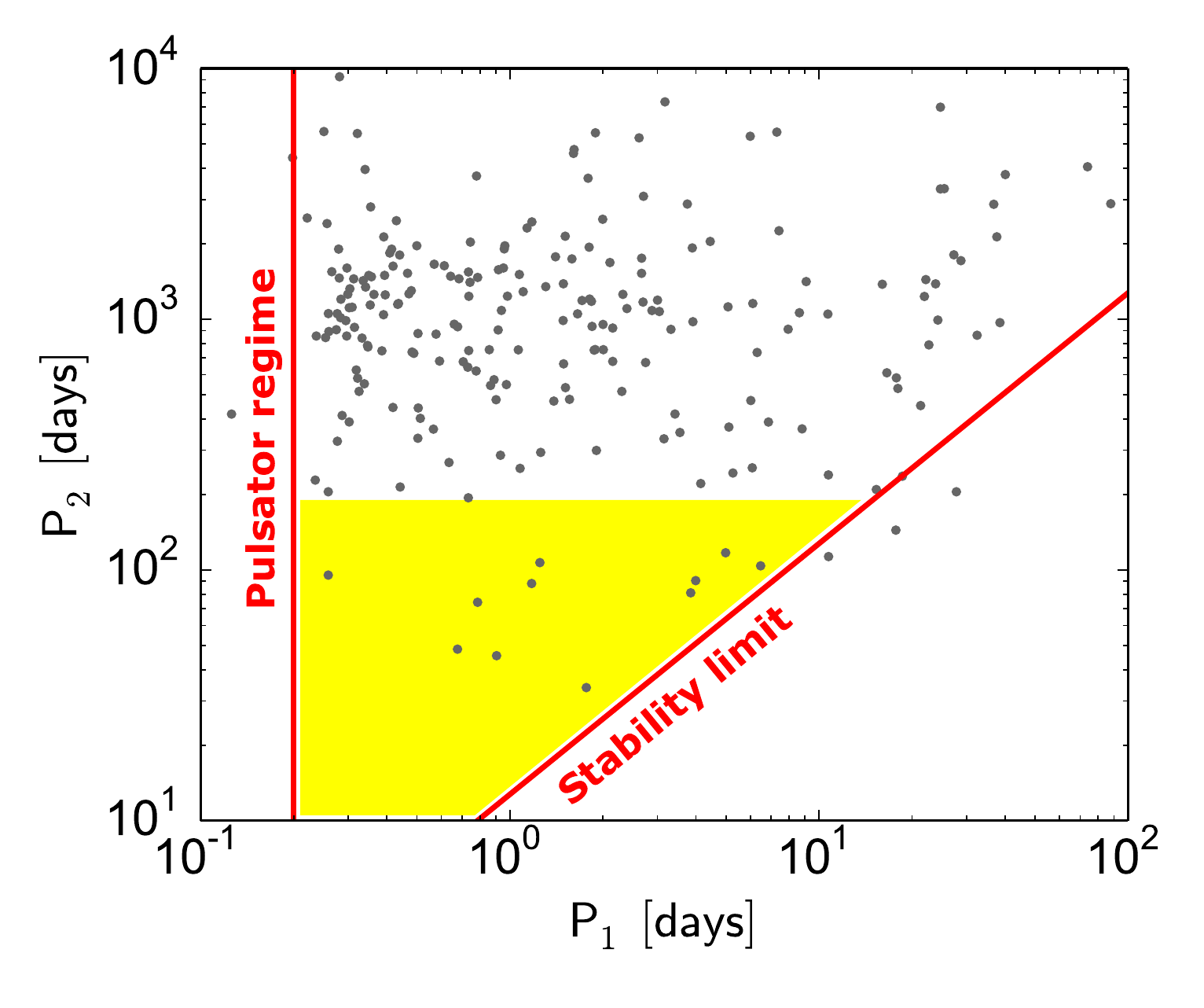} & \includegraphics[width=88mm]{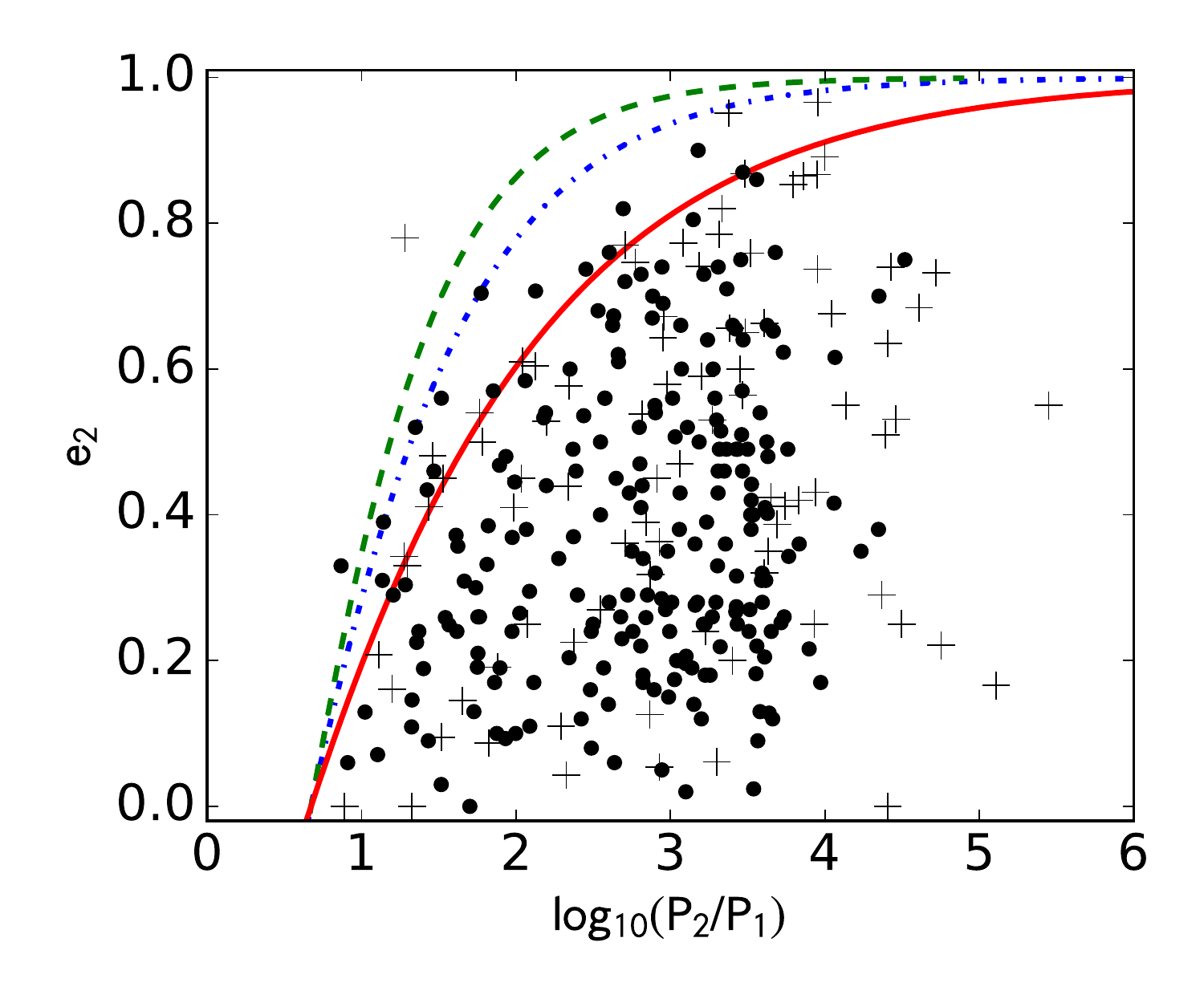} 
\end{tabular}
  \caption{\textit{Left:} Orbital period of the third companion $P_2$ versus orbital 
  period of the inner binary $P_1$ for 222 triple systems within 
  the original \textit{Kepler} field \citep{borkovits2016}. Area with the lack of triples 
  is highlighted with yellow. The stability limit is calculated for 
  the parameter $s = 1.8$ in Eq. (\ref{rov.stabilita_mardling}), with the assumption that $m_0 = m_1 = m_2 = 1$,
  for the mean value of orbital eccentricity of the outer orbit $\langle e_2\rangle \simeq 0.35$.
  \textit{Right:} The eccentricity of the outer orbit as a function of the 
  orbital period ratio. Dash-dot line is stability criterion according to \citet{mardling_aarseth}, dashed line \citep{sterzik_tokovinin2002} and red solid line is empirical criterion
  according to observations of \citet{tokovinin2007}. Eighty-eight triples from Multiple Star Catalog are plotted with black crosses \citep{tokovinin1997} and 222 triples in the original {\it Kepler} 
  field are plotted with black dots \citep{borkovits2016}.}
         \label{fig.kepler}
\end{figure*}

Another uncertainty persists in the problem of dynamical stability. 
\citet{mardling_aarseth} derived a theoretical limit on the stability of coplanar hierarchical 
triple systems as
\begin{equation}
P_2 \gtrsim 4.7 \biggl(\frac{m_0+m_1+m_2}{m_0+m_1}\biggr)^{1/10} \frac{(1+e_2)^{3/5}}{(1-e_2)^{S}}P_{1},
\label{rov.stabilita_mardling}
\end{equation}
where $m_0$ and $m_1$ are masses of inner binary, $m_2$ is a mass of
the third body, $e_2$ is an eccentricity of the outer orbit, and the parameter $s = 1.8$.
We follow the same notation hereafter. \citet{sterzik_tokovinin2002} 
later improved this limit according to numerical simulations and showed that 
exponent $s$ in relation (\ref{rov.stabilita_mardling}) has a different value, $s = 1.35,$ 
to the best conformity between model and simulated data. 
However, it turned out later that there are only a 
few systems close to this limit and that the vast majority of observed 
triple systems fulfill an even stricter empirical criterion with the value of the exponent $s = 3.0$
\citep{tokovinin2004, tokovinin2007}. The lack of observed systems 
with high $e_2$ might be caused by unmodeled dynamical effects \citep{tokovinin2004} but generally 
the empirical stability criterion is still not satisfactorily explained. Nevertheless, recent analysis 
of 222 compact triples located in the original {\it Kepler} field of view performed by \citet{borkovits2016} showed that the previously derived 
empirical criterion may rather be a result of some observational effects and all observed triples 
fulfils the theoretical criterion by \citet{sterzik_tokovinin2002} (see Fig.~\ref{fig.kepler}).
We point out that relation (\ref{rov.stabilita_mardling}) is very useful for an estimation of maximal $e_2$ 
from the period ratio $P_2/P_1$, even if one has no information about individual body mass, 
because the period ratio dependence on masses is rather weak.

In spite of the intense research, there are still only a few 
multiple systems with precisely determined orbits and physical 
parameters of all components, especially out of the Milky Way galaxy. Therefore, 
studying the known multiple systems, 
as well as a search for new ones (especially those systems, which are close to the 
stability limit) is crucial for obtaining sufficient statistics and comparison of 
theoretical simulations with observational data.

Special cases, when the inner pair of a multiple system is an eclipsing binary, 
are excellent laboratories for stellar evolution. 
Light curve analysis together with thorough 
study of radial velocities are able to reveal physical parameters 
(such as luminosities, masses, and radii) of 
components of the EB, as well as its orbit \citep{southworth, torres}. One can also estimate some 
parameters of a third body which can manifest itself via various physical effects. These effects are listed below:
\begin{itemize}
\item{eclipse time variations (ETVs) \citep{irwin, mayer1990, borkovits2016},}
\item{presence of a third light in the LC solution or a third spectrum in the overall spectrum,}
\item{visual resolution of the third body and its orbit,}
\item{variation of $\gamma$-velocity of the binary,}
\item{additional eclipses in a few rare cases \citep{carter2011, 
borkovits2015, alonso2015},}
\item{variation of LC amplitude as a result of orbital plane precession 
\citep{soderhjelm, borkovits2016, borkovits2011, breiter2015}.}
\end{itemize}
In the most favorable cases (e.g., high $L_3 / (L_1 + L_2)$ ratio or almost coplanar orbits of EB 
and the third component) a combination effect can occur, which 
makes the estimation of the third body parameters more accurate.
All these effects, except for the last one, are widely used for detection of 
new multiple systems. Effects of orbital precession led to discovery 
of a new triple only in relatively few cases mentioned in Sect. \ref{sec.known}. 
Timescale of the orbital precession $P_\mathrm{nodal} \sim P_2^2/P_1$ is usually very long compared 
to the time span of the most extended observations that are available (about 100\,years). 
Consequently, only those triples with sufficiently short periods of the outer orbits 
can be detected in available data sets. Because of this, there are still 
relatively few known systems showing orbital precession, despite the fact 
that all binaries with $P_1 \lesssim 1$~day probably have a third companion 
which caused shrinkage of an initially wide orbit via combination of Kozai cycles 
with tidal friction \citep{kozai, kiseleva, eggleton,fabrycky_tremaine2007} and orbital precession occurs 
in all cases when orbits are non-coplanar. An important step forward has been made recently 
by \citet{rappaport} and \citet{borkovits2016}, who found $42$ new compact triples showing
orbital precession in the original field of view of the \textit{Kepler} satellite. Unfortunately there are far fewer such systems
outside the Milky Way galaxy (see Sect. \ref{sec.known}).

Only the compact triple systems with low ratio $P_2/P_1$ and short period $P_2$
can be discovered by analysis of amplitude variation of LC with typical timespan of 
several dozens of years. Interestingly, those systems fall exactly into the area of unclear 
lack of triples in Fig.~\ref{fig.kepler}, which makes this method very suitable for 
detection these systems and extending of their poor statistics. In this 
work, we aim to develop suitable methods for detecting those systems and their thorough analysis.

\section{Orbital-plane precession} \label{sec.precession}

While the orbital-plane 
precession results in variation of inclination of both orbits, only changes of
inclination of the eclipsing binary are observable as changes of LC 
amplitude. Time dependence of observed inclination of the inner binary 
$i_0$ is given by the orbital geometry \citep{soderhjelm}
\begin{align}
\label{rov.soderhjelm}
        \cos i_0 &= \cos I \cos i_1 \pm \sin I \sin i_1 \cos {\dot \Omega}(t-t_0), \\
        \dot \Omega &= \frac{2 \pi}{P_\mathrm{nodal}}, \nonumber
\end{align}
where $I$ is an angle between system invariant plane and a plane tangent to 
the celestial sphere, $i_1$ is an angle between the invariant plane and the orbit 
of the EB, $t_0$ is a moment when the nodal line passes the plane tangent to 
the celestial sphere, $\dot\Omega$ is an angular velocity of the nodal 
line precession and $P_\mathrm{nodal}$ is a period of the nodal line precession\footnote{Description of all symbols used is summarized in Table \ref{tab.defsymb}.}. 
We recall that the invariant plane of the system is normal 
to its total and conserved orbital angular momentum.
It is clearly seen that the observed inclination 
oscillates in the interval $i_0 \in (I - i_1, I + i_1)$. Measuring 
time dependence of $i_0(t)$, we can use Eq.~(\ref{rov.soderhjelm}) to determine 
remaining parameters, which are deemed constant. These include 
$i_1$, $I$, ${\dot \Omega}$ and $t_0$. In what follows, we are mainly interested 
in solution of $i_1$ and  ${\dot \Omega}$. Obviously, because of their non-linear 
appearance in (\ref{rov.soderhjelm}), there are potentially severe correlations in their solution. We note
also that Eq. (\ref{rov.soderhjelm})
is invariant to the transformation $i_1 \rightarrow \pi - i_1$ a $I \rightarrow \pi - I$
and the orbital solution is ambiguous.
 
In case that eccentricity of the orbit of the inner pair is $e_1 \simeq 0$, which is 
fulfilled for a vast majority of short period binaries due to circularization, 
the precession rate of the nodal line in the invariant system may be derived 
analytically with a sufficient accuracy for most of the triple systems. We have
\begin{align}
\label{rov.vokrouhlicky}
\dot{\Omega} &\simeq \frac{2 \pi}{P_1} \frac{3}{4\eta_2^3}\frac{m_2}{M_2}\biggl(\frac{P_1}{P_2}\biggr)^2\cos j\sqrt{1-\gamma^2+2\gamma\cos j}, \nonumber \\
\gamma &= \frac{1}{\eta_2}\frac{l_1}{l_2} = \frac{1}{\eta_2}\frac{m_0 m_1 M_2 P_2 a_1^2}{m_2 M_1^2 P_1 a_2^2}, \; \eta_2 = \sqrt{1-e_2^2}, \\
M_1 &= m_0 + m_1, \; M_2 = m_0 + m_1 + m_2, \; j = i_1 + i_2,\nonumber
\end{align} 
where $i_2$ is an angle between the invariant plane and the orbital plane of the third body, $a_k$ is the semimajor axis and 
$l_k$ is the orbital angular momentum of the respective orbits \citep{breiter2015}. 
In these quantities, the indices $k = 1$ and $k = 2$ represent the inner and outer orbit, respectively. Angular velocity of the nodal 
line precession $\dot{\Omega}$ remains constant so long as the conditions $e_2 = \mathrm{const.}$ and 
$j = \mathrm{const.}$ are fulfilled. 

Modeling the variation of an EB inclination allows us to estimate the mass of the third body in the following way. Because $\sin i_2 = \gamma \sin i_1$, 
the mutual inclination of both orbits $j = i_1 + i_2$ can be rewritten as
\begin{equation}
\label{rov.j}
j = i_1 + \arcsin(\gamma \sin i_1).
\end{equation}
Eliminating $a_1$ and $a_2$ from Eq. (\ref{rov.vokrouhlicky}), we obtain for $\gamma$ the useful relation 
\begin{equation} 
\label{rov.gamma}
\gamma = \frac{1}{\eta_2}\frac{m_0 m_1}{M_1 m_2}\biggl(\frac{P_1 M_2}{P_2 M_1}\biggr)^{1/3}.
\end{equation}
The light curve solution together with spectra of components of the binary gives us the masses of
the inner binary components $m_0$ and $m_1$. Therefore, $\gamma$ is only a function of $m_2$, $P_2$, and 
 $e_2$, and thus ${\dot \Omega} = {\dot \Omega} (i_1,m_2,P_2, e_2)$. 
The equation (\ref{rov.soderhjelm}), applied to the observed dependence $i_0(t)$, gives 
a correlated least-squares fit of parameters $i_1$ and ${\dot \Omega}$. The size and
shape of the area with possible solutions in the $(i_1, {\dot \Omega})$ parameter space 
depend on the data time span and also on the period $P_\mathrm{nodal}$. 
In accordance with the functional dependence of ${\dot \Omega}$, 
each solution can be transformed to the $(m_2, P_2)$ parameter space,
which leads to restrictions on the third body mass and its orbital period.
Influence of unknown eccentricity $e_2$ on an estimation of $m_2$ and $P_2$ is rather 
small, because $\sqrt{1-e_2^2} \simeq 1$ even for $e_2$ values as large as $0.2-0.3$.
In addition, maximal $e_2$ can be estimated
according to the stability restrictions (see the right panel of Fig.~\ref{fig.kepler}).

A light curve solution leads to another restriction on the third body mass. A third body contributes with another light 
to the LC and it can be found during LC analysis in some cases. Then the third body mass can be estimated due to mass-luminosity 
relation. Even in cases when the third light is not detected, at least an upper limit on $m_2$ can be estimated.

Additional restriction on the third body mass can be obtained by analysis of ETVs. Orbital period $P_2$ is usually under the time resolution of 
long-term photometric surveys, because $P_2 \sim \sqrt{P_1 P_\mathrm{nodal}}$. Therefore, ETVs with period $P_2$ cannot 
usually be detected. Nevertheless, dispersion in eclipse timing residual diagram at least limits the maximal amplitude of ETVs.
ETVs include classical R{\o}emer's delay (so called light-time effect (LTE)), caused by finite travel time of light \citep{irwin,mayer1990}, and so called 
dynamical delay, which means a physical variation of $P_1$ as a result of the gravitational interaction of components of the inner binary 
with the third body \citep{rappaport, borkovits2003, borkovits2011, borkovits2015}. \citet{rappaport} showed that while R{\o}emer's delay
is dominant for systems with a long period of the third component and a short period of the inner binary, dynamical delay is dominant for compact 
systems with a short period of the third component. 

In this work, we deal with systems with $P_1$ of the order of days. Therefore, for period of a third body $P_2 < 200$~days 
\citep[see Fig.~7 in][]{rappaport}, dynamical delay dominates and the amplitude of ETVs is given by relation \citep{borkovits2003}
\begin{equation}
A_\mathrm{phys} = \frac{3}{8\pi}\frac{m_2}{m_0 + m_1 + m_2}\frac{P_{1}^2}{P_2}(1-e_2^2)^{-3/2},
\end{equation}
which allowed us to consider a restriction on relation $m_2 = f(P_2)$ because $A_\mathrm{phys}$ is bound from the dispersion of eclipse timing residual diagram 
and masses of the inner binary components can be estimated from LC solution. Conversely, for $P_2 > 200$~days classical R{\o}emer's delay dominates 
and ETVs amplitude is given as
\begin{equation}
A_\mathrm{LTE} = \frac{m_2 \sin i_3}{(m_0 + m_1 + m_2)^{2/3}} \frac{P_2^{2/3}\sqrt{1-e_2^2 \cos^2 \omega_2}}{173.15},  
\end{equation}
where the third body inclination $i_3$ oscilates within $(I - i_2, I + i_2)$ and its maximal values are known from the time dependence 
of $i_0$ \citep{irwin,mayer1990}\footnote{We note that the original Irwin's formula for the R{\o}emer's delay contains an extra 
term due to unusual convention of coordinate system. This term is constant as long as the elements of the orbit remains constant, 
which is clearly not the case of the systems analyzed in this work (see footnoote 2 in \citep{borkovits2016}). However, the amplitude 
$A_\mathrm{LTE}$ remains the same for the Irwin's as well as for the modern convention.}. We note that factor $173.15$ holds when $A_\mathrm{LTE}$ 
and $P_2$ are given in days and masses are in units of solar
mass. Outer orbit eccentricity $e_2$ is supposed to be unknown. However, \citet{tokovinin2007} showed that 
compact triple systems with high eccentricity of outer orbit $e_2$ becomes unstable and there is a natural limit on $e_2$ for each
 ratio of period $P_2/P_1$ (right panel of Fig.~\ref{fig.kepler}). According to selection effects of our methods, we can expect period ratios of analyzed systems within 
 the interval $P_2 /P_1 \in (5,100)$ and therefore $e_2 \leq 0.6$ in case of all systems analyzed below \citep[see fig. 3 in][]{tokovinin2006}. 


\section{Summary of known systems showing orbital precession} \label{sec.known}

Only 53 systems showing orbital precession have been discovered in the Milky Way galaxy so far. 
While 11 of them have been found by different observers within various fields of view in the sky (see Table~\ref{tab.known}),
42 systems have been identified within the original \textit{Kepler} field \citep{borkovits2016}. Away from the Milky Way galaxy, 17 more systems 
are located in the Large Magellanic Cloud (LMC) \citep{graczyk, zasche2} and there is only one known 
system in the Small Magellanic Cloud (SMC) \citep{pawlak2013}. 

Despite thorough analysis of the \textit{Kepler} systems in our Galaxy, only one system outside the Milky Way galaxy
has been carefully studied -- \object{MACHO 82.8043.171} \citep{zasche2}. Studying multiple systems in external galaxies and 
improving statistics of known systems
could help to reveal inter-galactic differences in star formation, which can be 
affected by different metallicity or other parameters and processes \citep{davies2015}.

\begin{table}
\caption{Known inclination changing EBs in the Milky Way galaxy (42 EBs within the \textit{Kepler} FOV are not included).}
\label{tab.known}
\centering
\begin{tabular}{cccccc}
\hline\hline
Designation &$P_{1}$& $P_{2}$ & $P_{\rm nodal}$ & Ref.\\
&(days) & & (years) & \\
\hline
\object{RW Per} & $13.1989$ & $68$~yr & \ldots & 1, 2, 3, 4\\  
\object{IU Aur} & $1.81147$ & $294.3$~days &  335 & 5, 6, 7, 8,\\
 &  &  &  & 9, 10 \\
\object{AH Cep} & $1.7747$ & $9.6$~yr\tablefootmark{*} & \ldots & 11, 12, 13\\
\object{AY Mus} & $3.2055$ & \ldots & \ldots & 14, 15\\
\object{SV Gem} & $4.0061$ & \ldots & \ldots & 16, 17, 18\\
\object{V669 Cyg} & $1.5515$ & \ldots & \ldots & 19, 20\\
\object{V685 Cen} & $1.19096$ & \ldots & \ldots & 21\\
\object{V907 Sco} & $3.77628$ & $99.3$~days & $68$\tablefootmark{*} & 22\\
\object{SS Lac} & $14.4161$ & $679$~days & $600$\tablefootmark{*} & 23, 24, 25\\
\object{QX Cas} & $6.004709$ & \ldots & \ldots & 26\\
\object{HS Hya} & $1.568024$ & $190$~days & $631$ & 27, 28\\
\hline
\end{tabular}
\tablefoot{
\tablefoottext{*}{Uncertain.}
}
\tablebib{
(1)~\citet{hall}; (2) \citet{mayer1984}; (3) \citet{schaefer_fried}; 
(4) \citet{olson1992}; (5) \citet{mayer1971}; (6) 
\citet{mayer_drechsel1987}; (7) 
\citet{mayer1987}; (8) \citet{mayer1991}; (9) \citet{drechsel1994}; (10) 
\citet{mason1998}; (11) \citet{mayer1980}; (12) \citet{drechser1989}; 
(13) \citet{kim2005}; (14) \citet{soderhjelm2}; (15) 
\citet{soderhjelm3}; (16) \citet{guilbault2001}; (17) 
\citet{paschke2006}; (18) \citet{paschke2006}; (19) 
\citet{azimov1991}; (20) \citet{lippky1994}; (21) \citet{mayer2004}; 
(22) \citet{lacy1999}; (23) \citet{milone2000}; (24) 
\citet{torres_stefanik2000}; (25) \citet{torres2001}; (26) \citet{guinan2012}; 
(27) \citet{zasche_paschke2012}; (28) \citet{torres1997}.
}
\end{table}

Large databases of medium quality lightcurves of EBs in the LMC and SMC, from photometric surveys like 
OGLE \citep{udalski5, udalski4, udalski2008, udalski6, udalski2015, szymanski} or MACHO \citep{macho1, 
macho2}, allow us to develop new methods to find candidate triple 
systems in the Magellanic Clouds. The goal of our study is to identify new LMC and SMC compact multiple 
systems via detection of amplitude variations in archival LCs. These amplitude variations could be caused by an orbital precession 
due to the presence of a third body. Because of rather short timescale of 
photometric surveys (in the order of ten~years) and typically long timescales of 
orbital precession it is possible to find only systems with short 
$P_2$ and small ratio $P_2/P_1$. However, those systems should be 
exactly in the area with lack of triples in the $P_1 - P_2$ distribution (yellow area in the Fig.~
\ref{fig.kepler}) near stability limit, which makes this simple method suitable for identifying new compact triples.

\section{Methods} \label{sec.methodics}

Light curves of eclipsing binaries observed by the OGLE III photometric survey
 have sufficient precision, low scattering and good homogeneity 
over the whole term of observation. This makes this database suitable for finding 
new compact multiple systems in the Magellanic Clouds.
 The OGLE III database contains LCs of $26\,121$ EBs in the LMC \citep{graczyk} and $6138$ EBs in the SMC 
\citep{pawlak2013} (mostly in Johnson \textit{I} photometric band) which are available 
online\footnote{\url{ftp.astrouw.edu.pl}} \citep{szymanski}.\footnote{We note that at the time
of the first part of our analysis, the OGLE IV data had not yet been released.} 

EB light curve amplitude variation can be caused by several effects, for example,  
physical variations of luminosity of the EB components,
stellar spots, changes 
of overlapping surfaces during eclipses as a result of apsidal motion, etc. 
Therefore, when the maximum luminosity of both components remains constant and 
the orbit is circular, the only possible explanation of amplitude changes is that the inclination 
angle varies. Both conditions can be easily checked from phased light curve (PLC).

In principle, there may be two basic methods for detection of LC amplitude 
variations:
\begin{itemize} 
        \item{Method A -- measurement of the eclipse depth of a PLC in different time intervals,} 
        \item{Method B -- measurement of the amplitude of a non-phased LC in different epochs.}
\end{itemize}
A typical number of data points $N$ in an OGLE III light curve is within the interval $(400, 600)$, 
from which only a small fraction is located near eclipses (in the case of the most 
frequent EB type -- detached binaries). Because of that,
the number of time intervals has to be small to get a reasonable fit of minima shapes. 
Therefore, we had no ambition to detect other but
linear amplitude changes on the timescale of OGLE III observations 
($\approx 8$~years), in the case of Method A.
  
\subsection{Data preparation}

Ephemerides and classification of each EB were taken from 
OGLE III catalog \citep{graczyk}. Only those LCs that contain sufficient amount 
of data points ($>250$) were taken into account. Those LCs, where the primary minima depth 
were smaller than the standard deviation of magnitudes, were discarded. It should be mentioned
that it obviously add an artificial selection effect to the analysis, because shallow minima may be 
caused by massive and luminous third companion. But this restriction is necessary 
to avoid false detection due to high scattering of some LCs. 

For the purpose of removing outliers, each LC was fitted with a linear 
function $f(t)$ and all data points outside the intervals $(f(t) - 3\sigma, 
f(t) + 7\sigma)$ in the case of detached binaries, $(f(t) - 3\sigma, 
f(t) + 6\sigma)$ in the case of semidetached binaries and $(f(t) - 3\sigma, 
f(t) + 3\sigma)$ in the case of overcontact binaries, were removed. We note that 
$\sigma$, in this case, means a standard deviation computed from the whole LC.
These asymmetrical intervals 
were chosen with respect to different characters of LCs of individual EB types so that 
the whole LC (except outliers) lies inside the interval. 

\subsection{Method A}

\begin{figure*}
\centering
        \begin{tabular}{@{}cc@{}}
    \includegraphics[width=88mm]{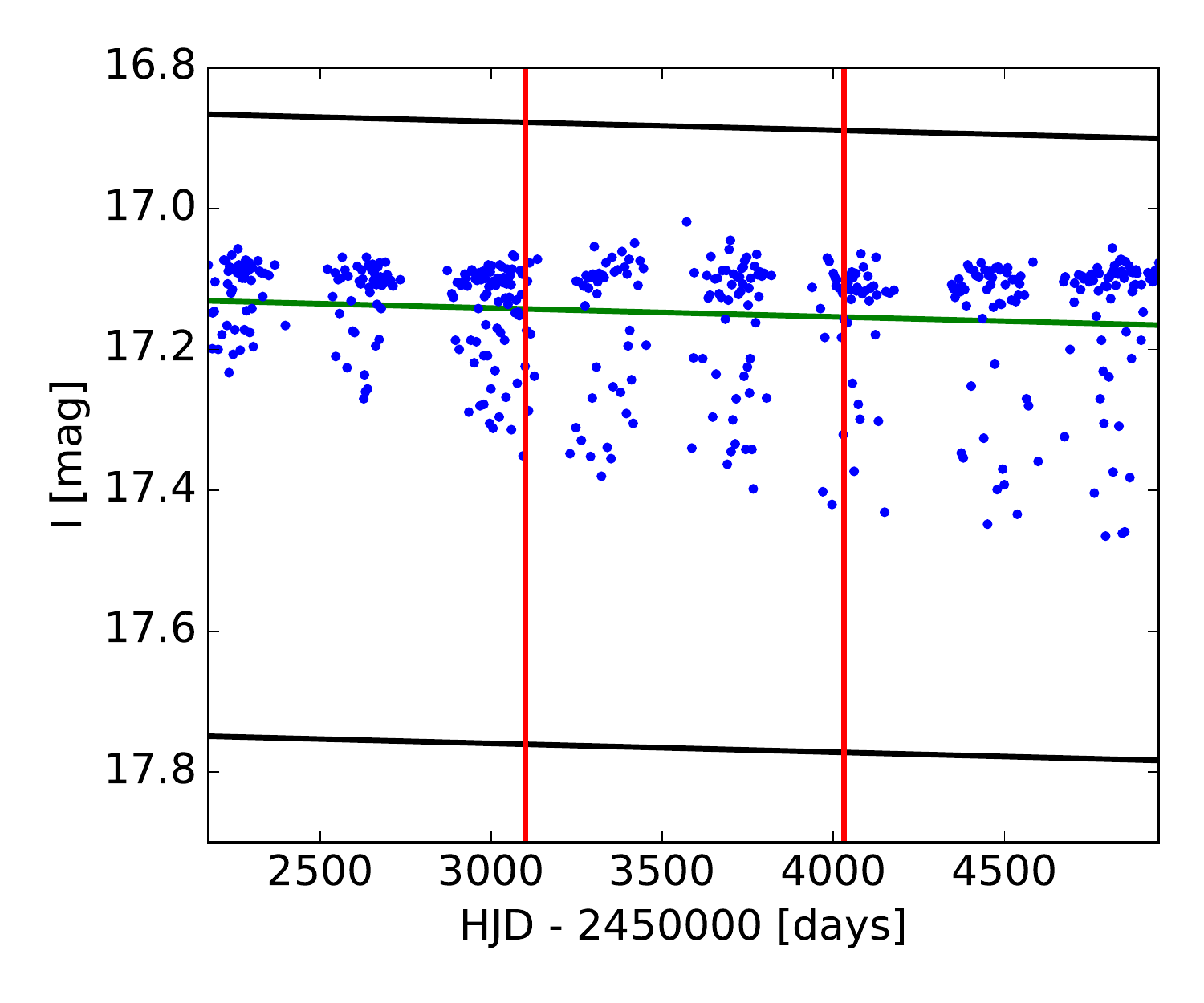} & 
    \includegraphics[width=88mm]{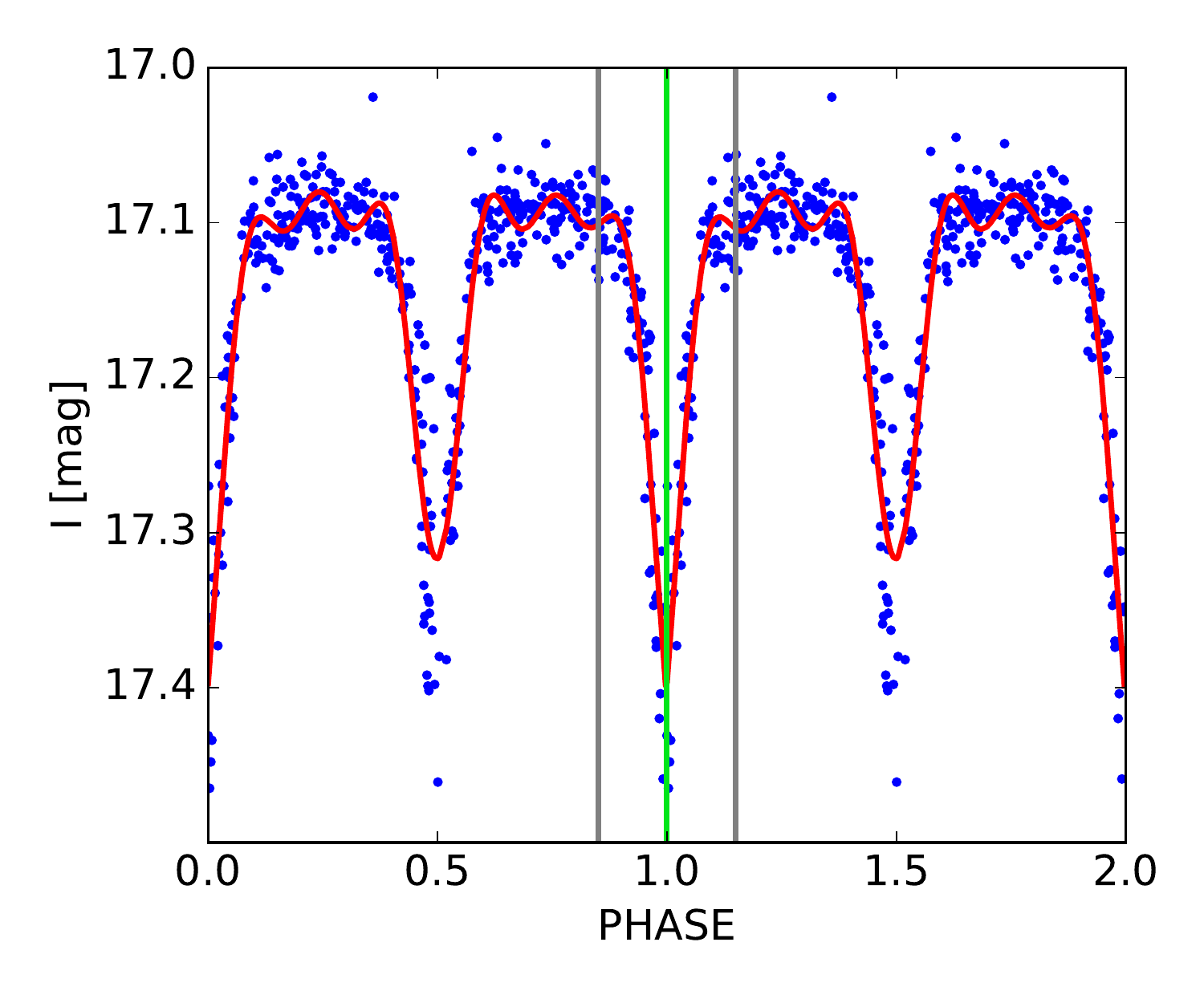} \\ 
        \end{tabular}
  \caption{Demonstration of Method A. \textit{Left:} Example LC of the system \object{OGLE-LMC-ECL-01350} 
  divided to three intervals. Green line represents linear fit, black lines show the limits for removal of the outliers ($-3\sigma$, $+7\sigma$).
  \textit{Right:} Phase curve for the whole LC fitted with a Fourier series of the
  $5^\mathrm{th}$ order (red line). Green line marks detected primary minimum, 
  gray lines represent detected region of the drop in brightness.}
   \label{fig.method_a1}
\end{figure*}

As the first step of this method, the whole LC of a given EB is divided into 
several intervals which contain approximately $150$ data points. A number of 
intervals depends on the number of data points in the whole LC.
 As shown in the left panel of Fig.~\ref{fig.method_a1},
a typical LC has been divided into three or four intervals. 

\begin{figure*}
\centering      
        \includegraphics[width=\textwidth]{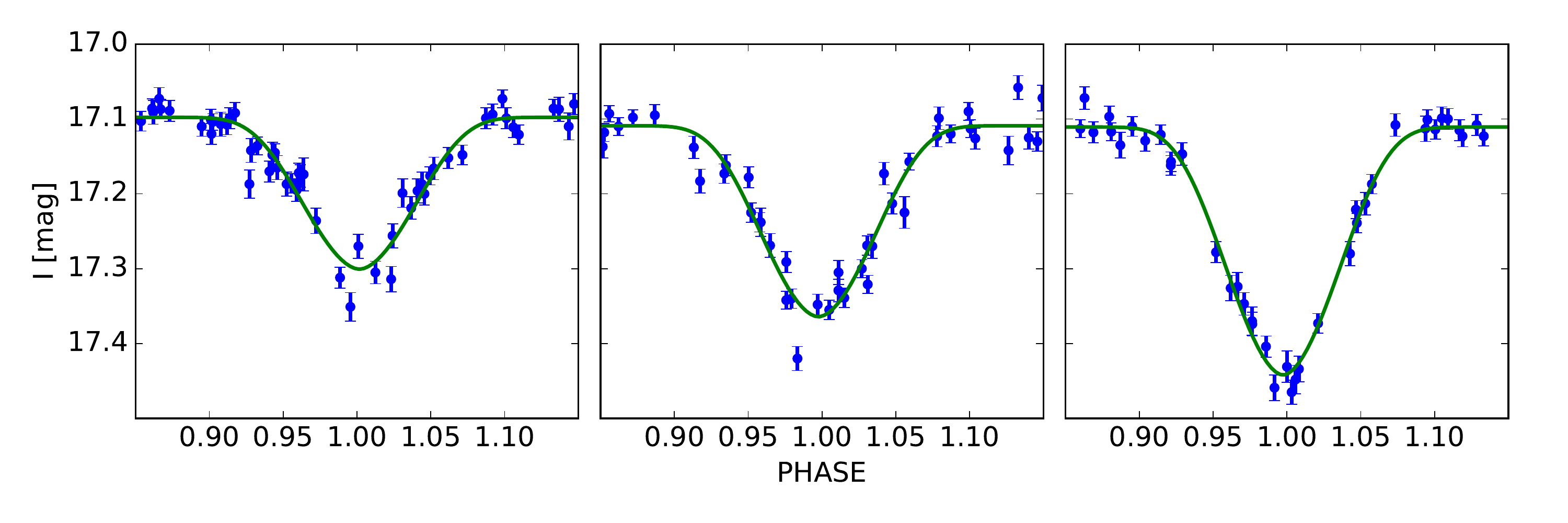}
  \caption{Demonstration of Method A. Fitting of primary eclipses of 
  \object{OGLE-LMC-ECL-01350} with function (\ref{eq.minimum}) on three 
  intervals of LC, corresponding to the left panel of Fig.~\ref{fig.method_a1}. The increase of the minima depth is apparent.}
  \label{fig.method_a2}
\end{figure*}

In the next step, the LC in
each interval is phased according to the catalog ephemerides, in order to 
obtain a dependence of magnitude in the Johnson $I$ band on the phase 
$I(\Phi)$. The core 
of this method is based on fitting of primary minimum on each phased 
interval of the LC with a proper phenomenological function, which is able 
to describe well the depth of an eclipse. There is a class of 
mathematical functions which generate similar dependencies such as shapes of real LCs around eclipse
\citep{andronov2012a,andronov2012b,mikulasek2012,mikulasek2015}.
 For our purposes we used the following formula \citep[taken from][]{andronov2012a}
\begin{equation} \label{eq.minimum}
        I(\Phi) = I_0 + A \left \{ 1 - \left[\left(\Phi - 
        \Phi_0\right)/d\right]^2\right \}^C,
\end{equation} 
where $I_0$ is a vertical shift in magnitude, $A$ is the depth of the eclipse, $\Phi_0$ 
is a phase of the time of minimum, $d$ represents minima width and $C$ 
is a parameter determines a shape of the eclipse. An example of fitted minima on each interval of the whole LC is shown in 
the Fig.~\ref{fig.method_a2}.

However, Eq. (\ref{eq.minimum}) describes well 
only a small part of the PLC around minimum and each PLC had to be properly reduced 
before fitting. As a reasonable compromise between robustness of the 
algorithm and its computing speed, preliminary fitting of the PLCs with Fourier 
series of the fifth, fourth, and second orders were performed 
in the case of detached, semidetached and overcontact binaries, respectively (see the right 
panel of Fig.~\ref{fig.method_a1}). The minimum of the Fourier model corresponds to the
minimum of given PLC and the first maxima of this model on both 
sides of an eclipse define the whole part of drop in 
brightness\footnote{An optimal solution was
found by repetition of the fitting procedure with several multiples of width 
obtained from the Fourier model.} (see the right panel of Fig.~\ref{fig.method_a1}).

Minima depths (the parameter $A$ from the Eq. (\ref{eq.minimum})) were registered on each interval 
of the whole LC and finally, the linear model of its time dependence was computed. For each EB,
the slope of the linear fit and $R^2$ parameter\footnote{Defined as $R^2 = 1 - \sum\limits_i w_i(y_i - f_i)^2 / \sum\limits_i w_i(y_i - \bar{y})^2$, 
where $y_i$ is the observed data value, $f_i$ is the predicted value from the fit, $\bar{y}$ is the mean 
of the observed data. $w_i = \sigma_i^{-2}$ is the weight for each data point, where $\sigma_i$ is uncertainty of 
each minima depth determined by the Levenberg-Marquardt curve-fitting algorithm.} were tested and if some specific values were exceeded,  
the EB was marked as suspicious of the  inclination change. Further details about setting these values are 
described in Sect. \ref{sec.thresholds}.

\subsection{Method B}

\begin{figure}[!t]
\centering
\includegraphics[width=88mm]{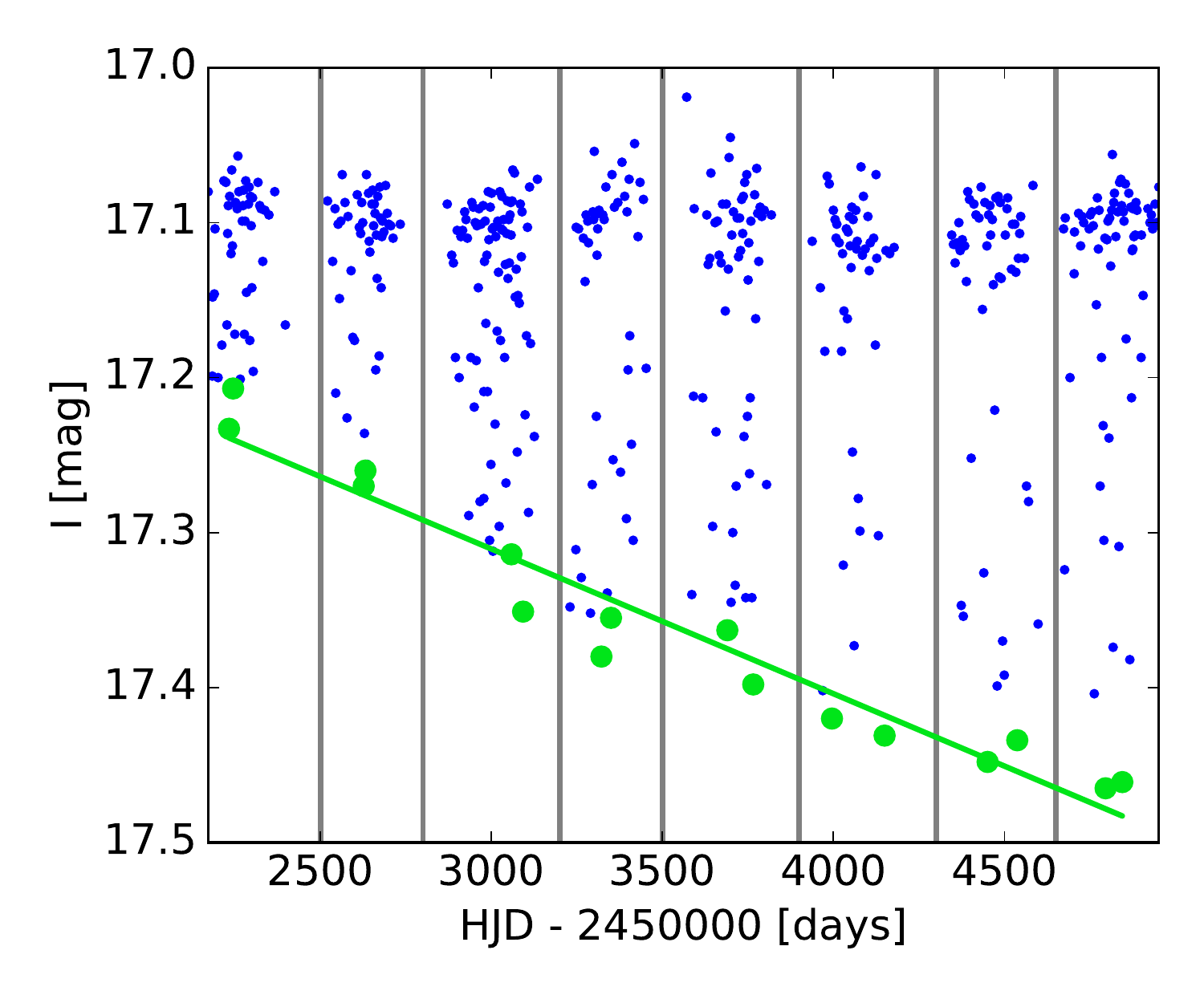}
  \caption{Demonstration of Method B. Example LC of the system OGLE-LMC-ECL-01350 
  splitted into 8 intervals according to observational seasons. 
  The lowest two points inside each interval are plotted with green circles, 
  the green line shows the linear fit.}
         \label{fig.method_b}
\end{figure}

In the second method, the whole LC was split into eight fixed intervals 
with respect to the seasons of observation (see Fig.~\ref{fig.method_b}). 
In contrast with Method A, Method B is less demanding as far as 
the number of data points in a given interval is concerned, 
because it is not necessary to have a lot of points around the eclipse and that 
allows division of the LC to more intervals. After LC splitting,
 two data points\footnote{Selection of more then one data points with 
 extremal value is for better robustness of the method.} corresponding to the lowest brightness were selected 
from each of the intervals of the LC and than all chosen data points were fitted 
with the linear model according to the $\sigma$-clipping method with a  
rejection of each point above the $2\sigma$ limit. As in the case of Method A, the detection criteria are the certain values of the slope of linear fit and the
$R^2$ parameter. The setting of those values is 
described in Sect. \ref{sec.thresholds}. 

The advantage of this simple method is that there is no need to know a 
precise orbital period of given EB and also there is no need to 
delimit a part of the LC with an eclipse. That makes this 
method faster than the Method A. On the other hand, what is actually fitted in 
this case, is not exactly the minimum depth, but only local LC amplitude, which makes this method more sensitive to outliers than slower fitting minima in each part of LC.

\subsection{Parameter thresholds} \label{sec.thresholds}

Specific values of slope and $R^2$ in the case of both methods served as thresholds for preliminary distinction whether 
changes of inclination occur in case of given EB or not. Setting of certain thresholds on these quantities is difficult 
because of the high variability of LC shape and their scattering that strongly depends on the brightness of 
a particular EB. However, the estimation of the threshold can be made in the parameter space defined 
by the 17 known systems in the OGLE III LMC database \citep{graczyk}. 

\begin{figure*}
\centering
        \begin{tabular}{@{}ccc@{}}
    \includegraphics[width=88mm]{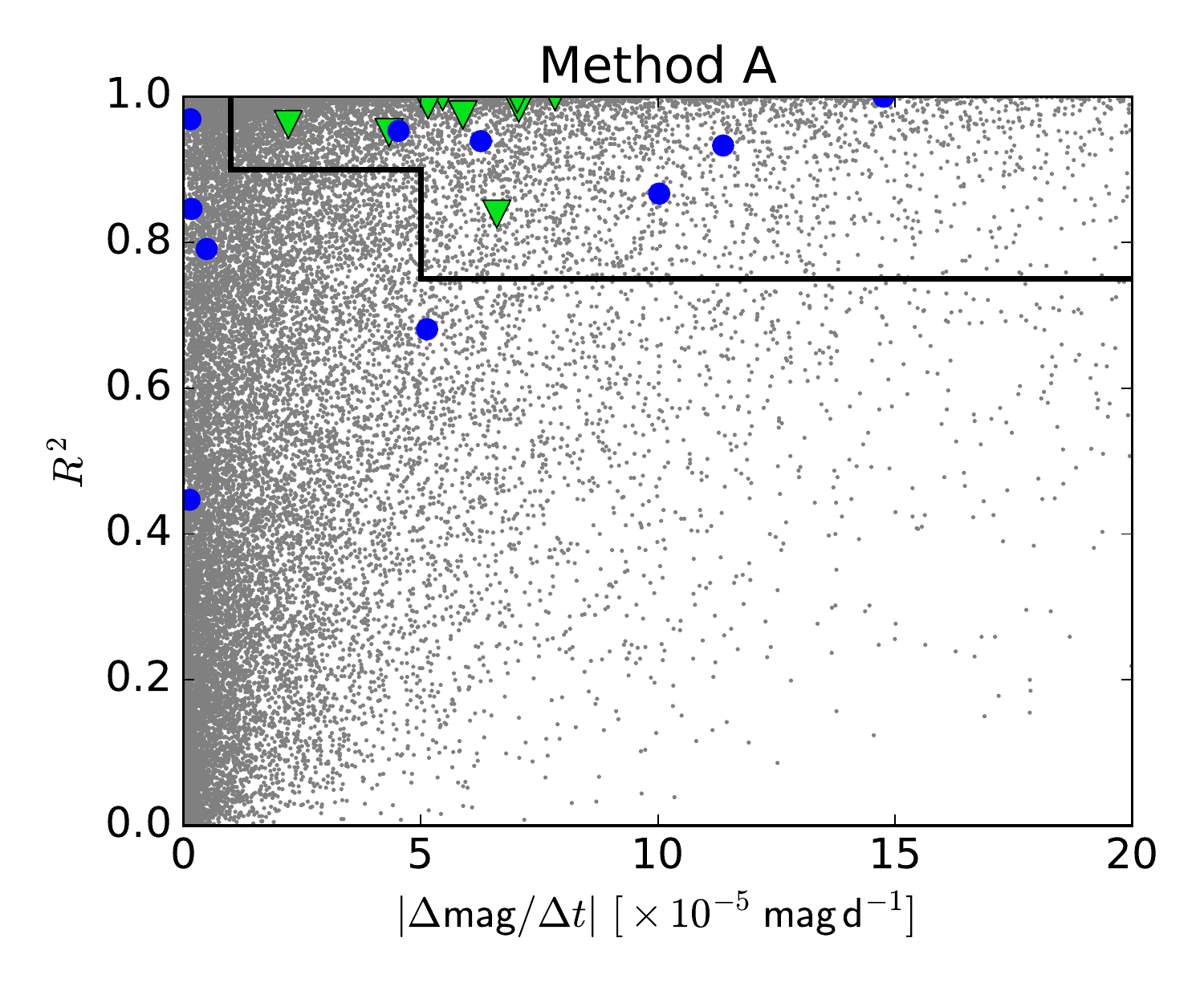} & 
    \includegraphics[width=88mm]{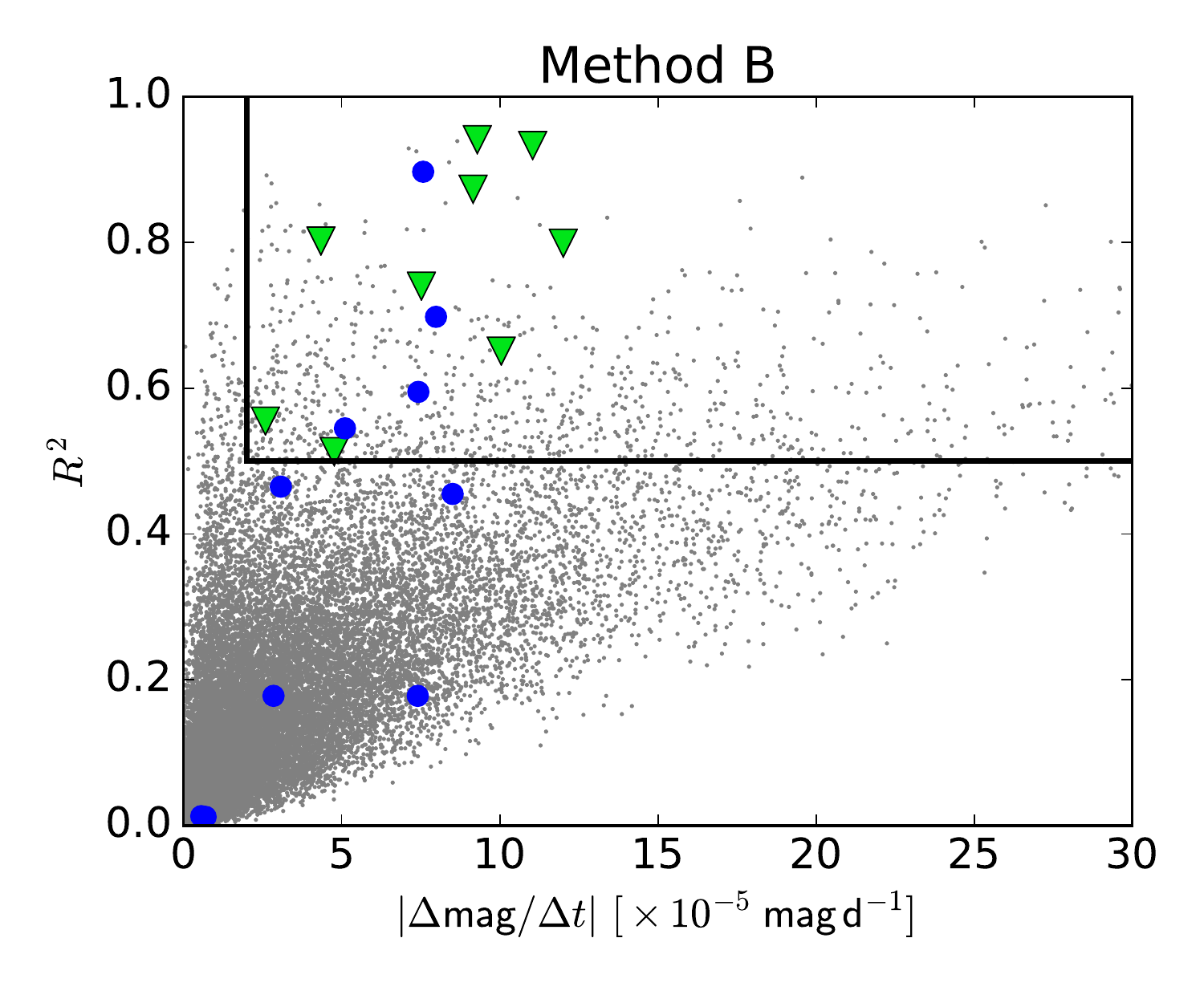} \\ 
        \end{tabular}
  \caption{All eclipsing binaries from the OGLE III LMC database (gray dots) plotted 
  in the (slope, $R^2$) parameter space of both methods. 
  Nineteen previously known systems showing variations of the LC amplitude are marked with blue circles
  (artefacts or systems with rapid changes of amplitude)
  and green triangles (systems with approximately linear amplitude change on the timescale of 
  observation). Our setting of thresholds is shown as a black line.}
   \label{fig.thresholds}
\end{figure*}

These systems, together with two others, which have been found independently by the 
second author of this paper, are shown in the parameter space of both 
methods in the Fig.~\ref{fig.thresholds}. However, careful examination of the previously
known systems from \citet{graczyk} showed that in some cases the change of the 
amplitude of LC could rather be an artefact than a real feature.
Furthermore, in some cases, amplitude 
variation is so fast that a significant part of LC is completely without 
eclipses. As described above, our 'linear' methods are usually not able to detect such systems (e.g., 
\object{OGLE-LMC-ECL-17212}, 17972, see Fig.~
\ref{fig.thresholds}) and parameters thresholds were set        
irrespective of these systems\footnote{However, several such systems have still been detected with described 
linear methods or with the modified method described below (see Figs.~\ref{fig.lmc_detected_systems} 
and \ref{fig.smc_detected_systems} showing all detected systems).}. 

Systems with approximately linear time dependence of eclipse depth on the OGLE III observational time scale 
(see Fig.~\ref{fig.thresholds}) served for setting the parameters thresholds.
The area of interesting systems in the parameter space 
was chosen to exclude as many EBs as possible without any amplitude change. 
Finally, our empirical thresholds for 
the Method A were $|\Delta \mathrm{mag} / \Delta t| \geq 5 
\times 10^{-5}\;\mathrm{mag\,day^{-1}}$ and $R^2 \geq 0.75$. However, some well 
covered LCs with very slow changing of amplitude could 
exist and for these systems the slope threshold is too strong. For this reason the restrictions were alleviated, and
 systems with $|\Delta \mathrm{mag} / 
\Delta t| \geq 1 \times 10^{-5}\;\mathrm{mag\,day^{-1}}$ and 
$R^2 \geq 0.90$ were marked as suspicious (see the Fig.~\ref{fig.thresholds}). For Method B the 
following thresholds were set: $|\Delta \mathrm{mag} / \Delta t| \geq 2 
\times 10^{-5}\;\mathrm{mag\,day^{-1}}$ and $R^2 \geq 0.50$.

In order to identify systems with shorter timescales of orbital 
precession ($P_\mathrm{nodal} \sim 10$~years) it was neccessary to modify our 
methods so that they could fit time dependence of LC amplitude by polynomials of higher 
orders. Method A cannot be modified in this way because it demands a relatively 
high number of data points around eclipses. But in Method B, LC is divided 
to eight intervals and the number of fitted 
data points is mostly 16, which allowed us to apply this modification.

Considering a scattering and coverage of LCs, polynomials up to the fourth order were used. 
Systems fulfilling the following criteria were marked as suspected of being of  higher order amplitude variations:
\begin{itemize}
        \item{At least one higher-order polynomial fit (second, third, or fourth) was significantly 
        better\footnote{Meaning that the reduced $\chi^2$ of polynomial fit was at least 
        two times smaller than $\chi^2$ of the linear fit.} than a linear fit.}
        \item{At least one higher-order polynomial fit had to have AIC 
        \citep{akaike1974}, 
        AICc \citep{sugiura1978} and BIC \citep{schwarz1978} information 
        criteria value lower than a linear fit.}
        \item{$\chi^{2}$ of the higher-order polynomial fit was smaller than 50
        to avoid systems with highly scattered data points around eclipses.}
\end{itemize}

\section{Results of detection of new systems with amplitude variation}

Our methods described in Sect. \ref{sec.methodics} were applied to 
all light curves of eclipsing binaries from the OGLE III LMC+SMC.
Thresholds set on the Method A and the linear regime of the Method B were exceeded in 
$8.1$~\% and $3.1$~\% of the LMC EBs, respectively. In the SMC it was in $3.9$~\% and 
$2.8$~\%. In the higher-order polynomial regime of Method B, $1.5$~\% of LMC EBs 
and $1.4$~\% of SMC EBs were marked as suspected of LC amplitude variations. All of 
the positive detections were subjects of thorough visual inspections and most of them 
were rejected as false positives.

False detections were mostly caused by too few 
data points around eclipses, which affects results of both 
methods. This is, for example, the case of \object{OGLE-LMC-ECL-10172}, which is a nice example
 of the system with rapid variation of LC amplitude during the term of OGLE III observations. 
 However, the orbital period of the eclipsing binary is $P_1=1.993644$~d, which makes 
impossible to observe eclipses in some seasons from one observing 
site and an artificial change of the LC amplitude occurs.

Results of Method A strongly depend on a precision of the catalog value of a binary orbital period, 
and also on time stability of the light curve.
In some cases, the mid-eclipse times of an EB vary in time due to apsidal motion or ETVs. 
That leads to 'blurriness' of a PLC, and results obtained by Method A may have been irrelevant and had to be rejected.

In addition, some systems had to be rejected because of variations of maximal brightness (magnitude 
at quadratures), which may be caused by a motion of surface spots, physical variability of one or both components of a
binary, or by a drift of zero points in the measuring equipment.  

After visual inspection and combining results of both methods
51 and 21 candidates remained in the LMC and SMC samples, respectively. These systems 
are listed in Tables~\ref{tab.lmc} and \ref{tab.smc} together with 
cross-identifications with the OGLE II and MACHO surveys and their lightcurves are shown in appendices \ref{fig.lmc_detected_systems} and \ref{fig.smc_detected_systems}. 
Apart from the 13 previously known systems\footnote{Only 13 from a whole sample of 17 known systems in LMC are 
detectable with our methods.}, there are 38 new systems in LMC, clearly showing 
amplitude variations while maximal brightness remains constant. From the 21 systems detected within the SMC, only one
was previously known \citep{pawlak2013}. 

\begin{table}[t]
\small
\caption{Detected systems with change of LCs amplitudes in the LMC.}
\label{tab.lmc}
\centering
\begin{tabular}{cccc}
\hline\hline
OGLE III LMC & OGLE II LMC& MACHO & Ref.\\
\hline
  01350 &      \ldots           &   \ldots            & 1\\
  02064 &      \ldots           &  17.1985.175  & this work\\  
  02269 &      \ldots           &  45.2118.679  & this work\\                        
  02480 &      \ldots           &   \ldots            & 1\\
  02641 &      \ldots           &   \ldots            & this work\\
  03163 &      \ldots           & 17.2472.176   & 1\\
  03183 &      \ldots           & 17.2473.43   & this work\\
  03747 &      \ldots           &   \ldots            & this work\\
  04002 &      \ldots           &   \ldots            & this work\\
  06385 & SC14\_175132 &     \ldots           & 1\\
  06464 & SC14\_160905 &1.3930.1050     & this work\\
  08377 &      \ldots           &    \ldots           & this work\\
  08628 &      \ldots           &   \ldots            & this work\\
  09770 & SC10\_255818 &      \ldots         & this work\\
  10063 &      \ldots           &   \ldots           &  this work\\
   10338 & SC9\_127779  &79.5378.336    & this work\\
  10369 & SC9\_128347  &       \ldots        & this work\\
   10413 & SC9\_127854  &79.5378.26     & this work\\
  11068 &     \ldots            &    \ldots           & this work\\
  11658 & SC8\_205379  &78.5859.237    & this work\\
  11168 &       \ldots          &  49.5774.50   & this work\\
  13150 &      \ldots           &  3.6485.107   & this work\\
  13399 &      \ldots           &      \ldots         & this work\\
  14394 & SC5\_169494  &       \ldots        & 1\\
  15256 &       \ldots          &     \ldots          & this work\\
  15520 & SC4\_227322  &77.7433.183    & this work\\       
  15993 & SC4\_391534  &     \ldots          & this work\\          
  16023 & SC4\_468053  &77.7554.200    & this work\\
  16323 &      \ldots           &     \ldots          & this work\\  
  16495 &      \ldots           &     \ldots          & this work\\
  16896 &      \ldots           &     \ldots          & this work\\
  17209 &      \ldots           &     \ldots          & this work\\
  17212 &      \ldots           &     \ldots          & 1\\
  17359 &      \ldots           &  82.8043.171  & 1, 2\\
  17890 &      \ldots           &    \ldots           & 1\\
  17972 &               \ldots            &             \ldots            & 1\\
  18240 &      \ldots           &    \ldots           & 1\\
  18686 &      \ldots           &    \ldots           & 1\\
  19066 & SC1\_152384  &       \ldots          & this work\\
  20389 &     \ldots            &      \ldots         & this work\\
  20742 &     \ldots            &      \ldots         & this work\\
  21083 &     \ldots            &      \ldots         & this work\\
  21928 &      \ldots           &      \ldots         & 1\\
  22555 &      \ldots           &       \ldots            & 1\\
  22686 &     \ldots            &  33.9990.38   & this work\\
  22885 &     \ldots            &      \ldots         & this work\\
  22918 &     \ldots            &      \ldots         & this work\\
  23148 &     \ldots            &  50.10240.867 & this work\\
  24123 &     \ldots            &      \ldots         & this work\\
  25108 &     \ldots            &      \ldots         & this work\\ 
  25373 &     \ldots            &      \ldots         & this work\\ 
\hline
\end{tabular}   
\tablebib{ (1)~\citet{graczyk}; (2)~\citet{zasche2}}
\end{table}

\begin{table}[t]
\small
\caption{Detected systems with change of LCs amplitudes in the SMC.}
\label{tab.smc}
\centering
\begin{tabular}{cccc}
\hline\hline
OGLE III SMC & OGLE II SMC& MACHO & Ref.\\
\hline
0648 &    \ldots        &       \ldots           & this work\\          
0718 &    \ldots        & 208.15571.181 & this work\\                                                                                    
0863 & SC3\_193792 &      \ldots         & this work\\
0917 & SC4\_14872 &  212.15677.1029 & this work\\   
1532 & SC5\_11681 &       \ldots         & this work\\   
1649 & SC5\_123484 &      \ldots         & this work\\ 
1872 & SC5\_160326 & 212.15957.454 & this work\\    
1946 & SC5\_230499 &     \ldots          & this work\\ 
1989 & SC5\_311575 &     \ldots          & this work\\
2212 & SC6\_18013  &     \ldots          & this work\\
2436 & SC6\_94470  &     \ldots          & this work\\
3317 & SC7\_115374 & 211.16311.196 & this work\\
3473 & SC7\_169045 &      \ldots         & this work\\
3613 & SC8\_46187  & 207.16431.1821 & this work\\  
3833 & SC8\_107524 &  207.16490.174 & this work\\  
4935 & SC10\_65845 & 206.16888.123  & this work\\
4952 &     \ldots       &       \ldots        & this work\\
5096 & SC10\_134445 &      \ldots         & 1\\
5662 &     \ldots       &      \ldots         & this work\\
5943 &    \ldots        &       \ldots        & this work\\
6118 &    \ldots        &        \ldots       & this work\\
\hline
\end{tabular}   
\tablebib{(1)~\citet{pawlak2013}}
\end{table}

\section{Analysis of individual systems}

Several detected systems showing LC amplitude variation were selected and analyzed thoroughly. The analysis was 
performed especially for systems with additional archival data (MACHO or OGLE II/IV) available.

For four selected systems, new Charge-coupled device (CCD)
 photometry was obtained at the La Silla Observatory in Chile,
with 1.54-m Danish telescope in the Johnson $I$ photometric band to secure consistency with the OGLE data. 
For one system -- \object{OGLE-SMC-ECL-1532} -- archival CCD frames, taken in the Johnson $R$ photometric band with the Danish telescope, 
were used. For aperture photometry we developed and used \texttt{Python 2.7} scripts with usage of the \texttt{photutils} package 
and differential magnitudes were obtained. The individual data (HJD vs. $\Delta m$) are listed in Table~\ref{tab.data}.

\begin{table*}[t]
\small
\caption{Photometric observations of individual systems.}
\label{tab.data}
\centering      
\begin{tabular}{cc| cc| cc| cc| cc| cc}
\hline\hline
\multicolumn{2}{c}{LMC01350} & \multicolumn{2}{c}{LMC13150} & \multicolumn{2}{c}{LMC16023} & \multicolumn{2}{c}{LMC18240} & \multicolumn{2}{c}{LMC23148} & \multicolumn{2}{c}{SMC1532} \\
\hline
JD & $\Delta m$ & JD & $\Delta m$ & JD & $\Delta m$ & JD & $\Delta m$ & JD & $\Delta m$ & JD & $\Delta m$ \\
6262.64078 & -0.534 & 7376.54703 & 0.563 & 7323.57195 & 0.998 & 6263.55449 & 0.684 & 7318.55154 & 0.264 & 6211.79967 & 0.509 \\ 
6262.64211 & -0.538 & 7376.54982 & 0.545 & 7323.57893 & 1.009 & 6263.55583 & 0.690 & 7318.55294 & 0.270 & 6211.80180 & 0.516 \\ 
6262.64345 & -0.481 & 7376.56837 & 0.575 & 7323.58033 & 1.026 & 6263.55718 & 0.711 & 7318.55436 & 0.282 & 6211.80391 & 0.545 \\ 
6262.64478 & -0.495 & 7376.57117 & 0.558 & 7323.59018 & 1.037 & 6263.62203 & 0.622 & 7318.55575 & 0.296 & 6211.80602 & 0.506 \\ 
6262.64612 & -0.467 & 7376.58991 & 0.559 & 7323.62295 & 1.044 & 6263.62336 & 0.651 & 7318.56420 & 0.263 & 6211.80815 & 0.532 \\ 
6262.64746 & -0.468 & 7376.59133 & 0.546 & 7323.68933 & 1.073 & 6263.62469 & 0.641 & 7318.56559 & 0.280 & 6211.81027 & 0.498 \\ 
6262.64879 & -0.511 & 7376.59275 & 0.549 & 7323.69211 & 1.085 & 6263.62602 & 0.629 & 7318.56697 & 0.296 & 6211.81412 & 0.536 \\ 
6262.65012 & -0.492 & 7376.61125 & 0.557 & 7323.70622 & 1.070 & 6263.62736 & 0.645 & 7318.56837 & 0.266 & 6211.81624 & 0.493 \\ 
6262.65146 & -0.471 & 7376.61265 & 0.582 & 7323.72312 & 1.059 & 6263.62869 & 0.623 & 7318.56978 & 0.279 & 6211.81836 & 0.510 \\ 
6262.65279 & -0.484 & 7376.61405 & 0.574 & 7323.72450 & 1.033 & 6263.63002 & 0.640 & 7318.57811 & 0.289 & 6211.82045 & 0.529 \\      
\hline
\end{tabular}
\tablefoot{This table is available in its entirety in machine-readable form via CDS. The listed JDs stands for $\mathrm{HJD}-2450000$.}
\end{table*}

Analysis of LCs of individual systems was performed in \texttt{PHOEBE} \citep{prsa_zwitter} program. Obtaining of spectra 
 of the stars outside of the Milky Way galaxy is complicated and requires long exposure time even when using the largest telescopes. 
 For this reason, a radial velocity curve is not available for any studied system and the mass ratio was fixed as $q = 1$
 in the first iteration of the LC modeling. When the photometric model was inconsistent with this assumption, it
 was recalculated with new $q$ value estimated from bolometric magnitudes of components. Synchronous rotation $F_1 = F_2 = 1$ of both 
 components was assumed and limb darkening coefficients were obtained from the square-root model which is more suitable 
 for hot stars than the logarithmic model \citep{diaz}. Bolometric albedos and gravity darkening were fixed as 
 $A_{1,2} = g_{1,2} = 1$ which is fulfilled for stars with $T > 7200$~K whose subsurface layers are in
 radiative equilibrium. Without any specroscopic data for given stars, solar metallicity was assumed and fixed.

The basic model 
was computed on a subset with the lowest data scattering and the best coverage and on the other subsets inclinations $i_0$ and luminosities 
$L_{1,2}$ were fitted only. With the assumption that physical parameters of components remain constant, that led us to obtain a time 
dependence of binary inclination. 

To improve the orbital period of the binary, primary and secondary minima were computed with the use of slightly modified 
AFP method \citep[for the description of the original AFP method see][]{zasche} and eclipse timing residual diagram for each system was constructed. 
Modification of the original AFP method was neccessary due to the fact that LC amplitudes of our systems vary with time. 
Therefore, we had to fit not only a phase and magnitude shift of the model curve, but also its 'contraction', which represents amplitude variation of the LC.

The final fixed parameter was primary temperature $T_1$ which had to be estimated from photometric indices because of a lack 
of another spectral information about given star in the LMC and SMC. However, correct estimation of the temperature is 
usually quite tricky. There are many photometric catalogs of the Magellanic Clouds with various color indices for a given star, which sometimes lead to different 
temperatures. In the case of the hot stars in our sample, relative differences in temperatures are up to $20\,\%$. Therefore, temperatures and 
masses of an individual component of an EB can be computed wrongly, which can also affect an estimated mass of the third body. But precision of estimation of the nodal period remains 
unaffected. For each system, all available color indices were collected and for the $T_1$ estimate the most probable value was selected.
Each color index was also corrected from an effect of interstellar reddening, according to the relation 
$(B-V)_0 = f((B-V),(U-B))$ \citep{johnson_morgan}, a map of interstellar reddening in the LMC \citep{haschke} and mean reddening in the direction toward the SMC \citep{massey1995}.

Photometric solutions of LCs of individual systems in the LMC and SMC are in Tables~\ref{tab.lc_sol_lmc} and \ref{tab.lc_sol_smc}, respectively. 
In these tables, the computed and fixed parameters are marked. 
Luminosities in $V$ and $R$ Johnson photometric bands were obtained from the MACHO data. MACHO photometry was not originally obtained in standard Johnson 
passbands but with $B_\mathrm{MACHO}$ and $R_\mathrm{MACHO}$ filters instead, and it had to be transformed before the \texttt{PHOEBE} modeling
according to the calibration relations in \citep{bessell1999}. Examples of LCs of each system, from which the time variations of inclination 
is apparent, are shown in Fig.~\ref{fig.lc}.

Precise linear ephemerides of the inner eclipsing binaries in the LMC and SMC are listed in Tables~\ref{tab.ephem_lmc} and \ref{tab.ephem_smc}, respectively.
In some cases, eclipse timing residual diagram of the EB is not linear and ETVs become apparent. For precise modeling of LCs of such systems
several sets of linear ephemeris had to be calculated for each part of a given LC. In Tables~\ref{tab.ephem_lmc} and \ref{tab.ephem_smc},
there are both the best ephemerides on whole time interval of observation and the set of ephemerides for each interval in cases when it was needed. 
Eclipse timing residual diagrams for every studied system is shown in Figs.~\ref{fig.LMC16023_oc+LMC18240_oc} and \ref{fig.oc}.  

Photometric solution of each LC subset of each system enables to derive the time dependence of inclination $i_0(t)$ of every EB. In order to 
obtain $P_\mathrm{nodal}$, physical parameters of a third body and mutual orientation of orbits, each dependence $\cos(i_0) = f(t)$ 
was modeled with Eq. (\ref{rov.soderhjelm}). The results are listed in Tables~\ref{tab.pnodal_lmc} and \ref{tab.pnodal_smc} together with 
$68\%$ confidence intervals. 
Confidence interval of each fitted parameter was estimated from projection of $\chi^2$ of the fit of $\cos(i_0) = f(t)$ dependence. 
In Fig.~\ref{fig.incl}, there is $\chi^2$ of the fit shown in $\cos(i_1)$ - $\dot\Omega$ parameter space, which provides an insight into how the parameters are limited. In most cases the model is not well limited and
many possible solutions have very similar $\chi^2$. That is the reason why the results in the Tables~\ref{tab.pnodal_lmc} 
and \ref{tab.pnodal_smc} have extreme uncertainties in some cases. However, the most likely parameters of the third body 
can be estimated. For a given solution of $\cos(i_0) = f(t)$, one value of mass of a third body $m_2$ and its orbital period $P_2$
can be calculated according to the Eqs. (\ref{rov.vokrouhlicky}), (\ref{rov.j}), and (\ref{rov.gamma}). In Fig.~\ref{fig.m2p2}, there
are all $m_2$ and $P_2$ computed from our model with $68.3 \, \%$ probability in both possible orientations of orbits according to the
"$180^\circ$" degeneracy of the solution. The area of possible solutions is not covered homogeneously and the number of solutions in each bin is indicated. Each figure is also shown for two extremal third body orbital eccentricities $e_2 = 0$ and $e_2 = 0.5$ according to the argumentation
in Sect. \ref{sec.precession}. 

From the distribution of possible $m_2$ and $P_2$, angles $i_2$ and $i_3$ can also be estimated. For each solution of $m_2$ and $P_2$, one value 
of $i_2$ can be computed from Eqs. (\ref{rov.j}) and (\ref{rov.gamma}). The most probable value and its uncertainty based on distribution,
 are listed in the Tables~\ref{tab.pnodal_lmc} and \ref{tab.pnodal_smc} for each system. With the knowledge of $i_2$, the also 'observable' inclination 
of the third body $i_3$ can be computed from the Eq. (\ref{rov.soderhjelm}). In Tables~\ref{tab.pnodal_lmc} and \ref{tab.pnodal_smc}, we list
their values, but usually with very large uncertainties which make it impossible to estimate real amplitude of radial velocities, which would be extremely 
useful for planning spectroscopic observations for confirmation of the presence of the third body.

From the LC solution, masses of EB's components can be estimated with an assumption that both components lie on the main sequence, 
where the mass-luminosity ratio is relatively well defined. Therefore, analysis of the third light can lead to an estimate of the third body mass.
In cases of the most of our analyzed EB the third light did not contribute to the LC with more than $1\,\%$ of the total light, which is not significant 
with respect to a precission of photometry. In these cases, however, at least a limit on maximal possible $m_2$ can be estimated and is also marked in Fig.~\ref{fig.m2p2}.
As mentioned in Sect. \ref{sec.precession}, additional observational bound is given from the amplitude of ETVs which can be estimated from scattering of eclipse timing residual diagram. 
All presented systems are relatively compact and thus ETVs are dominated by dynamical term and its amplitude $A_\mathrm{phys}$ is only function of 
$m_0$, $m_1$, $m_2$, period ratio and $e_2$. An upper limit on the third body mass is also shown in Fig.~\ref{fig.m2p2}, but in cases of small $m_2$ and 
short $P_2$ the stability limit is even stricter. All limits on masses and period of third bodies are summarized in Tables~\ref{tab.results_lmc} and \ref{tab.results_smc}.

\begin{table*}[!ht]
\caption{Orbital orientations of LMC systems and fitted parameters of $i_0$ time dependence.}
\label{tab.pnodal_lmc}
\centering
\begin{tabular}{cccccccc}
\hline\hline
OGLE-LMC-ECL- & $t_0$ & $P_\mathrm{nodal}$ & $\dot\Omega$ & $I / 180 - I$ & $i_1 / 180 - i_1$ & $i_\mathrm{2}$ & $i_3 / i_\mathrm{3,inv}$ \\
 & [HJD] & [years] & [rad/year] & [$^\circ$] & [$^\circ$] & [$^\circ$] & [$^\circ$]\\
\hline
01350 & $2450684^{+1290}_{-4400}$ & $23^{+31}_{-6}$ & $0.273^{+0.106}_{-0.157}$ & $7^{+68}_{-62} / 187^{+62}_{-68}$ & $71^{+2}_{-62} / 109^{+62}_{-2}$ & $6^{+70}_{-3}$ & $7^{+53}_{-40} / 169^{+32}_{-26}$ \\[5pt]      
13150  & $2448006^{+1700}_{-28200}$ & $67^{+399}_{-22}$ & $0.094^{+0.045}_{-0.081}$ & $9^{+63}_{-87} / 171^{+87}_{-63}$ & $72^{+11}_{-64} / 108^{+64}_{-11}$ & $4^{+40}_{-2}$ & $7^{+33}_{-91} / 167^{+49}_{-202}$ \\[5pt]
16023 & $2452310^{+20}_{-38}$ & $48^{+4}_{-13}$ & $0.129^{+0.050}_{-0.011}$ & $27^{+34}_{-13} / 153^{+13}_{-34}$ & $49^{+14}_{-33} / 131^{+14}_{-33}$ & $3^{+52}_{-2}$ & $26^{+35}_{-13} / 152^{+18}_{-32}$ \\[5pt]
18240 & $2456663^{+2210}_{-580}$ & $43^{+134}_{-23}$ & $0.146^{+0.172}_{-0.11}$ & $176^{+2}_{-74} / 4^{+74}_{-2}$ & $83^{+2}_{-74} / 97^{+74}_{-2}$ & $4^{+51}_{-2}$ & $172^{+17}_{-1} / 2^{+83}_{-2}$ \\[5pt]
23148 & $2450809^{+276}_{-540}$ & $42^{+36}_{-16}$ & $0.15^{+0.09}_{-0.07}$ & $10^{+53}_{-5} / 170^{+5}_{-53}$ & $62^{+5}_{-53} / 118^{+53}_{-5}$ & $7^{+69}_{-4}$ & $4^{+53}_{-9} / 163^{+21}_{-22}$ \\[5pt]
\hline
\end{tabular}           
\end{table*}

\begin{table*}[!ht]
\caption{Orbital orientations of SMC systems and fitted parameters of $i_0$ time dependence.}
\label{tab.pnodal_smc}
\centering
\begin{tabular}{cccccccc}
\hline\hline
OGLE-SMC-ECL- & $t_0$ & $P_\mathrm{nodal}$ & $\dot\Omega$ & $I / 180 - I$ & $i_1 / 180 - i_1$ & $i_\mathrm{2}$ & $i_3 / i_\mathrm{3,inv}$ \\
 & [HJD] & [years] & [rad/year] & [$^\circ$] & [$^\circ$] & [$^\circ$] & [$^\circ$]\\
\hline  
1532 & $2457103^{+27853}_{-778}$ & $75^{+278}_{-56}$ & $0.084^{+0.253}_{-0.066}$ & $-40^{+23}_{-38} / 140^{+38}_{-23}$ & $28^{+38}_{-18} / 152^{+18}_{-38}$ & $3^{+43}_{-2}$ & $44^{+67}_{-38} / 142^{+83}_{-26}$ \\[5pt]
3317 & $2455733^{+53600}_{-315}$ & $62^{+351}_{-44}$ & $0.101^{+0.253}_{-0.086}$ & $29^{+42}_{-96} / 151^{+96}_{-42}$ & $36^{+54}_{-27} / 144^{+27}_{-54}$ & $2^{+57}_{-1}$ & $31^{+71}_{-97} / 153^{+125}_{-48}$ \\[5pt]
6118 & $2455763^{+112000}_{-154}$ & $92^{+1010}_{-78}$ & $0.068^{+0.382}_{-0.063}$ & $10^{+54}_{-46} / 170^{+46}_{-54}$ & $53^{+37}_{-43} / 127^{+37}_{-43}$ & $8^{+69}_{-6}$ & $18^{+87}_{-53} / 176^{+51}_{-293}$ \\[5pt]                                          
\hline
\end{tabular}           
\end{table*}

\begin{table*}[!ht]
\caption{Masses and periods of studied triples in the LMC.}
\label{tab.results_lmc}
\centering
\begin{tabular}{ccccccc}
\hline\hline
OGLE-LMC-ECL- & $P_1$ & $P_2$ & $P_\mathrm{nodal}$ & $m_0$ & $m_1$ & $m_2$ \\
 & [days] & [days] & [years] & [$\rm M_\odot$] & [$\rm M_\odot$] & [$\rm M_\odot$]\\
\hline
01350 & $1.0988325275$ & $<70$ & $23^{+31}_{-6}$ & $3.4$ & $3.4$ & $<1.4$ \\[5pt]         
13150 & $0.95597619$ & $< 140$ & $67^{+399}_{-22}$ & $2.8$ & $2.8$ & $< 1.5$  \\[5pt]
16023 & $0.78825135$ & $< 60$ & $48^{+4}_{-13}$ & $3.9$ & $3.9$ & $< 2.15$ \\[5pt]
18240 & $2.764104952$ & $< 200$ & $43^{+134}_{-23}$ & $5.4$ & $3.6$ & $< 2.2$ \\[5pt]
23148 & $1.28218324$ & $< 95$ & $42^{+36}_{-16}$ & $13.0$ & $11.0$ & $< 6.3$  \\[5pt]
\hline
\end{tabular}           

\end{table*}

\begin{table*}[!ht]
\caption{Masses and periods of studied triples in the SMC.}
\label{tab.results_smc}
\centering
\begin{tabular}{ccccccc}
\hline\hline
OGLE-SMC-ECL- & $P_1$ & $P_2$ & $P_\mathrm{nodal}$ & $m_0$ & $m_1$ & $m_2$ \\
 & [days] & [days] & [years] & [$\rm M_\odot$] & [$\rm M_\odot$] & [$\rm M_\odot$]\\
\hline  
1532 & $1.0283876$ & $<80$ & $75^{+278}_{-56}$ & $4.9$ & $1.26$ & $<1.8$ \\[5pt]
3317 & $0.70421558$ & $<170$ & $62^{+351}_{-44}$ & $4.7$ & $0.9$ & $<2.3$  \\[5pt]
6118 & $0.9372806$ & $<260$ & $92^{+1010}_{-78}$ & $11.0$ & $10.9$ & $<4.1$  \\[5pt]
\hline
\end{tabular}            
\end{table*}


\subsection{OGLE-LMC-ECL-16023}

\object{OGLE-LMC-ECL-16023} (05:27:04.86 $-$69:29:01.6, \textit{I} = 16.9 mag) is an overcontact binary with early spectral type components. The 
best estimate based on photometric indices leads to B7V spectral type and temperature of the primary component 
$T_1 = 14\,000 \, \mathrm{K}$. Quite a large amount of photometric data from several databases including our own observations 
(MACHO, OGLE II/III/IV, DK154) is available, which allowed us to calculate inclination of the binary in 28 time intervals 
over 23 years (see Fig.~\ref{fig.incl}). Therefore, together with relatively short period $P_\mathrm{nodal}$, parametric 
space is rather well limited. Estimated orbital period of the third body is very short 
$P_2 < 60 \, \mathrm{days}$, which makes this system to be very compact. 

\begin{figure*}
\centering
        \begin{tabular}{@{}cc@{}}
    \includegraphics[width=88mm]{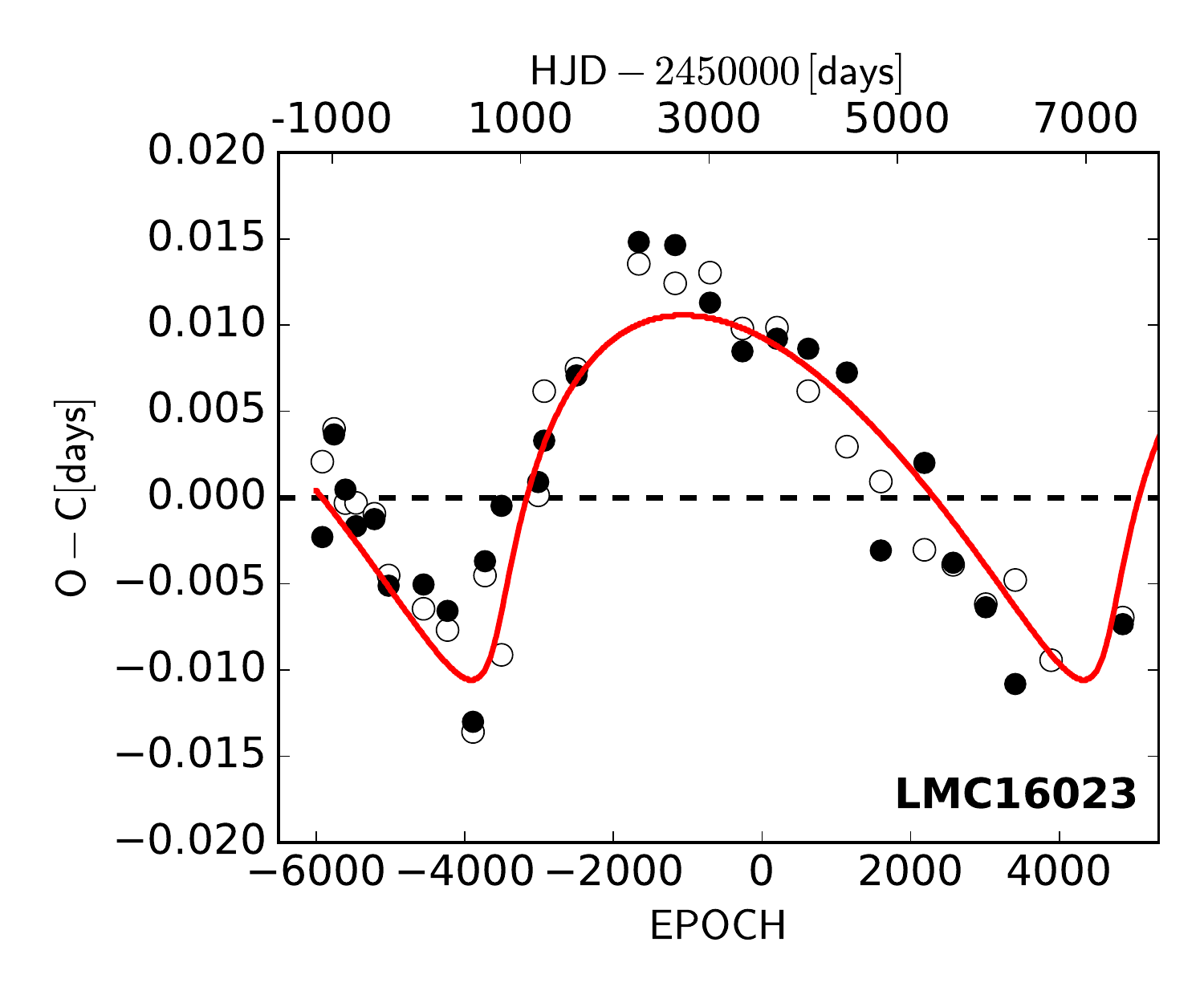} & 
    \includegraphics[width=88mm]{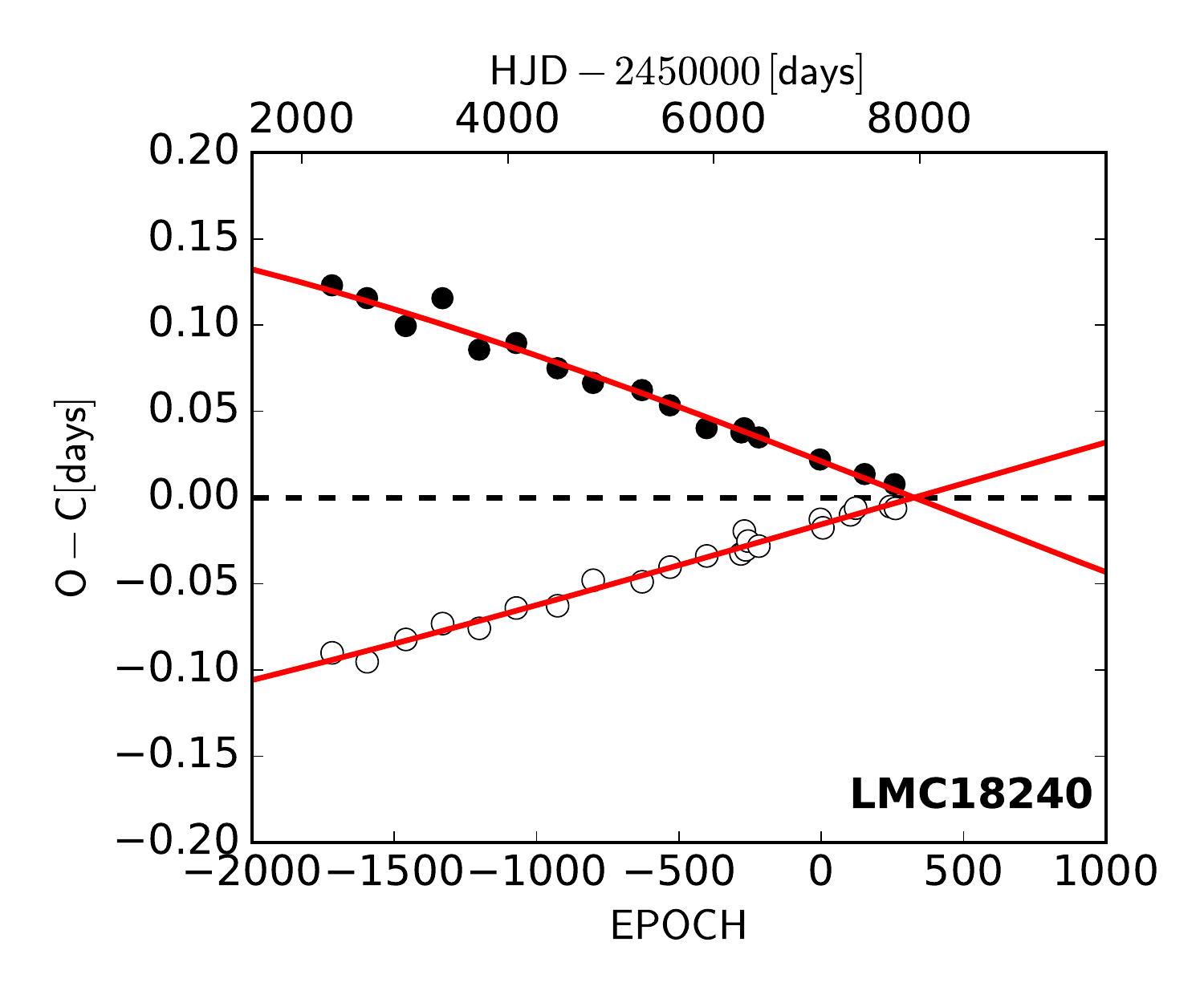} \\ 
        \end{tabular}
  \caption{Eclipse timing residual (observed $-$ computed) diagrams for \object{OGLE-LMC-ECL-16023} and 18240 with respect to linear ephemerides listed in Table 
  \ref{tab.ephem_lmc}. Black and white points represent primary and secondary minima, respectively. The red line is the best 
  fit of LTE and apsidal motion in the case of \object{OGLE-LMC-ECL-16023} and 18240, respectively.}
   \label{fig.LMC16023_oc+LMC18240_oc}
\end{figure*}

However, in eclipse timing residual diagram in the Fig.~\ref{fig.LMC16023_oc+LMC18240_oc}, ETVs with periods of approximately $ 6500 \, \mathrm{days}$ are 
apparent. That is about two orders of magnitude longer than the expected orbital period of a third body which therefore cannot be responsible
for this phenomena. Observed variation of minima timings may be due to LTE caused by another body in the system. 
With respect to this hypothesis, we have fitted the fourth body orbit and results are listed in Table~\ref{tab.LMC16023_lte},
where $T_0$ is Julian date of periastron passage of the hypothetical fourth body. 
Because of a large uncertainty of the third body mass, the mass function of the fourth component could not be calculated without
additional spectroscopic observation. 
 
We note that the third body mass limit, based on the limit of detectable third light in the LC, appears to be rather an
approximate limit on both, third and fourth body, masses. This would mean that the maximum orbital period of the third body is even shorter
than $60$ days.

\begin{table}[!ht] 
\caption{Fitted parameters of the LTE for \object{OGLE-LMC-ECL-16023}.} 
\label{tab.LMC16023_lte} 
\centering
\begin{tabular}{cc} 
\hline\hline 
$P_{\rm out}$ [days] & $6495 \pm 88$  \\ 
$T_0$  [HJD] & $2418279 \pm 482$   \\
$A_\mathrm{LTE}$ [days] & $0.01057 \pm 0.00066$ \\
$\omega_1$ [$^{\circ}$] & $323.0 \pm 8.3$     \\   
$e_1$ & $0.711 \pm 0.078$         \\         
\hline 
\end{tabular} 

\end{table}

\subsection{OGLE-LMC-ECL-18240}

\object{OGLE-LMC-ECL-18240} (05:31:33.66 $-$71:14:25.1, \textit{I} = 17.0 mag) is a detached eclipsing binary, which was found
to be also an eccentric one. For this reason, its analysis was a little
different. We also included the hypothesis of apsidal motion for
the detailed description of its eclipse timing residual diagram analysis \citep[e.g.,][]{gimenez1983}. The effect of apsidal motion also affects
the depths of both minima, however, the nodal precession is the
most dominant contribution. Moreover, the effect of changing
depth in eccentric binaries was properly modelled in our solution
using the PHOEBE code. The time coverage is still rather
poor and the apsidal motion slow, but the change of the periastron is
apparent in the data covering about 15 years. The eccentricity of the inner
 orbit of the system was found
to be of about $0.2$ and the apsidal motion period of 149 yr. The complete solution of apsidal motion is listed in Table~\ref{tab.LMC18240_apsid}, where $P_\mathrm{s}$ stands for sidereal period.
Detailed photometric monitoring in the upcoming years would help us to derive
its apsidal parameters with higher confidence. Once this result is achieved, 
the constrained apsidal motion of the inner binary and the orbital precession may 
provide more severe limits on the third star mass and its period.

\begin{table}[!ht] 
\caption{Fitted parameters of the apsidal motion for \object{OGLE-LMC-ECL-18240}.} 
\label{tab.LMC18240_apsid} 
\centering
\begin{tabular}{cc} 
\hline\hline
$\mathrm{HJD}_0$ [HJD] & $2457044.469 \pm 0.010$ \\
$P_\mathrm{s}$ [days] & $2.764105 \pm 0.000014$ \\
$e_1$ & $0.201 \pm 0.076$ \\
$\omega_1$ [$^{\circ}$] & $267.5 \pm 3.18$ \\
$\mathrm{d}\omega_1/\mathrm{d}t$ [$^{\circ}$/cycle] & $ 0.0183 \pm 0.0088$ \\
\hline 
\end{tabular} 
\end{table}

\section{Discussion and conclusions}

Multiple stellar systems are important astrophysical laboratories which could help us to understand general mechanisms 
of mutual N-body dynamic interactions between components, as well as the process of their formation. 
However, the number of relatively well studied multiple systems still remains low, especially the compact ones
manifesting orbital precession. Besides that, results from the \textit{Kepler} mission show an interesting distribution of orbital 
periods of triple systems, which is not explained theoretically and more investigation is needed. 


In this work, we focused on changing of inclination of eclipsing binaries, and developed new methods, 
which appear to be suitable for detection of
new triples with a small $P_2/P_1$ ratio. The presented methods led to an identification of 58 new compact triple candidates within the LMC and 
SMC which is, together with 14 previously known systems, the largest published sample of inclination changing compact triple
candidates out of the Milky Way galaxy. 

\begin{figure}[t!]
  \includegraphics[width=88mm]{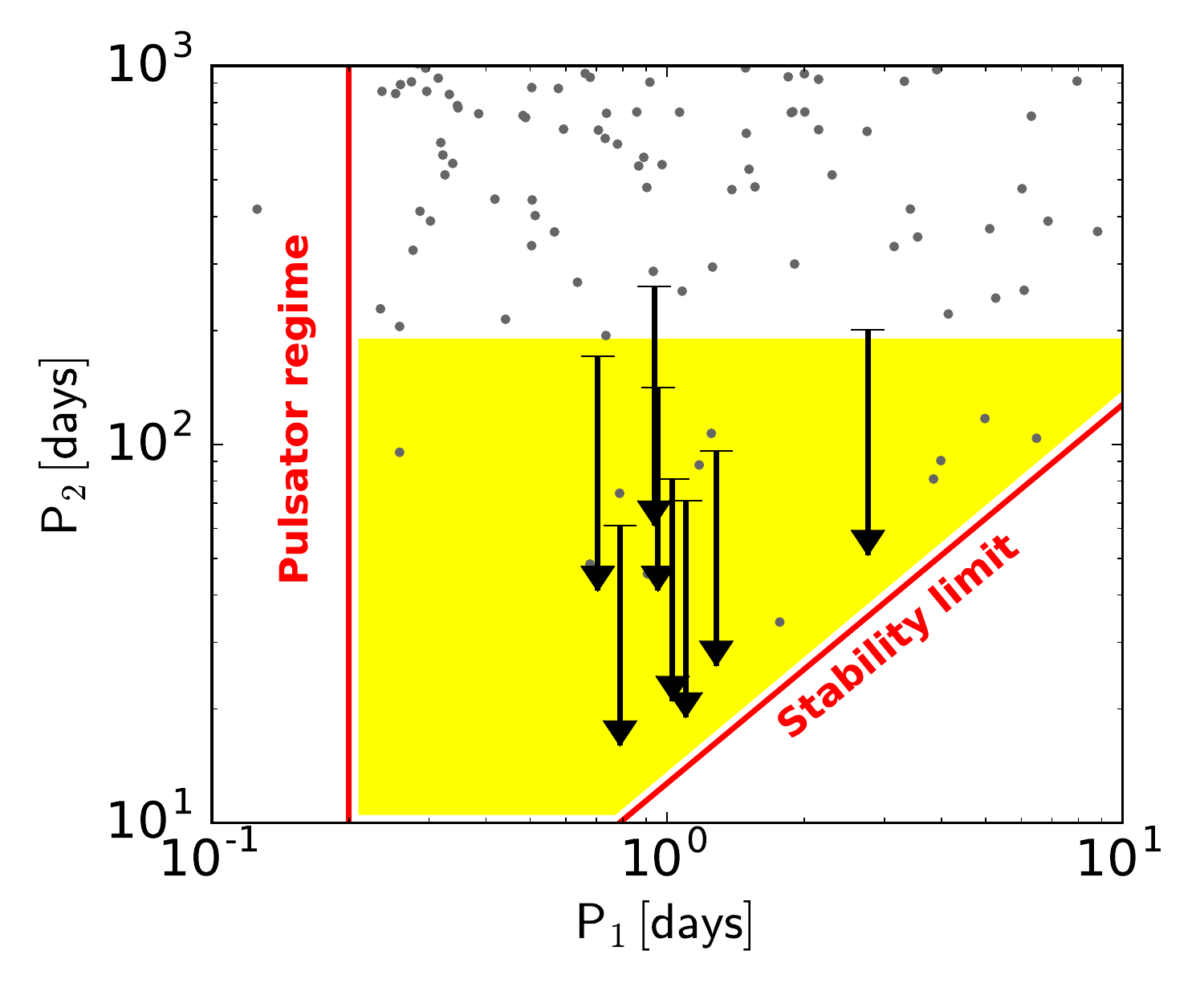}
  \caption{Area with the lack of triples in $P_1 - P_2$ distribution with the systems studied in this work. As only upper limit on $P_2$ has been estimated, each system is depicted like an arrow with upper limit.
  Triples found within the \textit{Kepler} field are marked with a gray dots.}
   \label{fig.P1P2}
\end{figure}

Eight of detected systems were studied thoroughly to determine the basic physical parameters of the eclipsing pair, the third
component, and mutual orientation of the orbits.
Unfortunately, we found that for precise determination of mutual inclinations a time base of given observations 
(more than $\approx 20$ years in some cases) is still too short and obtained angles were computed with very large uncertainties. 
In some cases also $P_\mathrm{nodal}$ were computed with large uncertainty, but despite that the systems in our sample still belongs to 
those with rather small $P_\mathrm{nodal}$ (compare 
Tables~\ref{tab.results_lmc} and \ref{tab.results_smc} in this work with Table~10 in \citet{borkovits2016}, 
with a large sample of compact triples discovered by \textit{Kepler} mission). Our analysis also led to 
restrictions on the upper limit of possible orbital period $P_2$ of the third body. The distribution of the $P_1$ and 
$P_2$ with the results of our study are shown in the Fig.~\ref{fig.P1P2}. 
One can clearly see that our sample is almost completely located within the area of the lack of triples,
as reported by \citet{borkovits2016}.
Moreover, periods of some systems in our sample might be close to the limit of stability, which is not 
determined unambiguously \citep[see][]{mardling_aarseth, sterzik_tokovinin2002, tokovinin2004, tokovinin2007}. 
But new observations with longer time bases, in the ideal case both photometric and spectroscopic,
 are needed to obtain parameters for all identified triples and to improve the precision of the 
 determined parameters for the eight analyzed systems. 

It should be noted that the upper limits on $m_2$ and consequently $P_2$ are estimated with relatively rough assumptions based on absence of third 
light in the light curve solutions and the real upper limits might be slightly different. However, it cannot change the fact that systems presented in this paper are most 
probably within the region with the lack of triples in the $P_1 - P_2$ distribution, and that makes listed systems as perfect targets for 
 a campaign of photometric observations targeting on minima timing with cadence of the order of weeks, which should lead to precise 
 determination of the third body orbital period. Absence of a third light in the light curve solution for all studied systems 
 could seem quite interesting, because it also means that $m_2 < m_0 + m_1$ but in the Milky Way galaxy the third body mass $m_2$ tends 
 to be similar to the mass of eclipsing pair $m_0 + m_1$ and for $81 \, \%$ of triples $m_2/(m_0 + m_1)$ ratio is greater than 
 $0.2$ \citep{tokovinin2008, correia2006}. Comparison of distributions of triple component masses between 
 the Milky Way galaxy and the Magellanic Clouds with different metalicities could be very interesting and useful for theory of multiple
 stellar systems formation. But in this case the third light absence is most probably result of selection effects of our methods.
 
\begin{acknowledgements}
This work was supported by the Czech Science Foundation grant no. GA15-02112S, and
also by the grant MSMT INGO II LG15010. We are also grateful to the ESO team at
the La Silla Observatory for their help in maintaining and operating the Danish telescope.
We do thank the {\sc MACHO} and {\sc OGLE} teams for making all of the observations easily public
available.
We would like to thank also the EROS-1 team -- Jean-Baptiste Marquette, Philippe Schwemling and Marc Moniez,
who kindly provided us archival photometric data. 
Marek Skarka acknowledges the support of the postdoctoral fellowship program of the Hungarian Academy of Sciences 
at the Konkoly Observatory as a host institution and the financial support of the Hungarian NKFIH Grant K-115709.
This research was carried out under the project CEITEC 2020 (LQ1601) with financial support from the Ministry of 
Education, Youth and Sports of the Czech Republic under the National Sustainability Programme II.
This research has made use of the SIMBAD and VIZIER databases, 
operated at CDS, Strasbourg, France and of NASA’s Astrophysics Data System Bibliographic Services.
\end{acknowledgements}

\bibliographystyle{aa} 
\bibliography{paper} 

\begin{appendix} 

\FloatBarrier
\onecolumn
\section{Light curves of EBs with amplitude variation located in the LMC and SMC.}


\begin{figure*}[ht!]
\centering
        \begin{tabular}{@{}ccc@{}}
    \includegraphics[width=58mm]{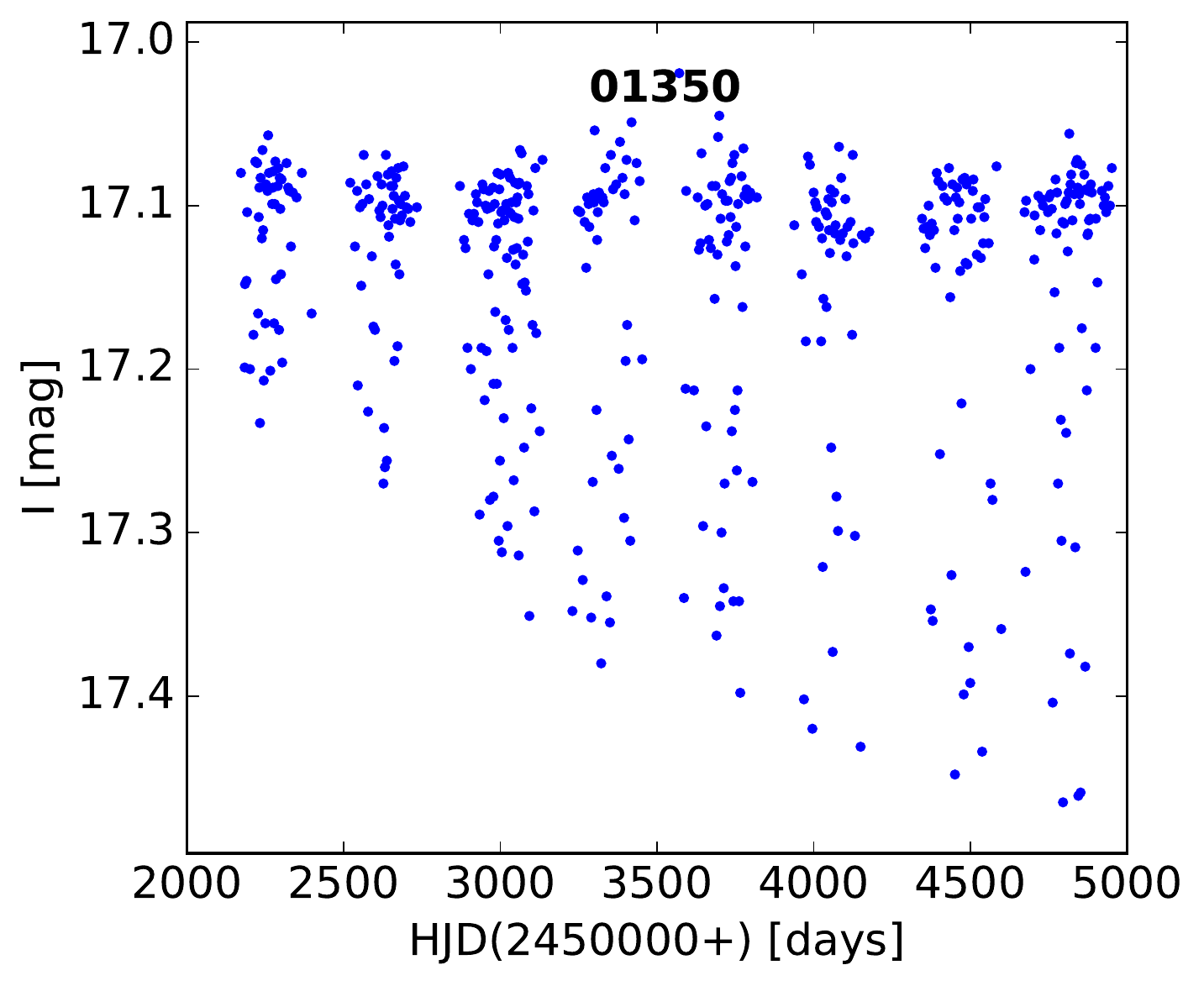} &
        \includegraphics[width=58mm]{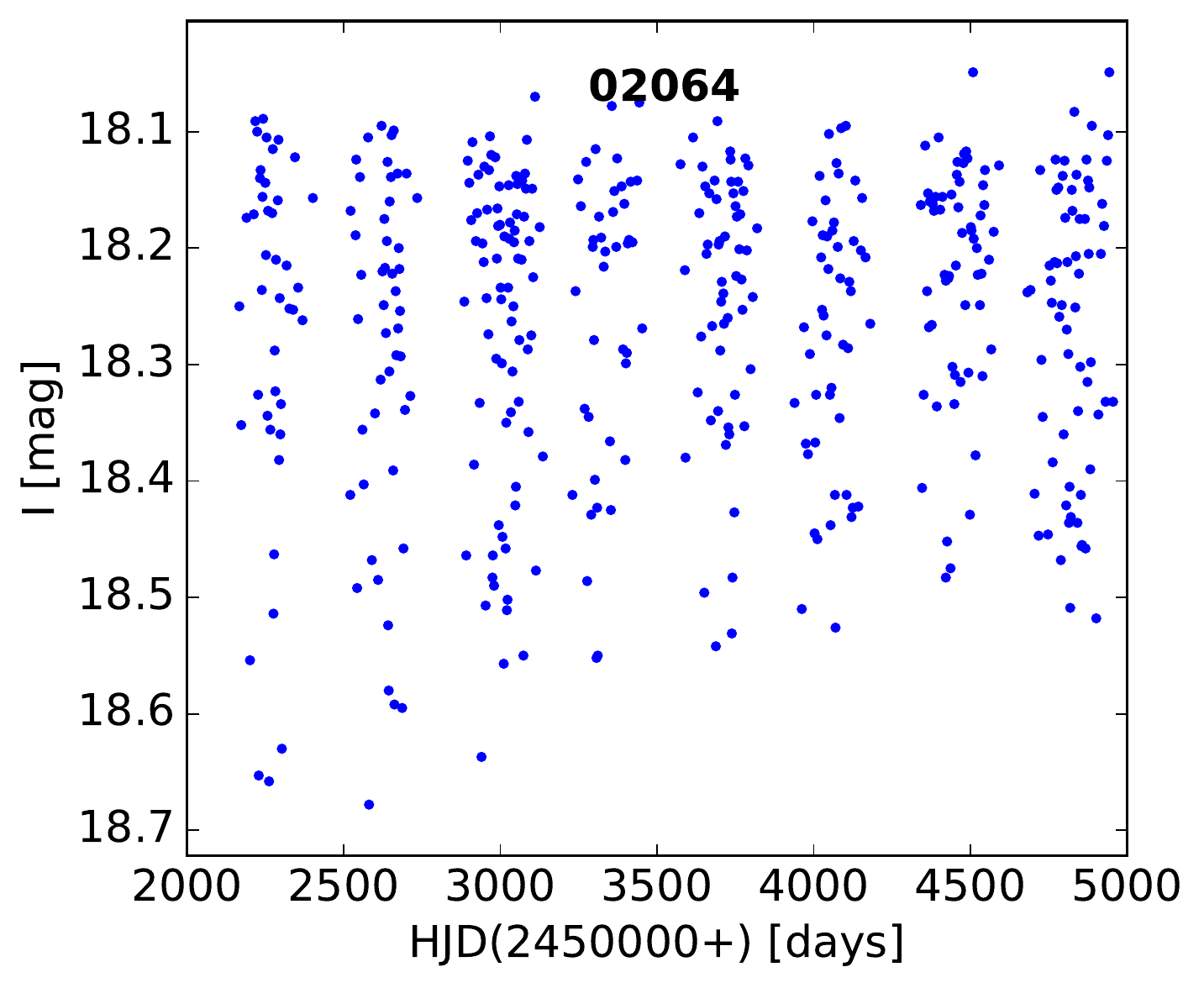} &
        \includegraphics[width=58mm]{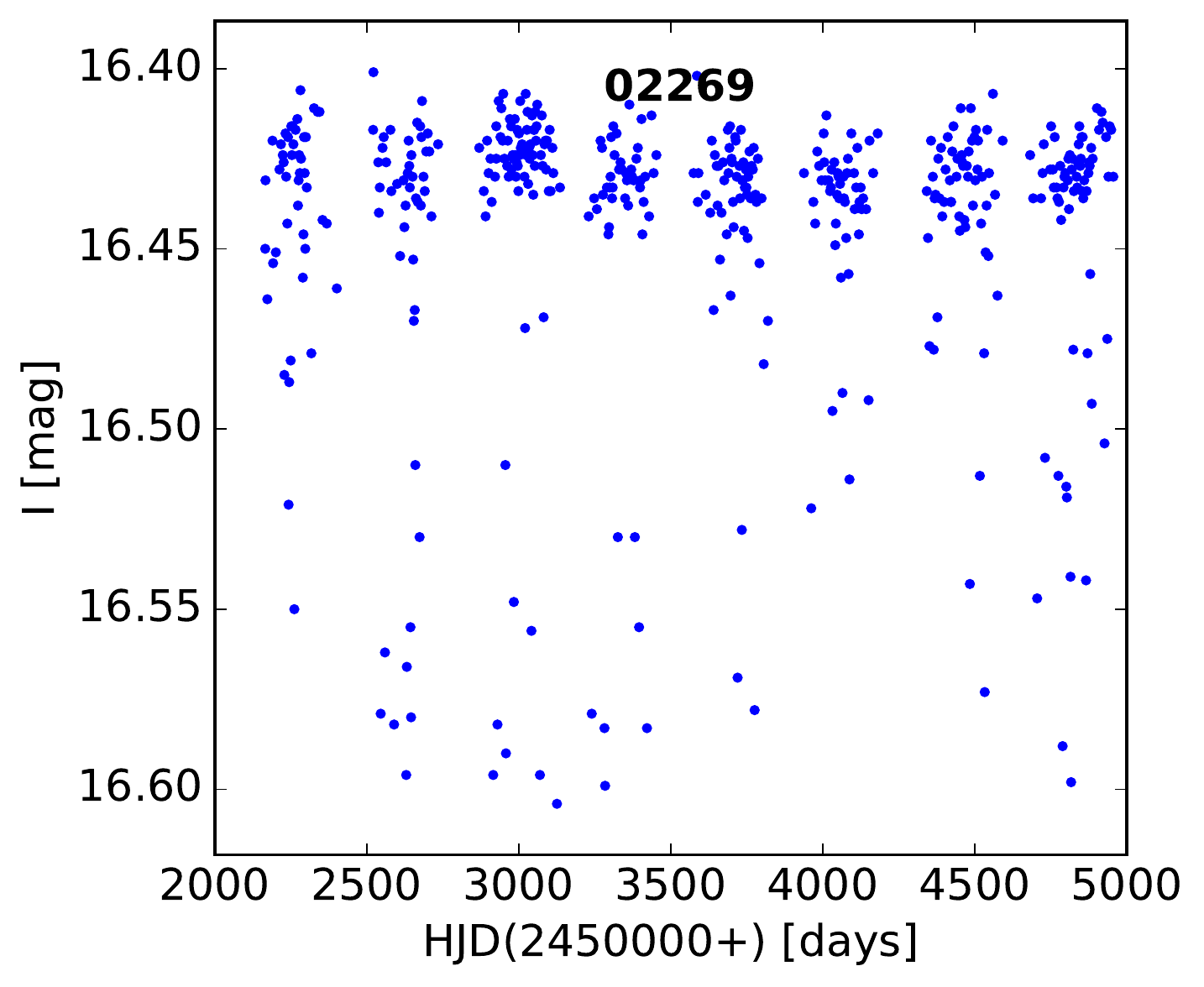} \\
        \includegraphics[width=58mm]{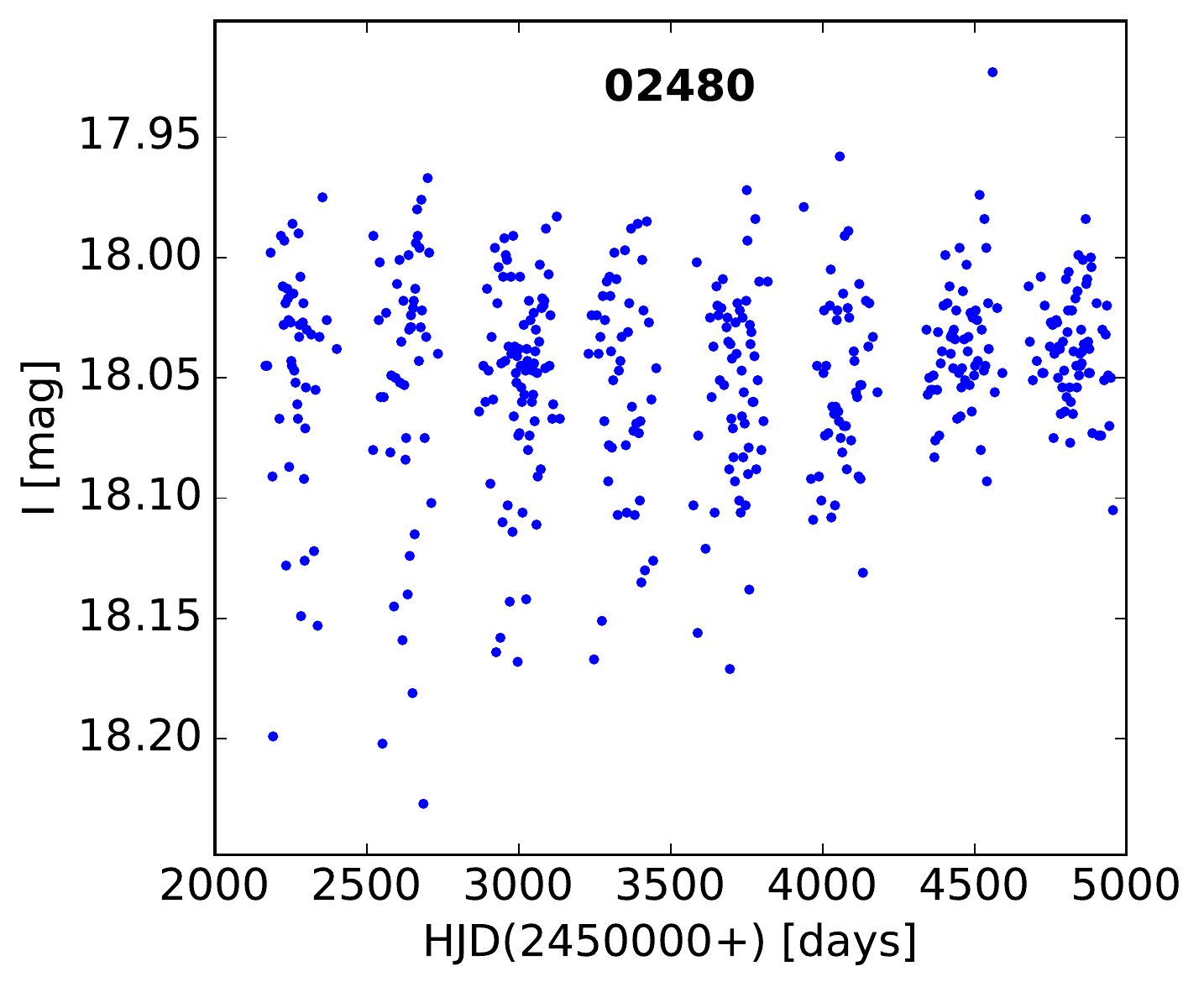} &
        \includegraphics[width=58mm]{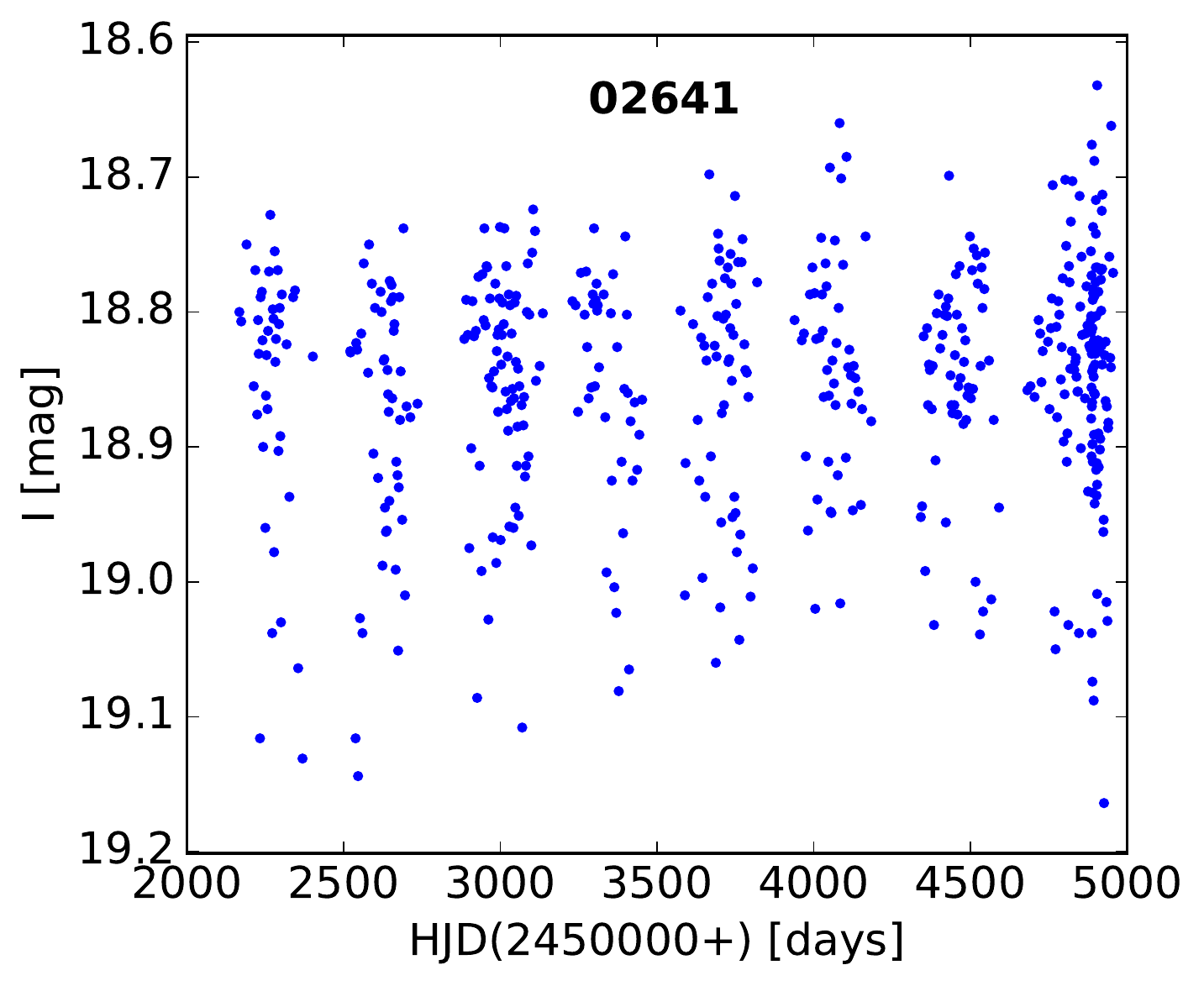} &
        \includegraphics[width=58mm]{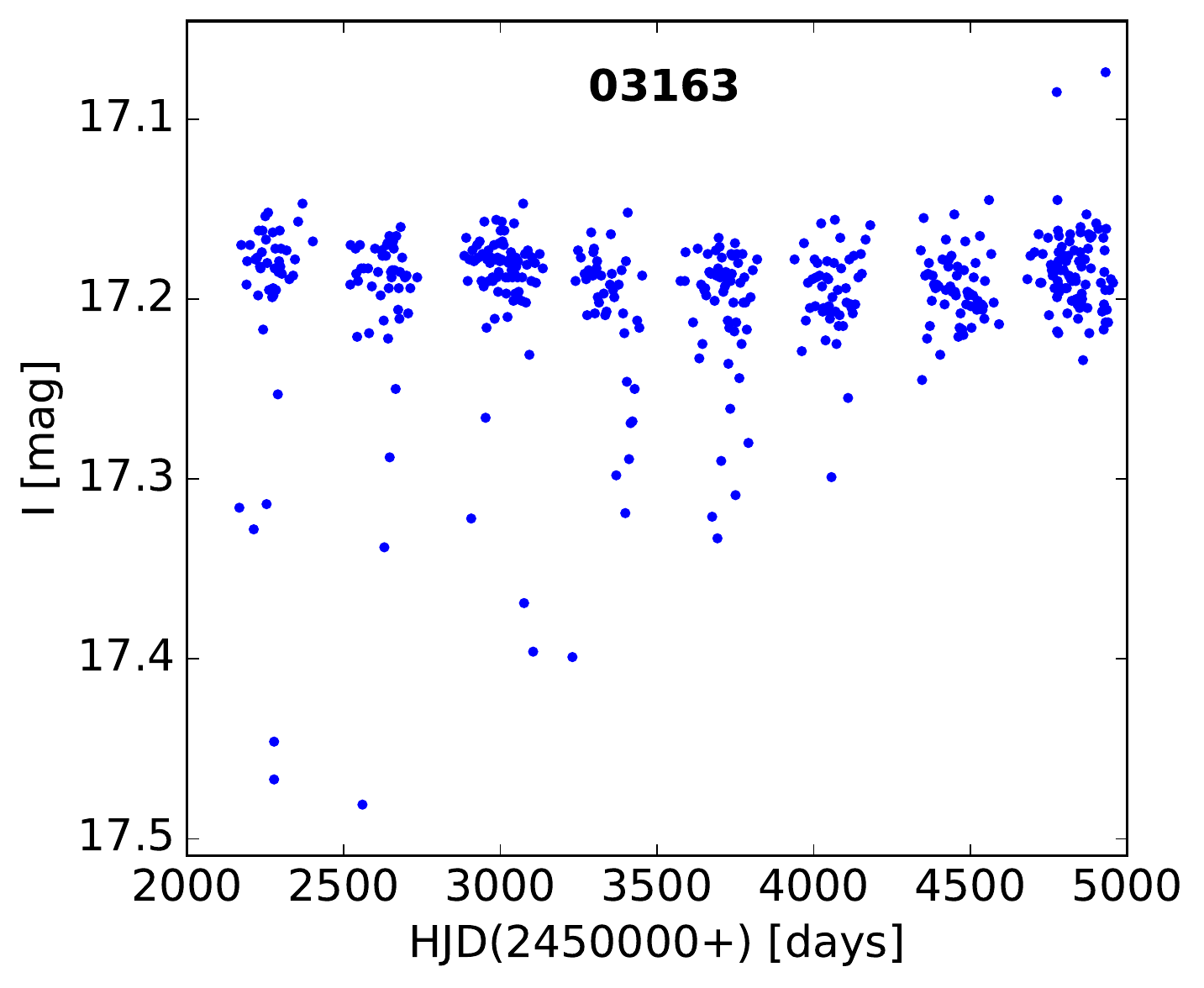} \\
        \includegraphics[width=58mm]{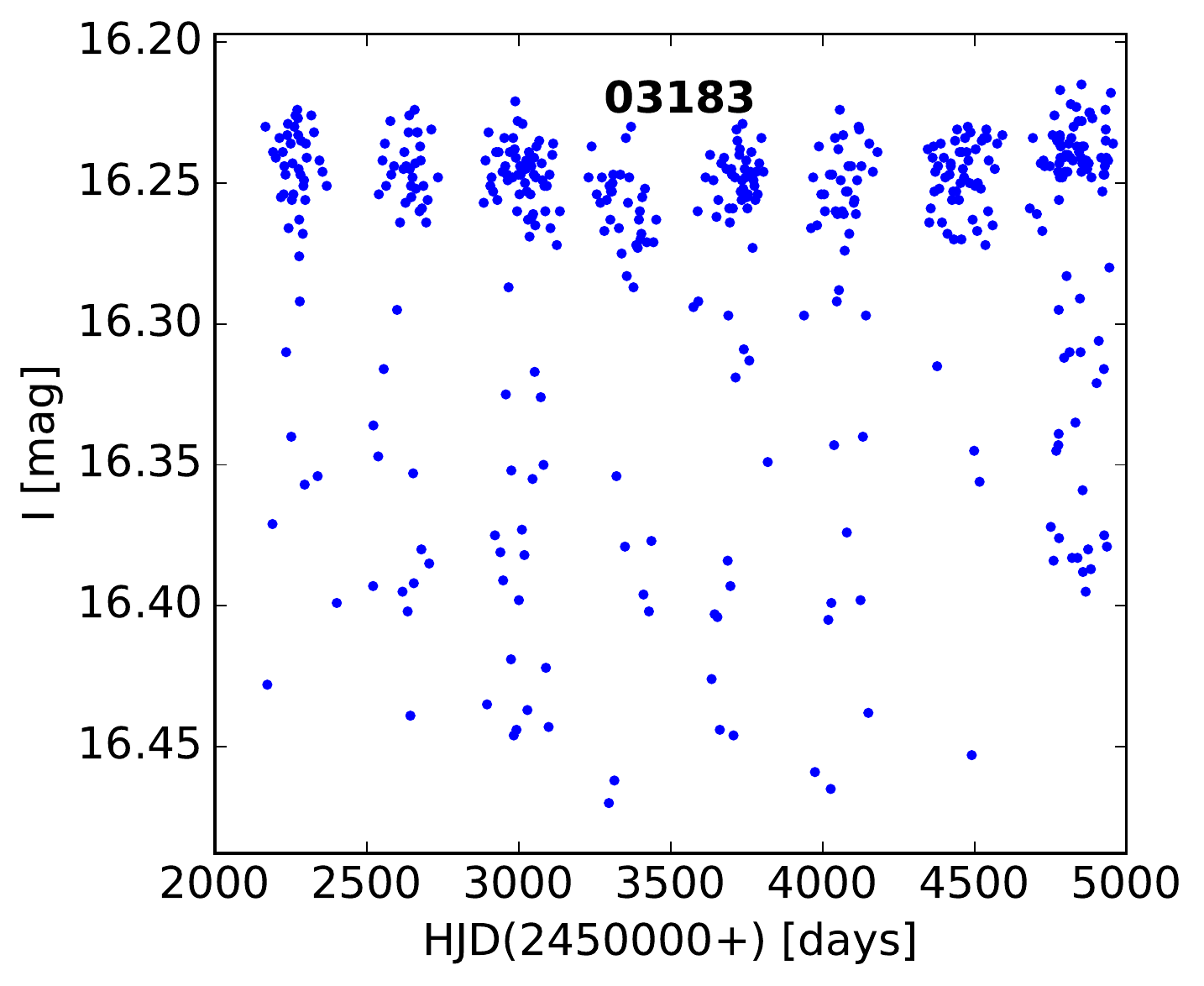} &
        \includegraphics[width=58mm]{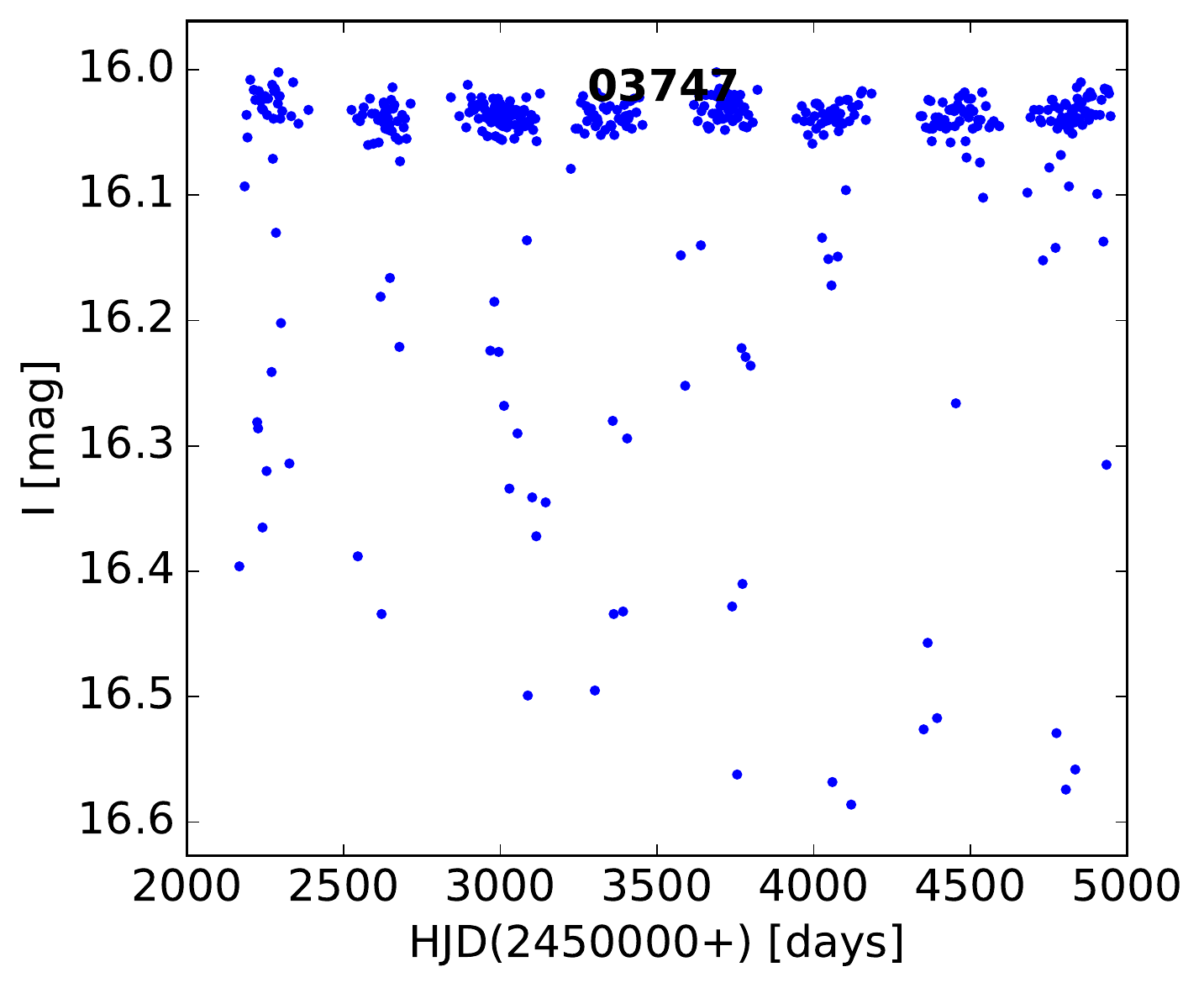} &
        \includegraphics[width=58mm]{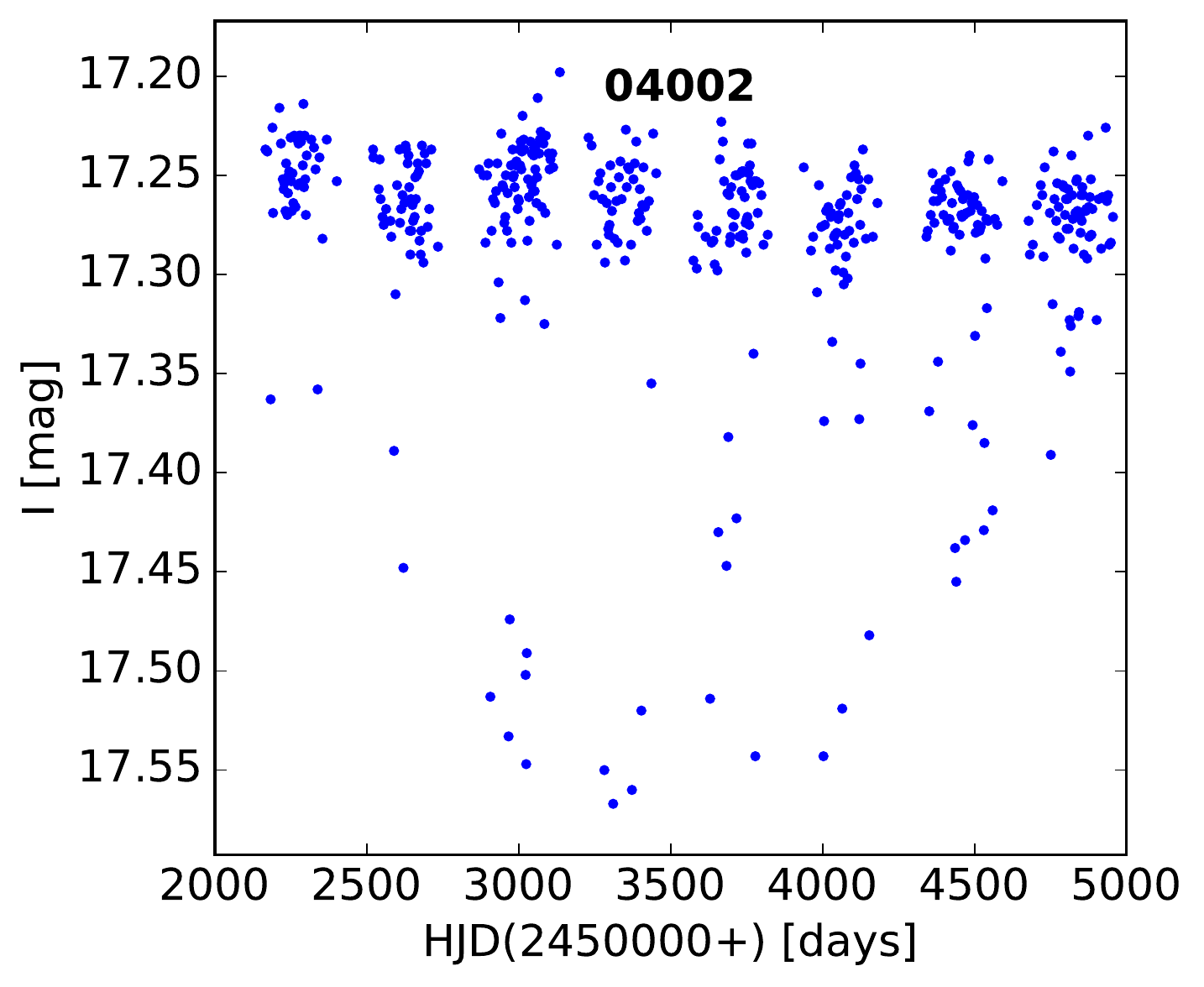} \\
        \includegraphics[width=58mm]{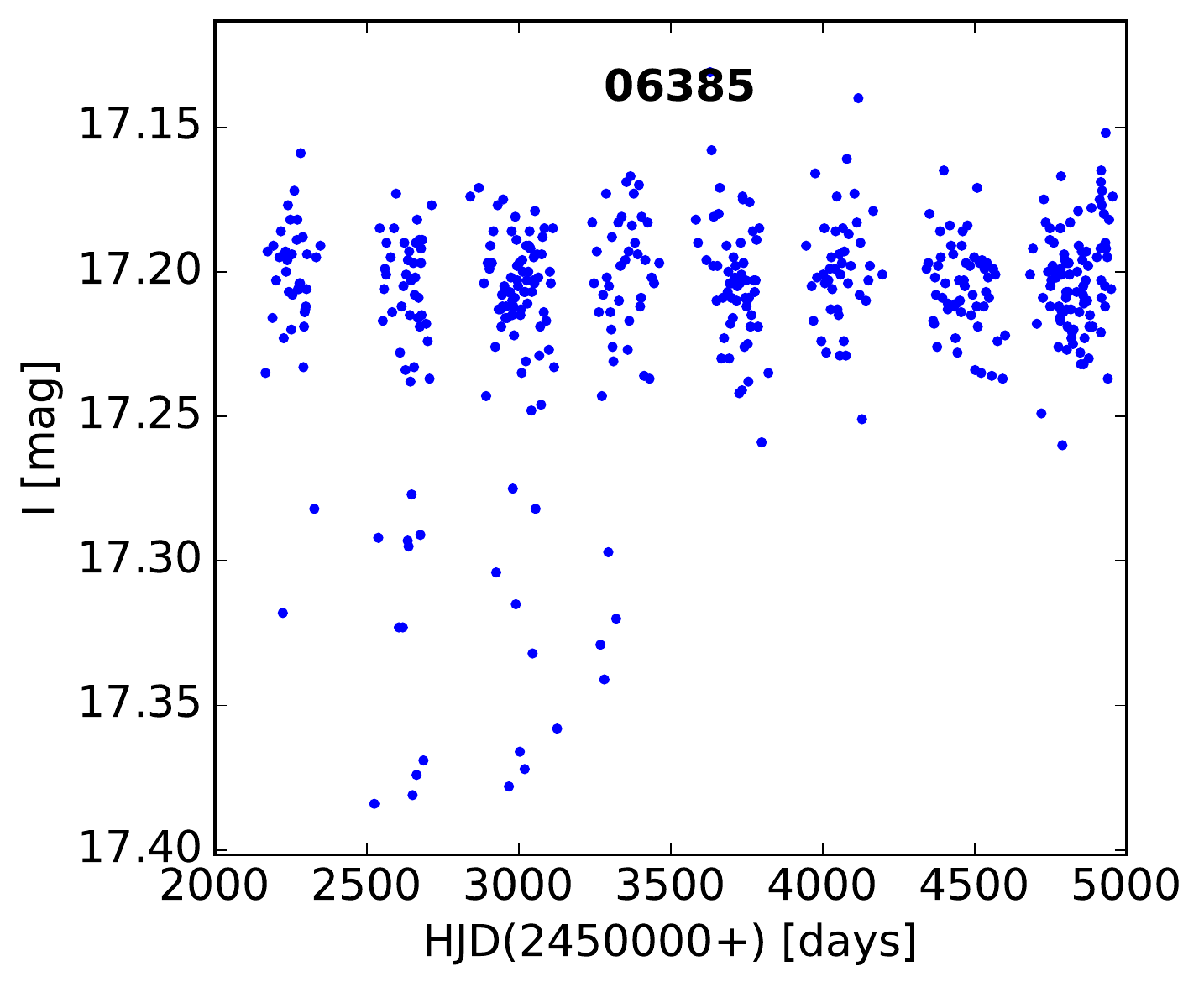} &
        \includegraphics[width=58mm]{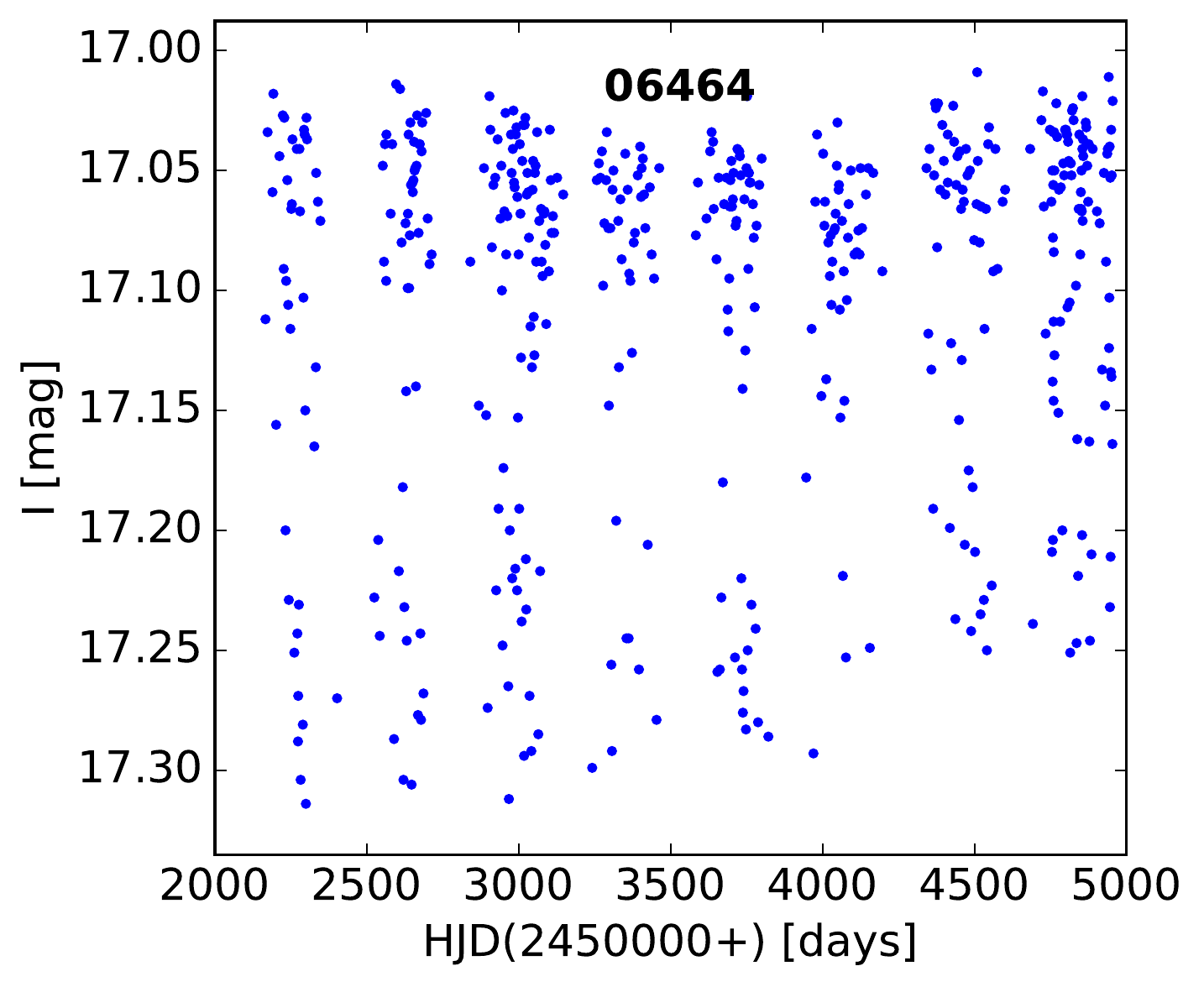} &
        \includegraphics[width=58mm]{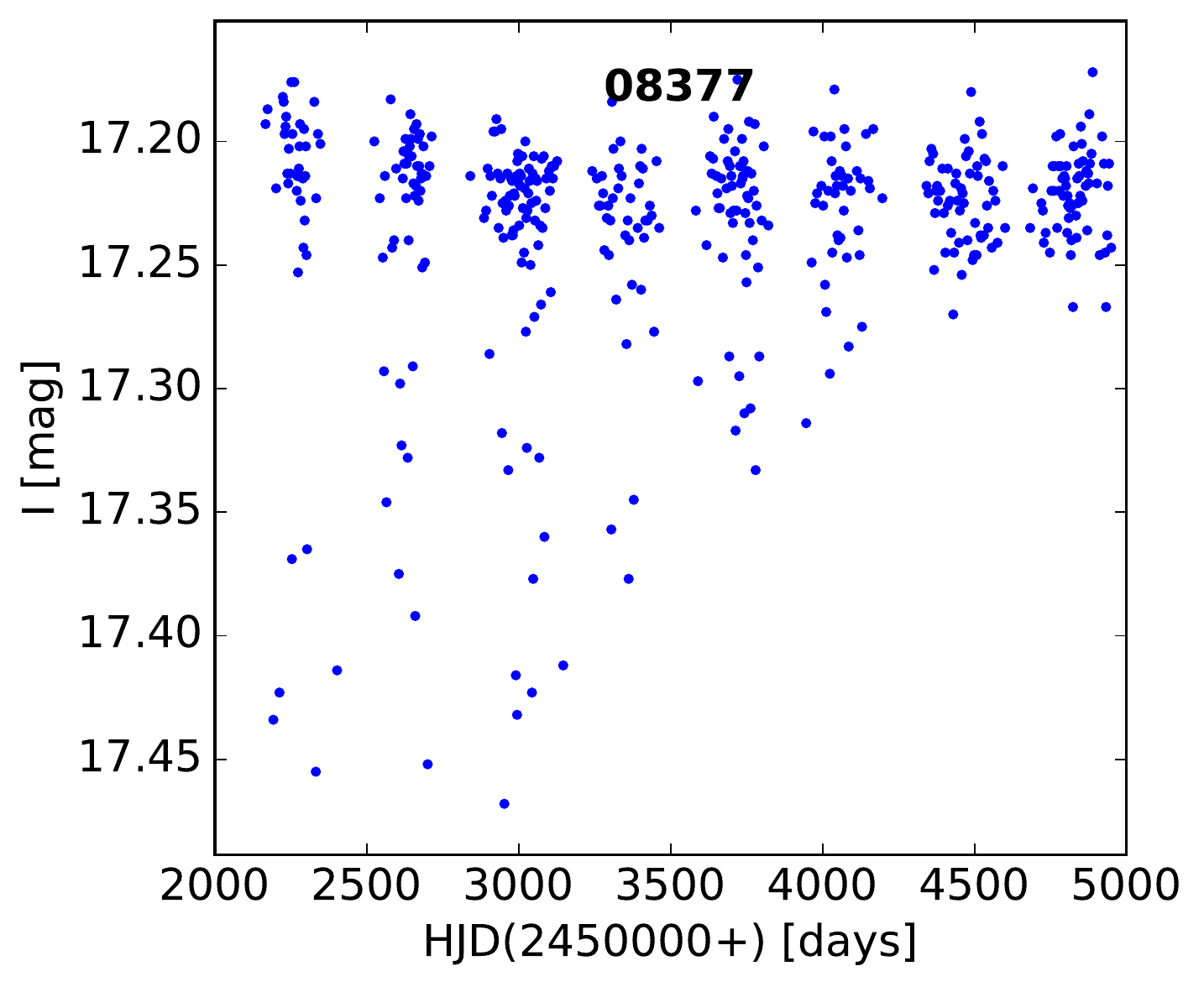} \\
        \end{tabular}
  \caption{Light curves of EBs with amplitude variation from the OGLE III LMC database.}
   \label{fig.lmc_detected_systems}
\end{figure*}

\begin{figure*}
\ContinuedFloat 
\centering
        \begin{tabular}{@{}ccc@{}}      
        \includegraphics[width=58mm]{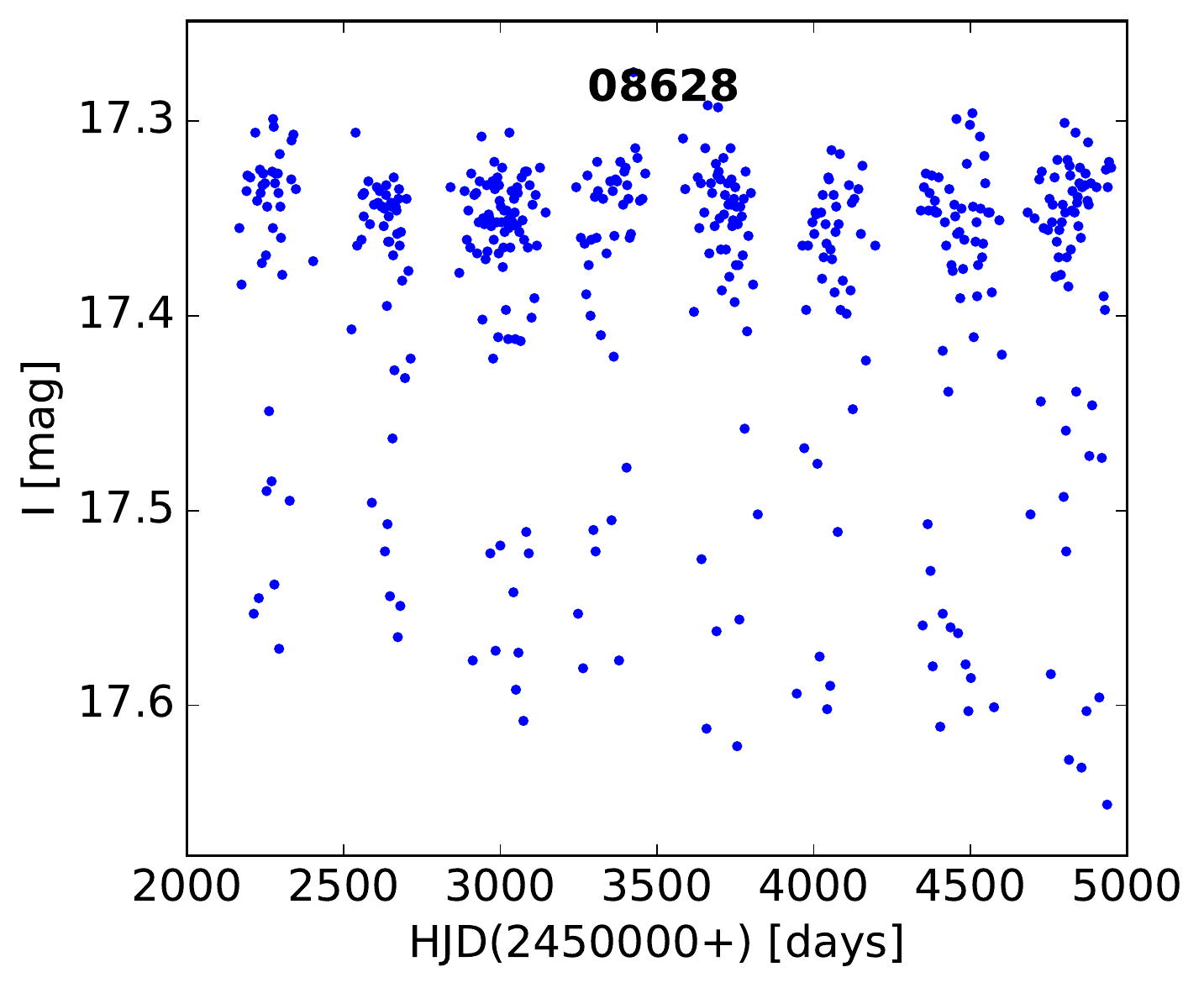} &
        \includegraphics[width=58mm]{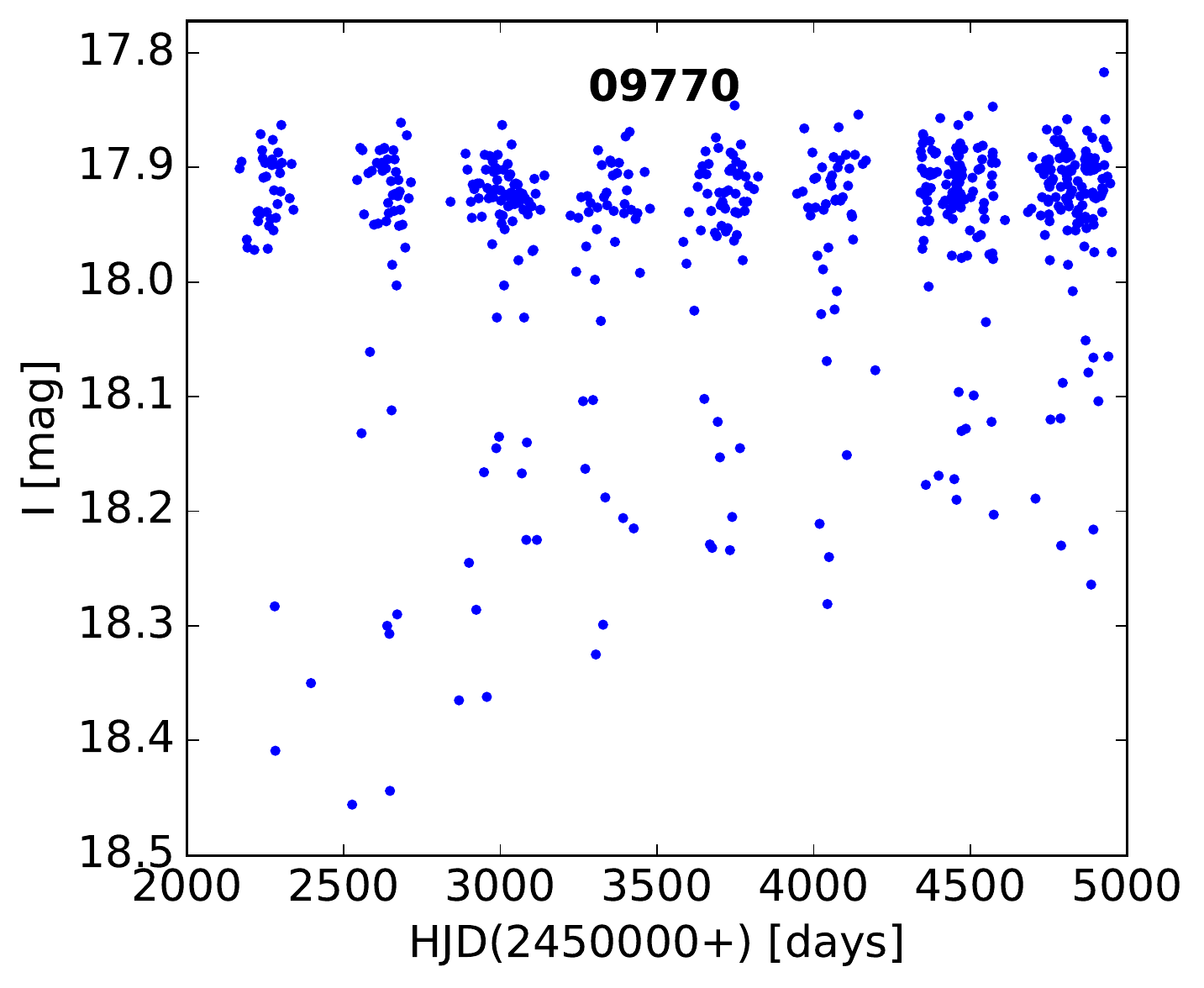} &
        \includegraphics[width=58mm]{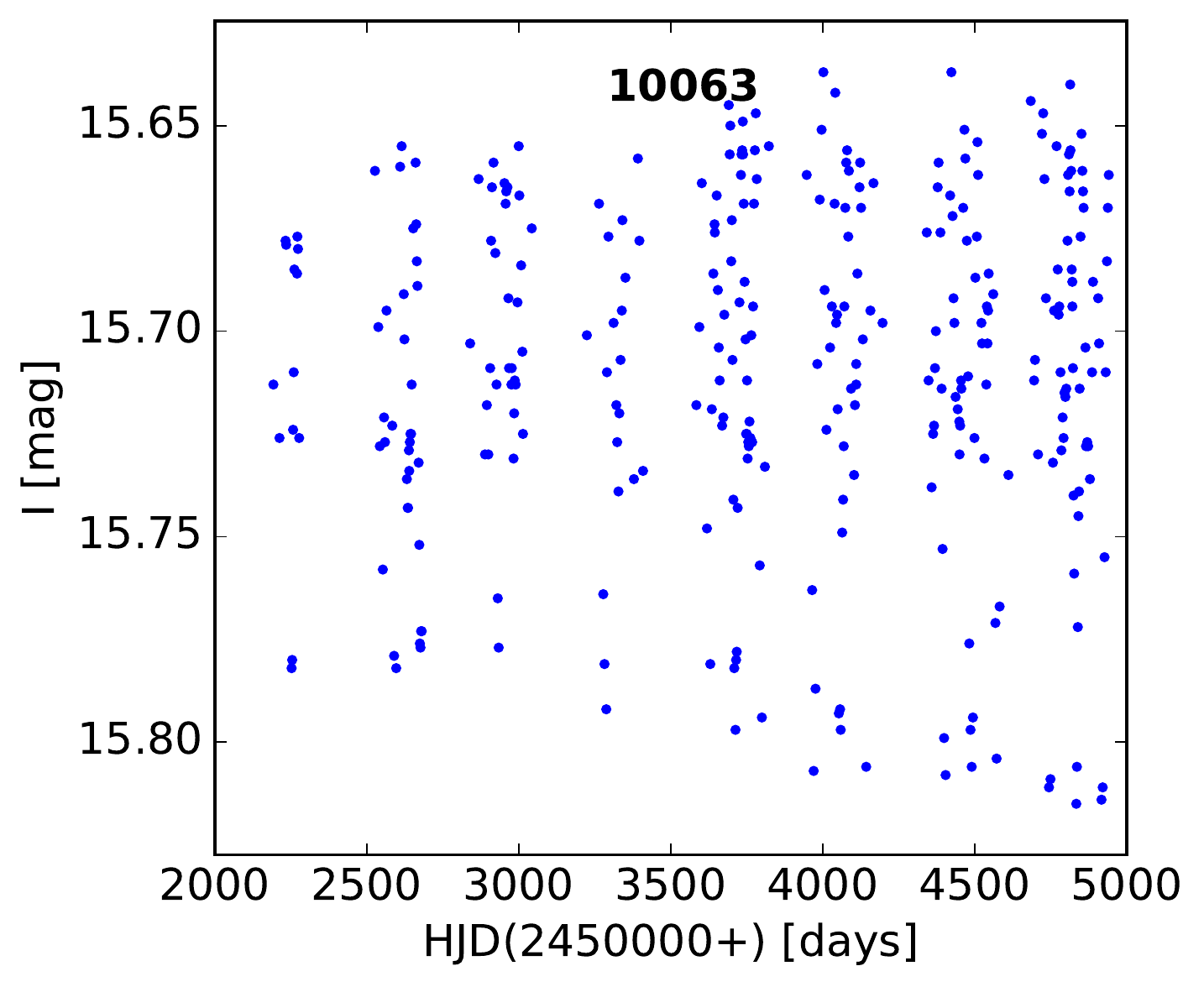} \\
        \includegraphics[width=58mm]{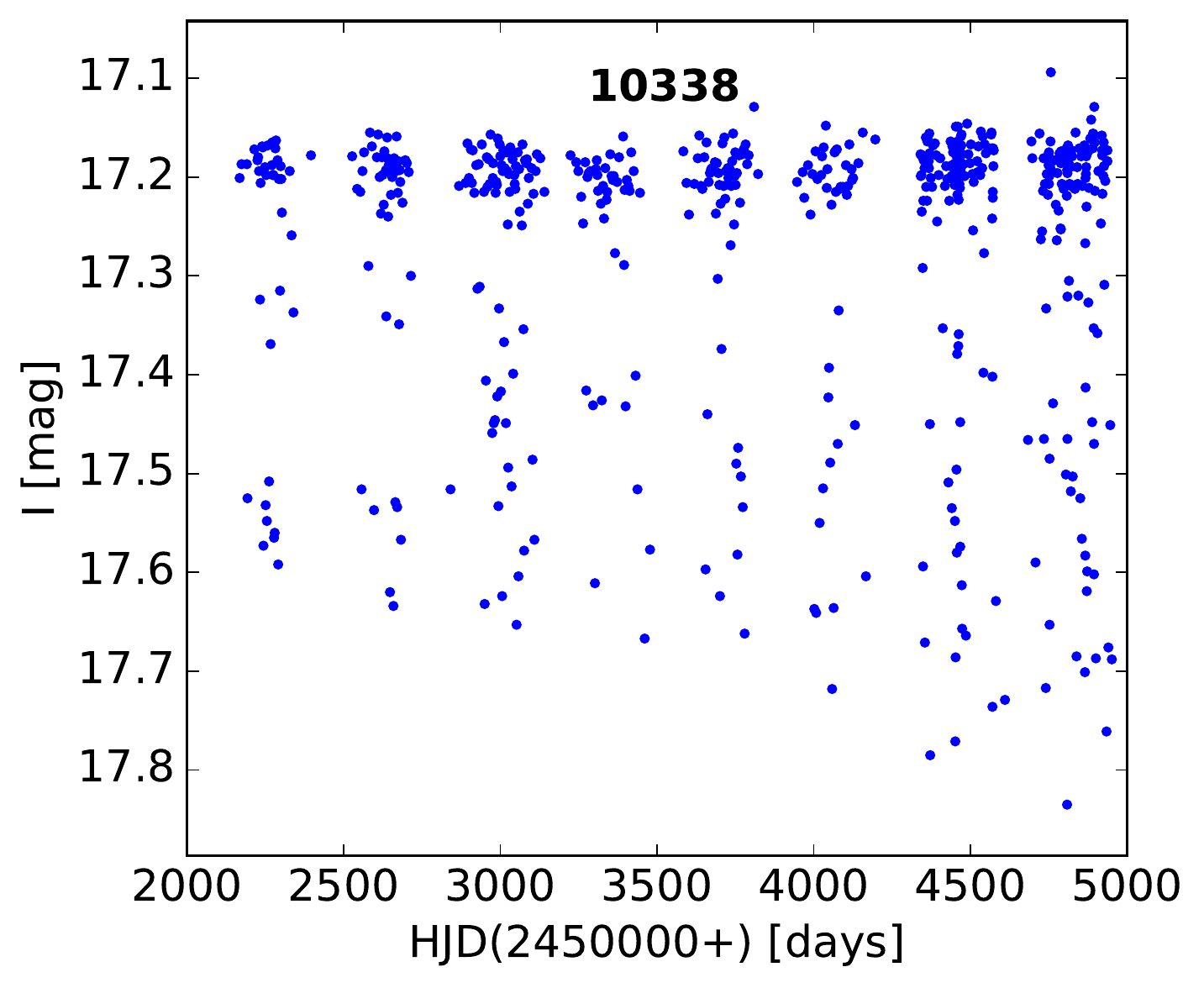} &
        \includegraphics[width=58mm]{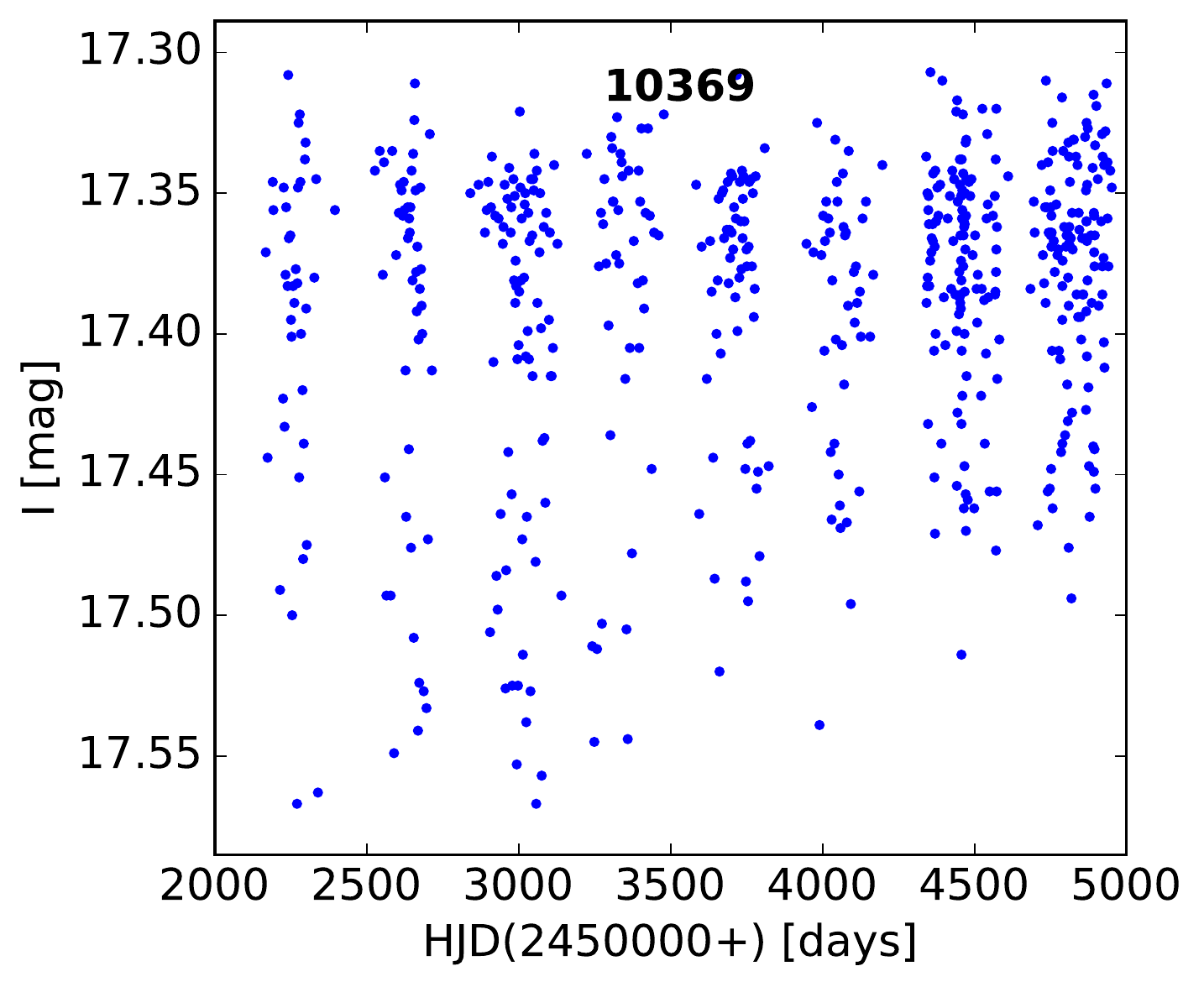} &
        \includegraphics[width=58mm]{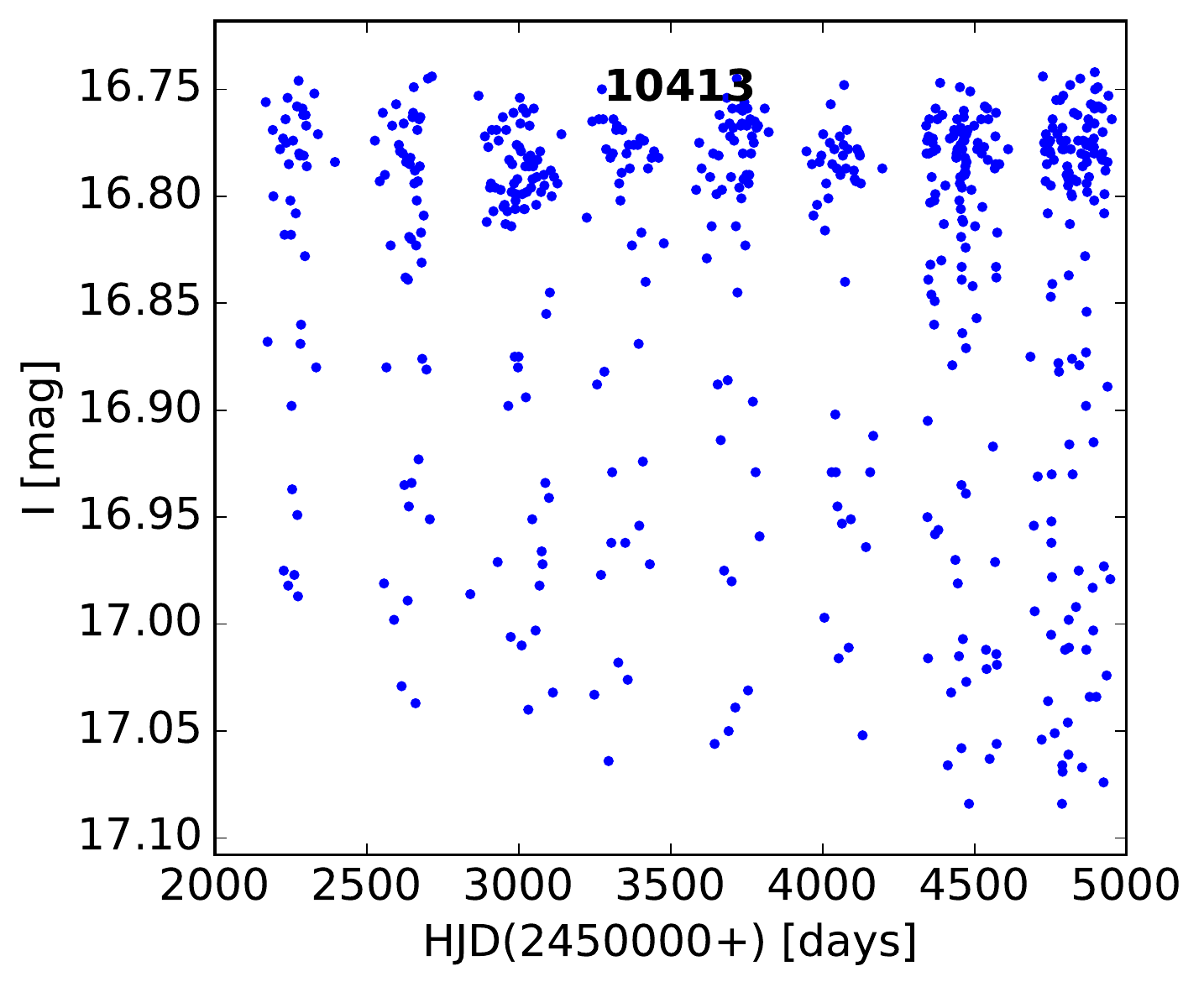} \\
        \includegraphics[width=58mm]{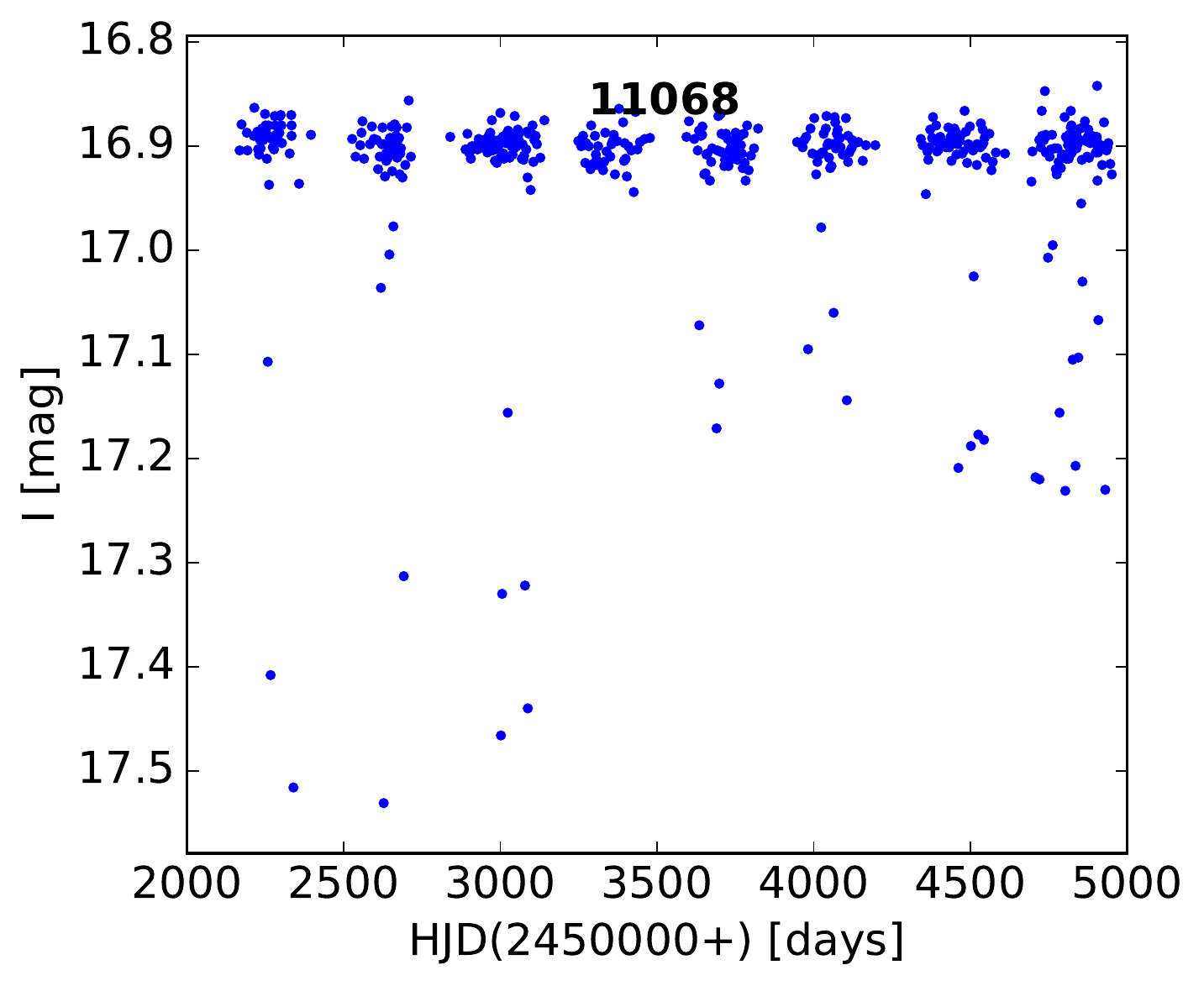} &
        \includegraphics[width=58mm]{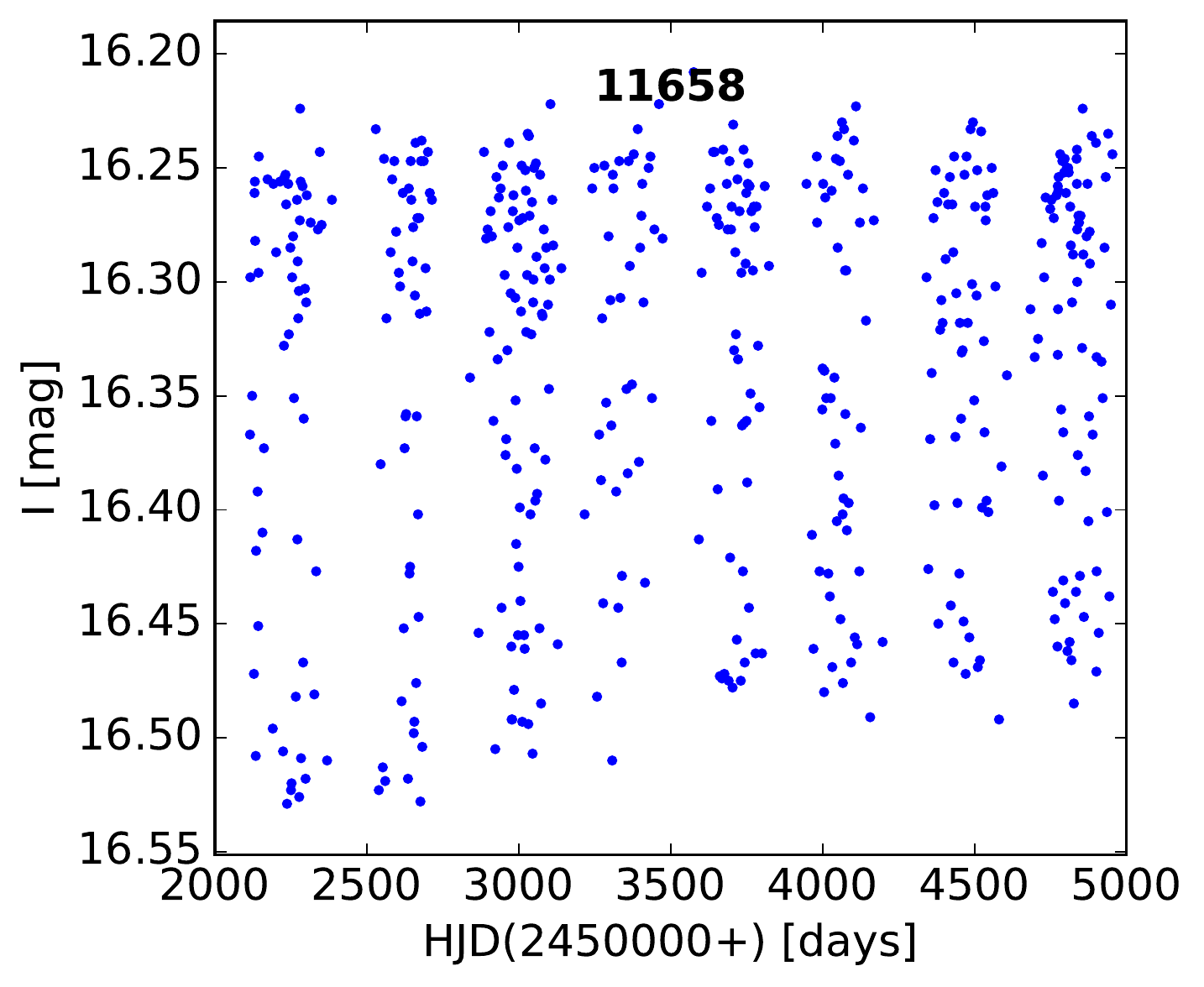} &
        \includegraphics[width=58mm]{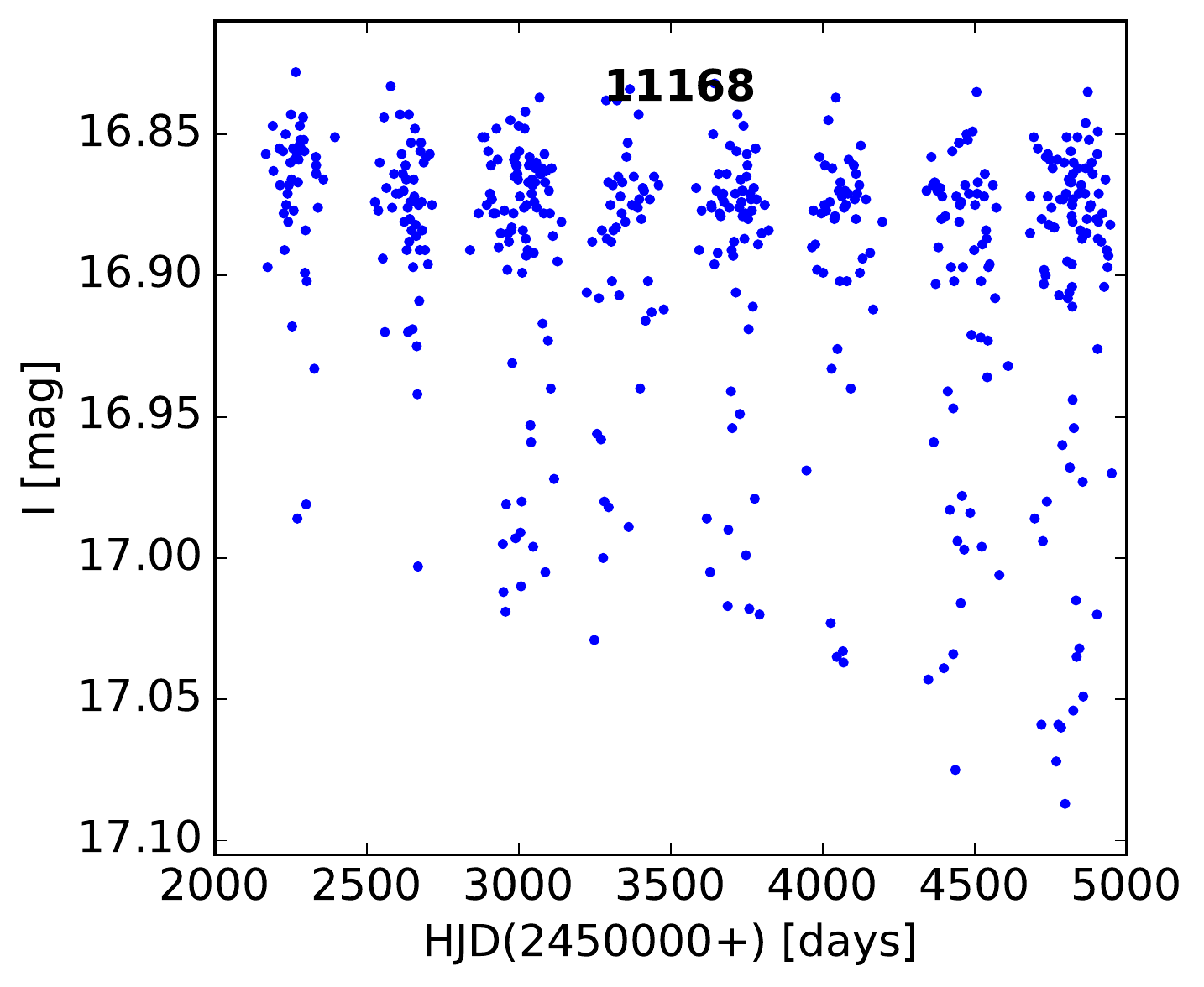} \\
        \includegraphics[width=58mm]{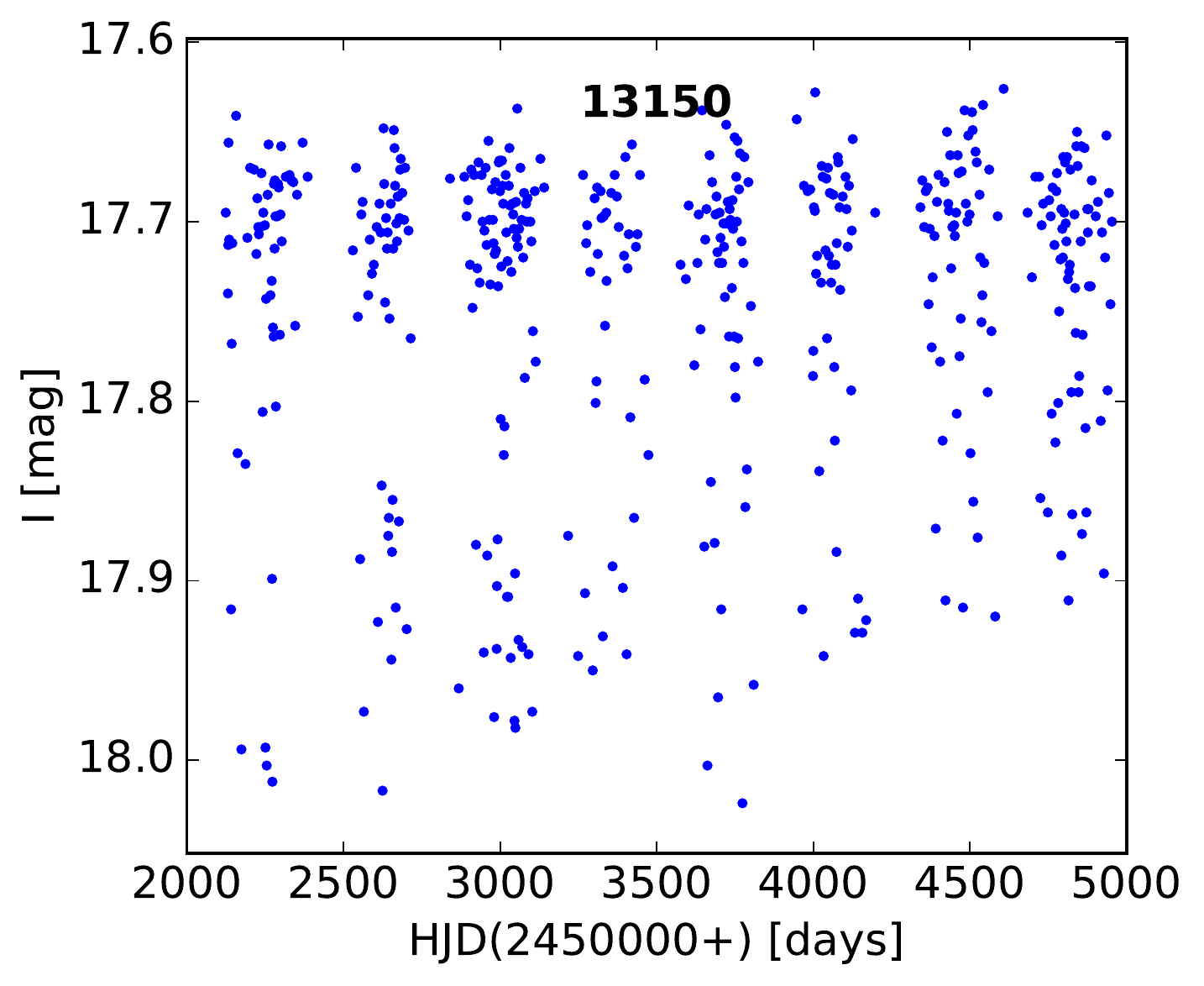} &
        \includegraphics[width=58mm]{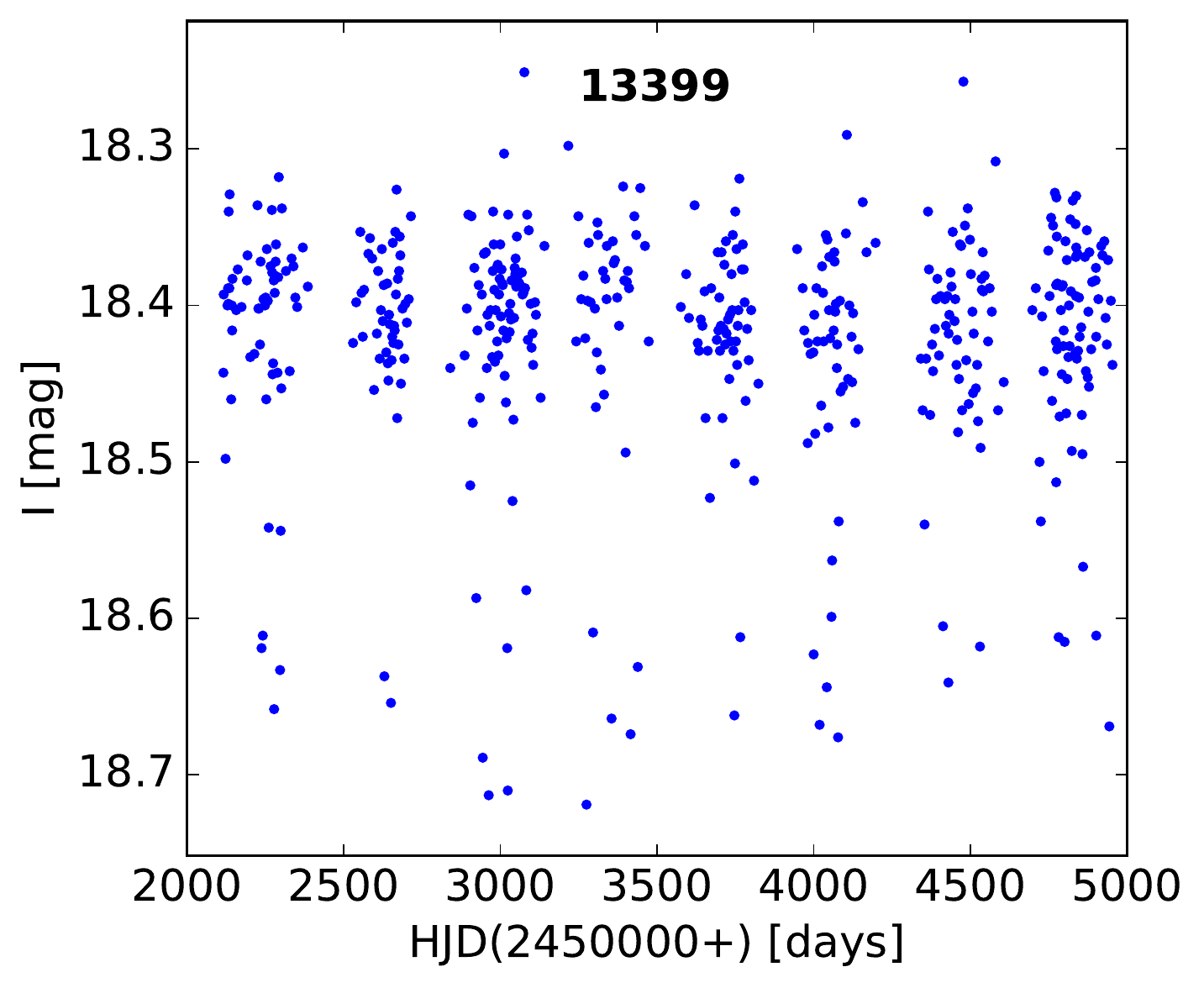} &
        \includegraphics[width=58mm]{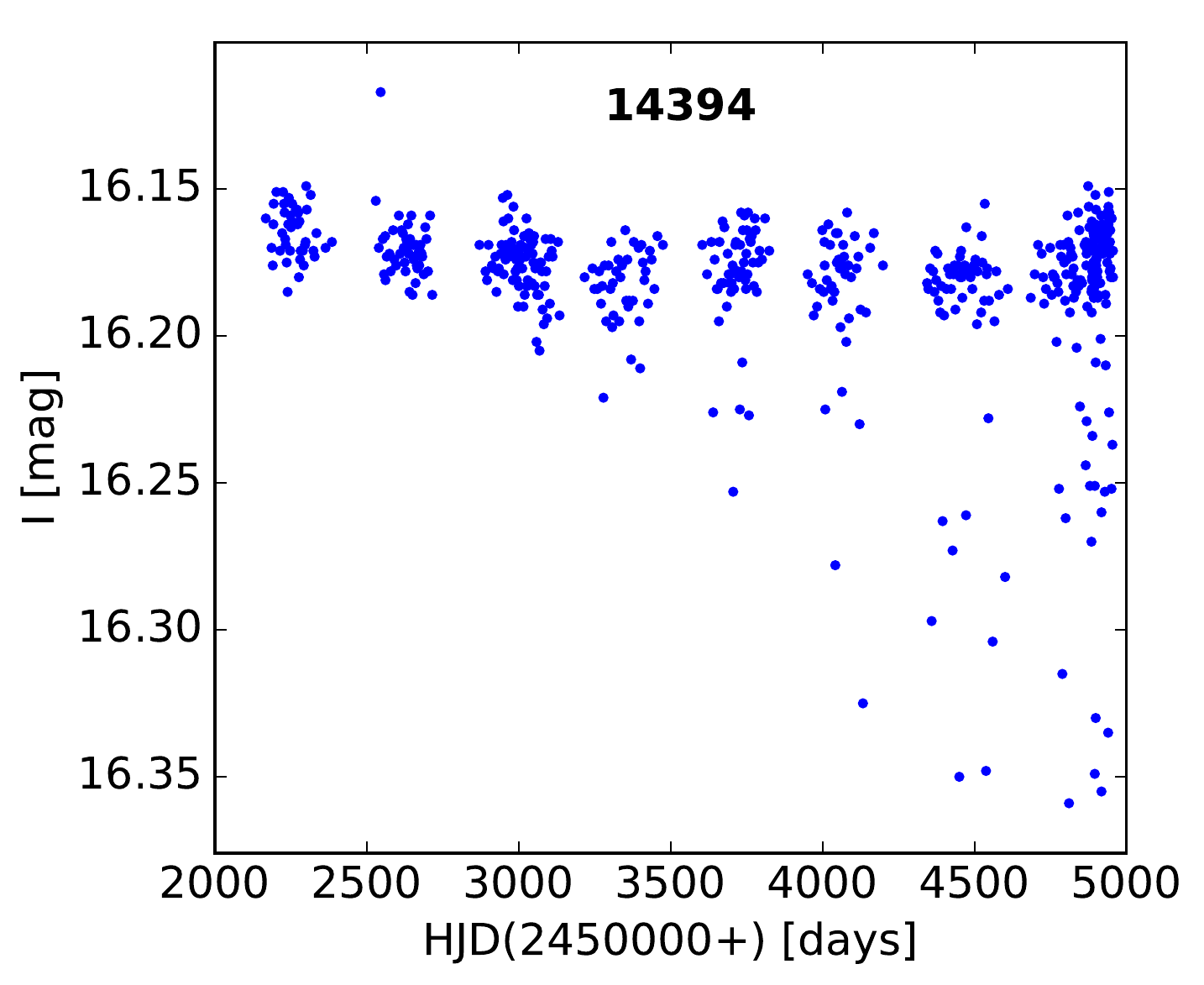} \\
        \includegraphics[width=58mm]{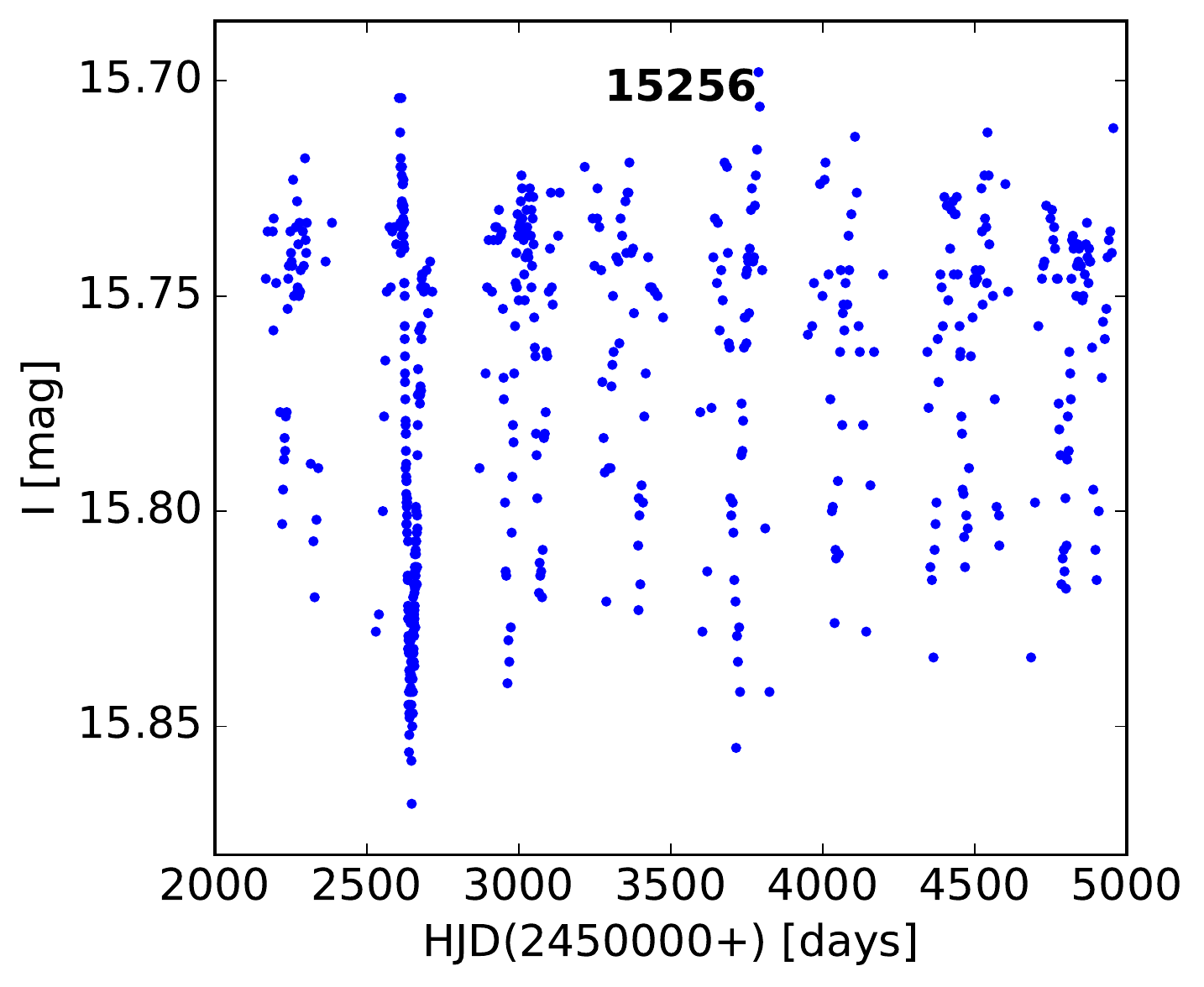} &
        \includegraphics[width=58mm]{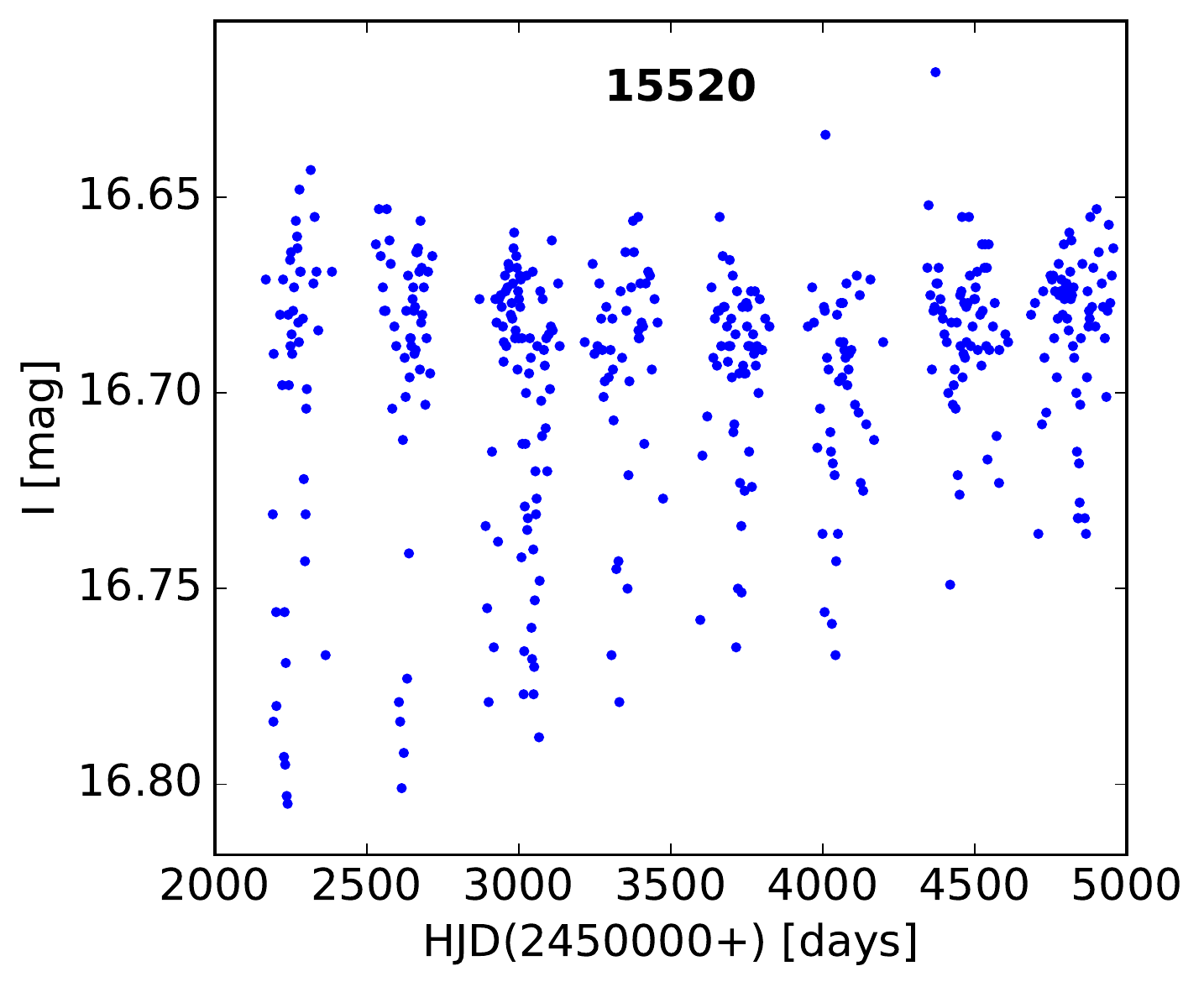} &
        \includegraphics[width=58mm]{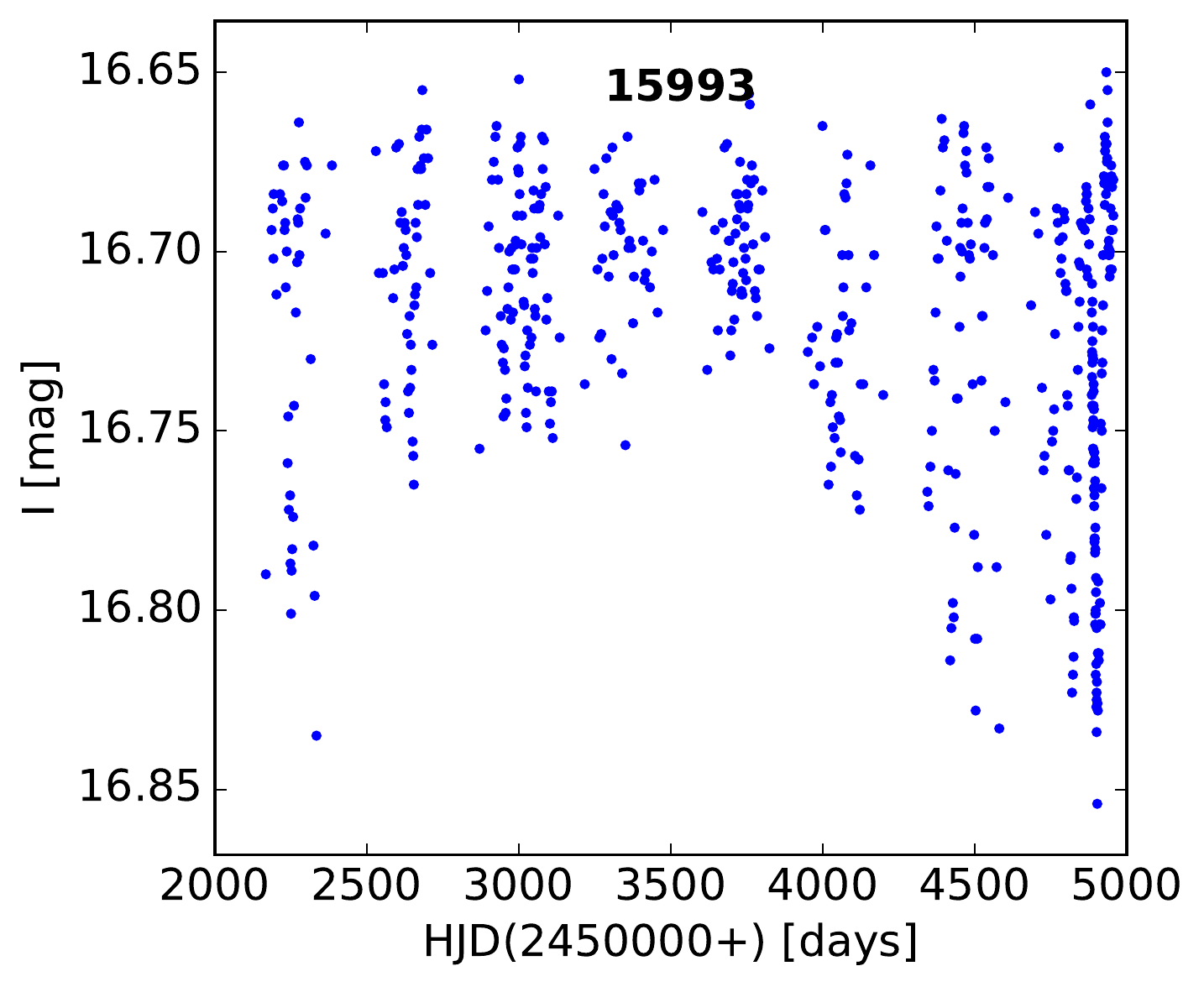} \\
        \end{tabular}
  \caption{\textit{continued}}
\end{figure*}

\begin{figure*}
\ContinuedFloat 
\centering
        \begin{tabular}{@{}ccc@{}}              
        \includegraphics[width=58mm]{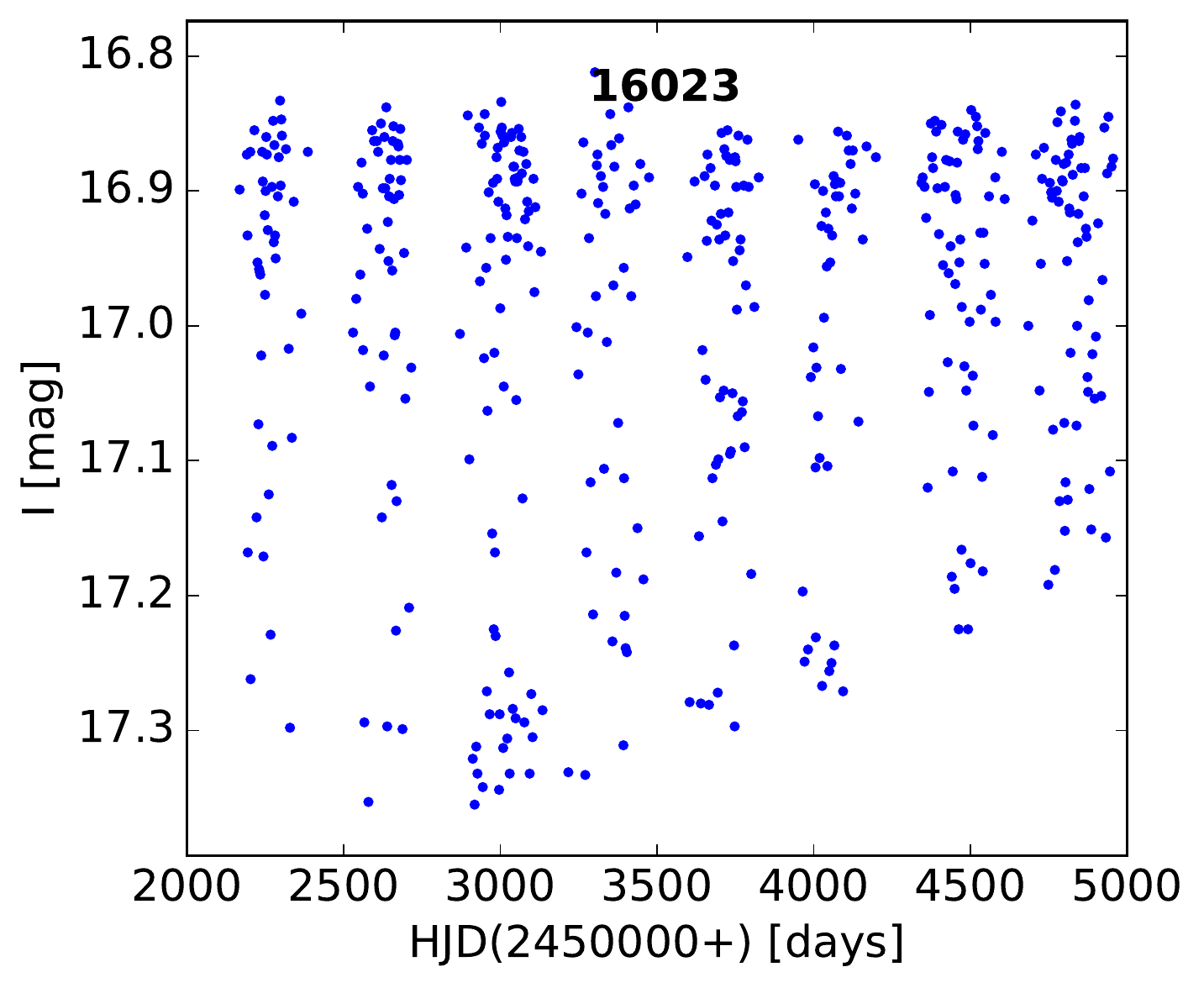} &
        \includegraphics[width=58mm]{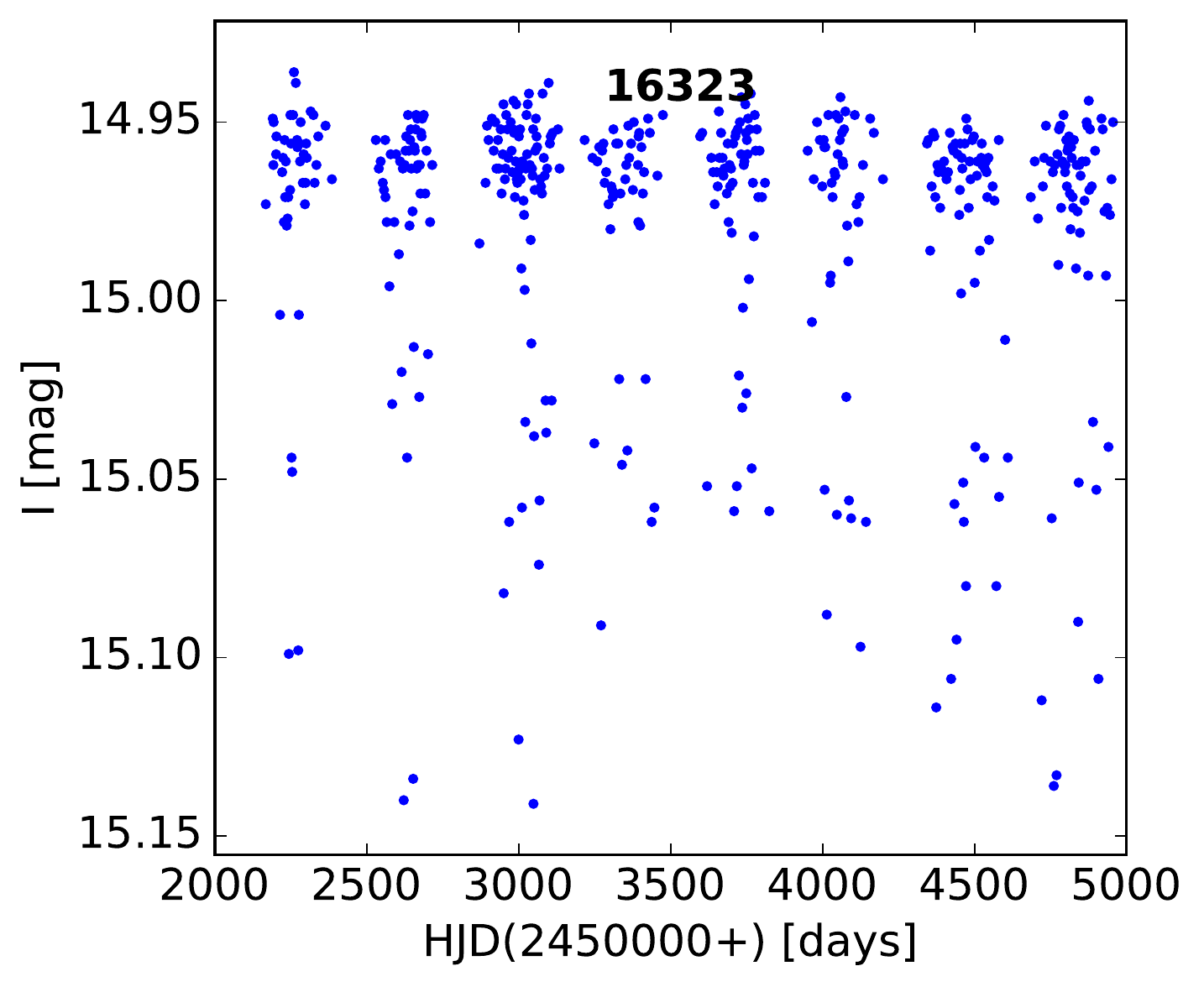} &
        \includegraphics[width=58mm]{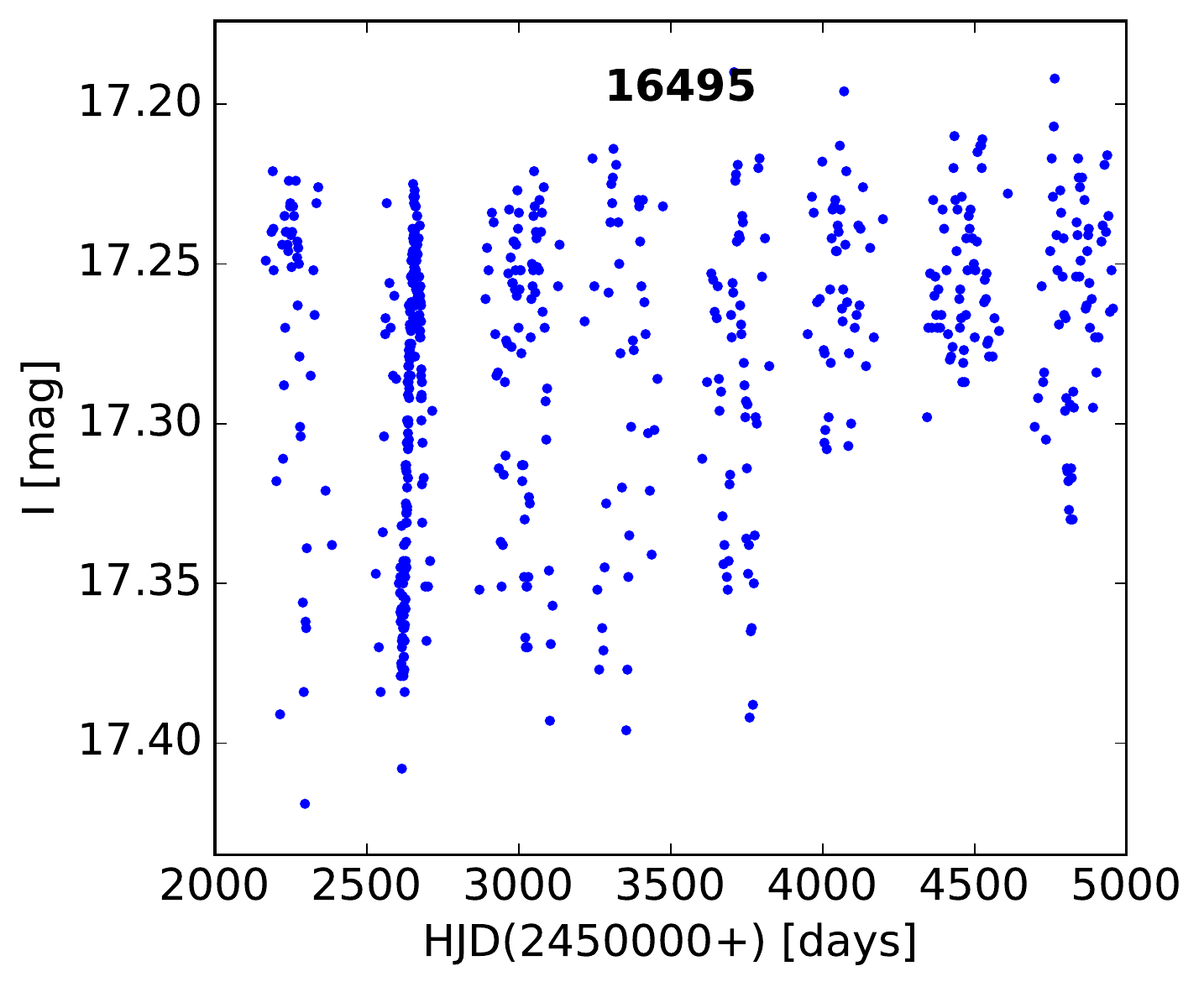} \\
        \includegraphics[width=58mm]{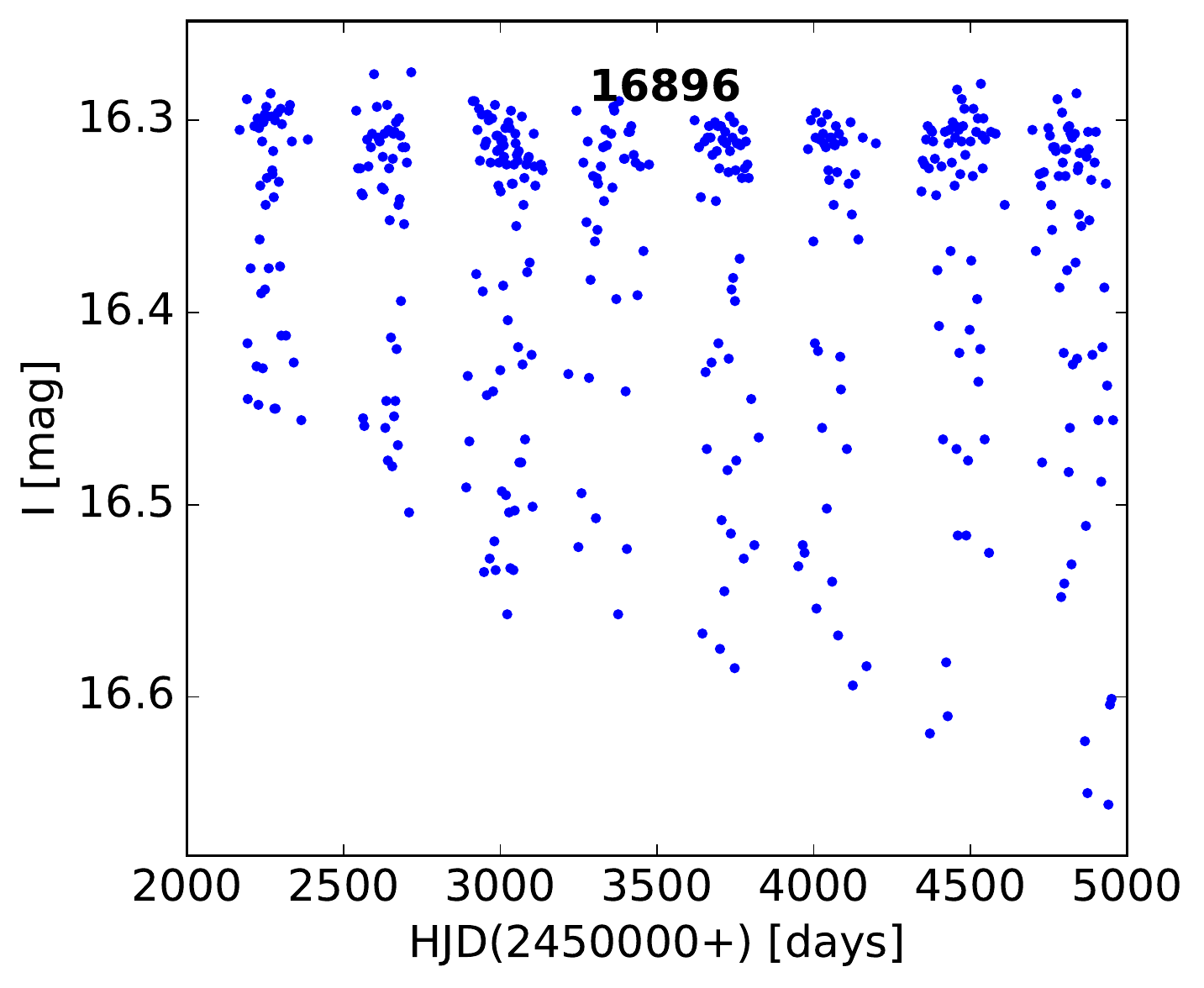} &
        \includegraphics[width=58mm]{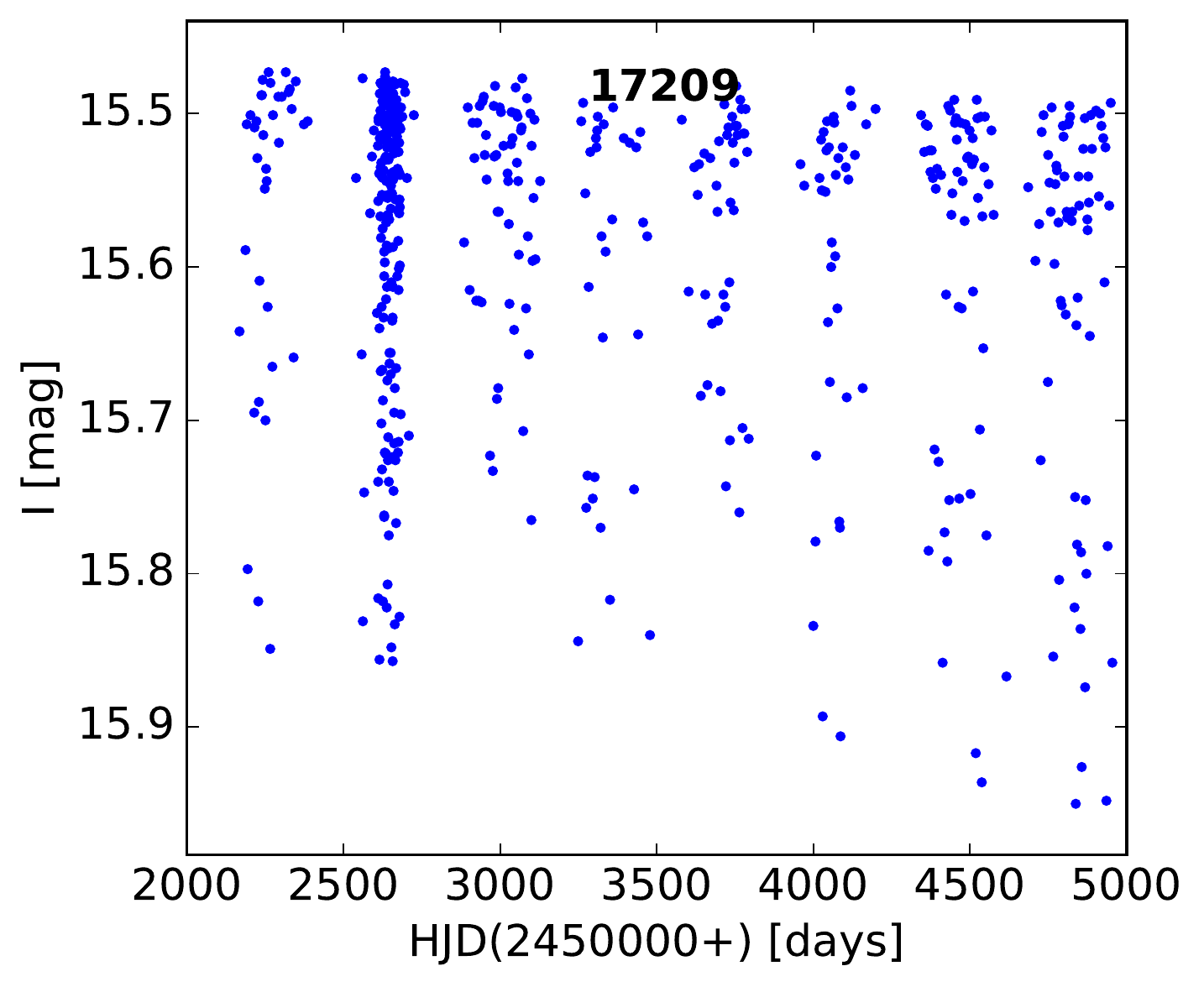} &
        \includegraphics[width=58mm]{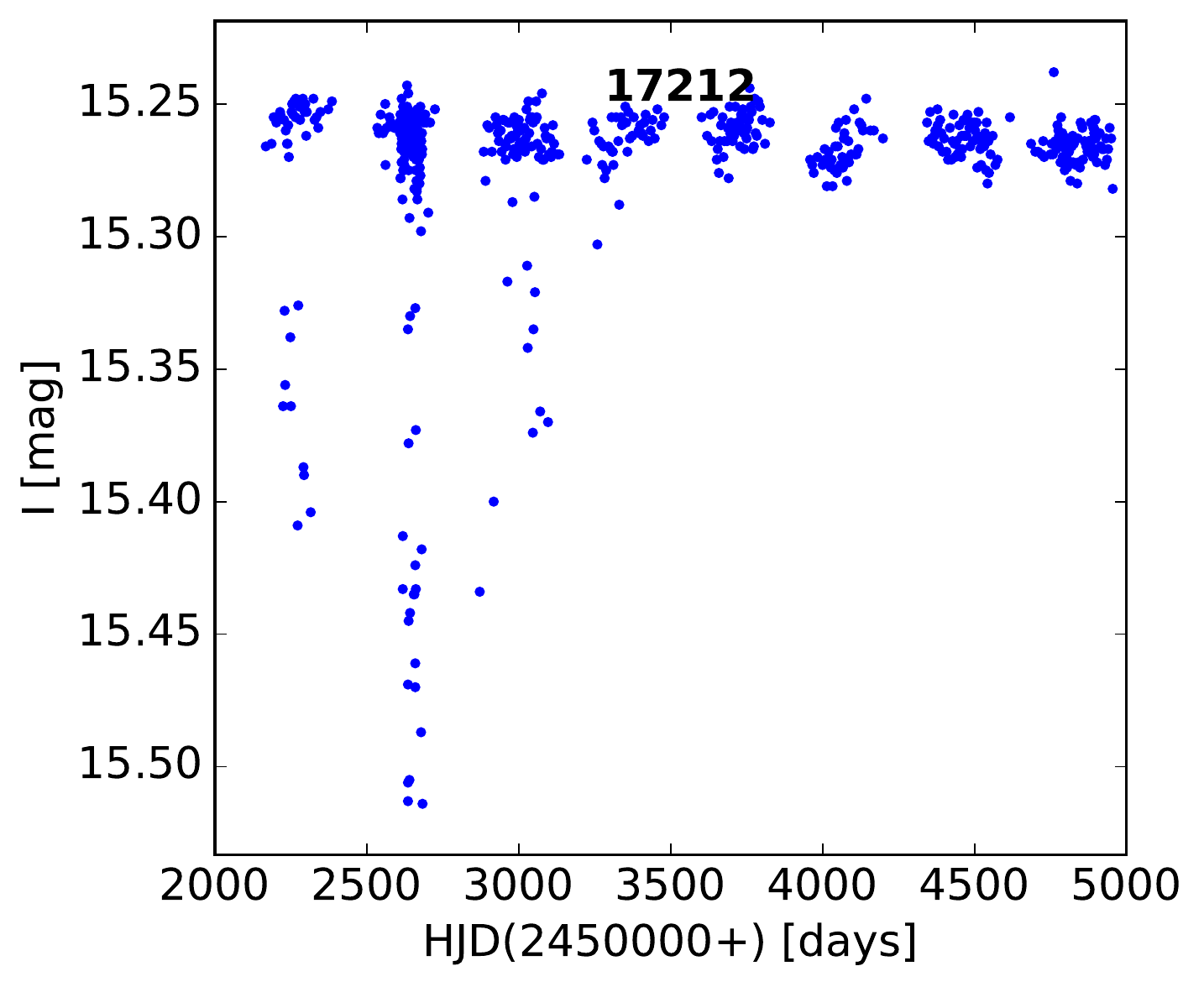} \\
        \includegraphics[width=58mm]{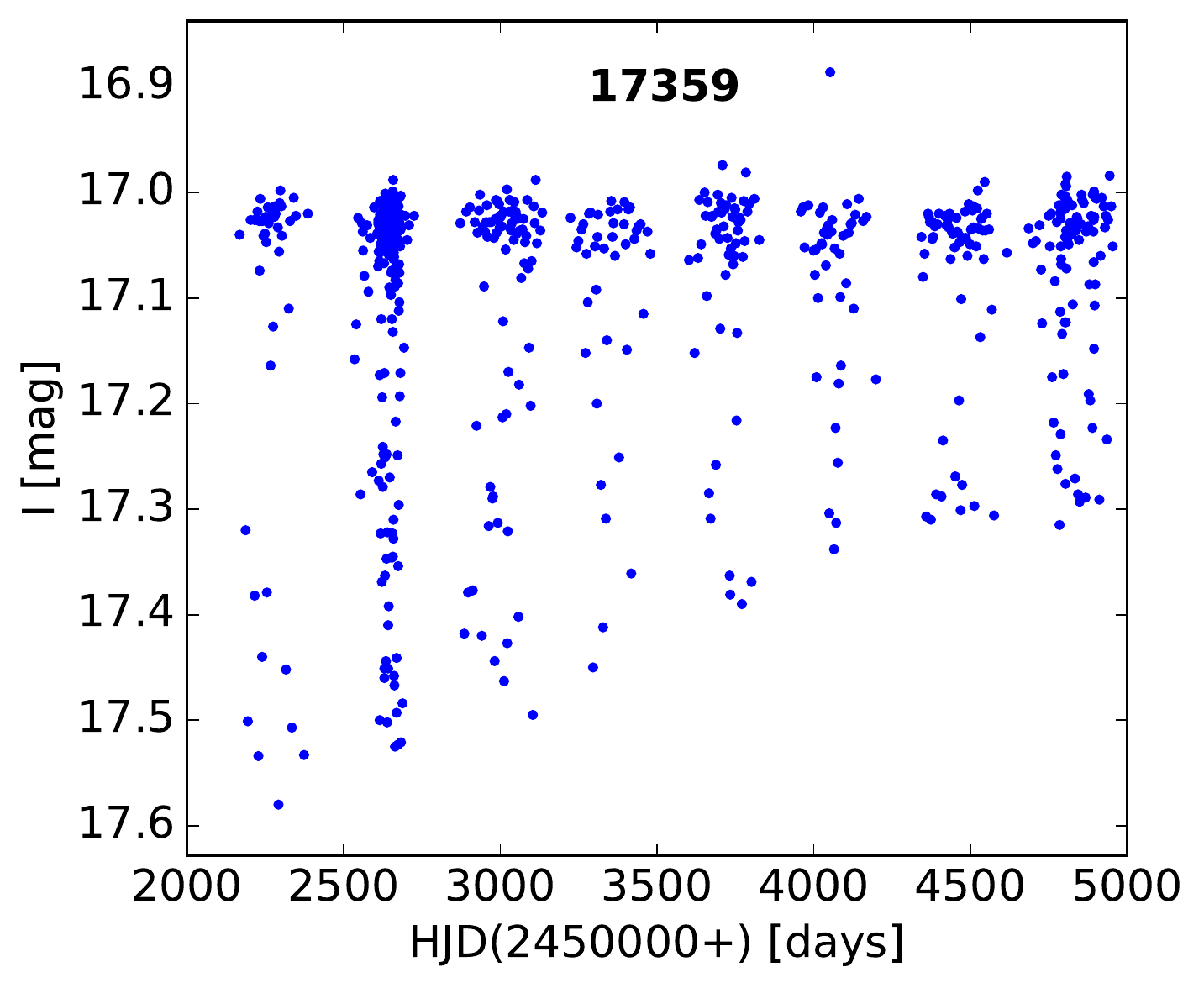} &
        \includegraphics[width=58mm]{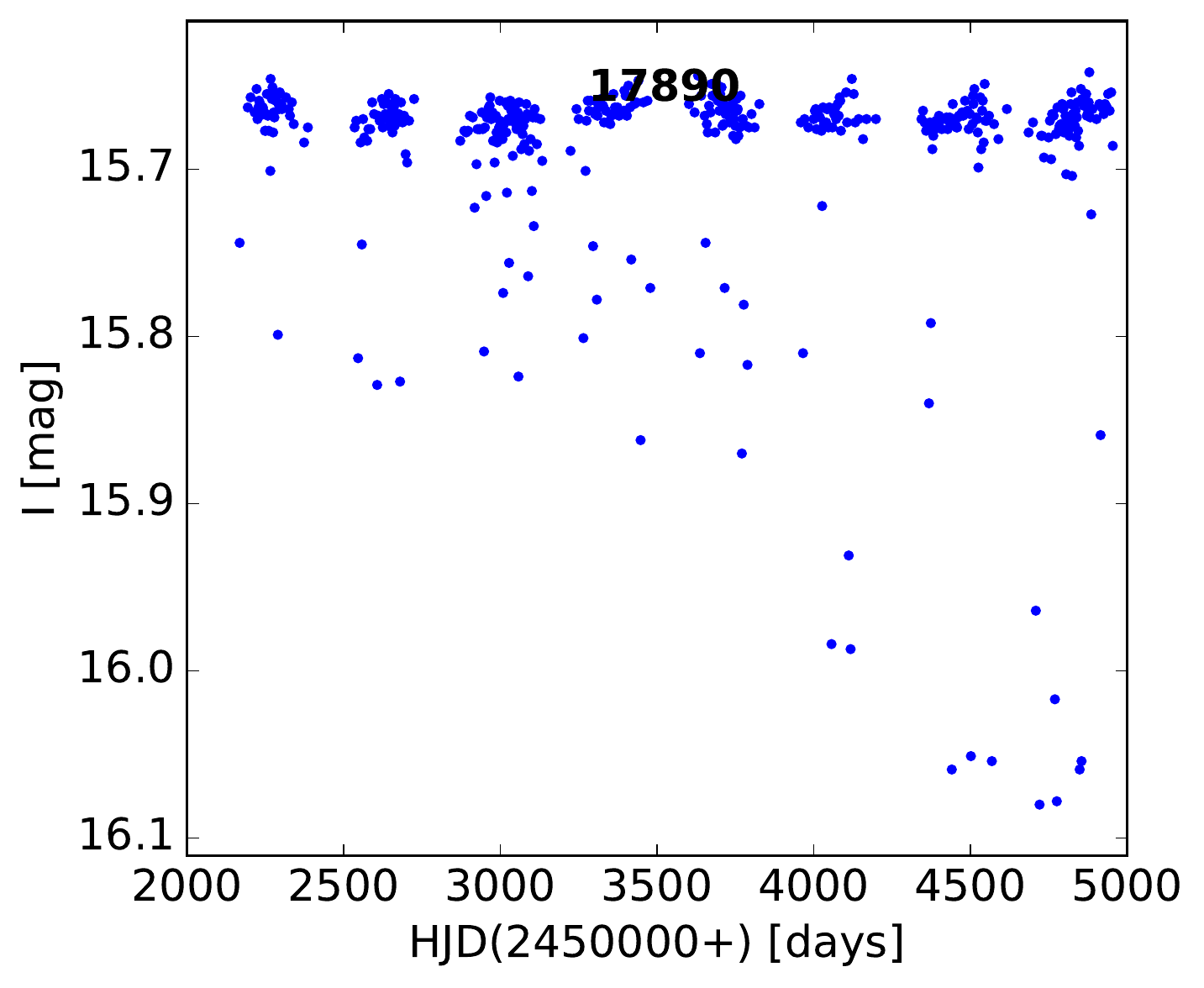} &
        \includegraphics[width=58mm]{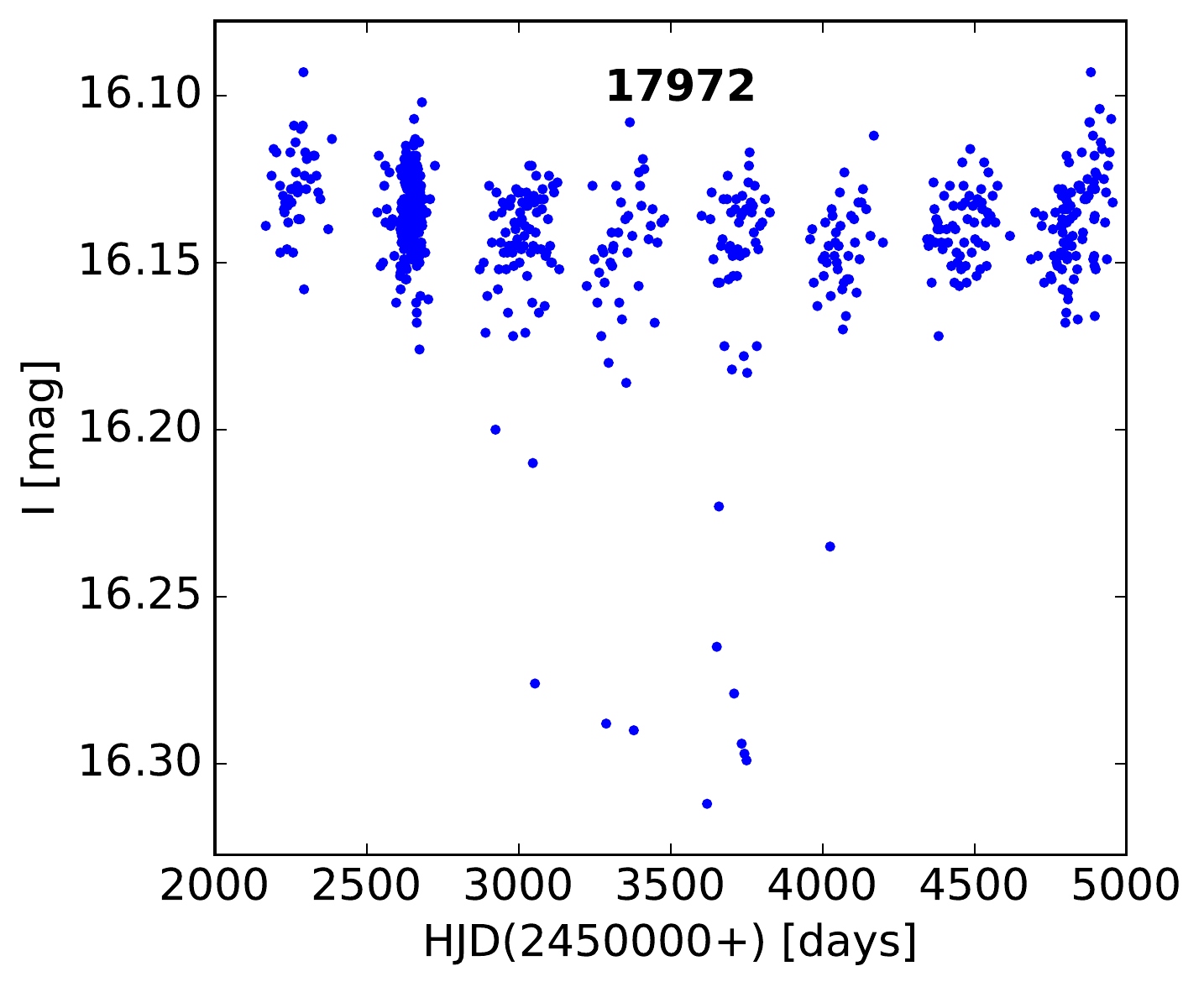} \\
        \includegraphics[width=58mm]{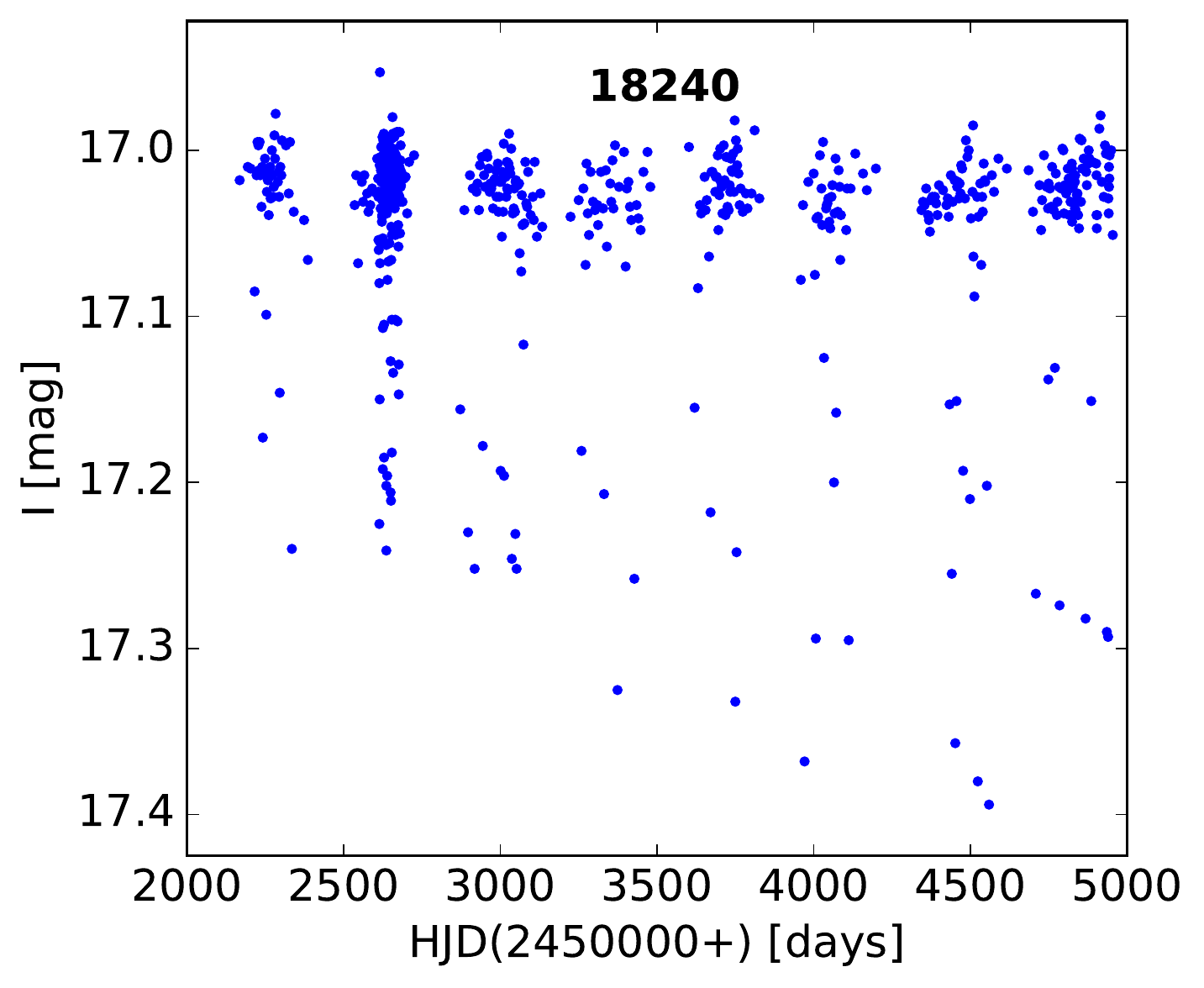} &
        \includegraphics[width=58mm]{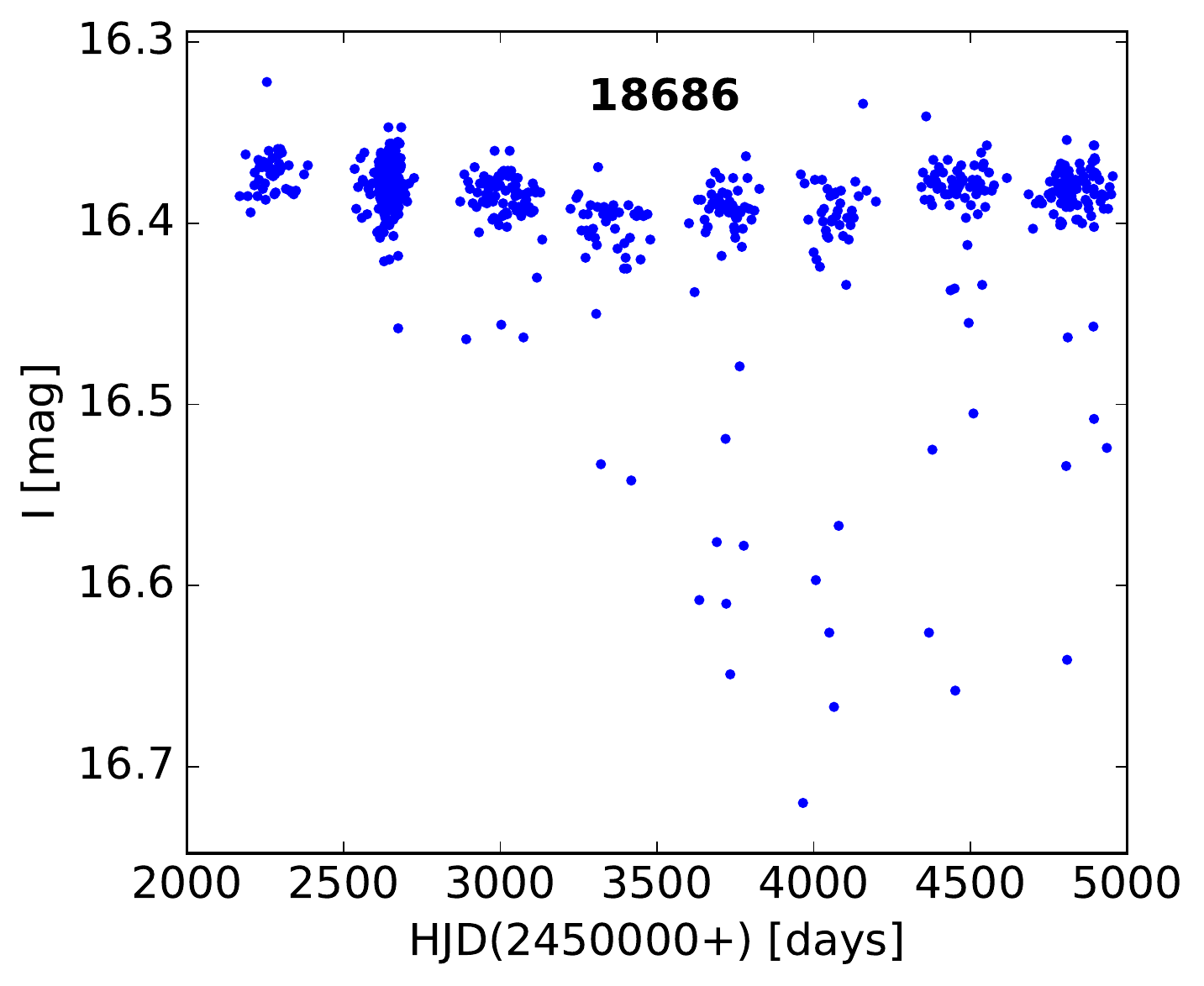} &
        \includegraphics[width=58mm]{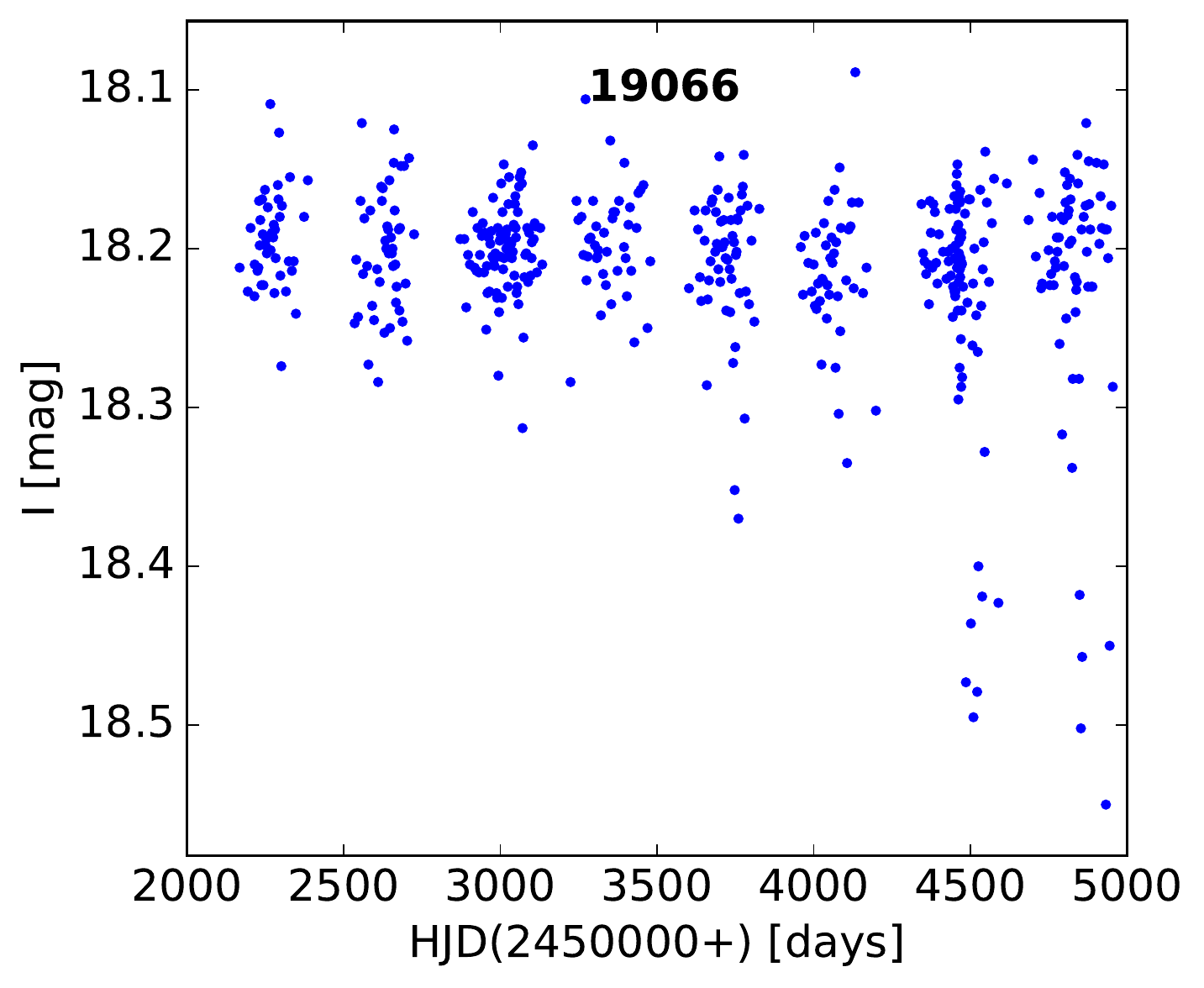} \\
        \includegraphics[width=58mm]{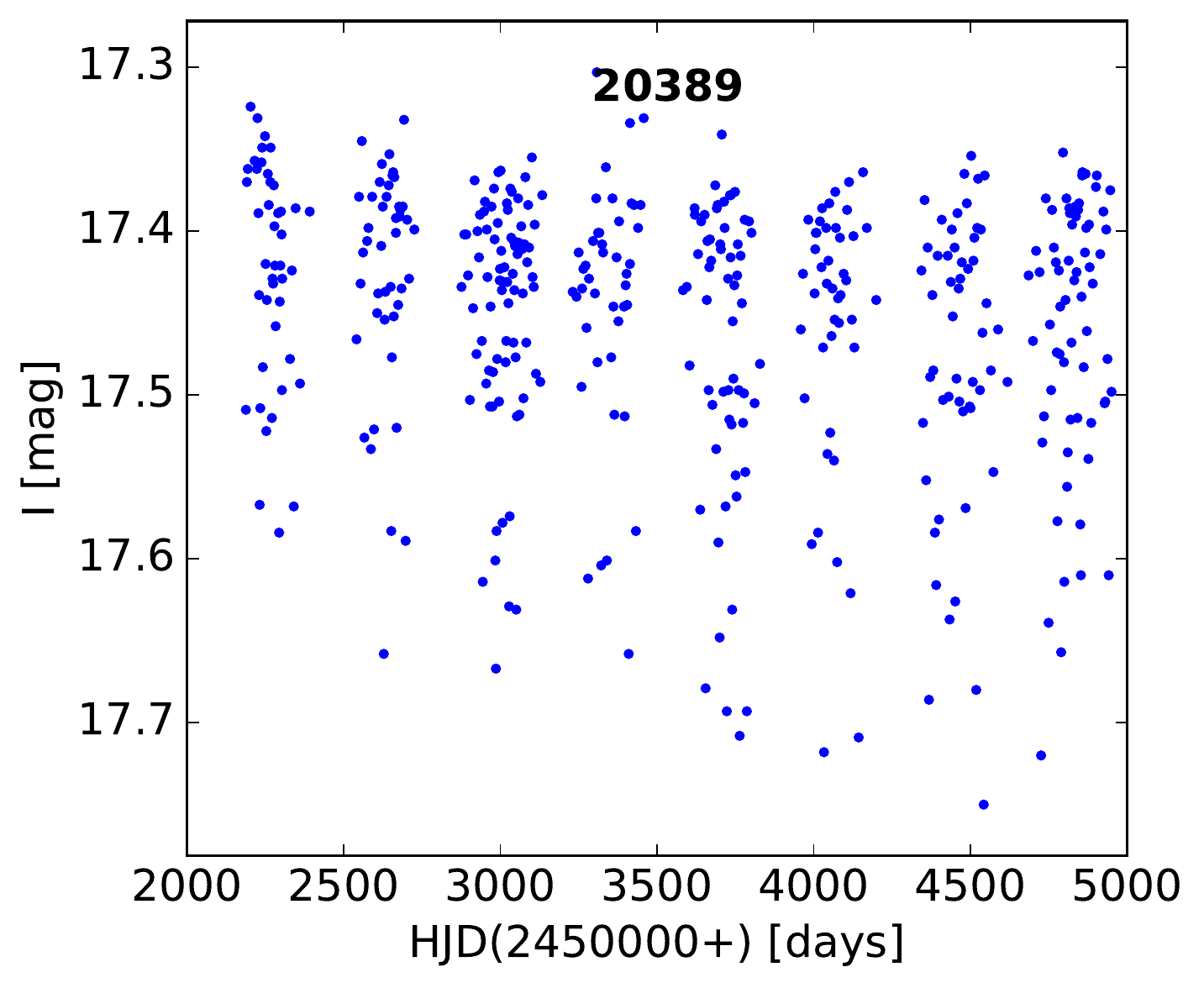} &
        \includegraphics[width=58mm]{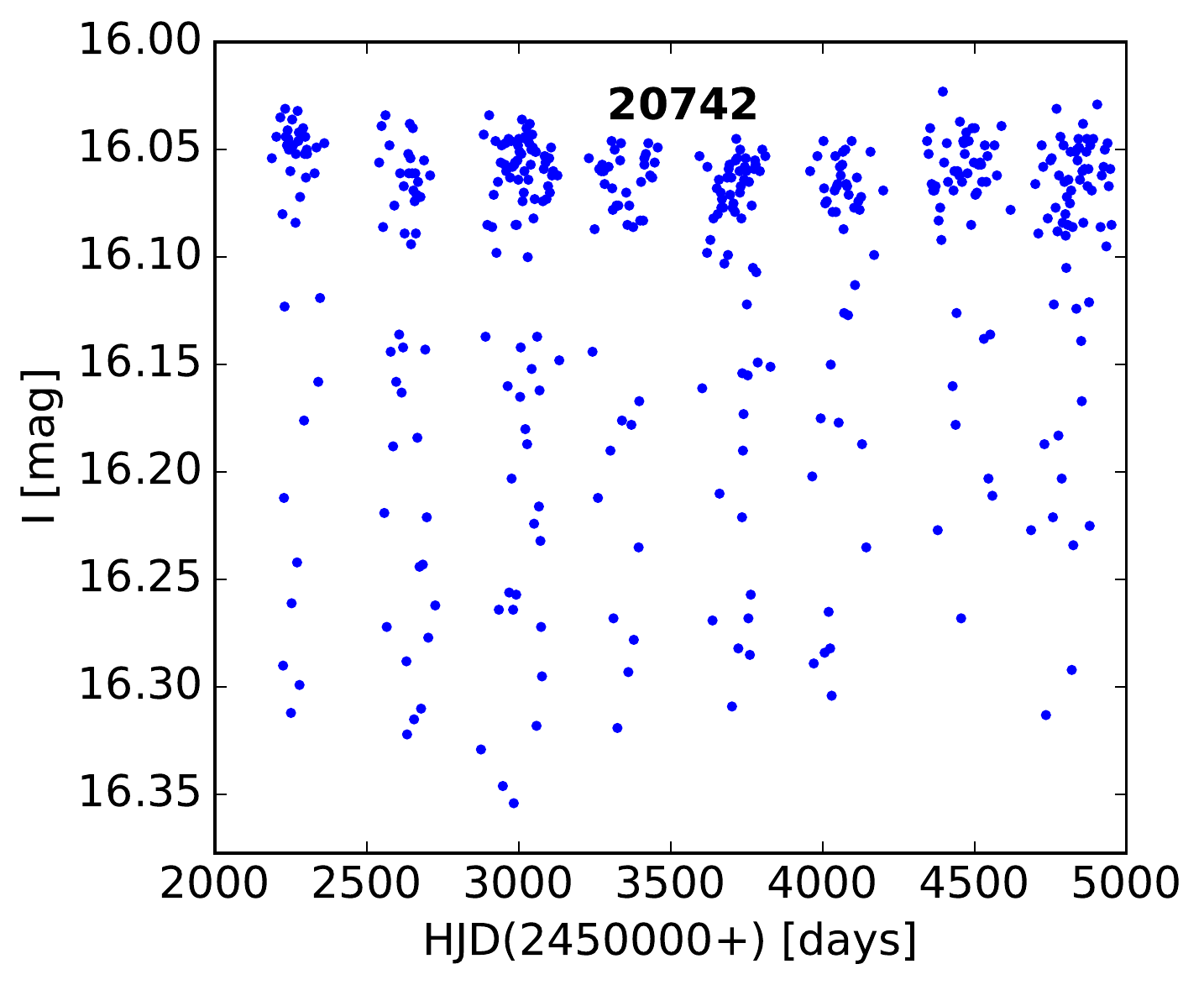} &
        \includegraphics[width=58mm]{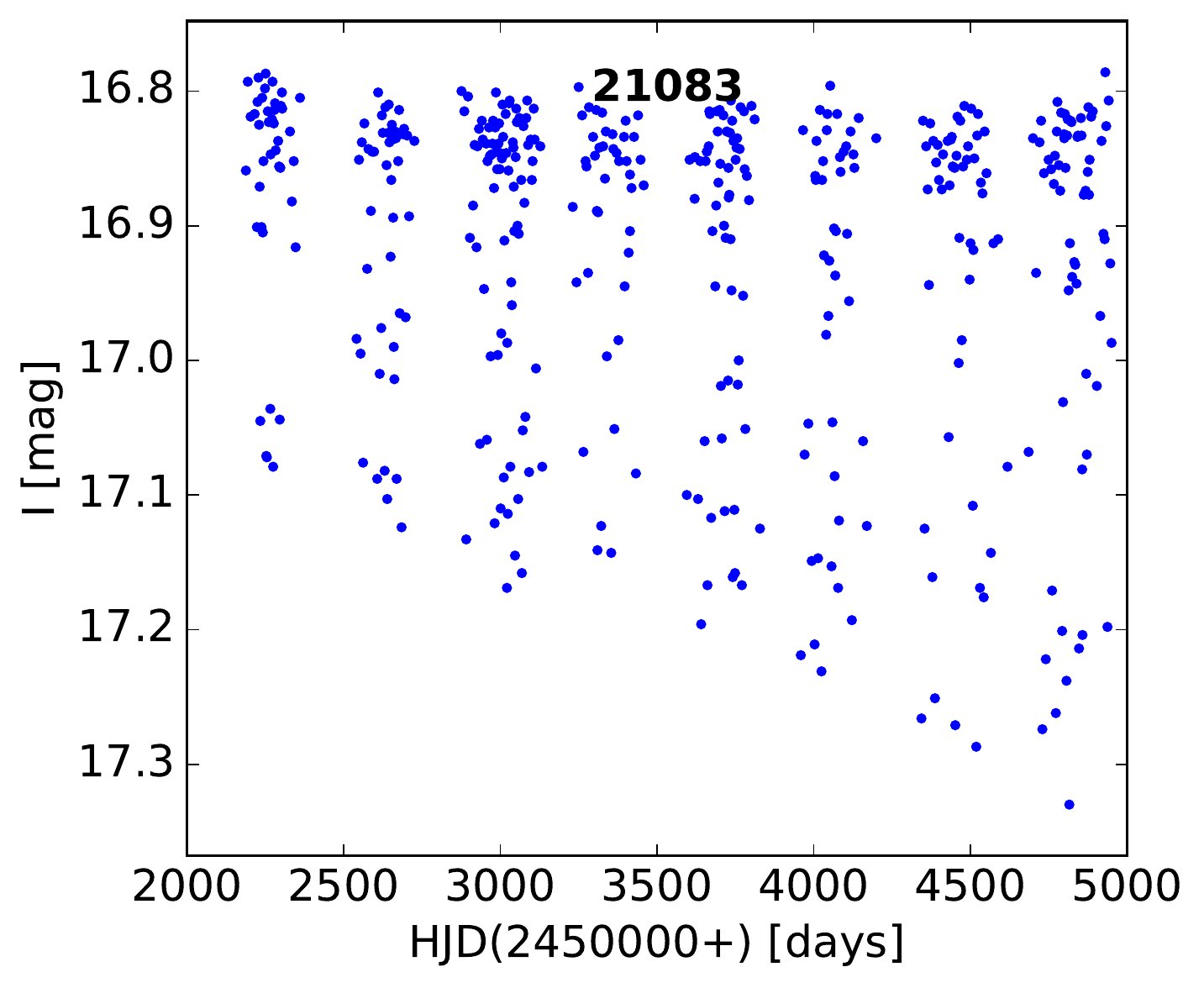} \\
        \end{tabular}
  \caption{\textit{continued}}
\end{figure*}

\begin{figure*}
\ContinuedFloat 
\centering
        \begin{tabular}{@{}ccc@{}}      
        \includegraphics[width=58mm]{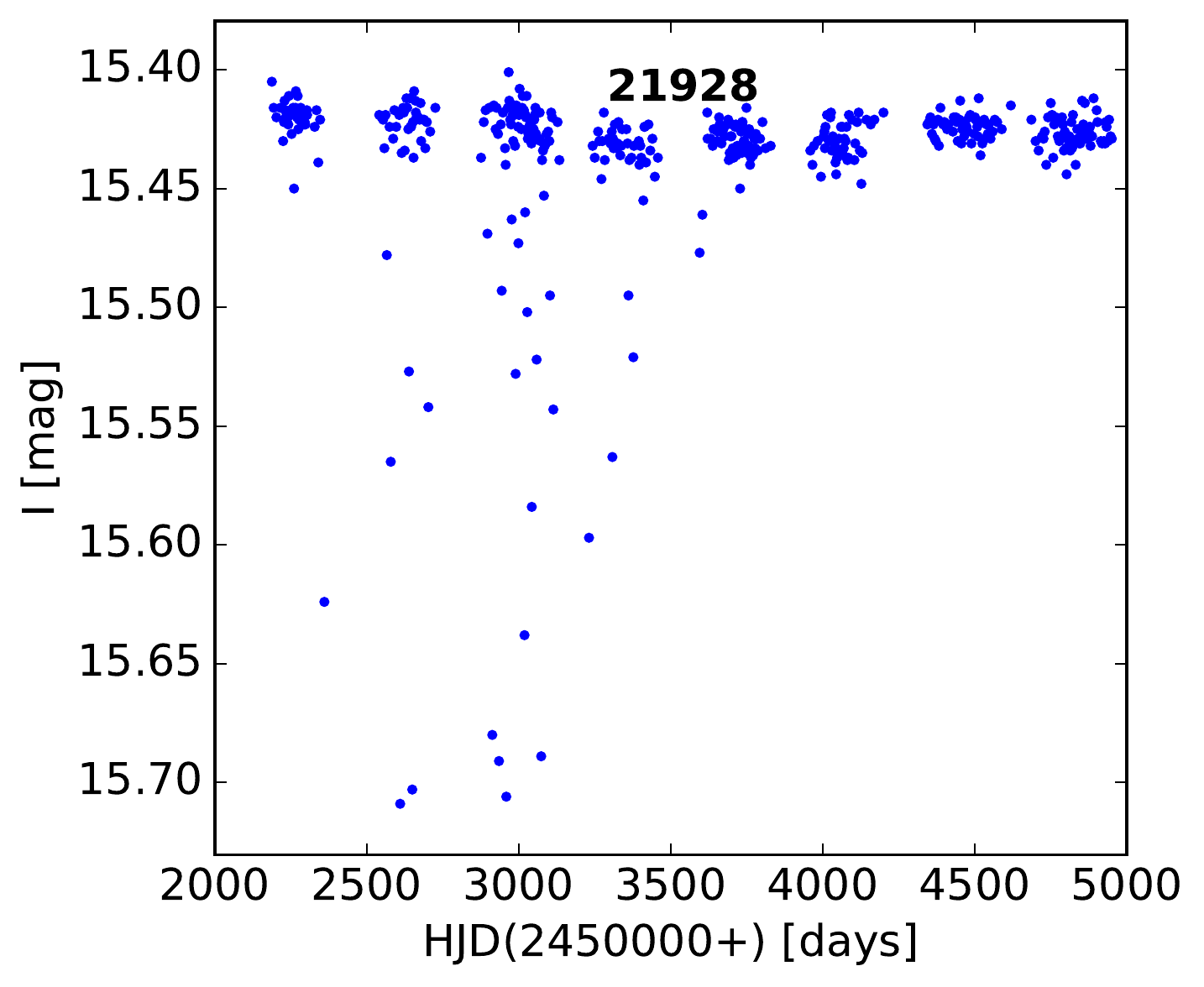} &    
        \includegraphics[width=58mm]{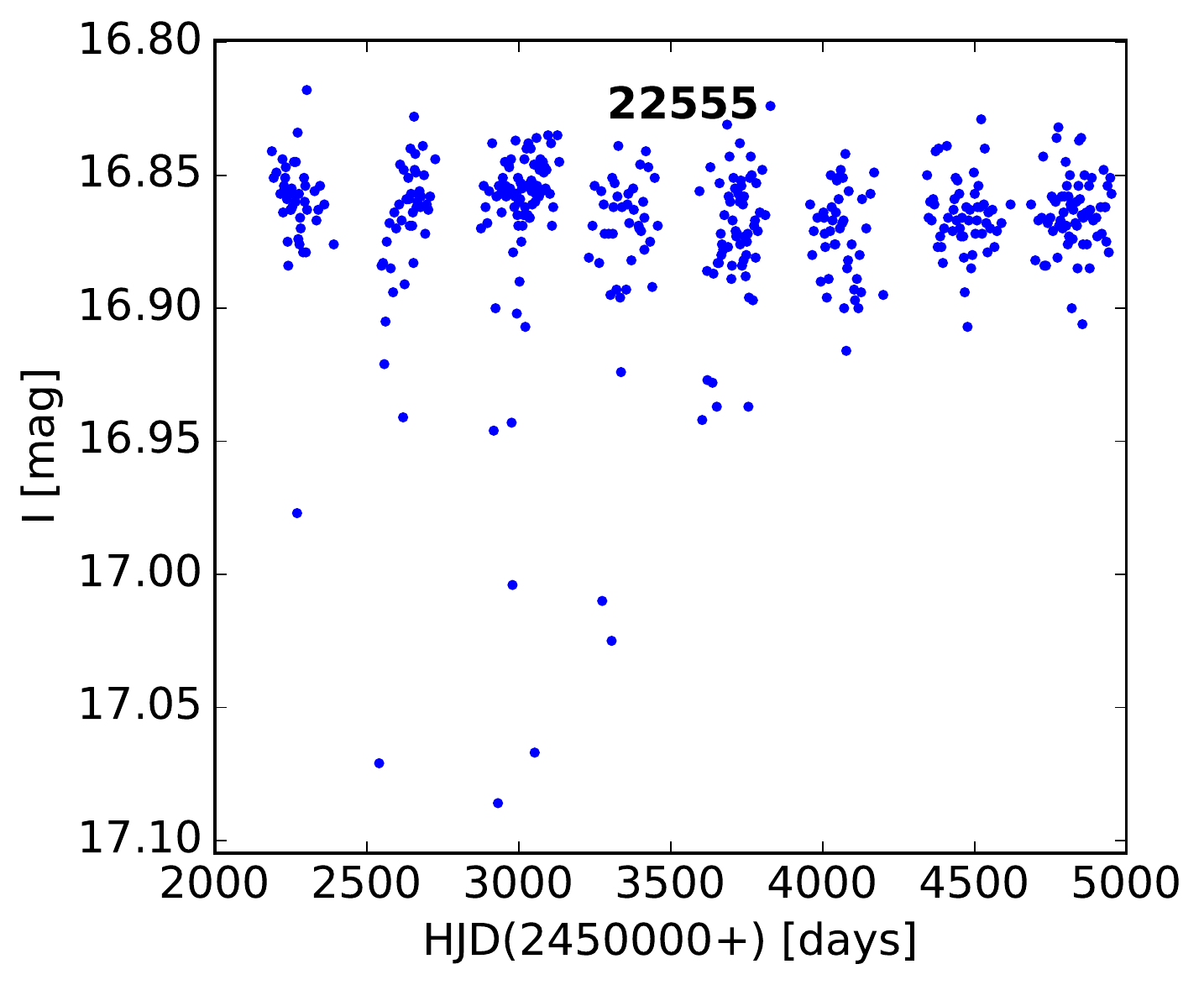} &
        \includegraphics[width=58mm]{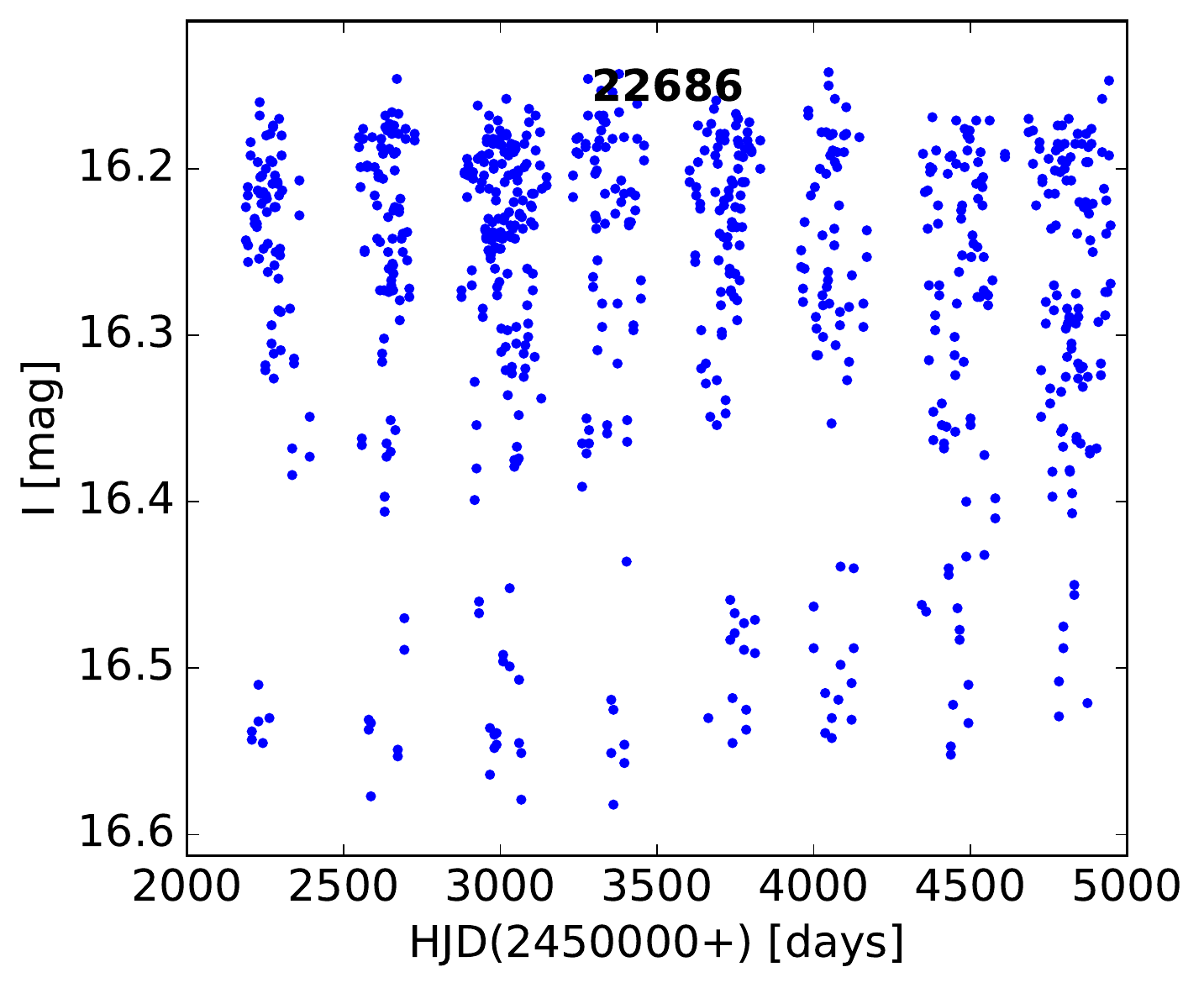} \\
        \includegraphics[width=58mm]{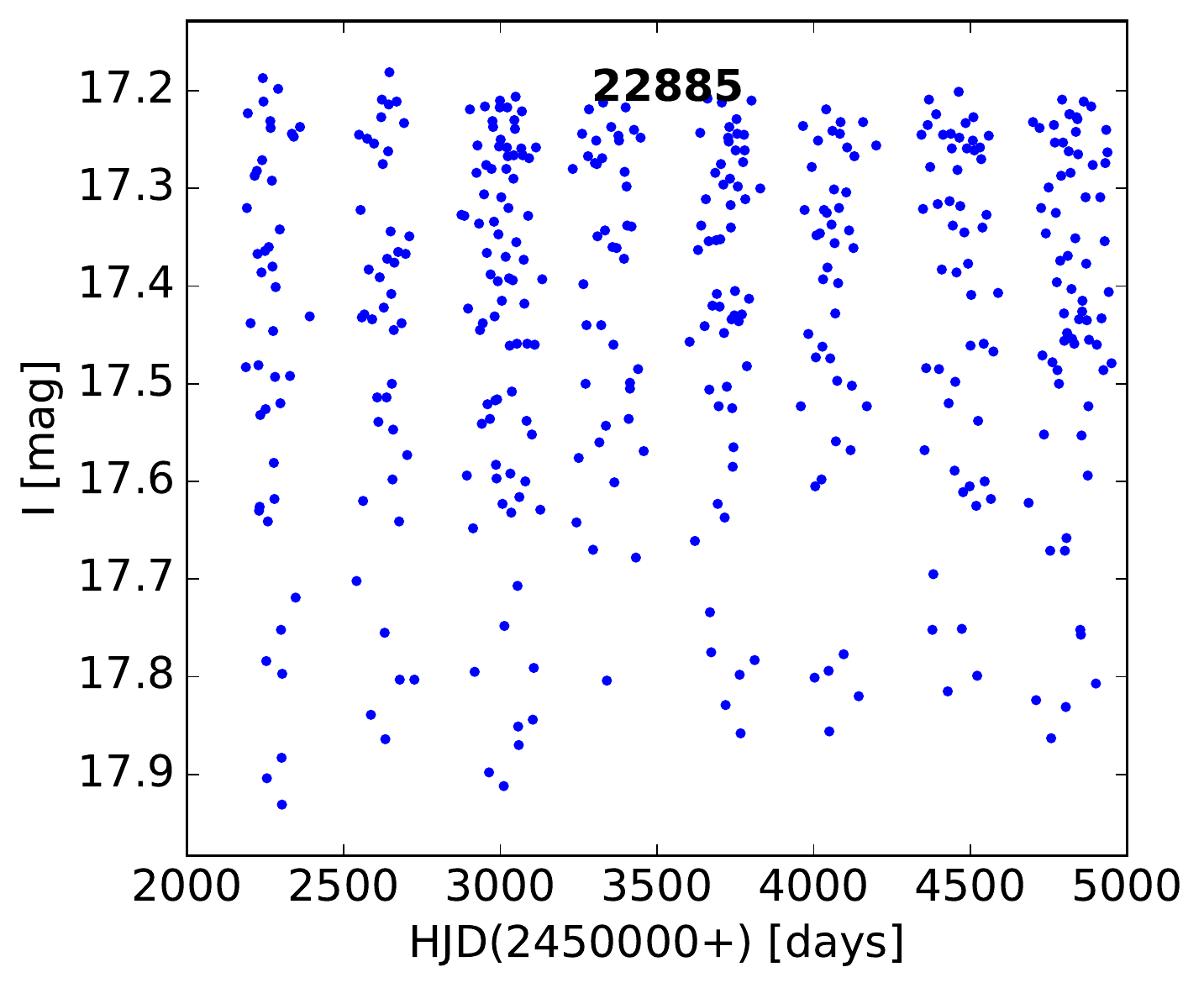} &
        \includegraphics[width=58mm]{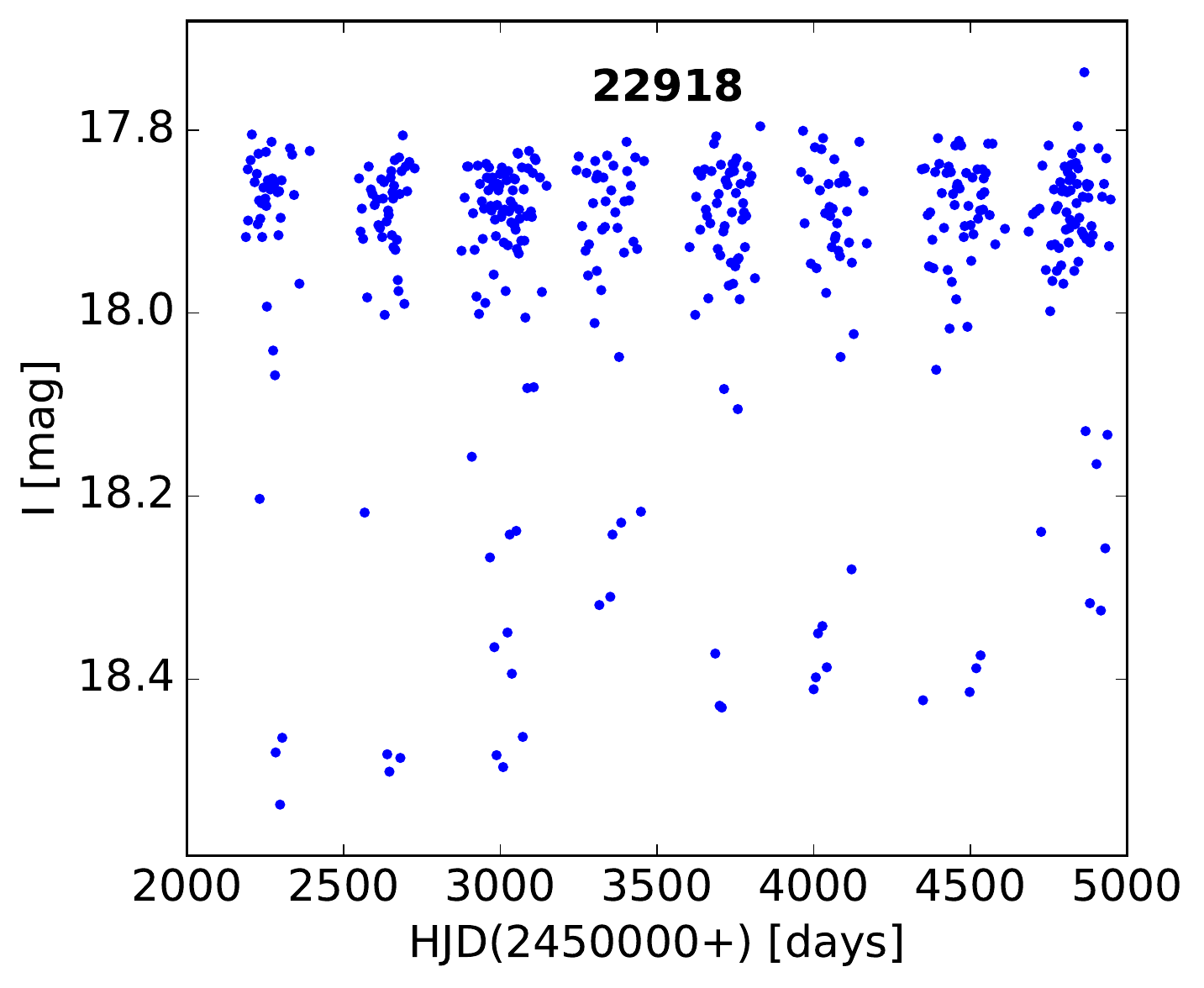} &
        \includegraphics[width=58mm]{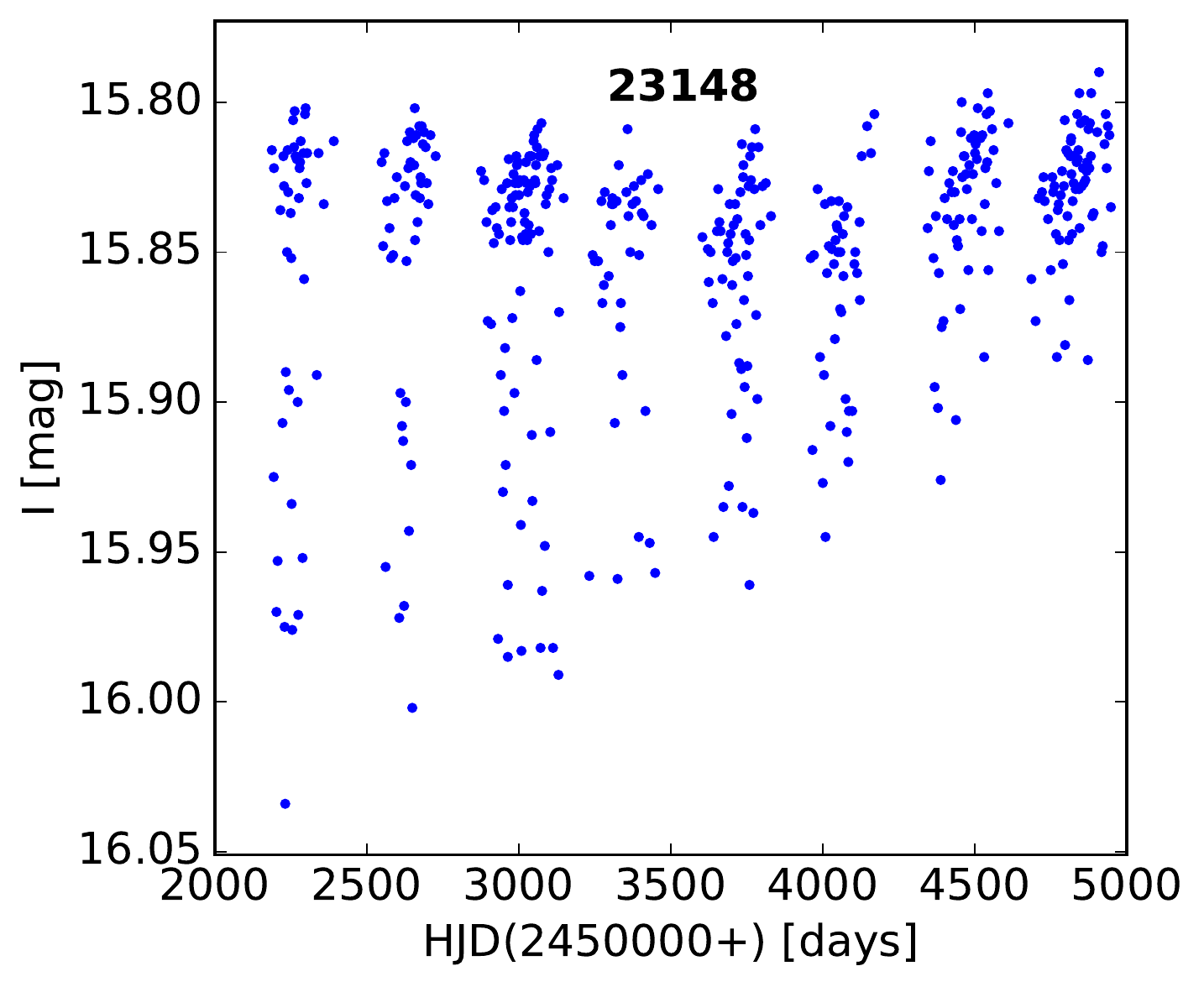} \\
        \includegraphics[width=58mm]{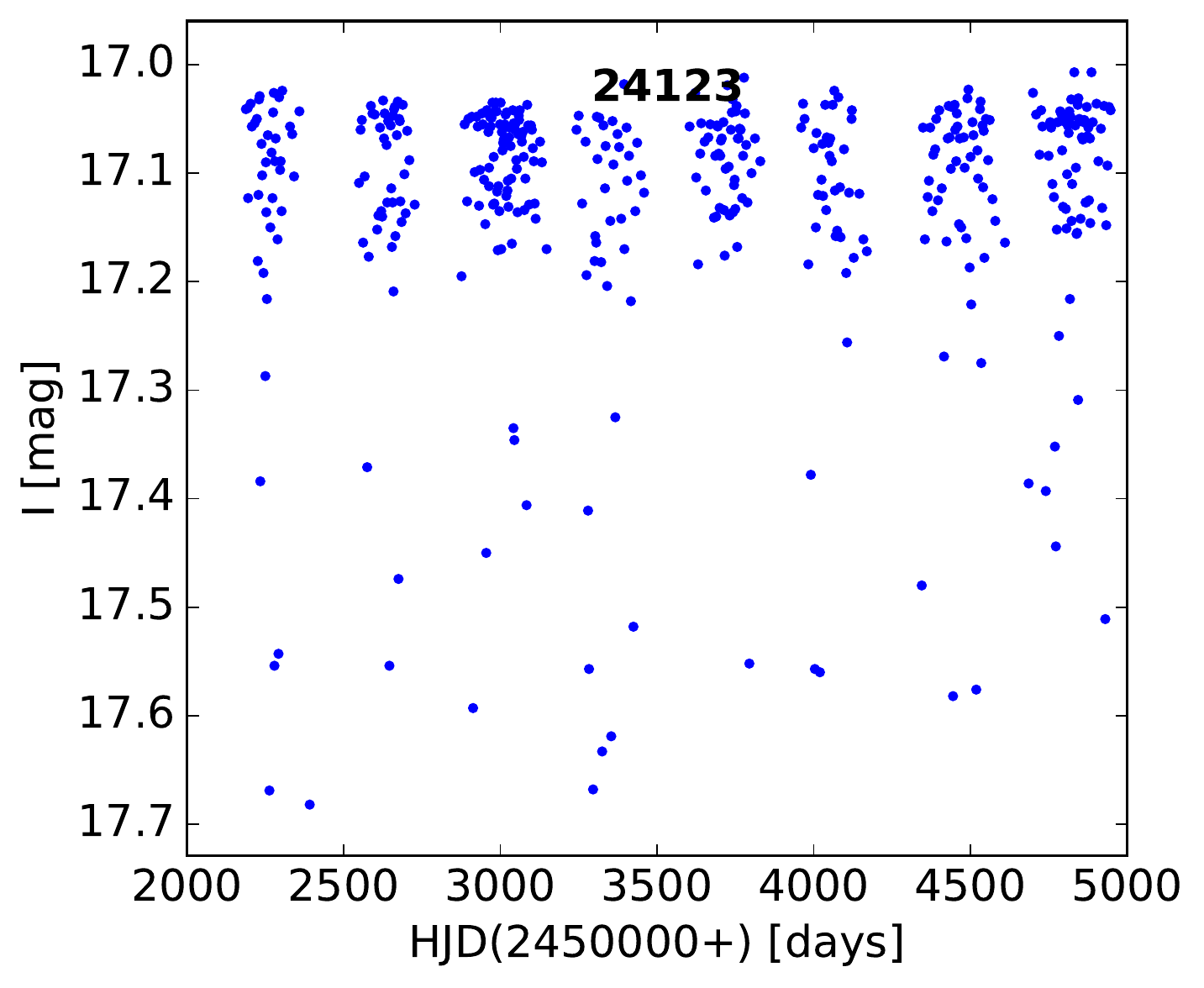} &
        \includegraphics[width=58mm]{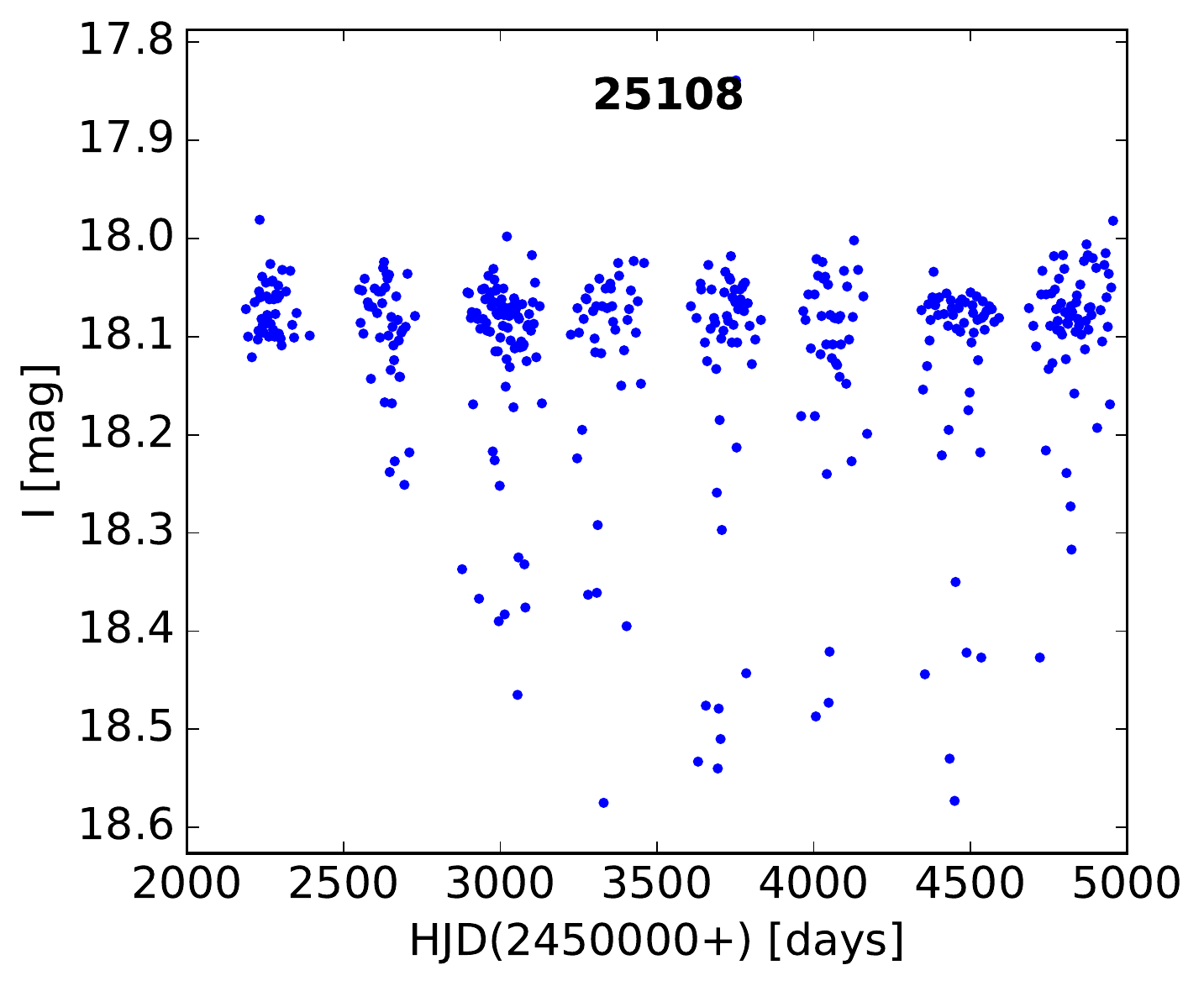} &
        \includegraphics[width=58mm]{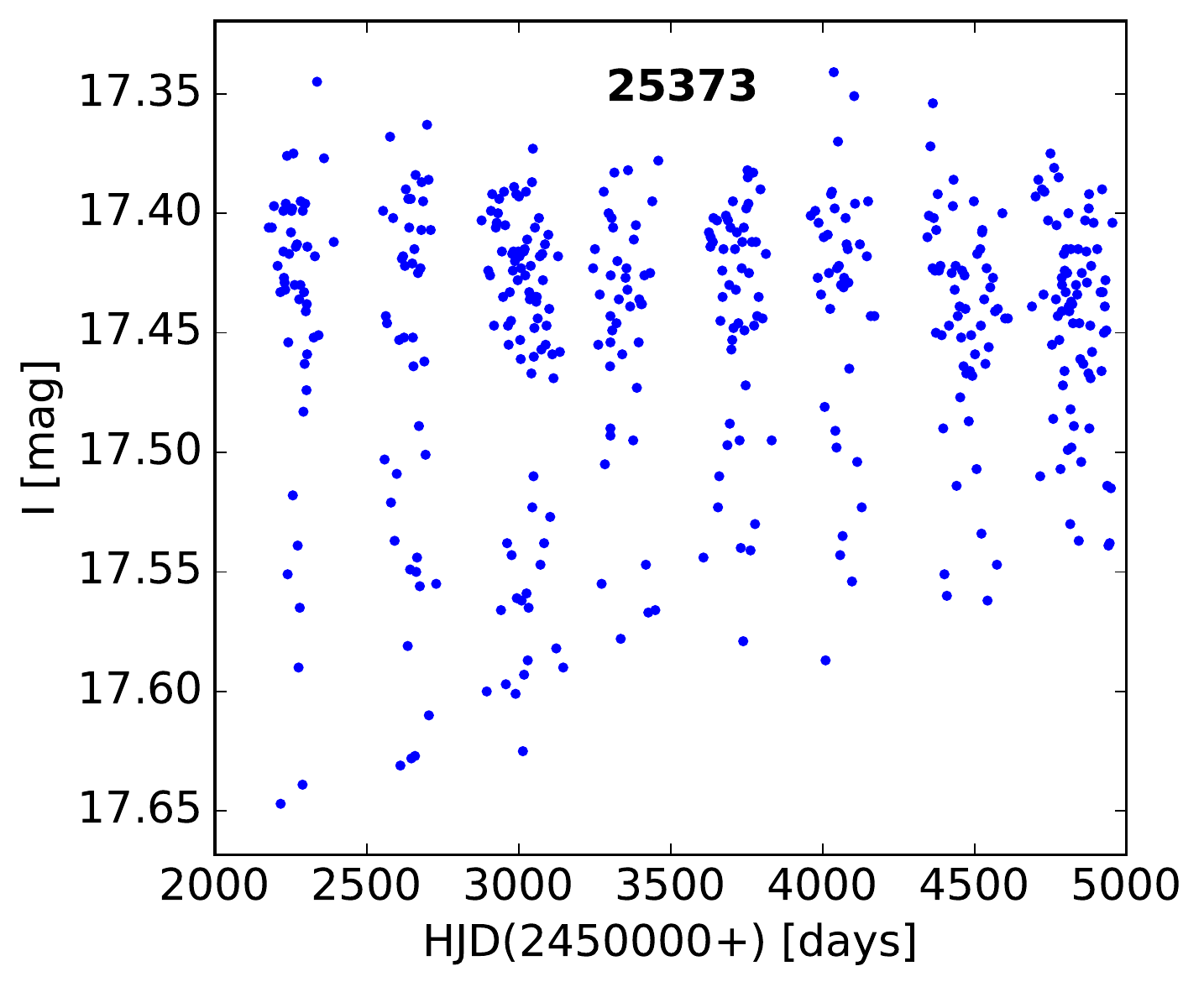} \\
        \end{tabular}
  \caption{\textit{continued}}
\end{figure*}

\begin{figure*}
\centering
        \begin{tabular}{@{}ccc@{}}
        \includegraphics[width=58mm]{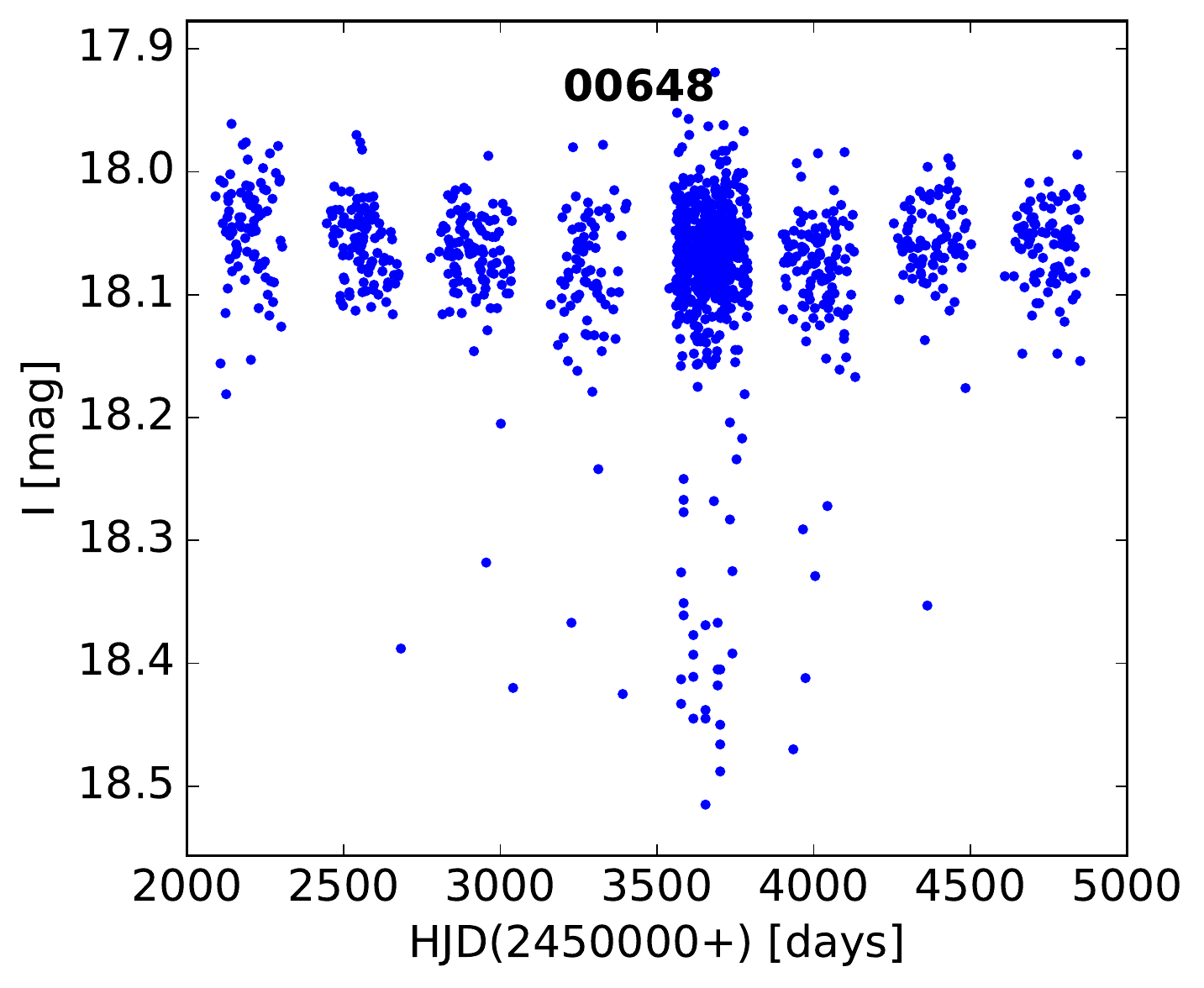} &
        \includegraphics[width=58mm]{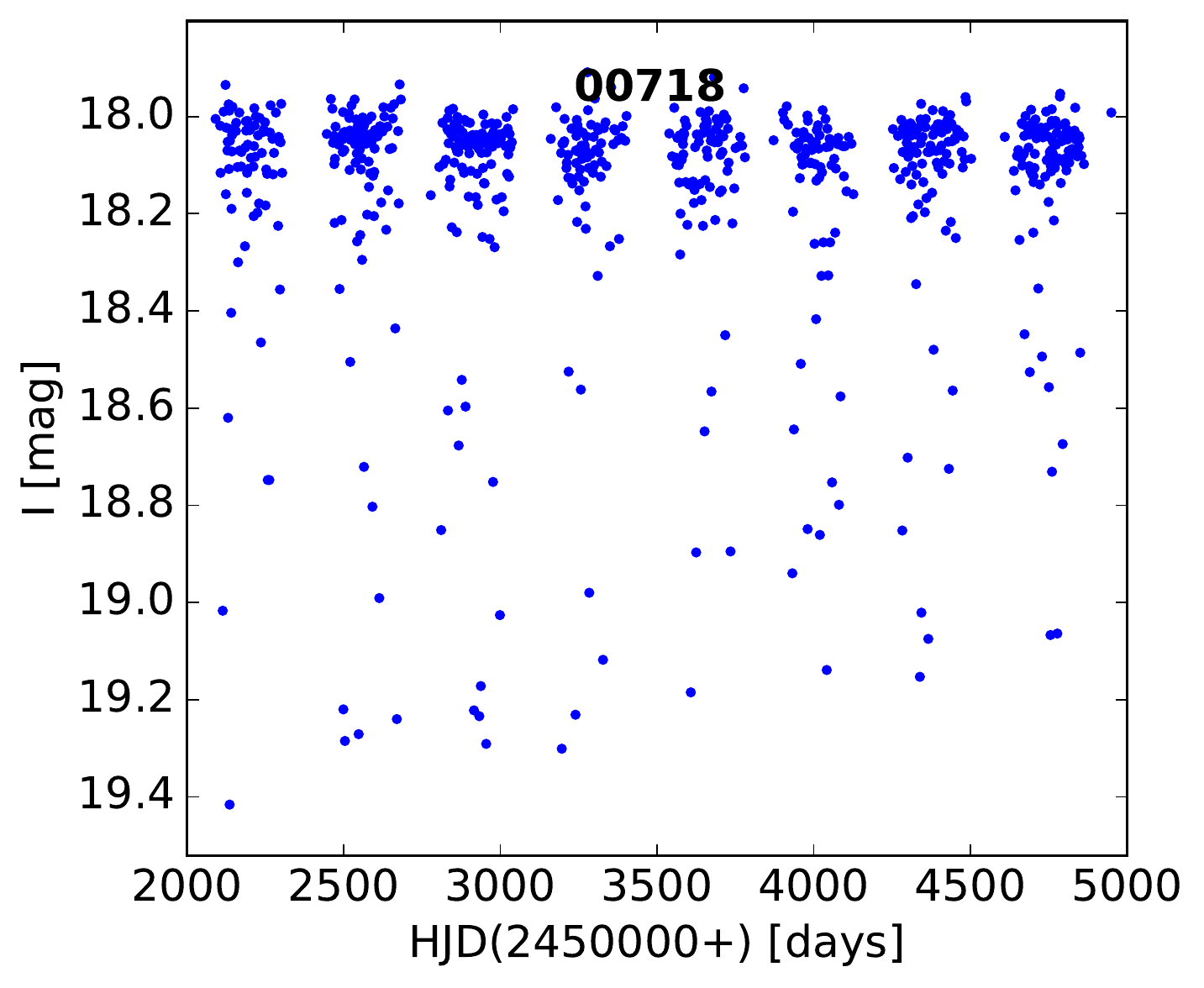} &
        \includegraphics[width=58mm]{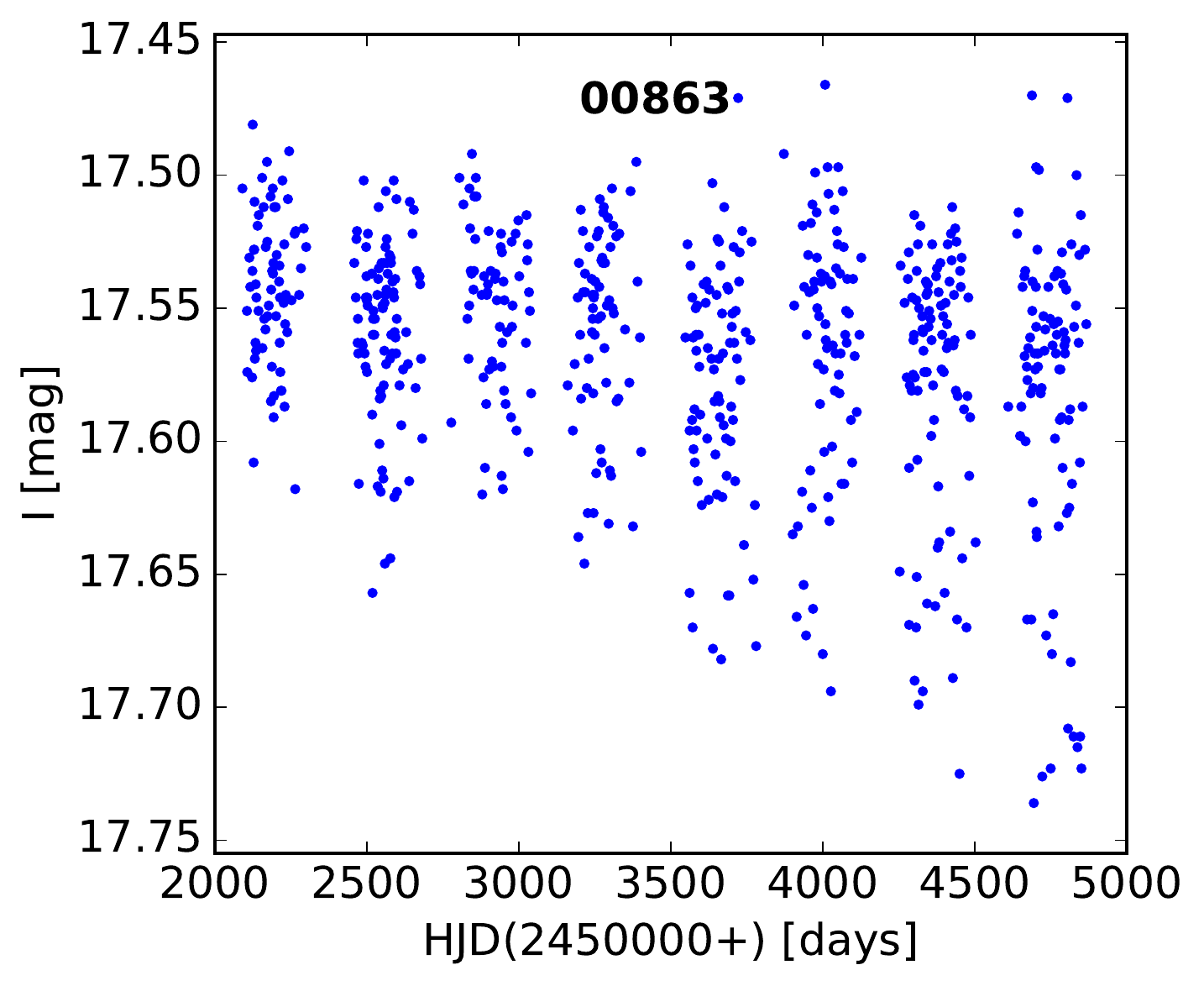} \\
        \includegraphics[width=58mm]{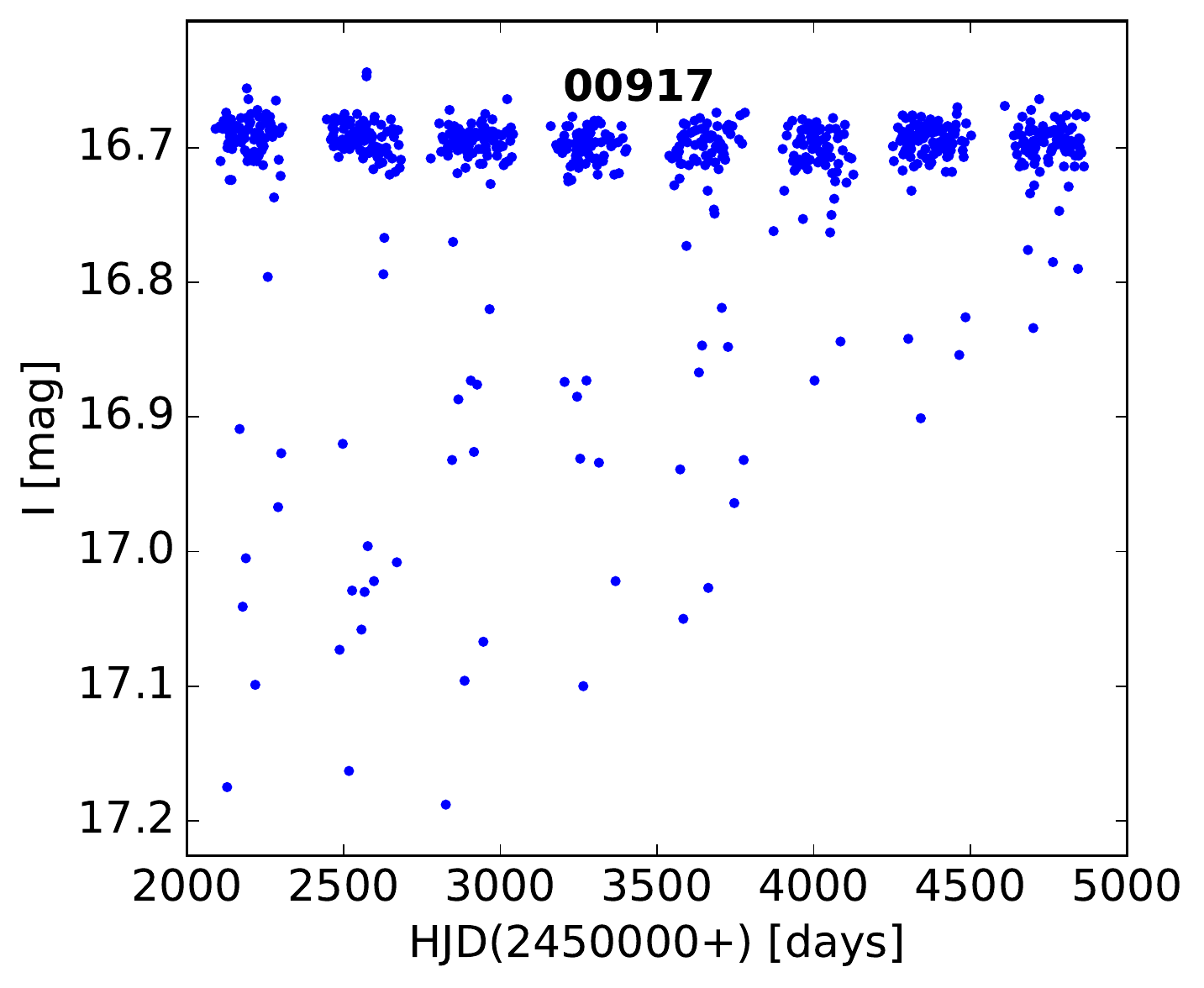} &
        \includegraphics[width=58mm]{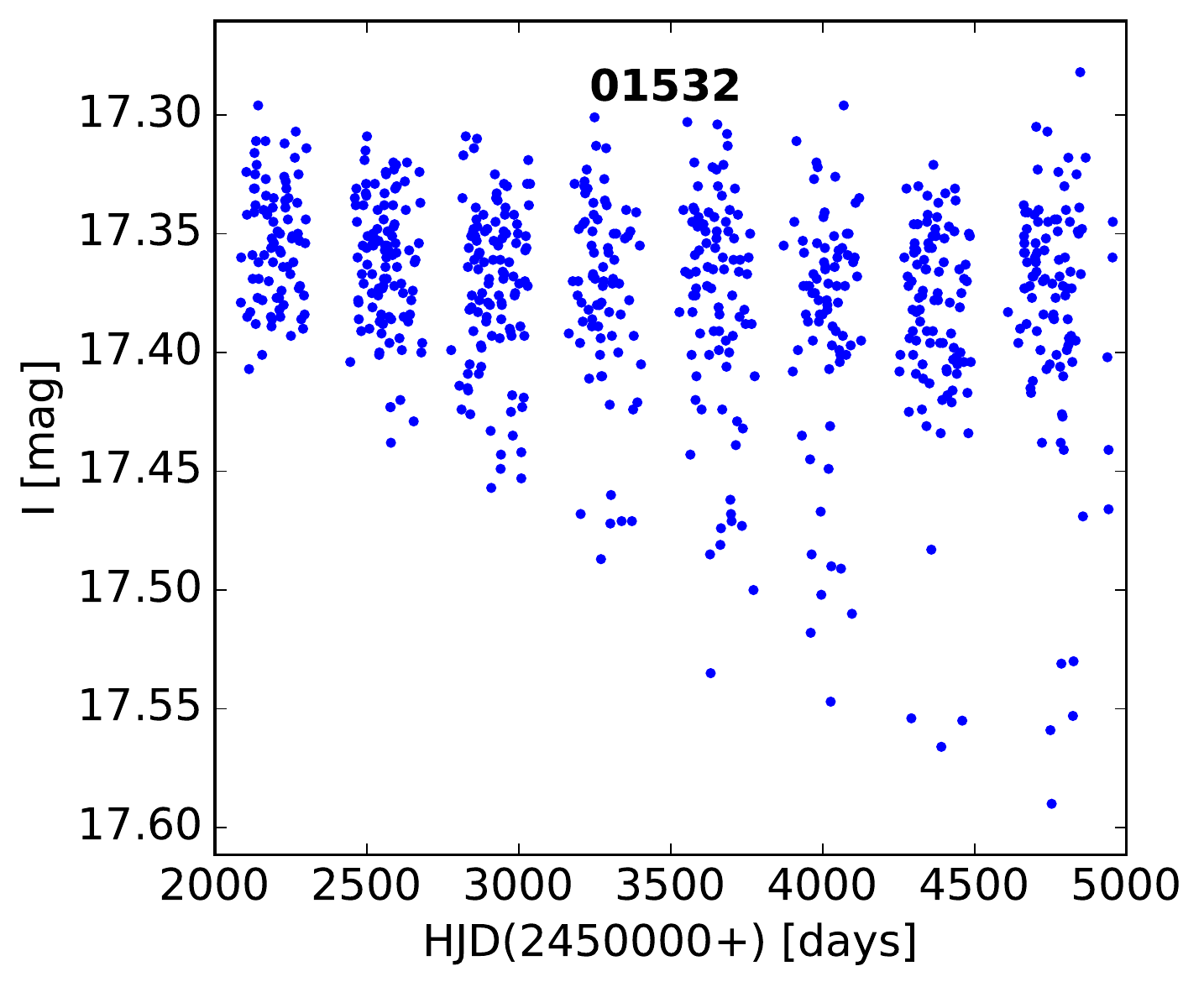} &
        \includegraphics[width=58mm]{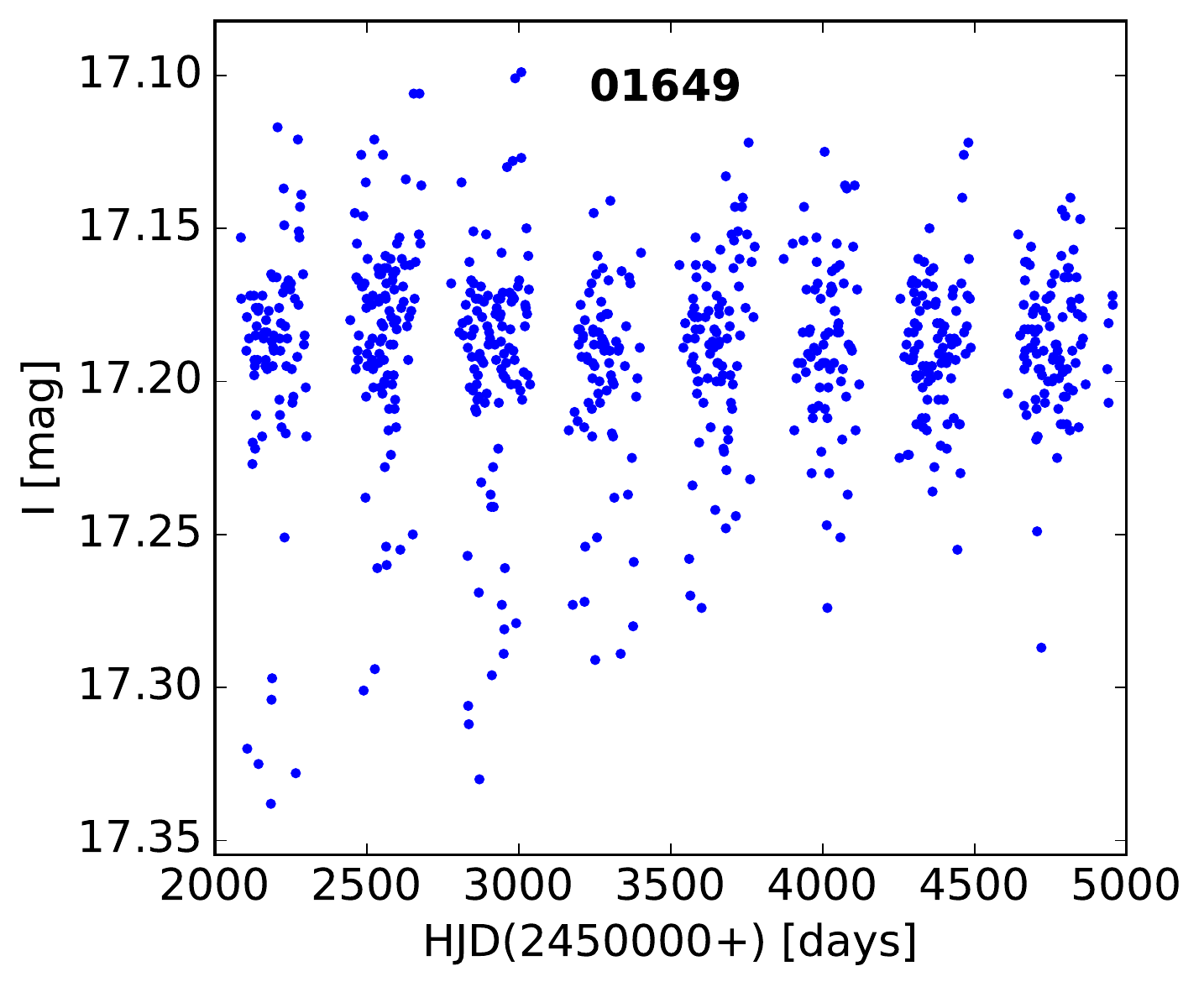} \\
        \includegraphics[width=58mm]{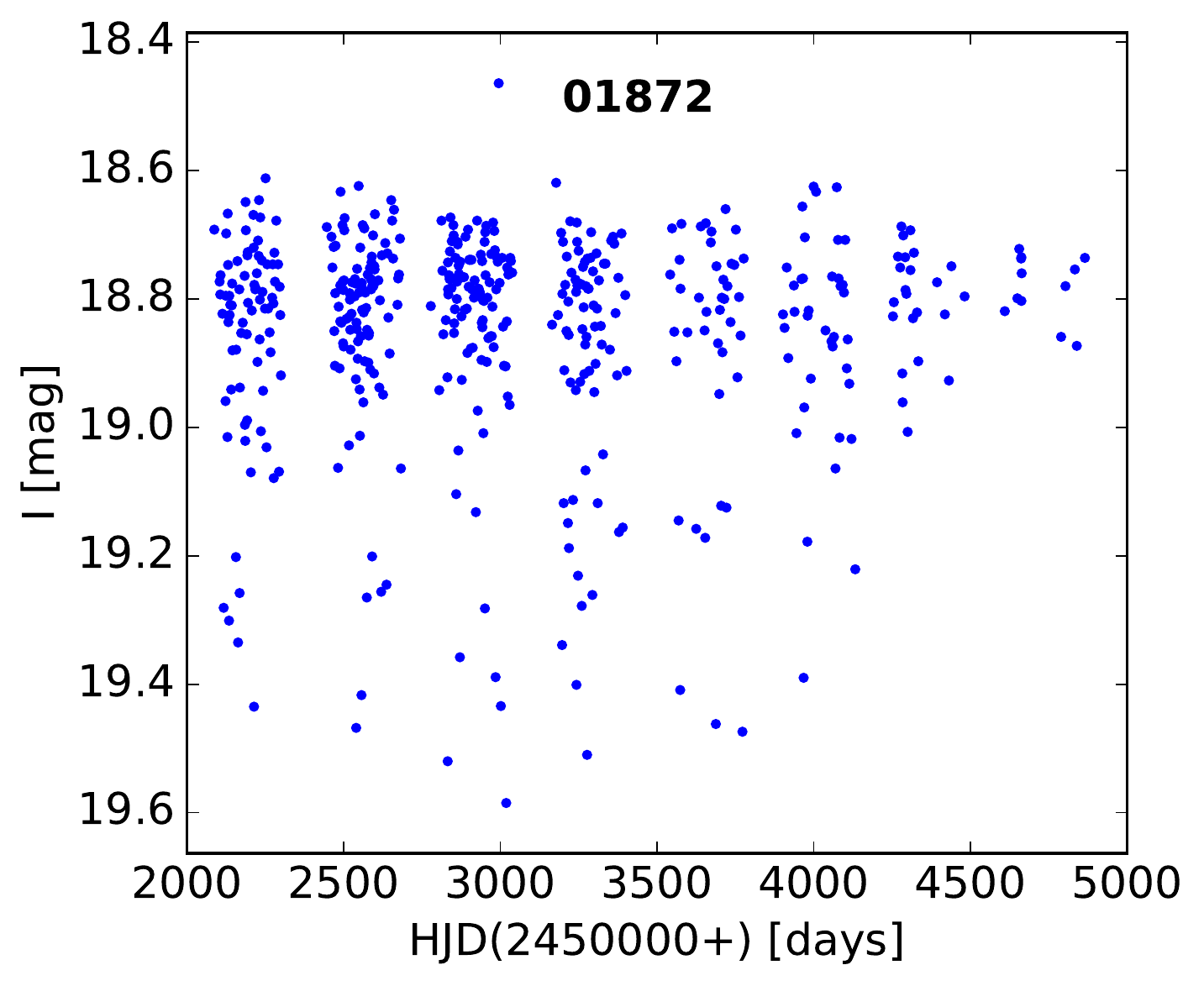} &
        \includegraphics[width=58mm]{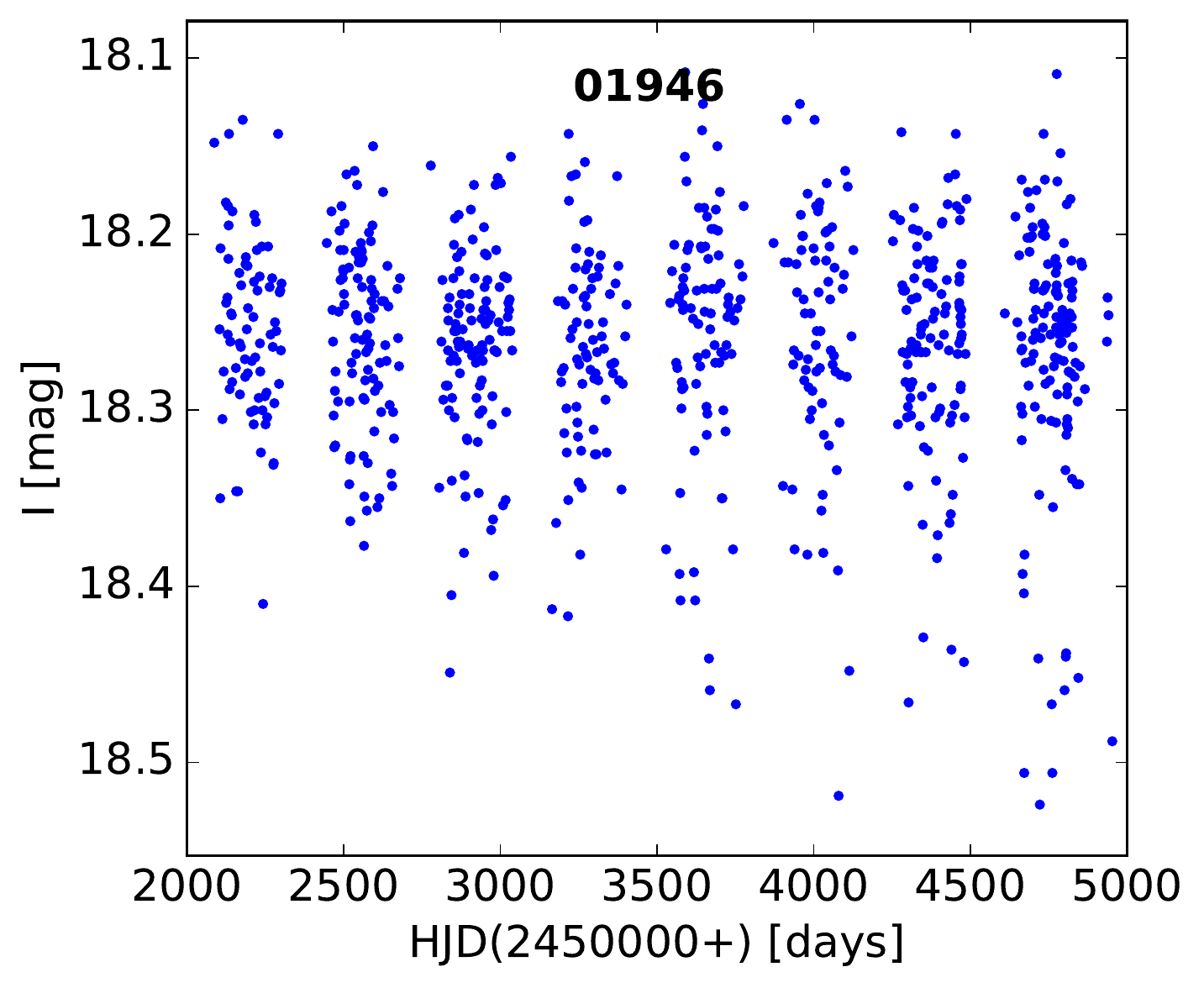} &
        \includegraphics[width=58mm]{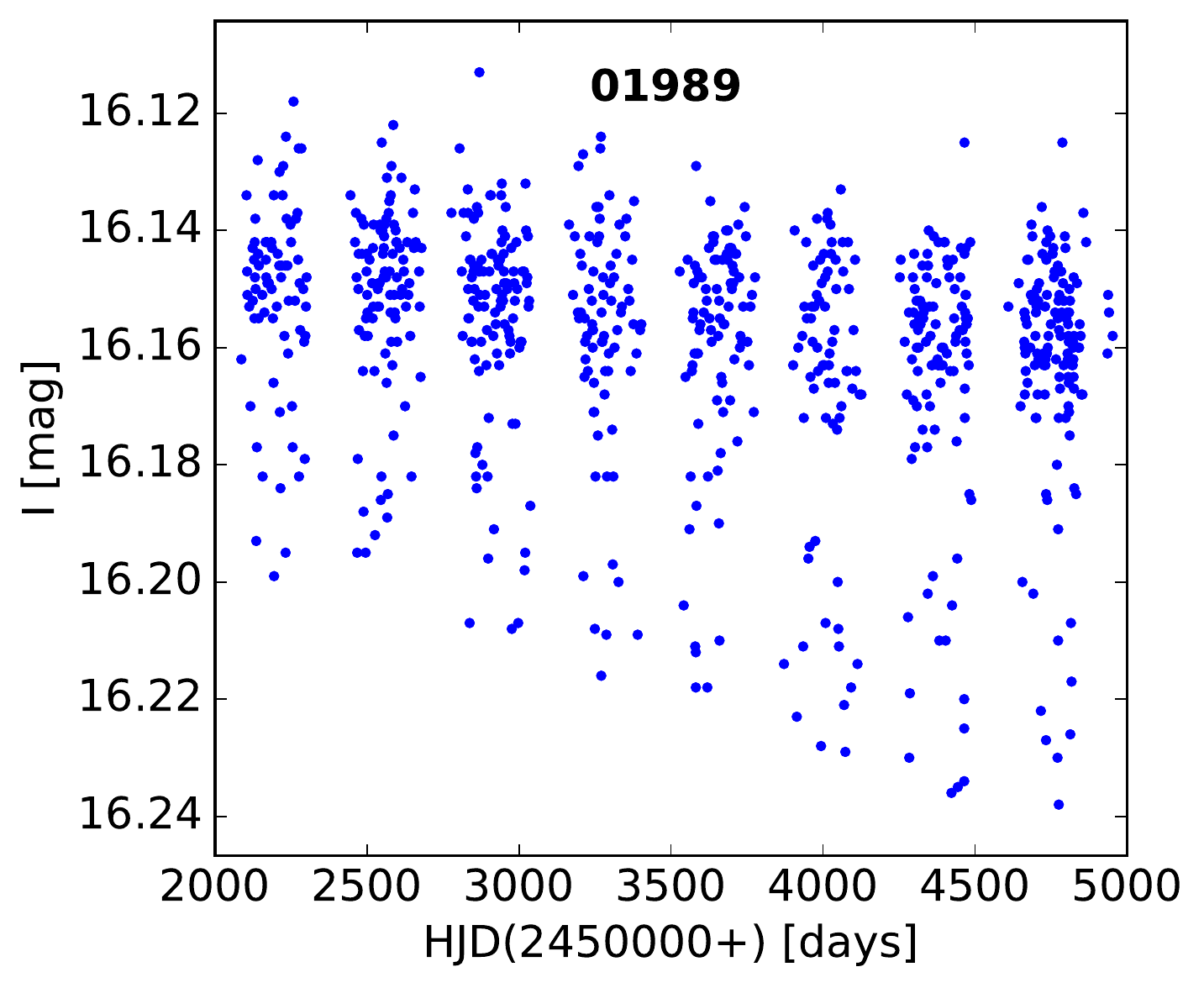} \\
        \includegraphics[width=58mm]{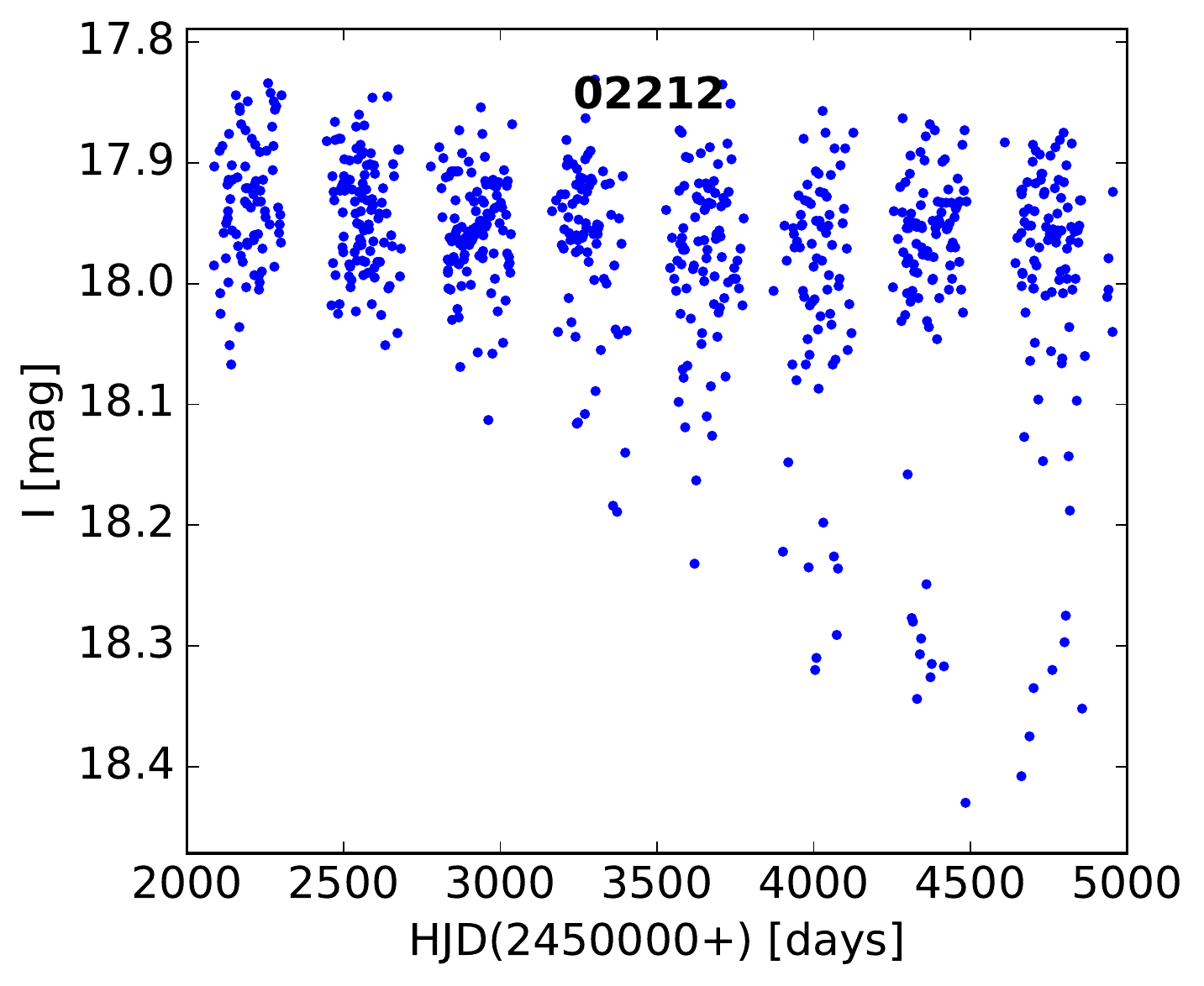} &
        \includegraphics[width=58mm]{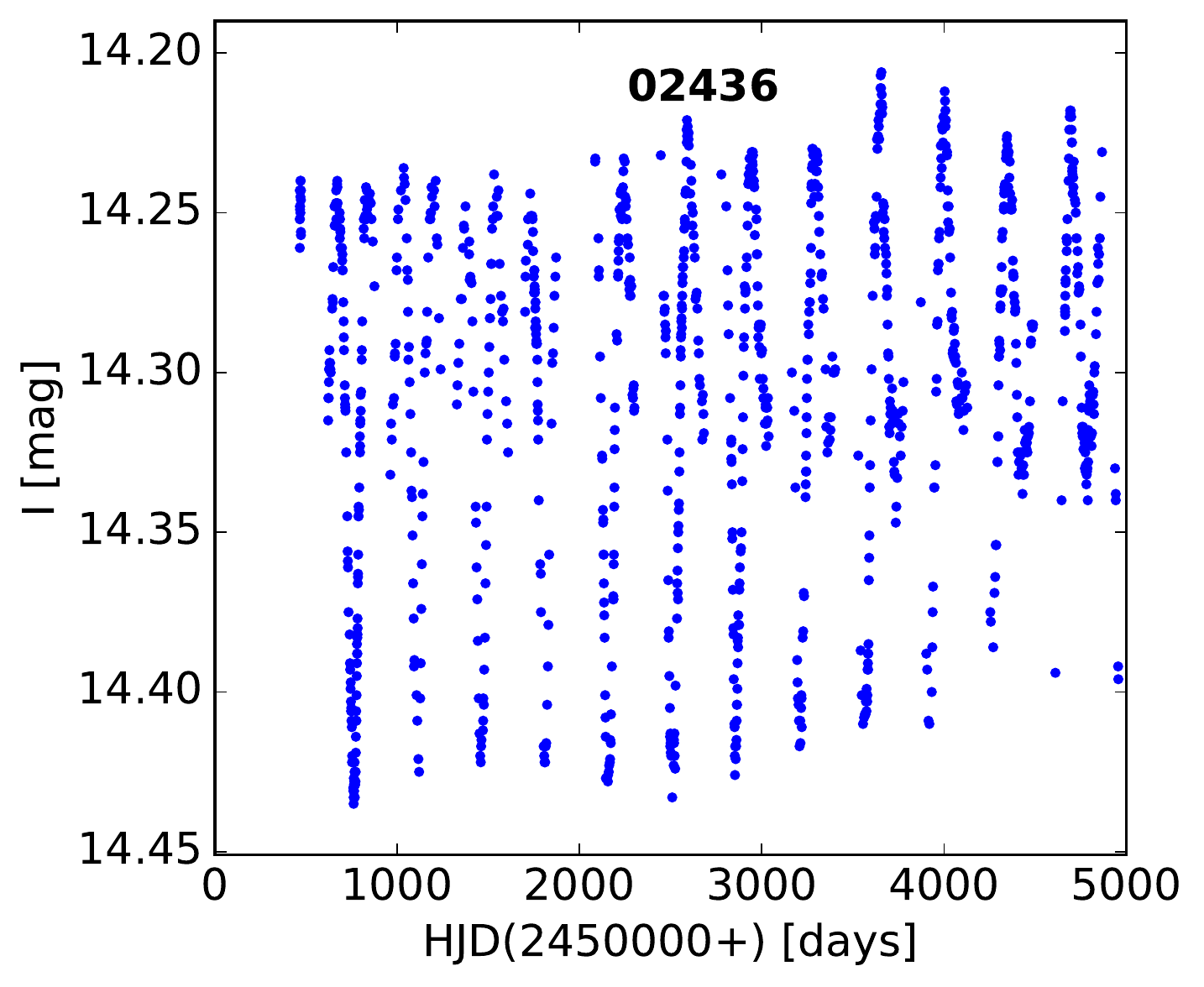} &
        \includegraphics[width=58mm]{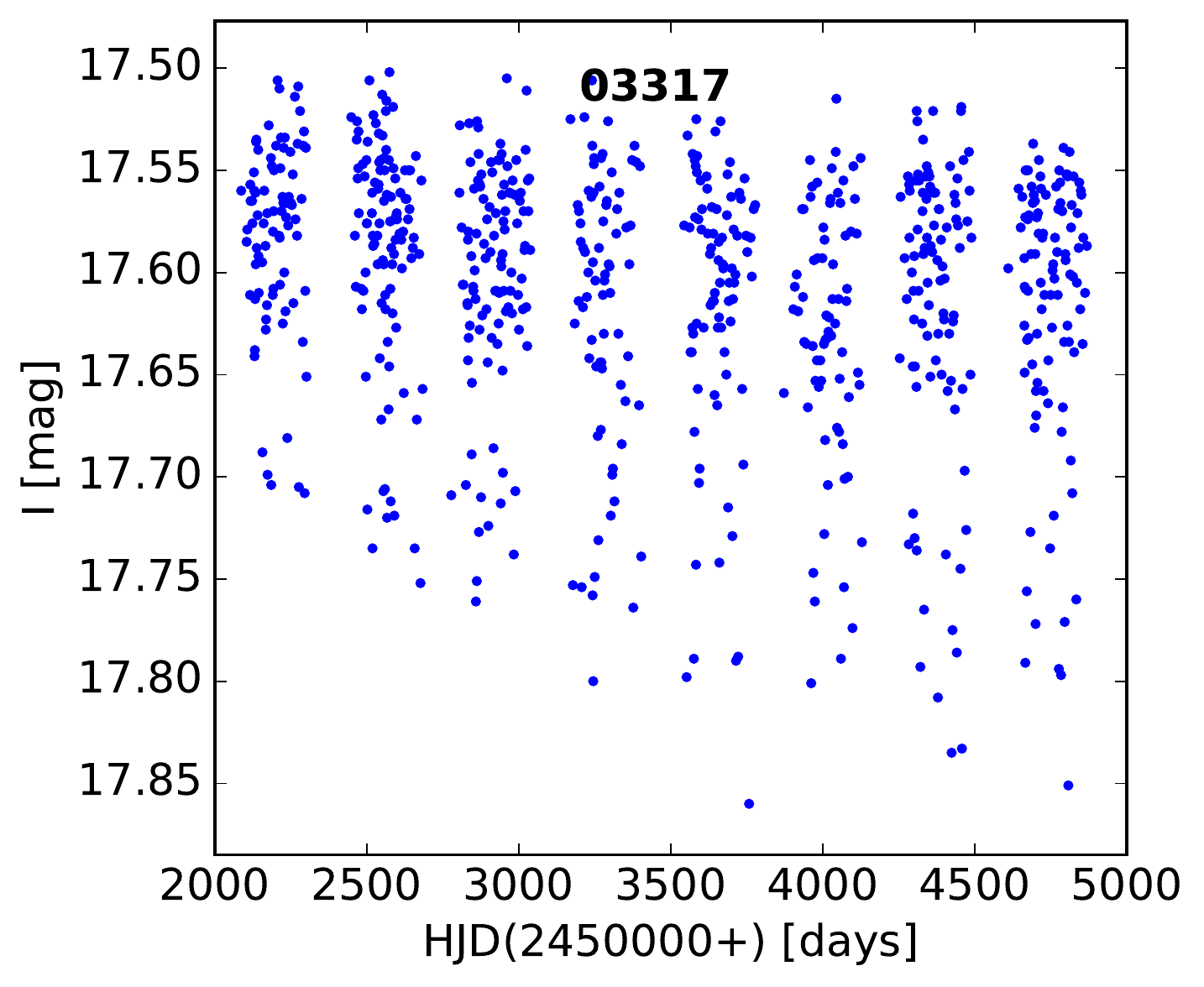} \\
        \includegraphics[width=58mm]{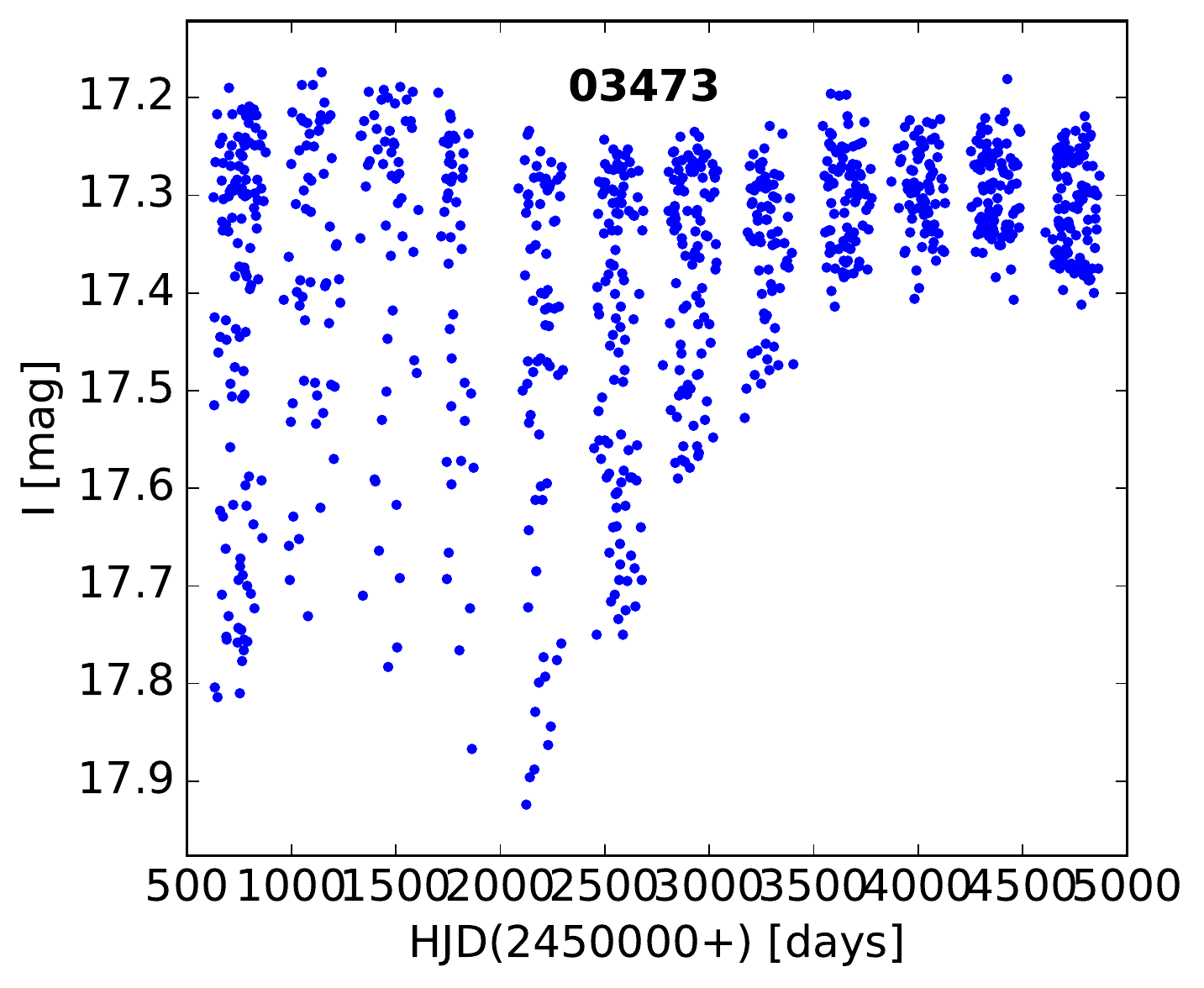} &
        \includegraphics[width=58mm]{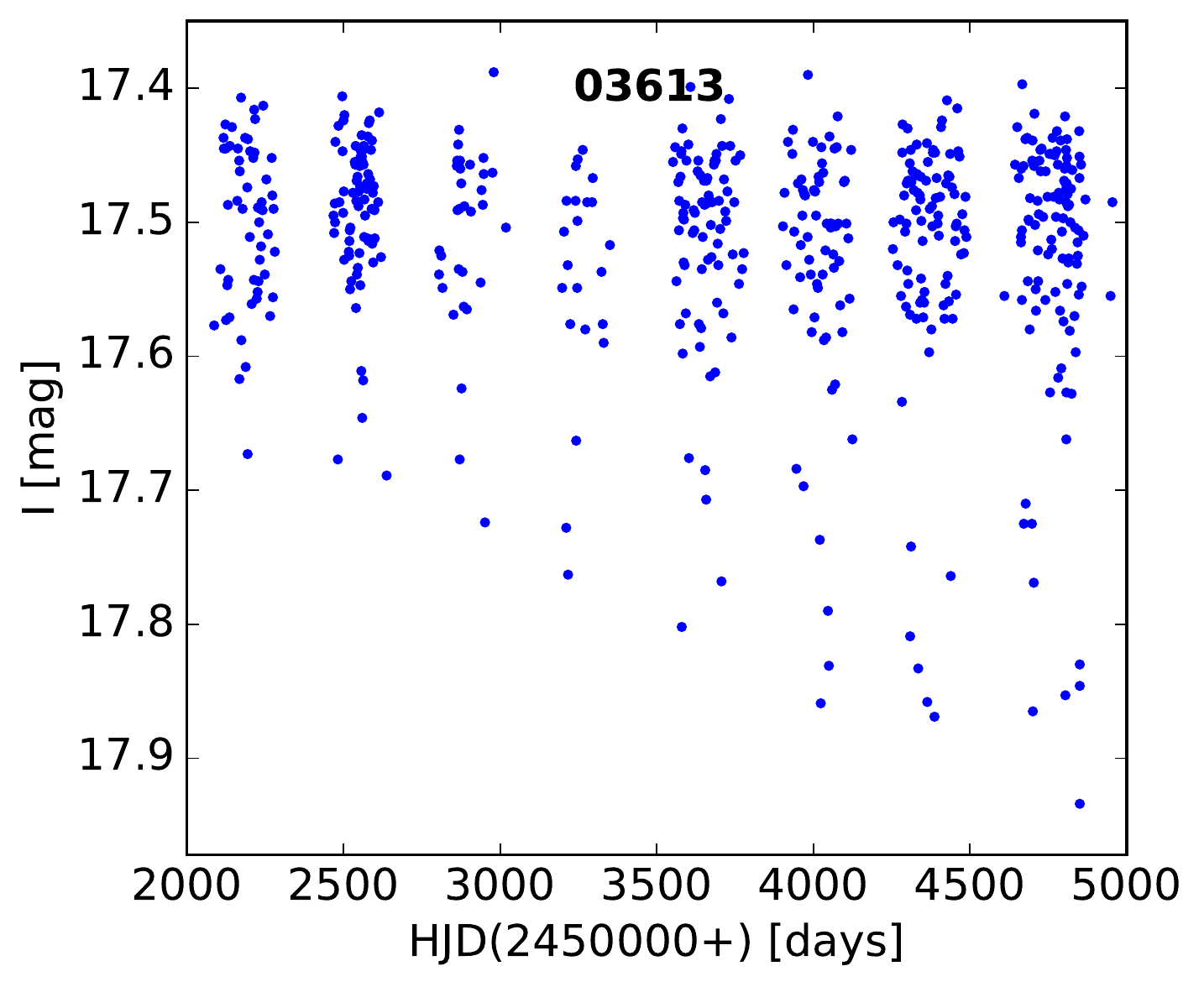} &
        \includegraphics[width=58mm]{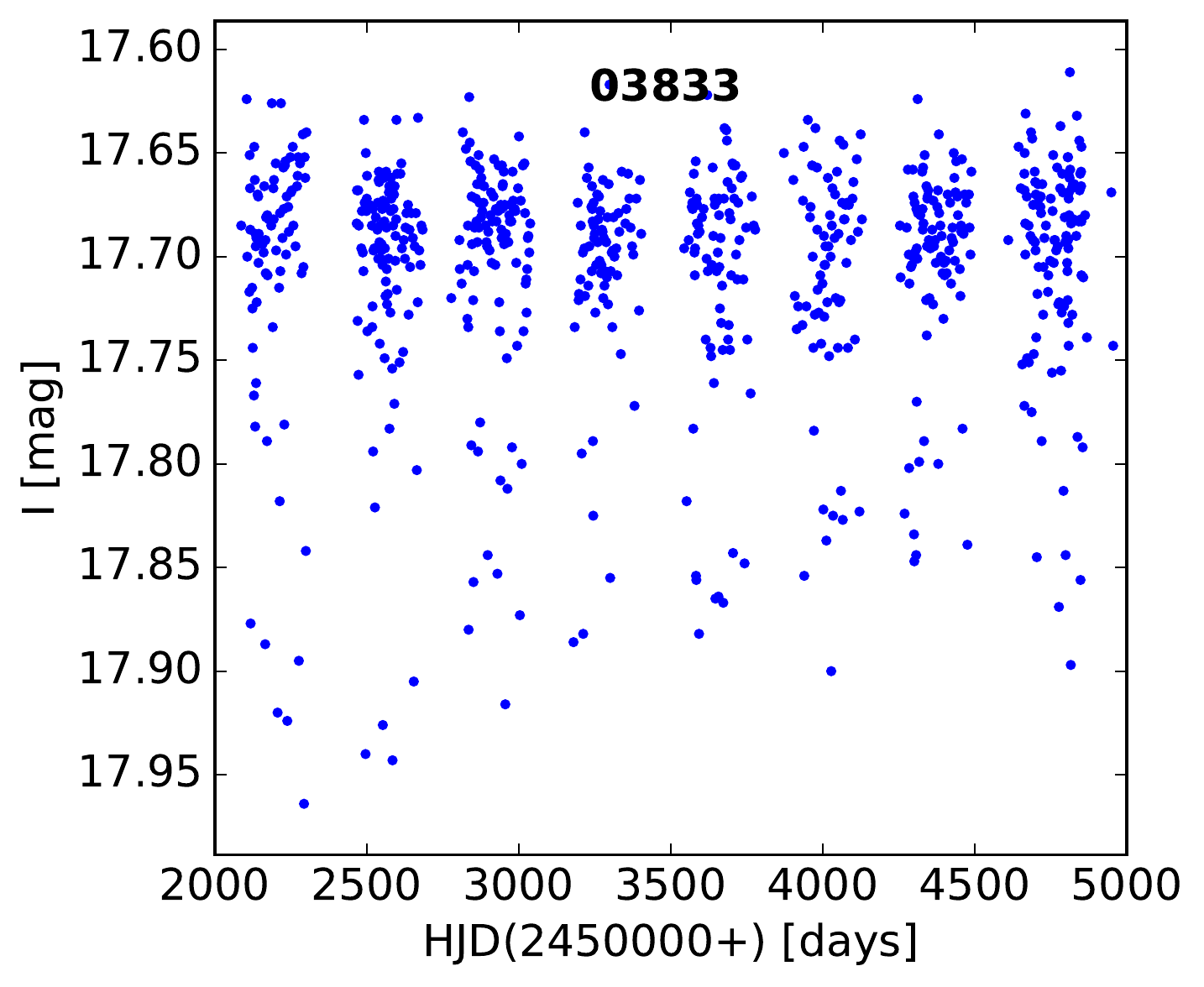} \\
        \end{tabular}
  \caption{Light curves of EBs with amplitude variation from the OGLE III SMC database.}
   \label{fig.smc_detected_systems}
\end{figure*}

\begin{figure*}
\ContinuedFloat
\centering
        \begin{tabular}{@{}ccc@{}}
        \includegraphics[width=58mm]{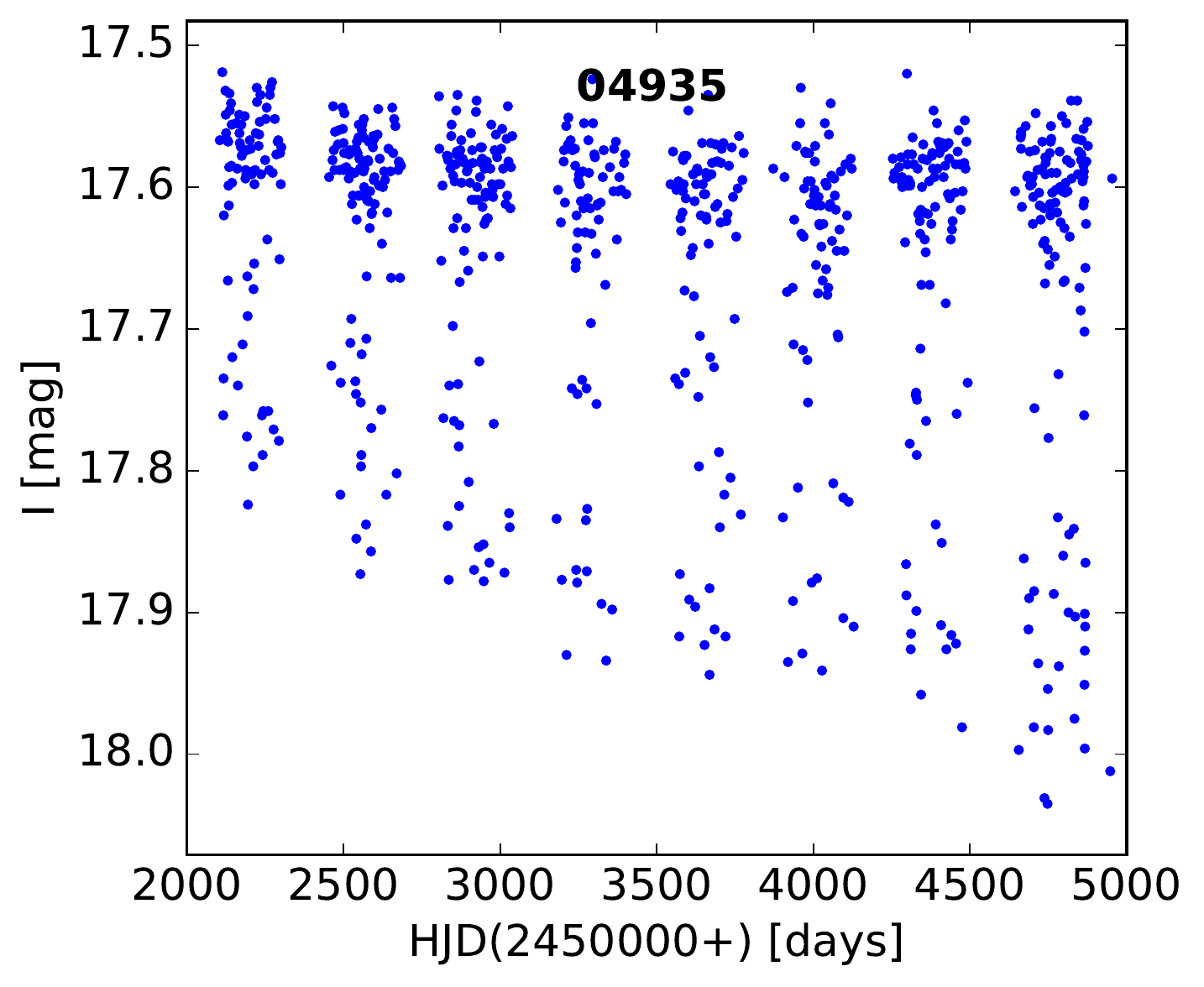} &
        \includegraphics[width=58mm]{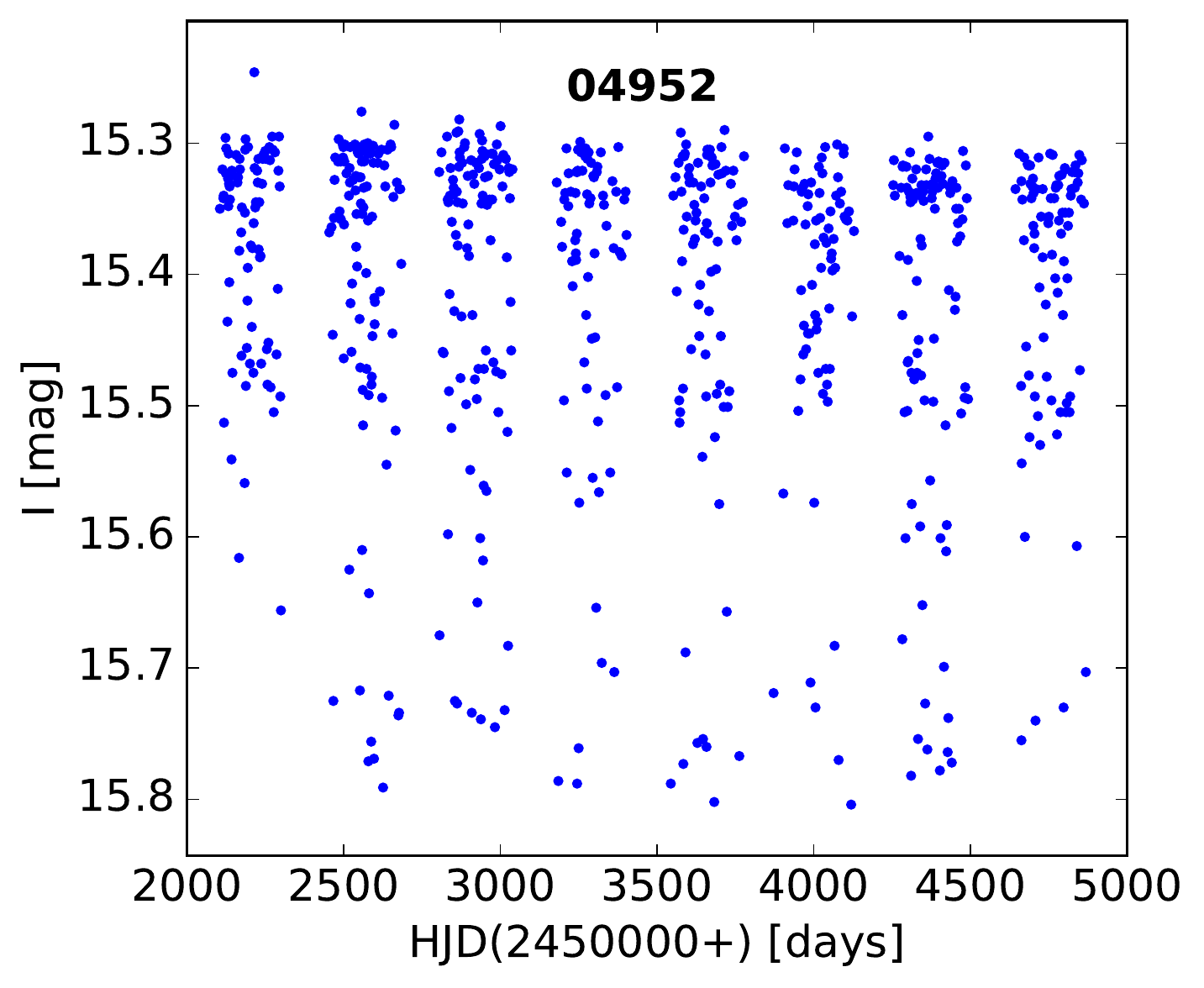} &
        \includegraphics[width=58mm]{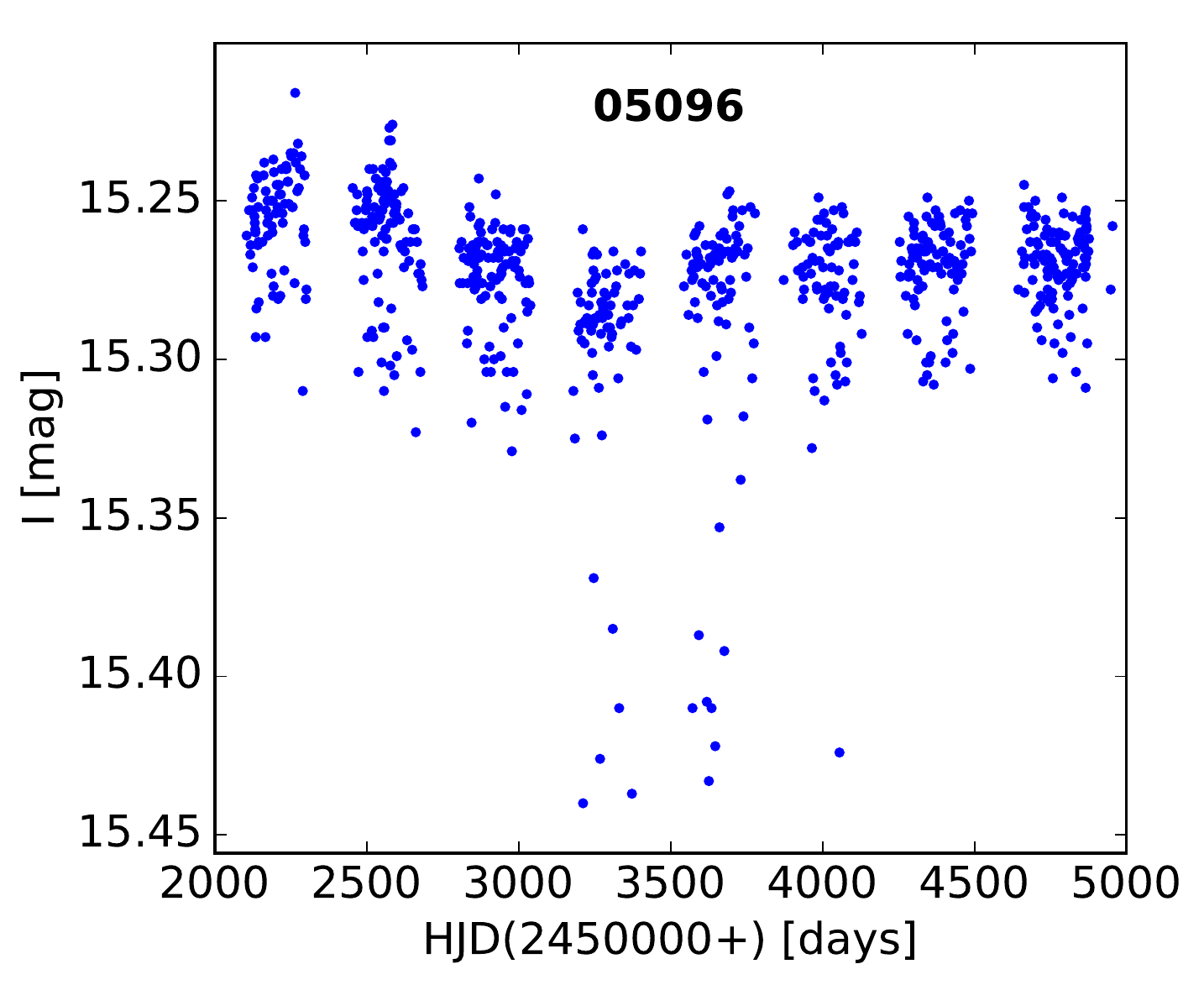} \\
        \includegraphics[width=58mm]{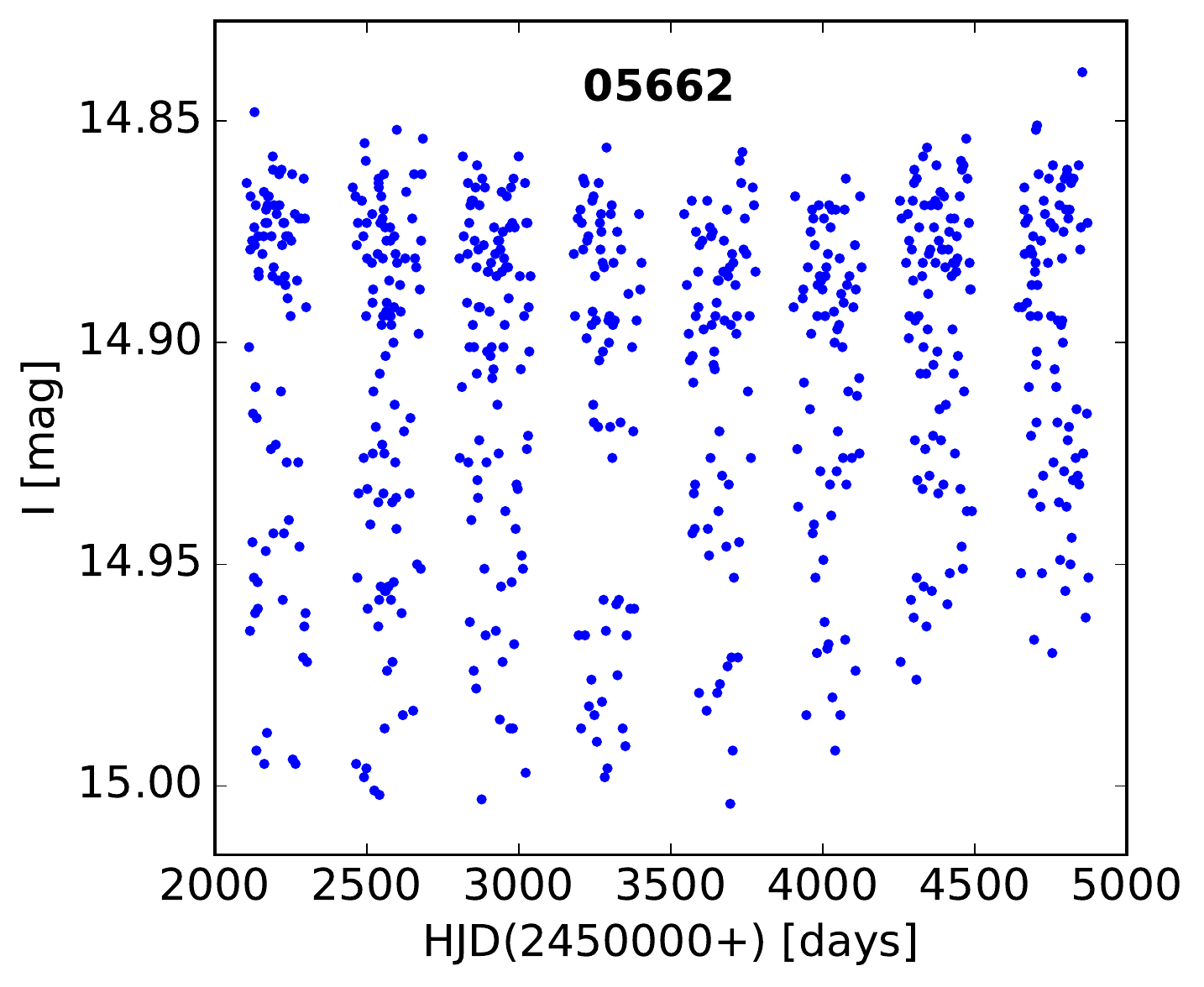} &
        \includegraphics[width=58mm]{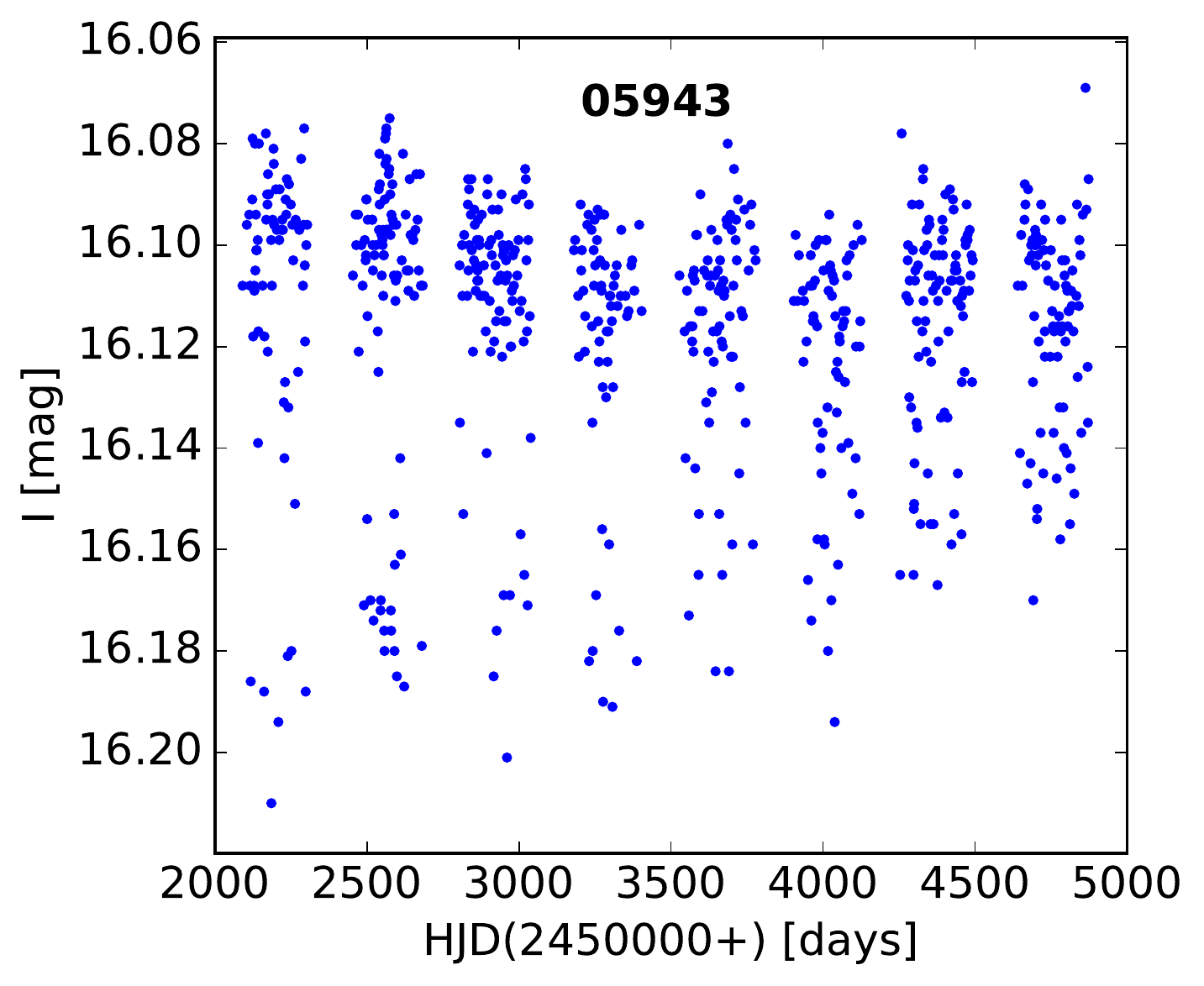} &
        \includegraphics[width=58mm]{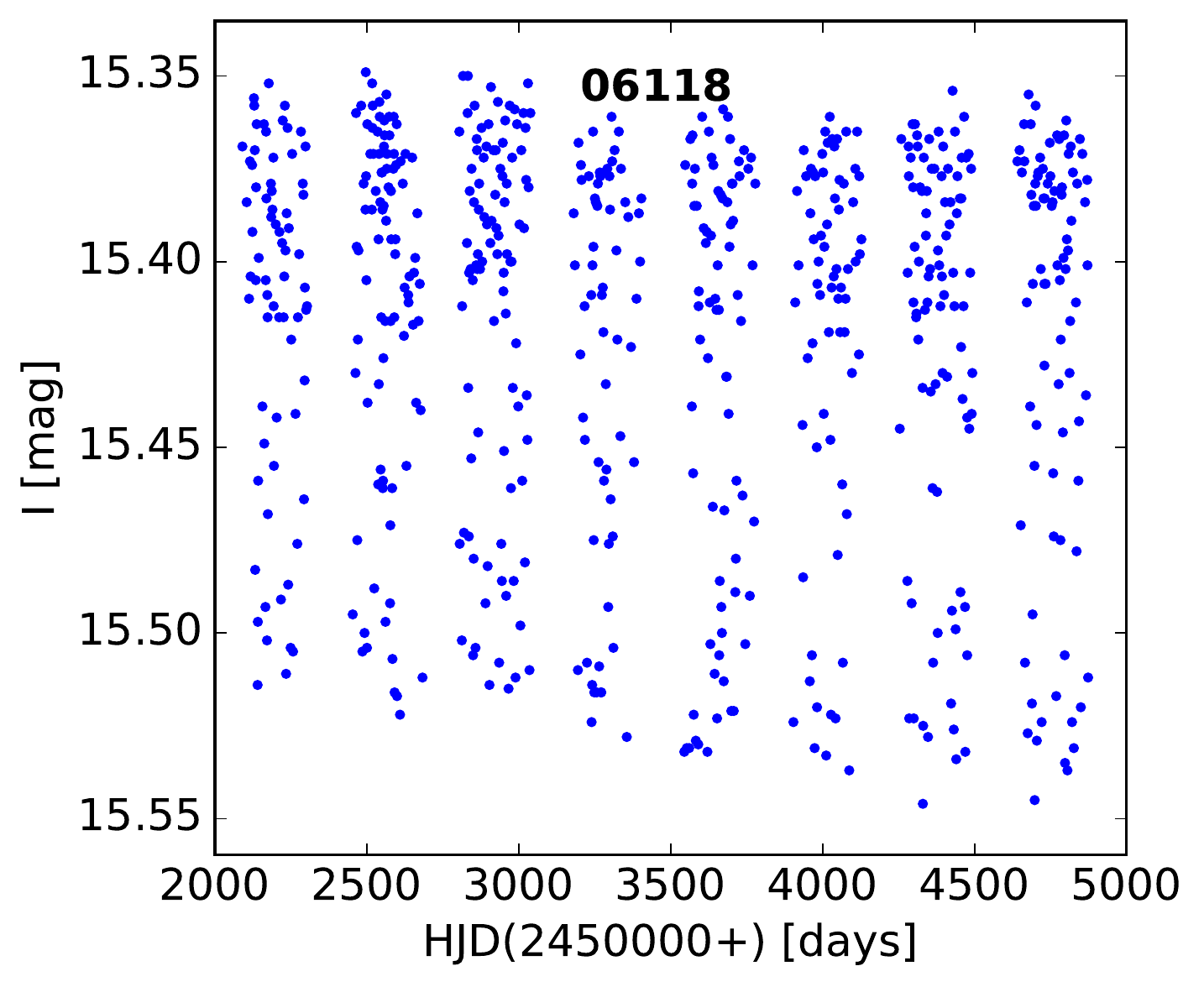} \\
        \end{tabular}
  \caption{\textit{continued}}
\end{figure*}


\FloatBarrier

\section{Photometric solutions for selected eclipsing binaries in the LMC and SMC}

\begin{table*}[ht!]
\caption{Eclipsing binaries in the LMC.}             
\label{tab.lc_sol_lmc}      
\centering          
        \begin{tabular}{c c c c c c }   
        \hline\hline   
        
        OGLE-LMC-ECL- & 01350 & 13150 & 16023 & 18240 & 23148 \\ \cline{0-0}
        Parameter & & & & & \\
        \hline
        $T_0 \, \mathrm{(K)}$ (fixed)& $13000$ & $11000$ & $14000$ & $17000$  & $28000$ \\
        $T_1 \, \mathrm{(K)}$ &  $12612$        & $10984$       & $13482$          & $13075$  & $22985$ \\
                                                                                 
        $i_0\,(^{\circ})$ & $67.2 - 77.4$ & $65.3 - 81.1$ & $48.8 - 75.7$ & $92.8 - 97.8$ & $<60.0 - 72.1$ \\
        
        $q = m_1/m_0$ & $1.0$ (fixed) & $1.0$ (fixed) & $1.0$ (fixed) & $1.0$ (fixed) & $0.78$ \\
        
        $R_0/a_1$ & $0.26$ & $0.27$ & $0.37$ & $0.17$ & $0.27$ \\
        $R_1/a_1$ & $0.30$ & $0.29$ & $0.37$ & $0.11$ & $0.27$ \\
        
        $M_{\rm bol 0}$ (mag) & $-0.83$ & $-0.2$ & $-1.9$ & $-1.1$ & $-4.3$ \\
        $M_{\rm bol 1}$ (mag) & $-1.01$ & $-0.4$ & $-1.8$ & $-1.0$ & $-3.4$ \\
        
        $L_{\rm V 0 }$ (\%) & \ldots & $45.4$ & $51.6$ & \ldots &  $58.1$\\
        $L_{\rm V 1 }$ (\%)     & \ldots & $54.7$ & $48.4$ & \ldots & $41.9$ \\
        
        $L_{\rm R 0 }$ (\%) & \ldots & $45.4$ & $51.4$ & \ldots &  $57.6$\\
        $L_{\rm R 1 }$ (\%)     & \ldots & $54.6$ & $48.6$ & \ldots & $42.4$ \\
        
        $L_{\rm I 0 }$ (\%) & $43.9$ & $45.3$ & $51.4$ & $78.0$ &  $57.0$\\
        $L_{\rm I 1 }$ (\%)     & $56.1$ & $54.7$ & $48.6$ & $22.0$ & $43.0$ \\                                              
                                                        
        \hline                  
        \end{tabular}
\end{table*}

\begin{table*}[ht!]
\caption{Eclipsing binaries in the SMC.}             
\label{tab.lc_sol_smc}      
\centering          
        \begin{tabular}{c c c c}   
        \hline\hline   
        
        OGLE-SMC-ECL- & 1532 & 3317 & 6118 \\ \cline{0-0}
        Parameter & & & \\
        \hline
        $T_0 \, \mathrm{(K)}$ (fixed) &  $16500$ & $16000$ & $26000$ \\
        $T_1 \, \mathrm{(K)}$ &                  $6373$  & $5397$  & $25906$  \\
                                                                                 
        $i_0\,(^{\circ})$ & $<55.0 - 68.1$ & $<53.0 - 66.1$ & $60.6 - 62.3$ \\
        
        $q = m_1/m_0$ & $0.3$ & $0.25$ & $1.0$ (fixed) \\
        
        $R_0/a_1$ & $0.36$ & $0.49$ & $0.33$ \\
        $R_1/a_1$ & $0.23$ & $0.25$ & $0.34$ \\
        
        $M_{\rm bol 0}$ (mag) &  $-2.5$ & $-3.13$ & $-4.39$ \\
        $M_{\rm bol 1}$ (mag) &  $2.18$ & $3.08$ &  $-4.42$ \\
        
        $L_{\rm V 0 }$ (\%) & \ldots & \ldots & \ldots \\
        $L_{\rm V 1 }$ (\%)     & \ldots & \ldots & \ldots  \\
        
        $L_{\rm R 0 }$ (\%) & $92.8$ & \ldots & \ldots  \\
        $L_{\rm R 1 }$ (\%)     &$2.2$ & \ldots & \ldots  \\
        
        $L_{\rm I 0 }$ (\%) & $90.2$ & $97.3$ & $49.0$ \\
        $L_{\rm I 1 }$ (\%)     & $9.8$ &  $2.7$ &  $51.0$ \\                                           
                                                        
        \hline                  
        \end{tabular}
\end{table*}

\pagebreak
\FloatBarrier
\section{Ephemerides for analyzed eclipsing binaries}

\begin{table*}[ht!]
\caption{Ephemerides for the LMC eclipsing binaries.}             
\label{tab.ephem_lmc}      
\centering          
        \begin{tabular}{c c l l}   
        \hline\hline   
         OGLE-LMC-ECL- & Time interval (HJD) & HJD0 & $P_1$ (days) \\
        \hline     
        01350   & whole & 2456308.2933(10) & 1.09883233(21) \\[10pt]            
                        
        13150   & whole & 2453541.39142(97) & 0.95597619(38) \\[10pt]
        
        \multirow{4}{*}{16023}  & 2448896.0883 < HJD < 2450389.1889     & 2453562.506(10) & 0.7882459(20) \\
                                                        & 2450389.1889 < HJD < 2452484.4118      & 2453562.5692(88) & 0.7882625(28) \\
                                                        & 2452484.4118 < HJD < 2457389.4966      & 2453562.5422(13) & 0.78824656(69) \\
                                                        & whole & 2453562.5333(23) & 0.78825063(63) \\[10pt]       
        
        18240   & whole & 2457044.4687(72) & 2.7641050(60) \\[10pt]
        
        23148   & whole & 2443568.0121(59) & 1.28218324(86) \\[10pt]                                                                                                            
                                                                                                                
        \hline                  
        \end{tabular}
\end{table*}

\begin{table*}[ht!]
\caption{Ephemerides for the SMC eclipsing binaries.}             
\label{tab.ephem_smc}      
\centering          
        \begin{tabular}{c c l l}   
        \hline\hline   
         OGLE-SMC-ECL- & Time interval (HJD) & HJD0 & $P_1$ (days) \\
        \hline     
        
        1532    & whole & 2455000.4915(34) & 1.0283876(18) \\[10pt]
        
        3317    & whole & 2455000.4583(21) & 0.70421606(63) \\[10pt]
        
        6118    & whole & 2455000.5435(25) & 0.9372806(15) \\[10pt]                                                                                             
                                                                                                                
        \hline                  
        \end{tabular}
\end{table*}

\pagebreak
\FloatBarrier
\section{Solutions of the time dependencies of inclination for selected systems}

\begin{figure*}[ht!]
\centering
        \begin{tabular}{@{}cc@{}}
                \includegraphics[width=87mm]{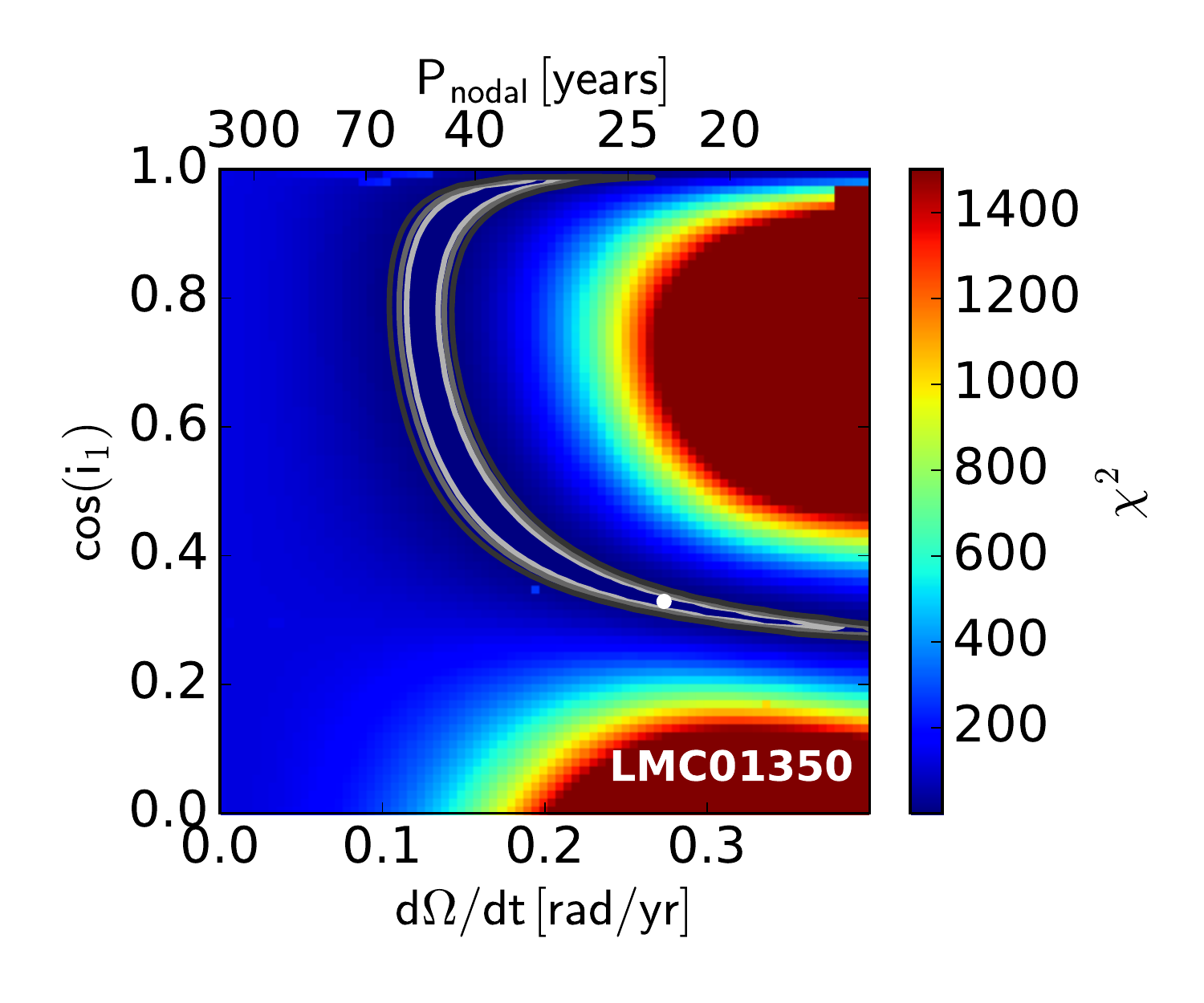} &
                \includegraphics[width=87mm]{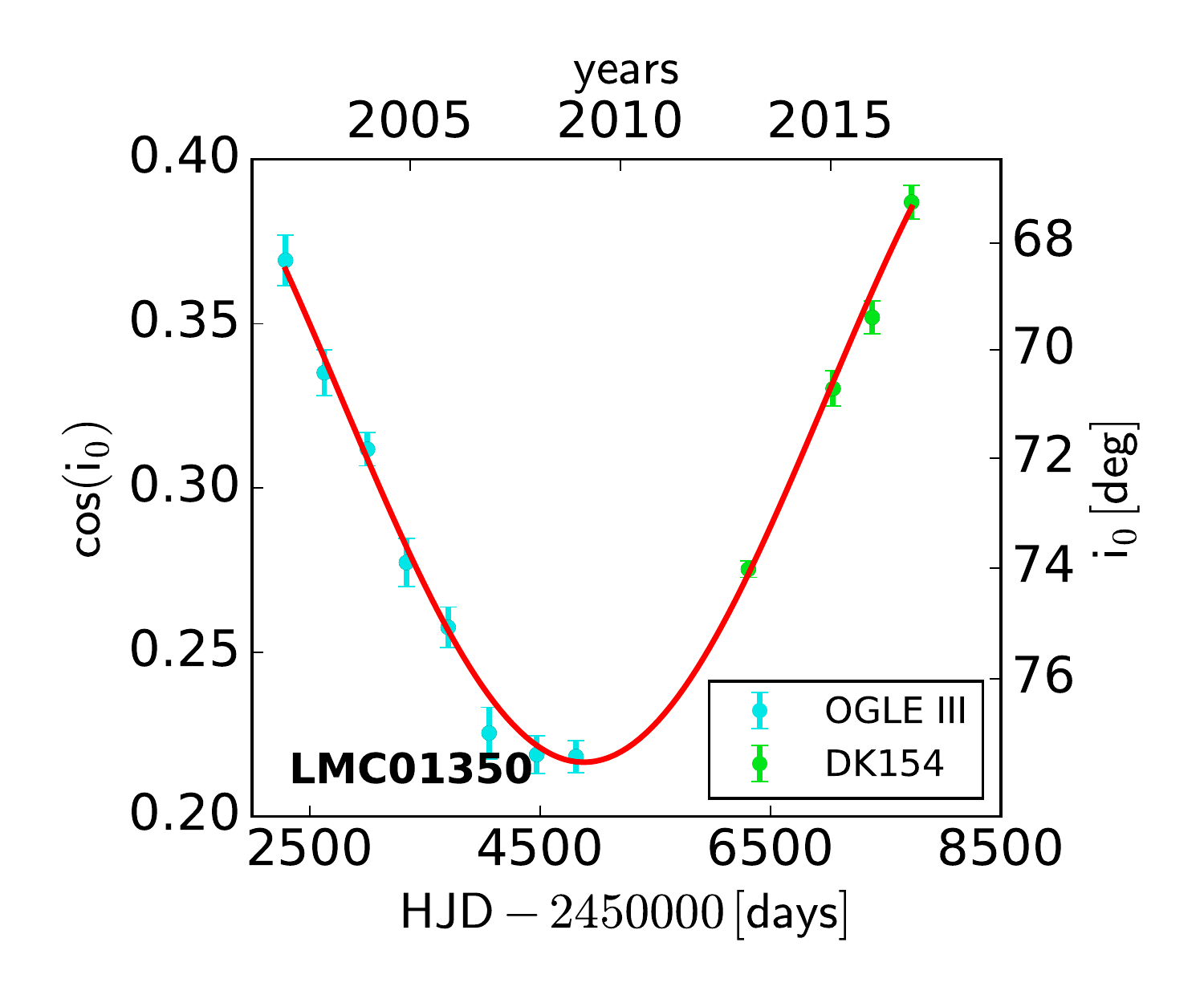} \\
                \includegraphics[width=87mm]{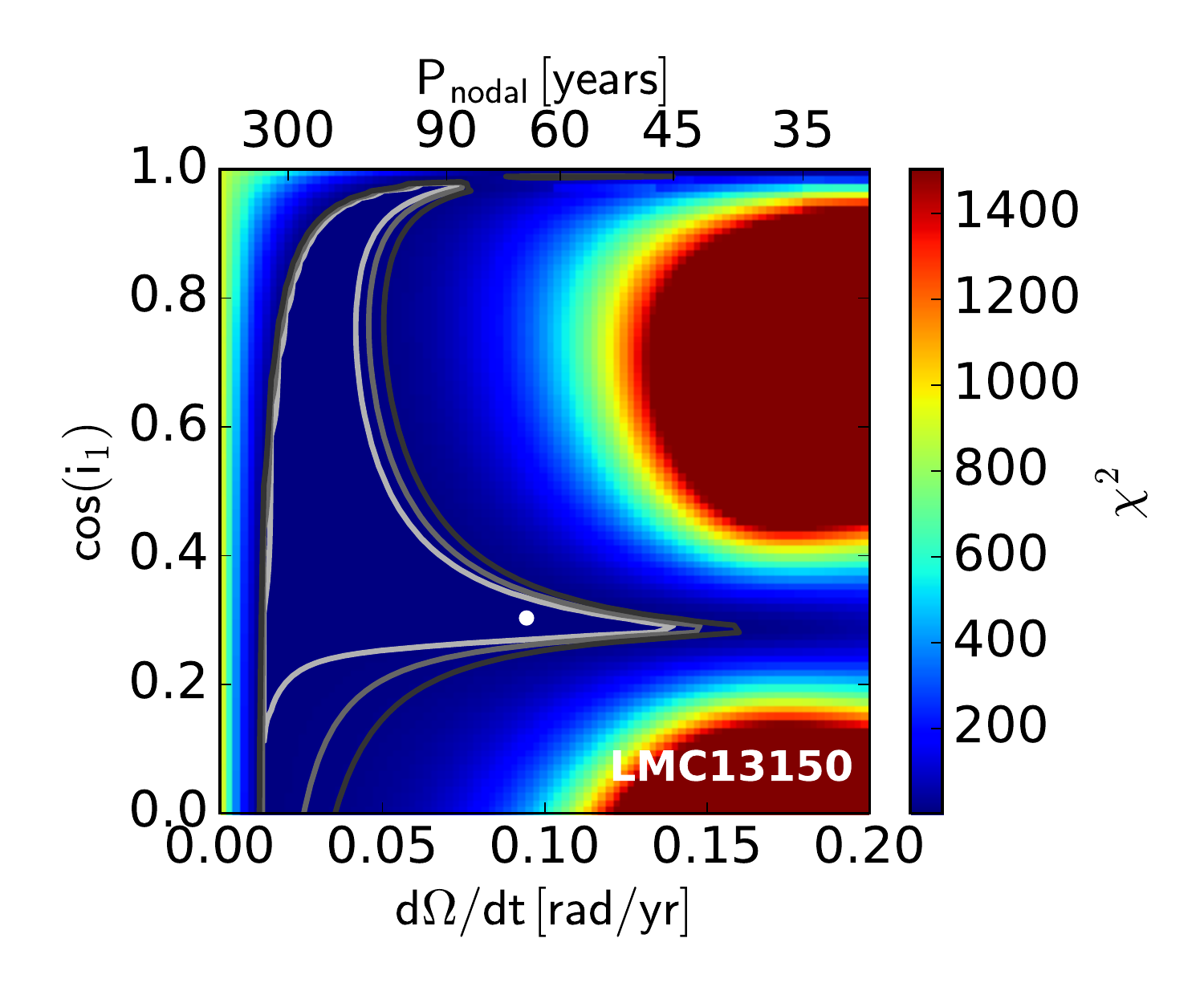} &
                \includegraphics[width=87mm]{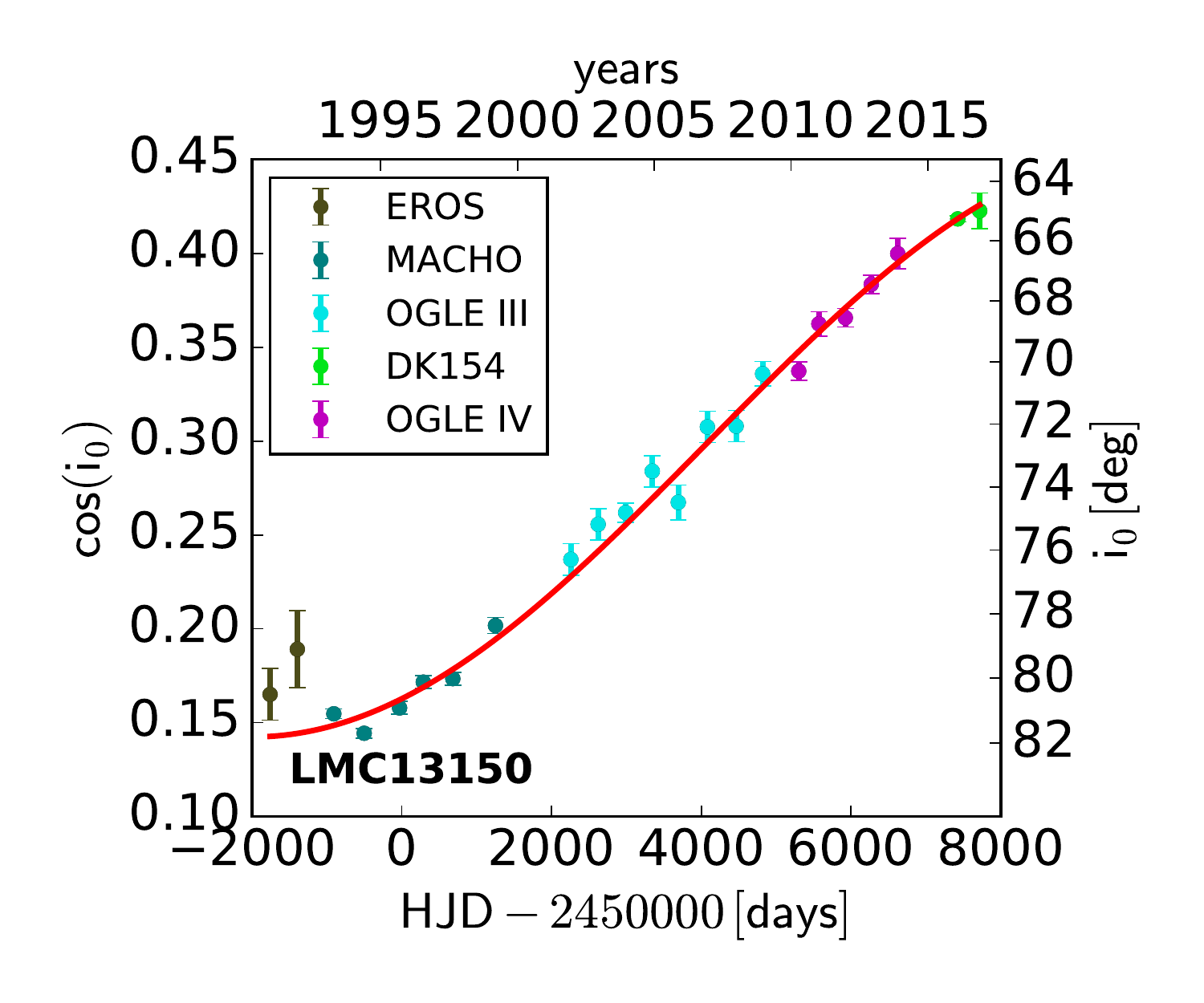} \\
                \includegraphics[width=87mm]{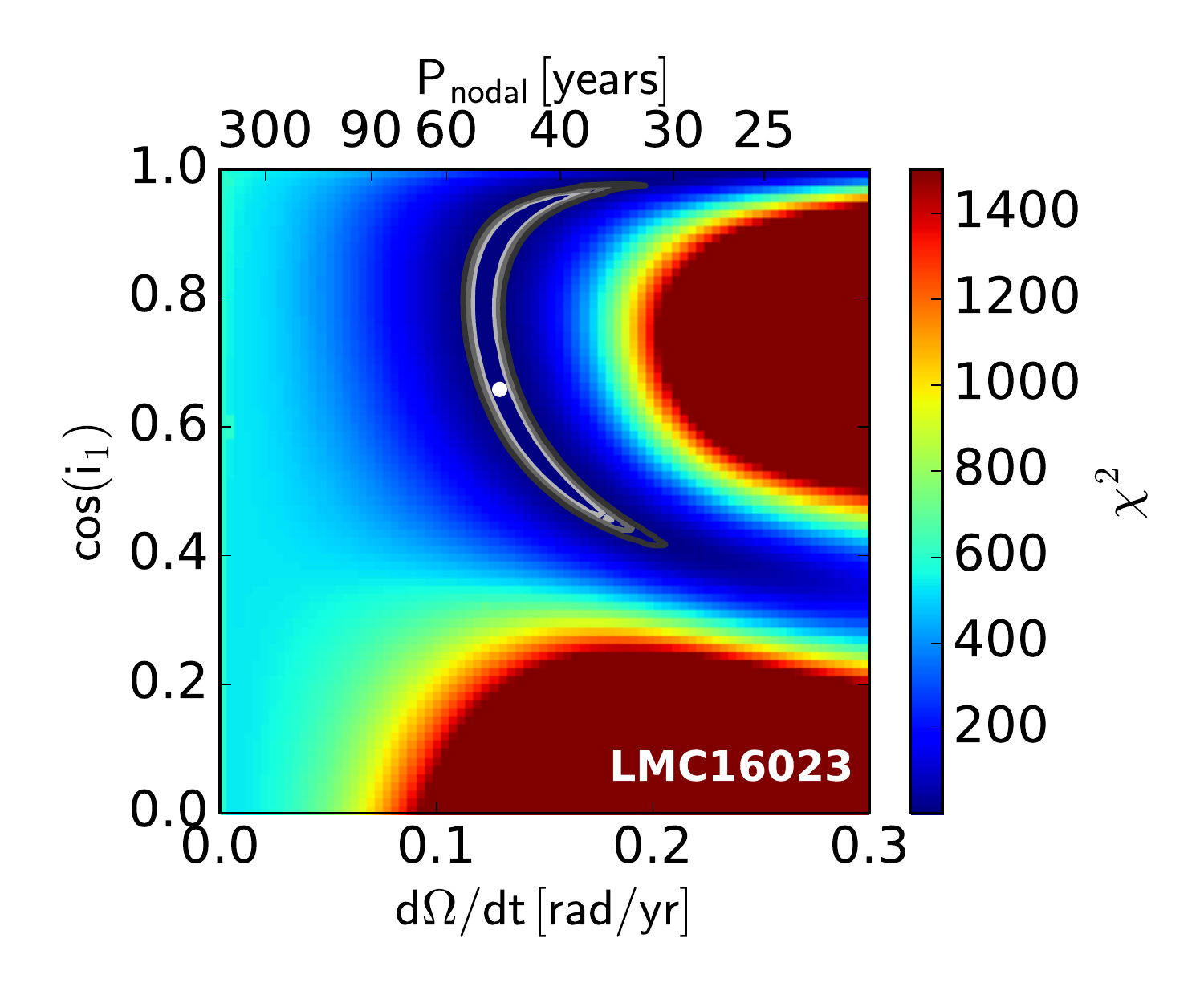} &
                \includegraphics[width=87mm]{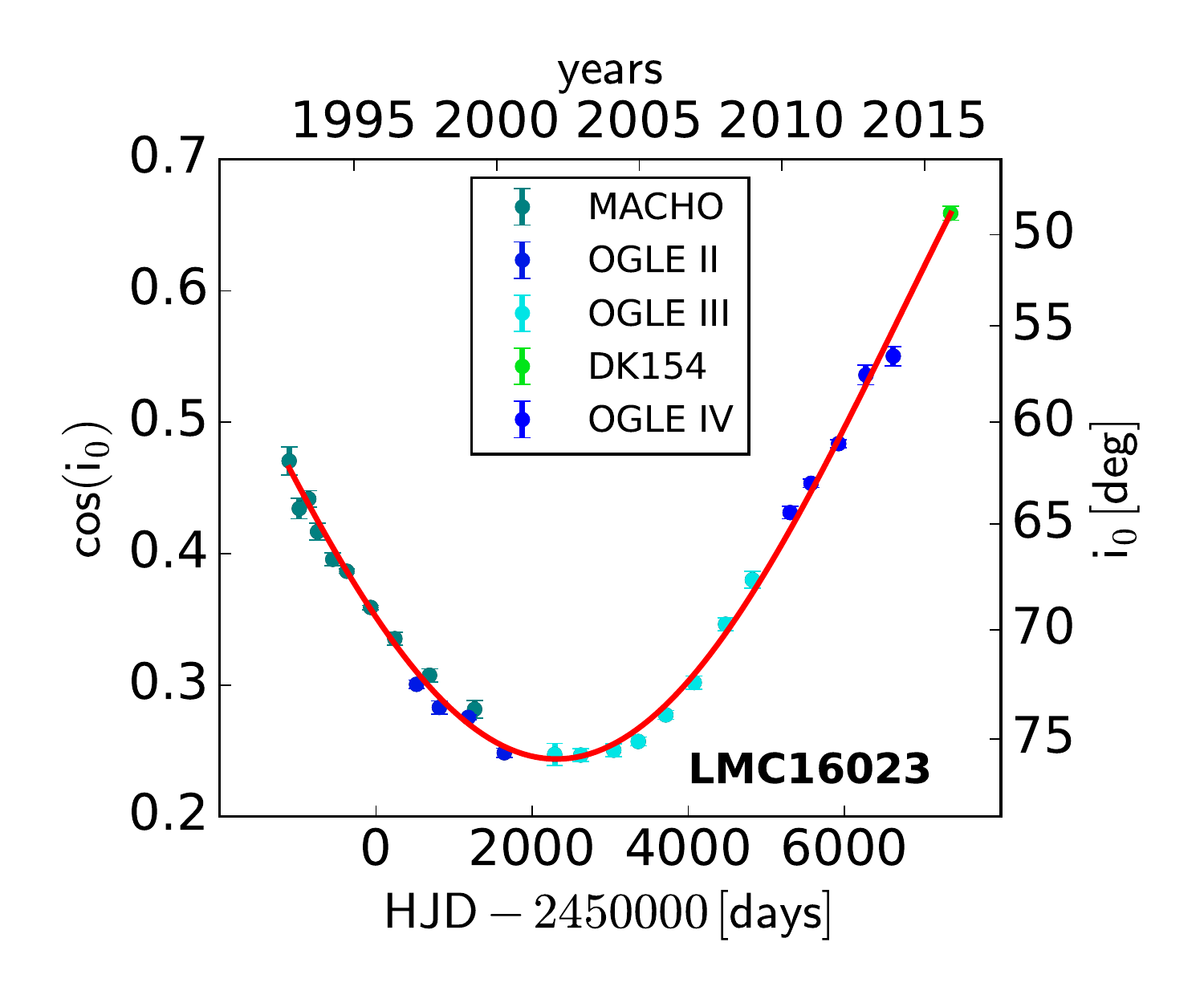} \\
        \end{tabular}
  \caption{\textit{Left:} Projections of $\chi^2$ of fitted dependence $\cos (i_0) = f(t)$ to $\cos(i_1)$ -- $\dot\Omega$ plane. 
  Confidence intervals $68.3 \, \%$, $90.0 \, \%$ and $99.0 \, \%$ are indicated with gray lines. White points show the best solutions whose
  parameters are listed in Tables~\ref{tab.pnodal_lmc} and \ref{tab.pnodal_smc}.
  \textit{Right:} The best solution of dependence $\cos (i_0) = f(t)$ according to relation (\ref{rov.soderhjelm}).}
   \label{fig.incl}
\end{figure*}

\begin{figure*}
\ContinuedFloat 
\centering
        \begin{tabular}{@{}cc@{}}
                \includegraphics[width=87mm]{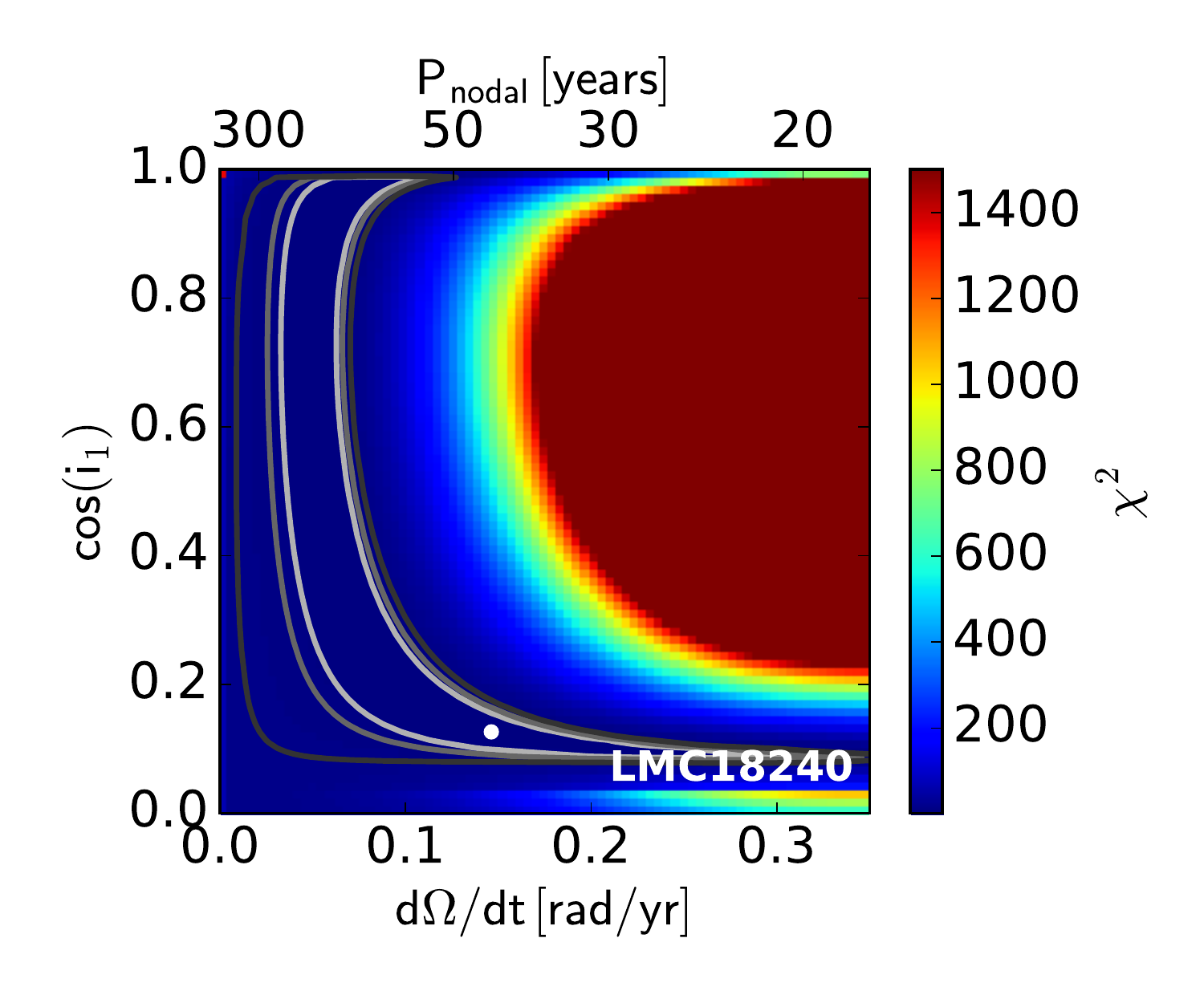} &
                \includegraphics[width=87mm]{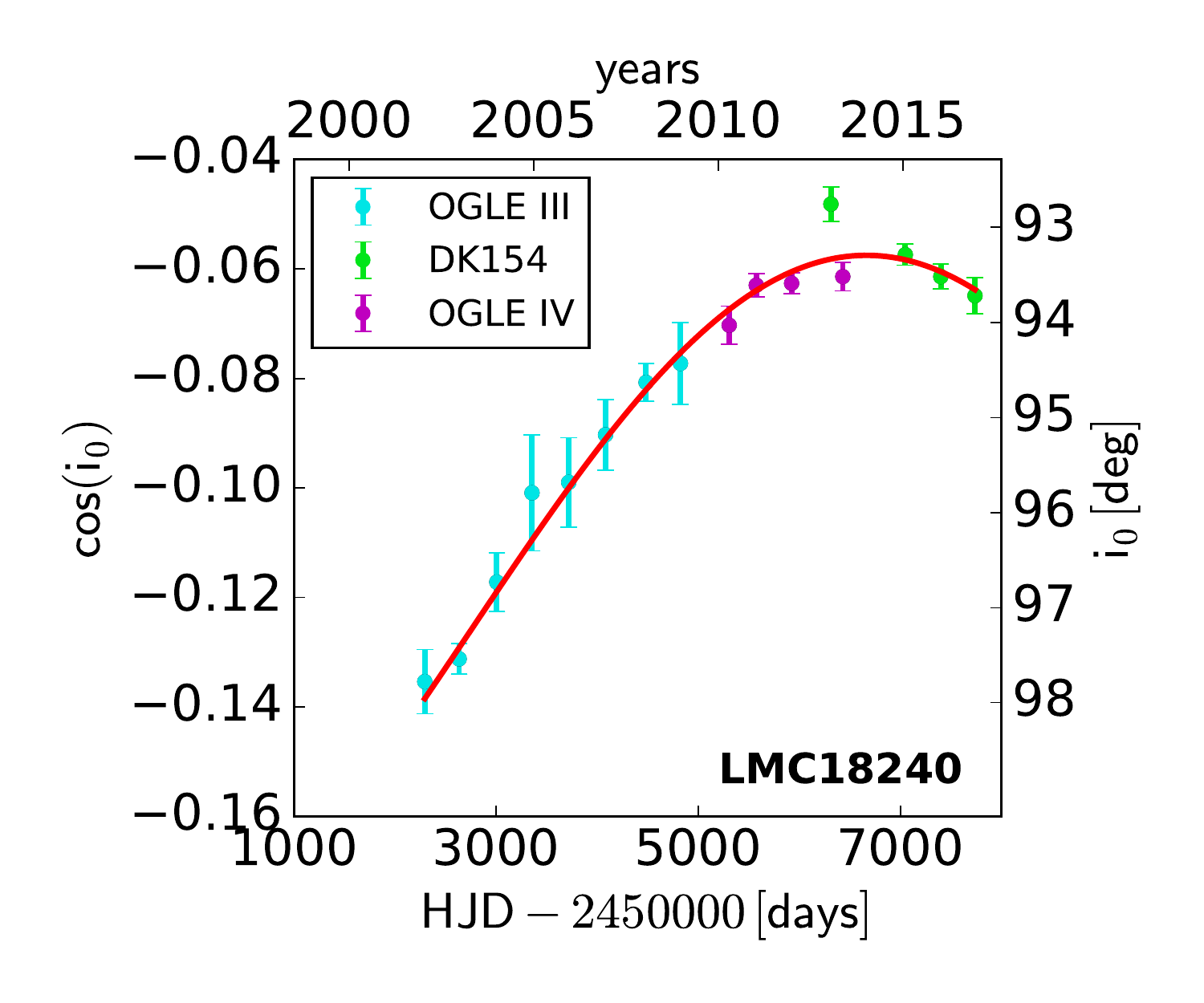} \\
                \includegraphics[width=87mm]{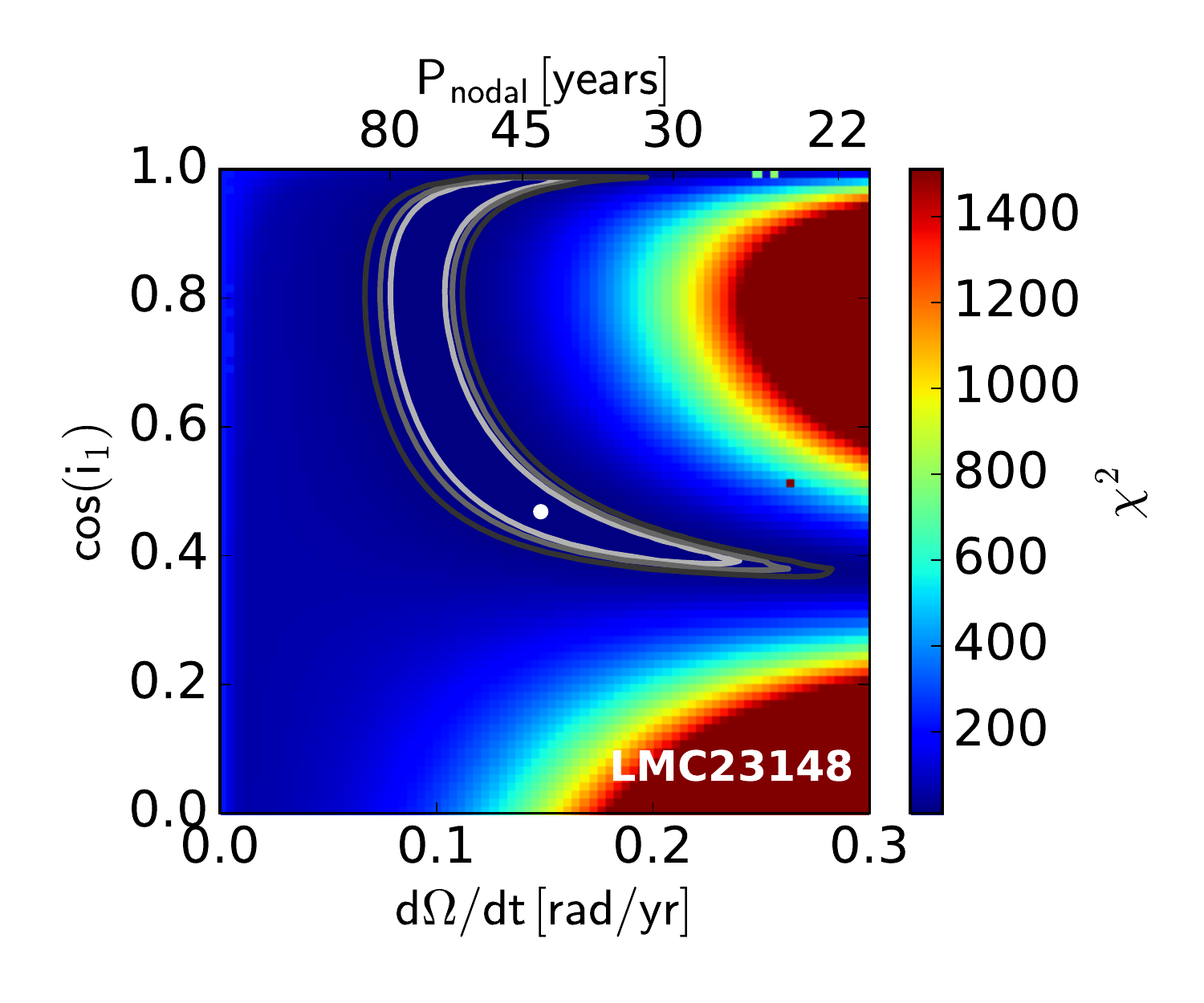} &
                \includegraphics[width=87mm]{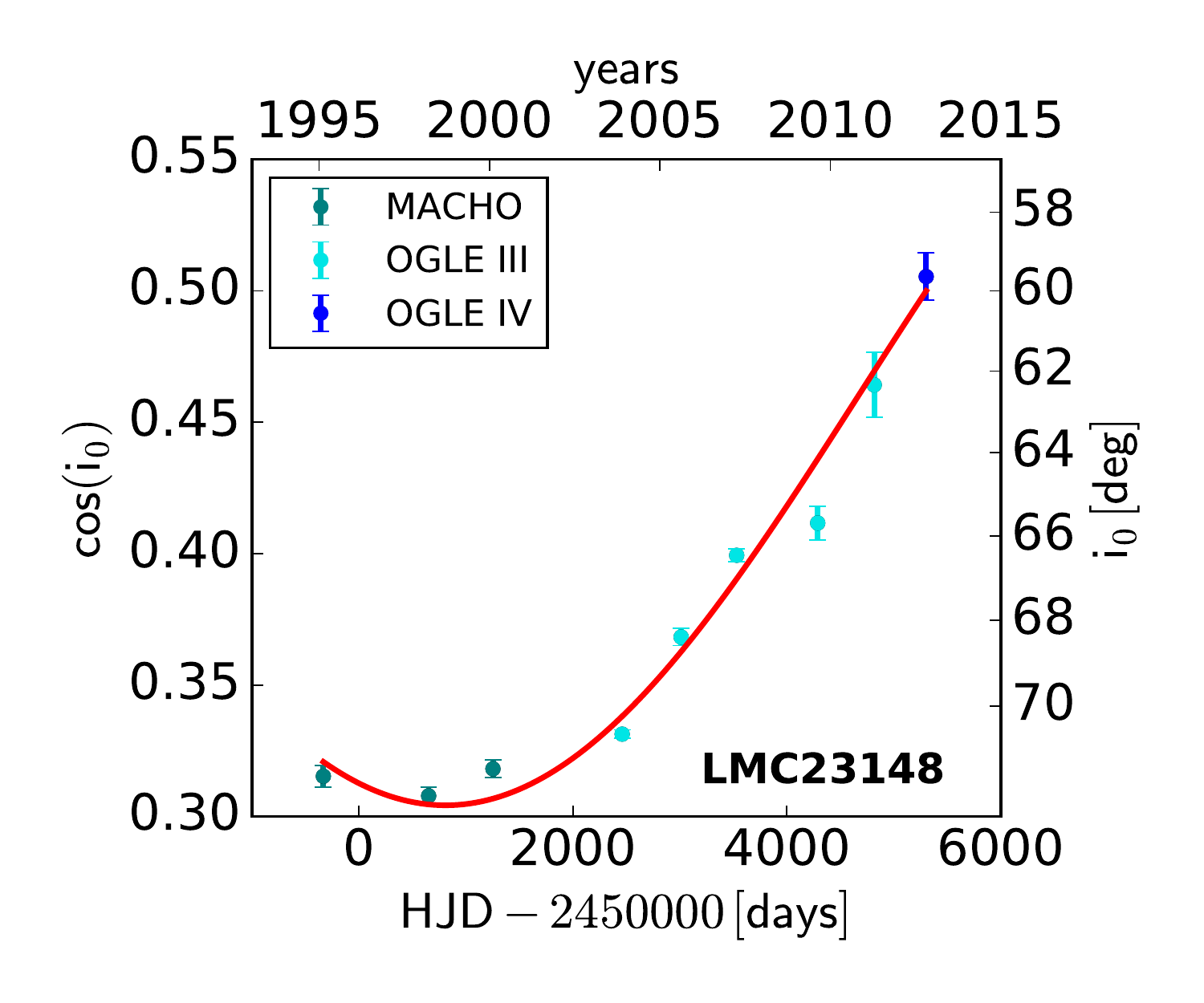} \\
                \includegraphics[width=87mm]{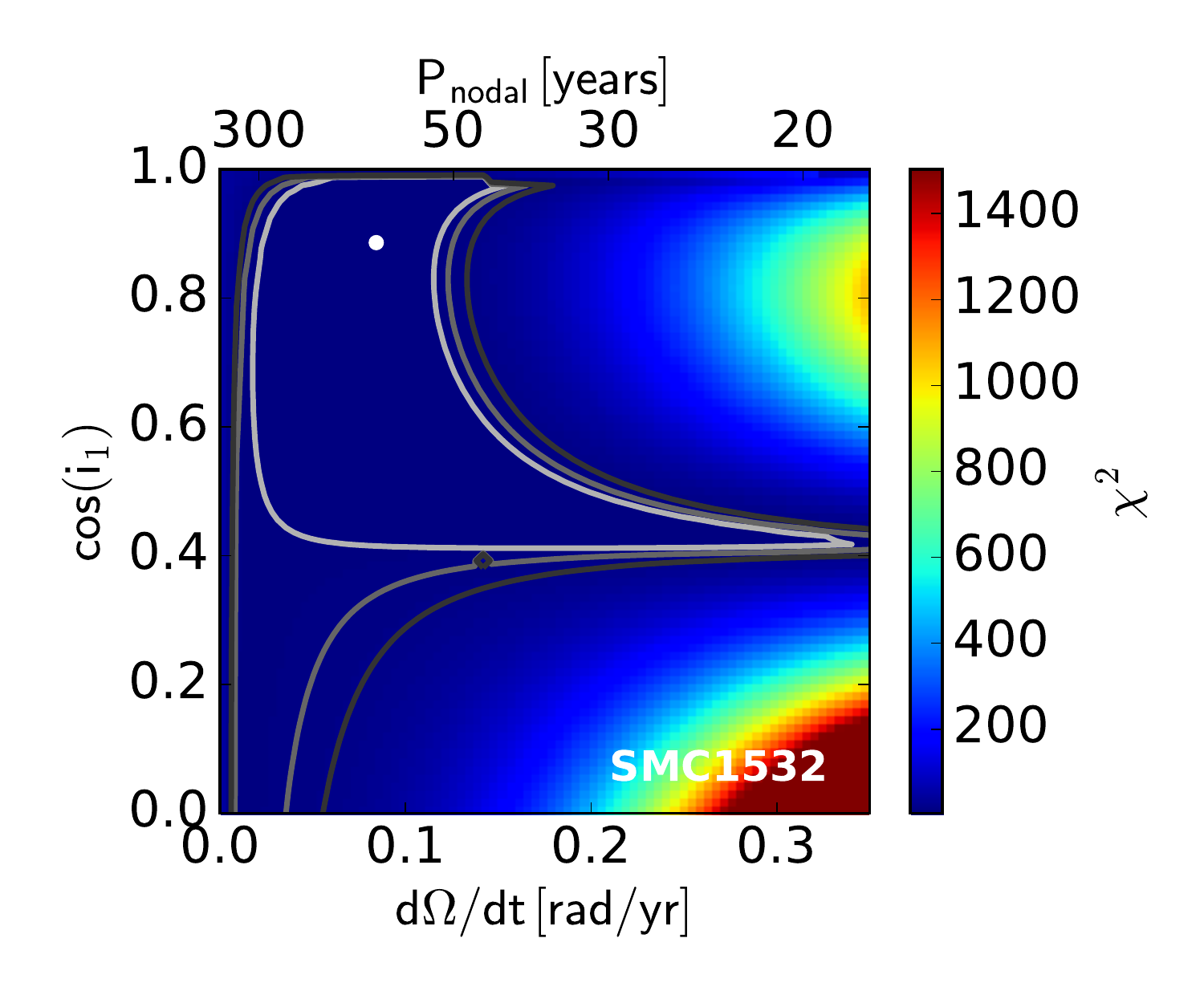} &
                \includegraphics[width=87mm]{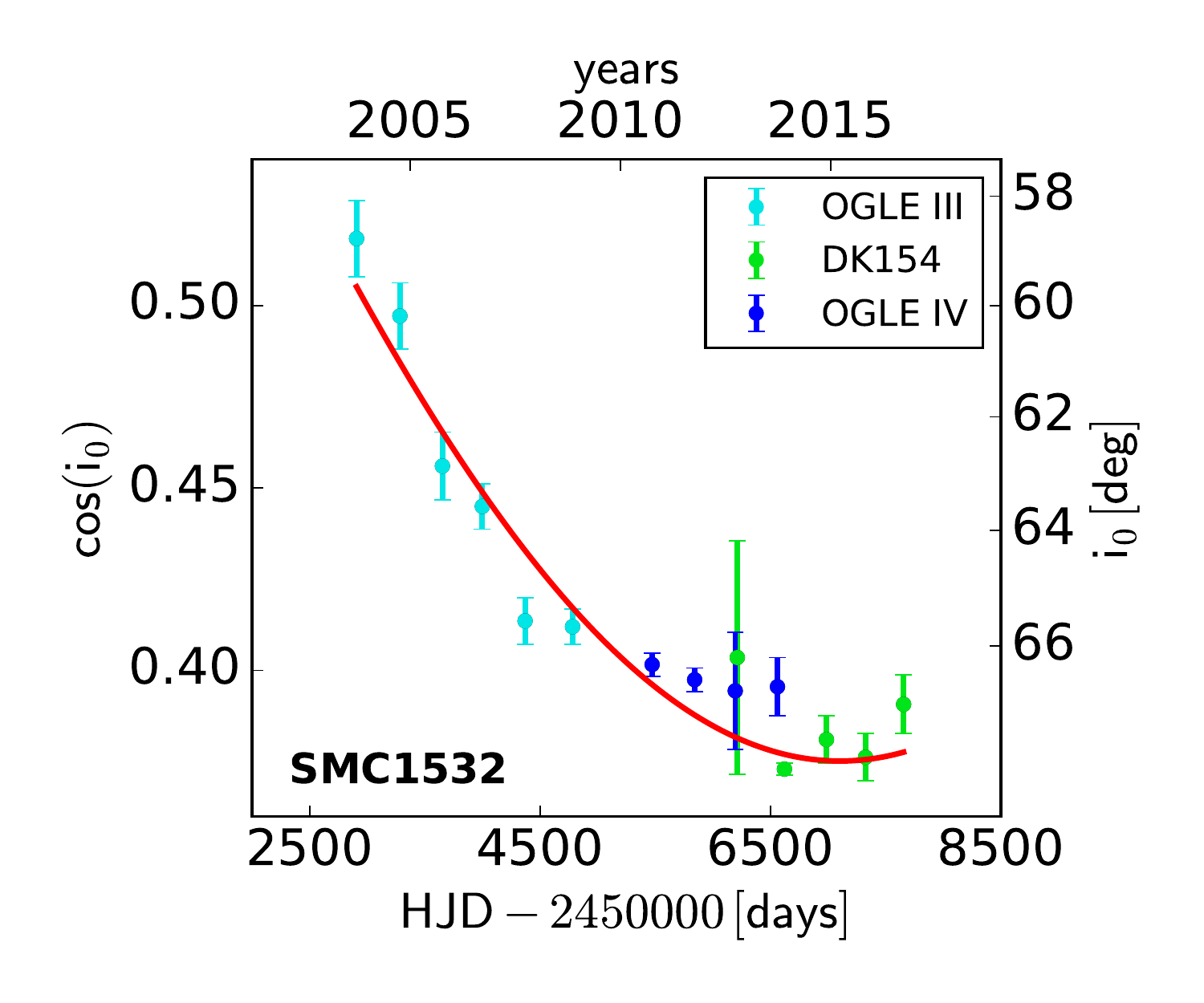} \\
        \end{tabular}
  \caption{\textit{continued}}
\end{figure*}

\begin{figure*}
\ContinuedFloat 
\centering
        \begin{tabular}{@{}cc@{}}
                \includegraphics[width=87mm]{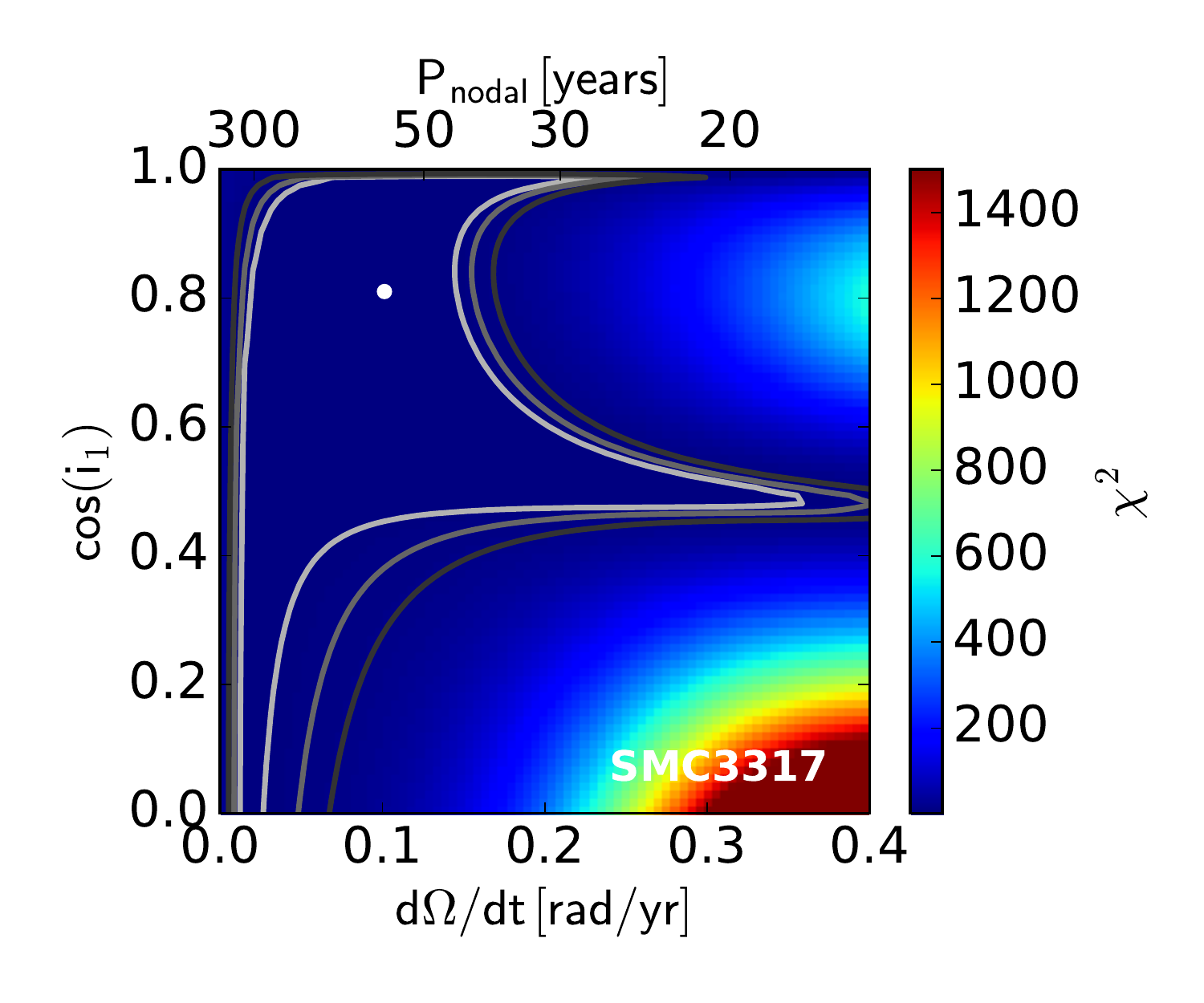} &
                \includegraphics[width=87mm]{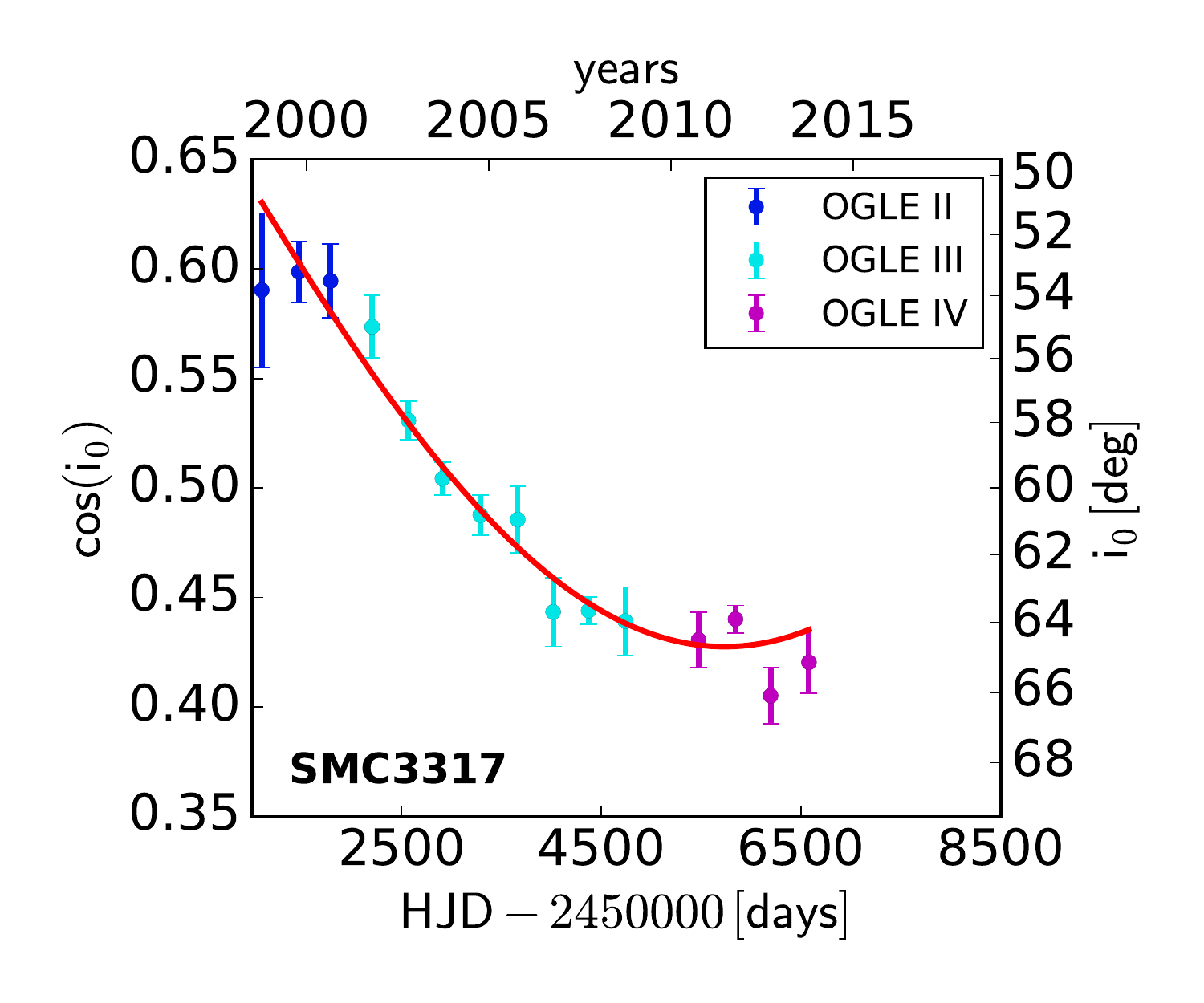} \\
                \includegraphics[width=87mm]{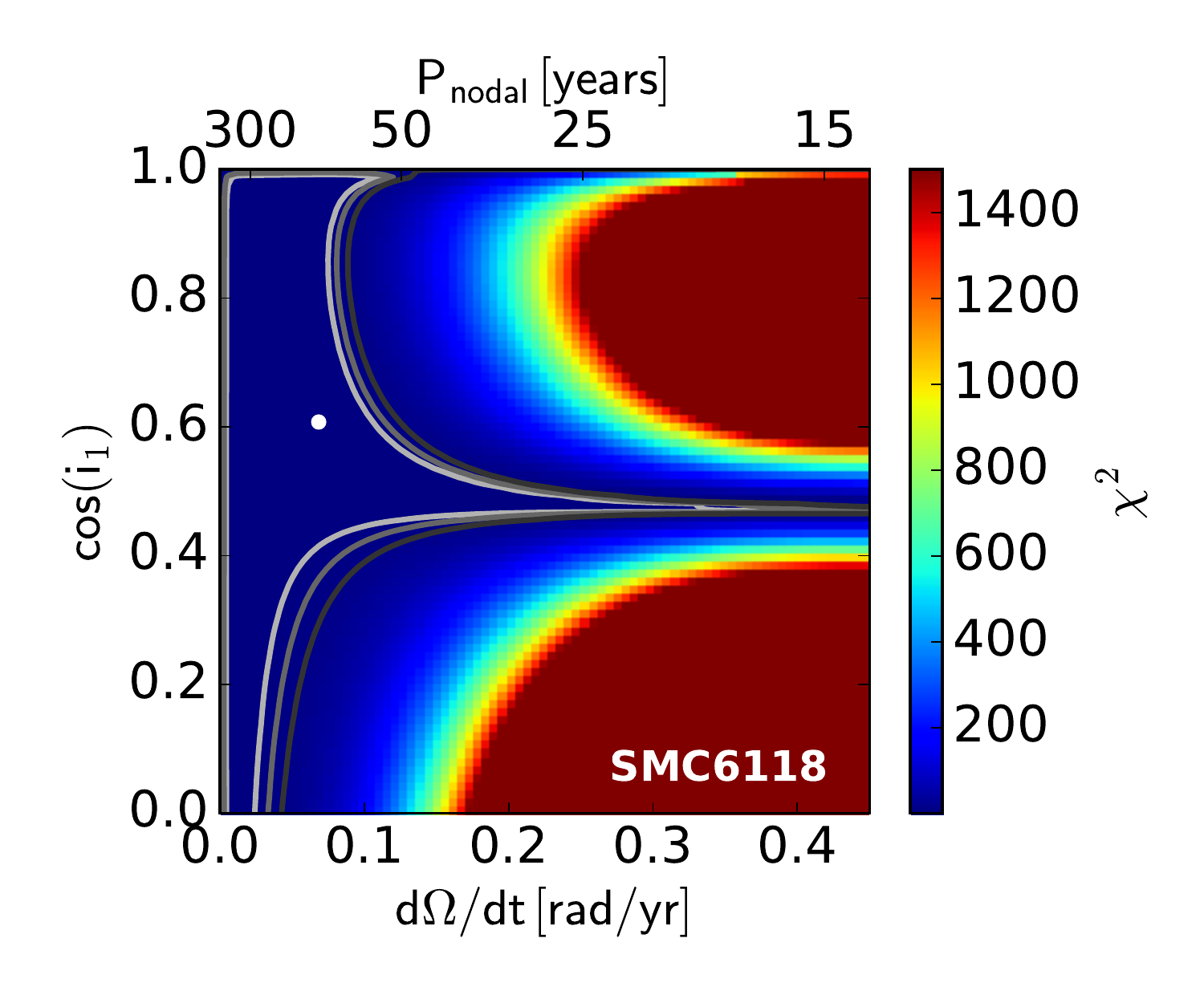} &
                \includegraphics[width=87mm]{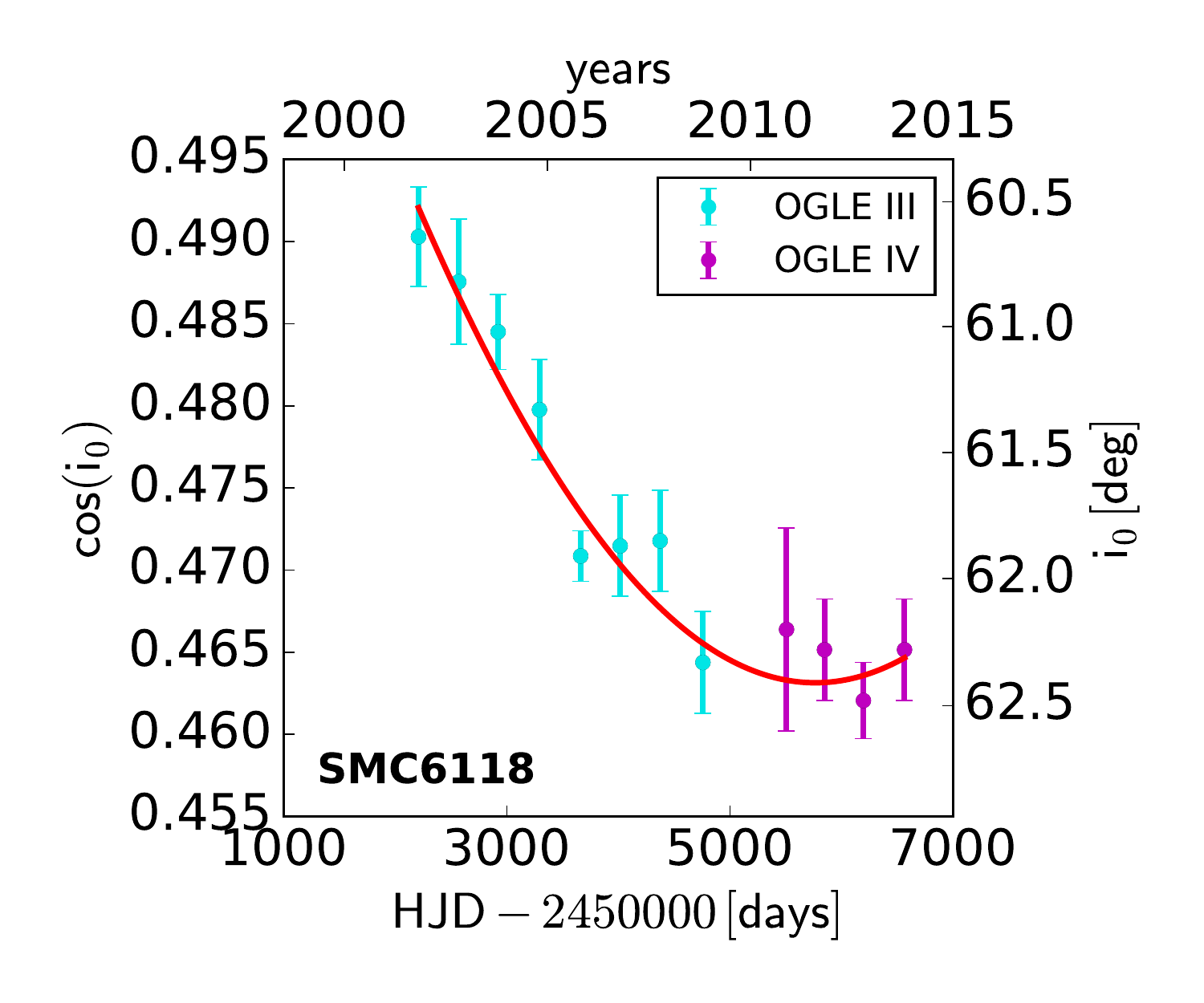} \\
        \end{tabular}
  \caption{\textit{continued}}
\end{figure*}

\FloatBarrier
\section{Possible masses and periods of third bodies}

\begin{figure*}[ht!]
\centering
        \begin{tabular}{@{}cc@{}}
                \includegraphics[width=87mm]{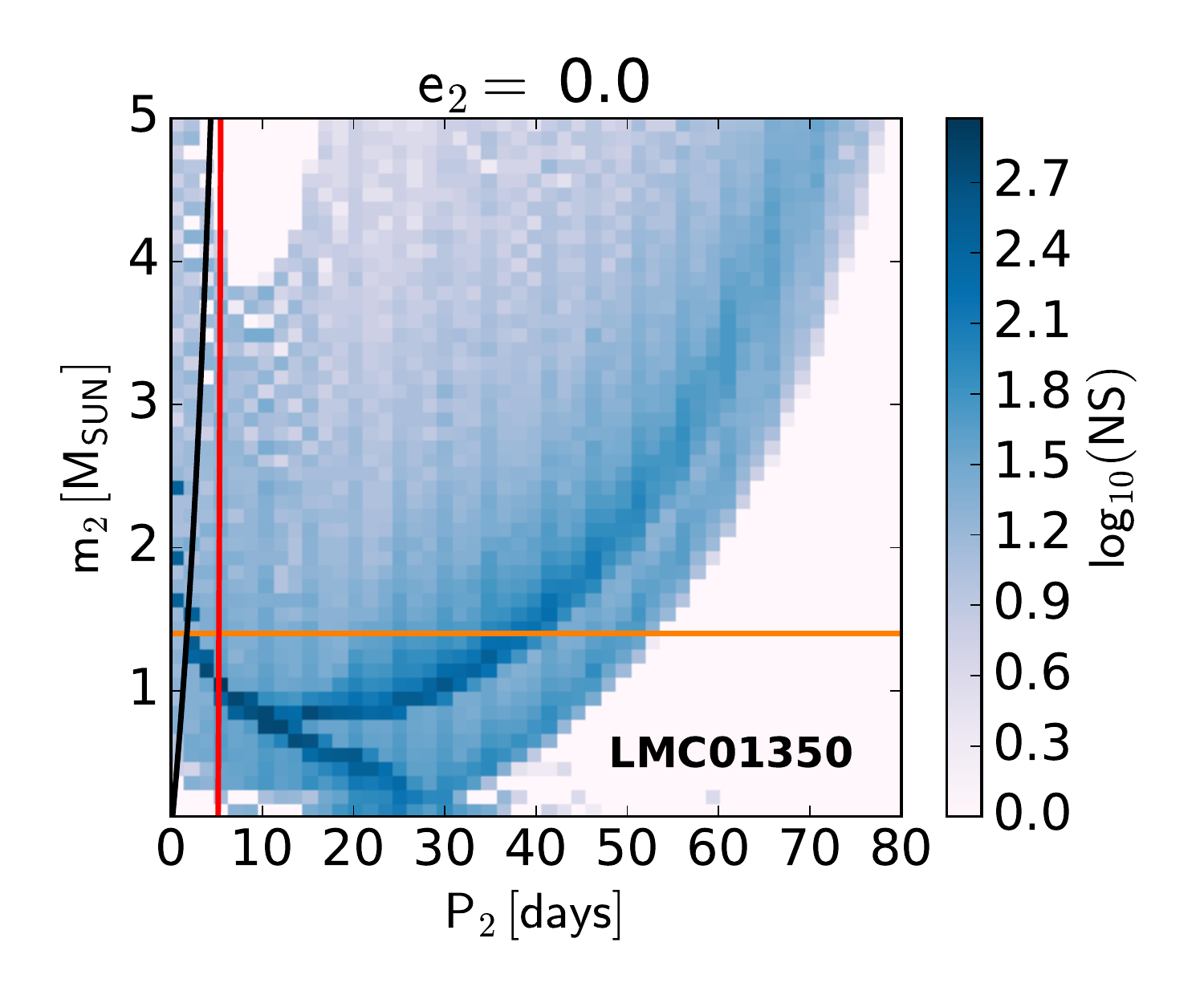} &
                \includegraphics[width=87mm]{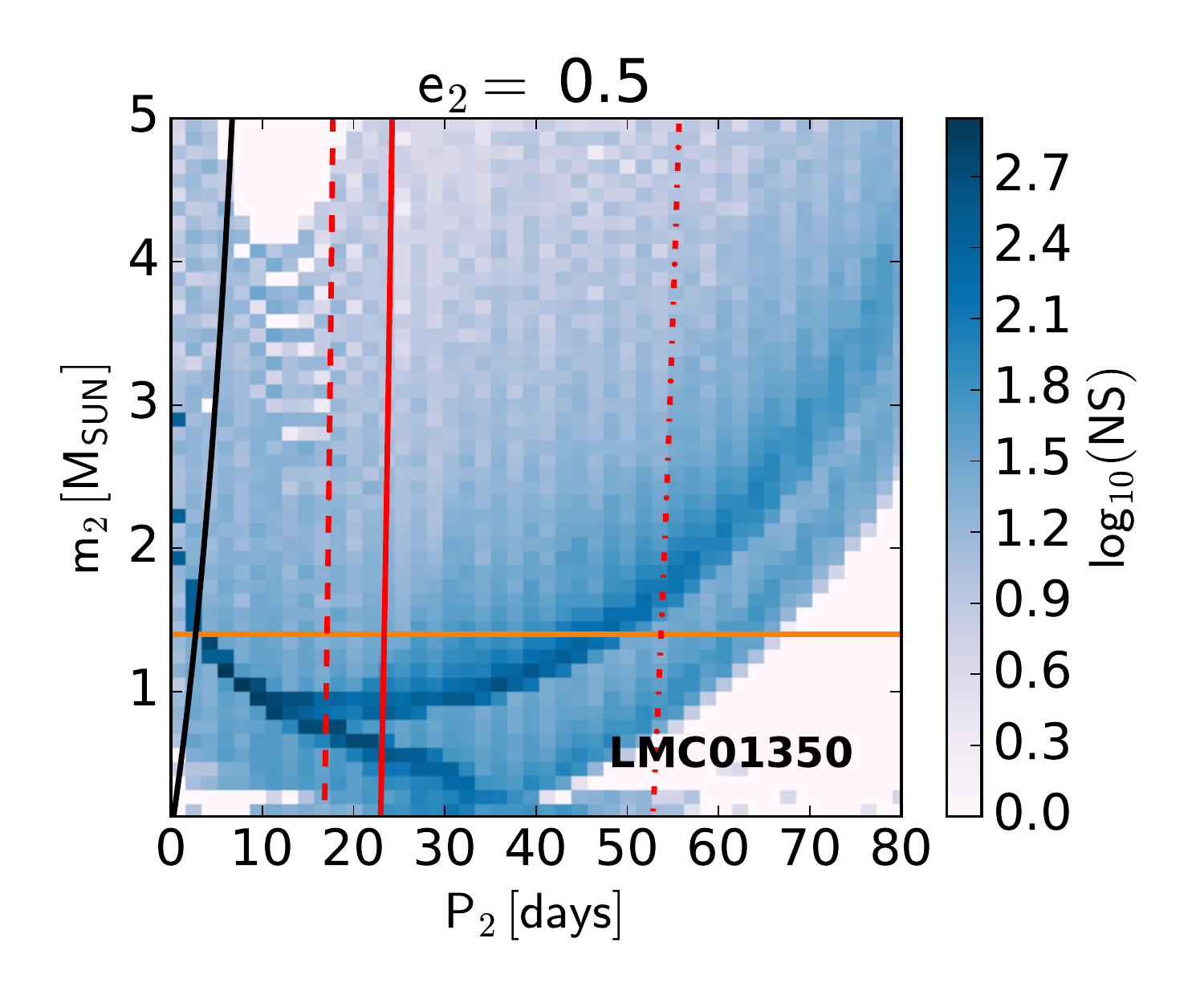} \\
                \includegraphics[width=87mm]{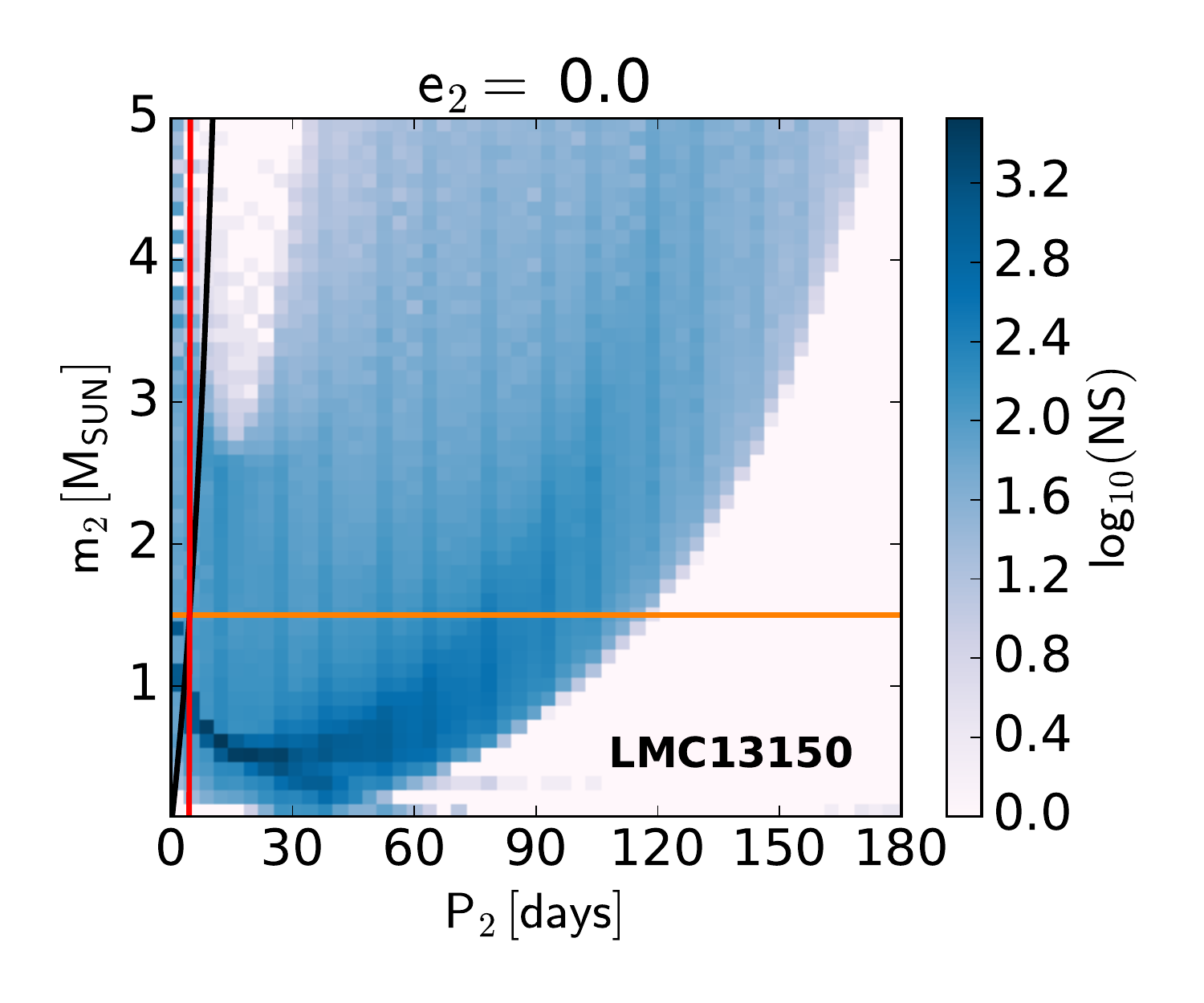} &
                \includegraphics[width=87mm]{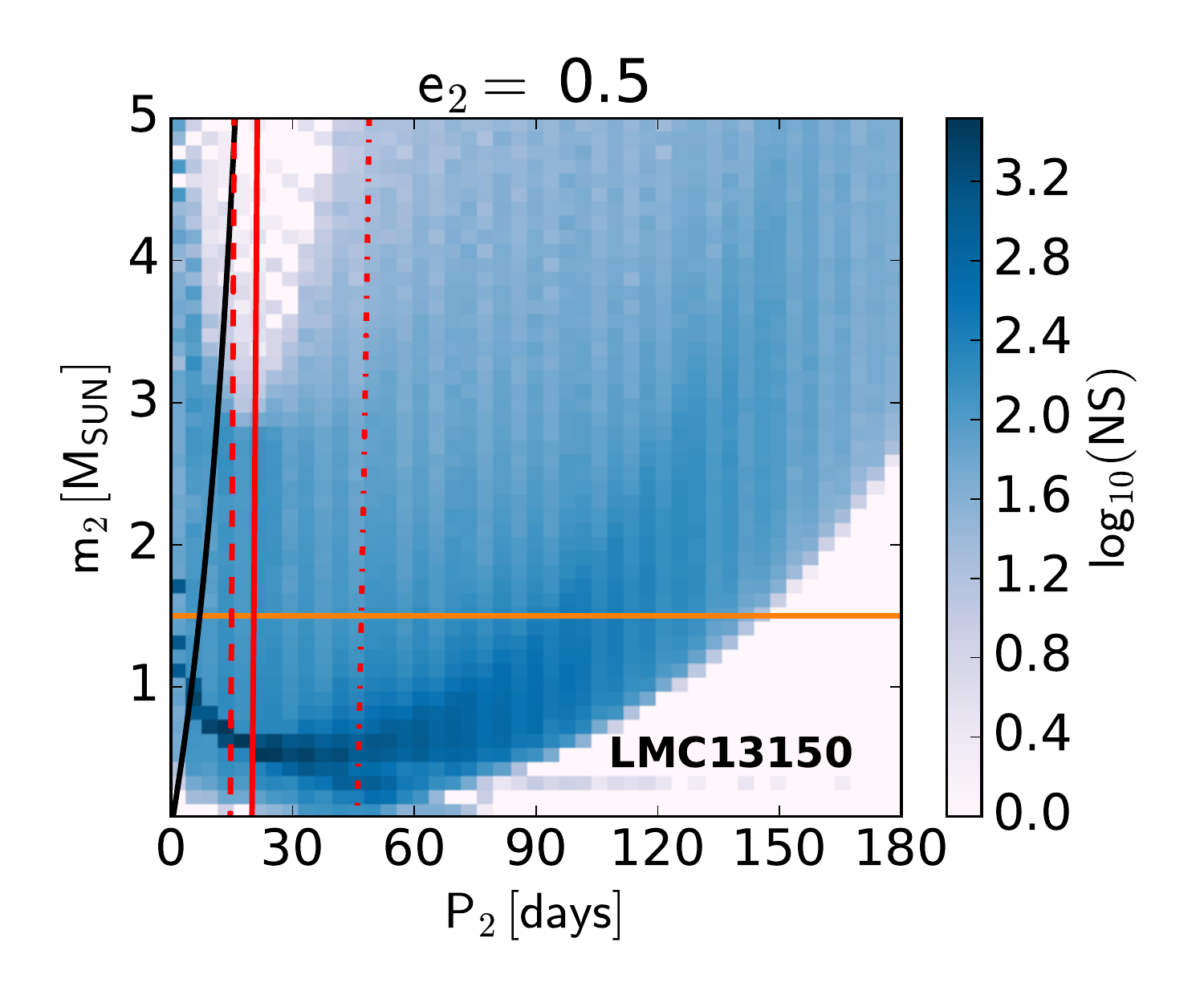} \\
                \includegraphics[width=87mm]{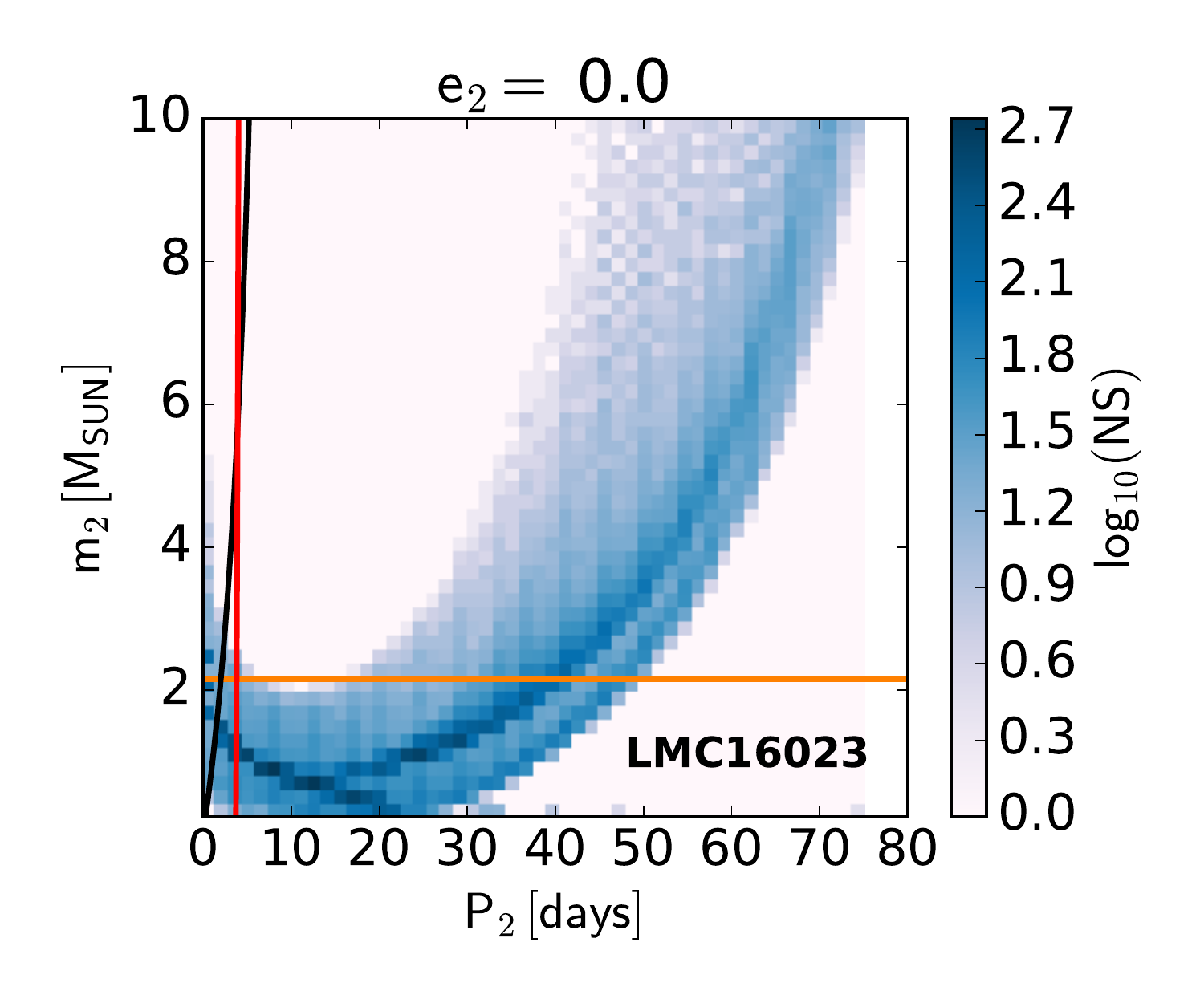} &
                \includegraphics[width=87mm]{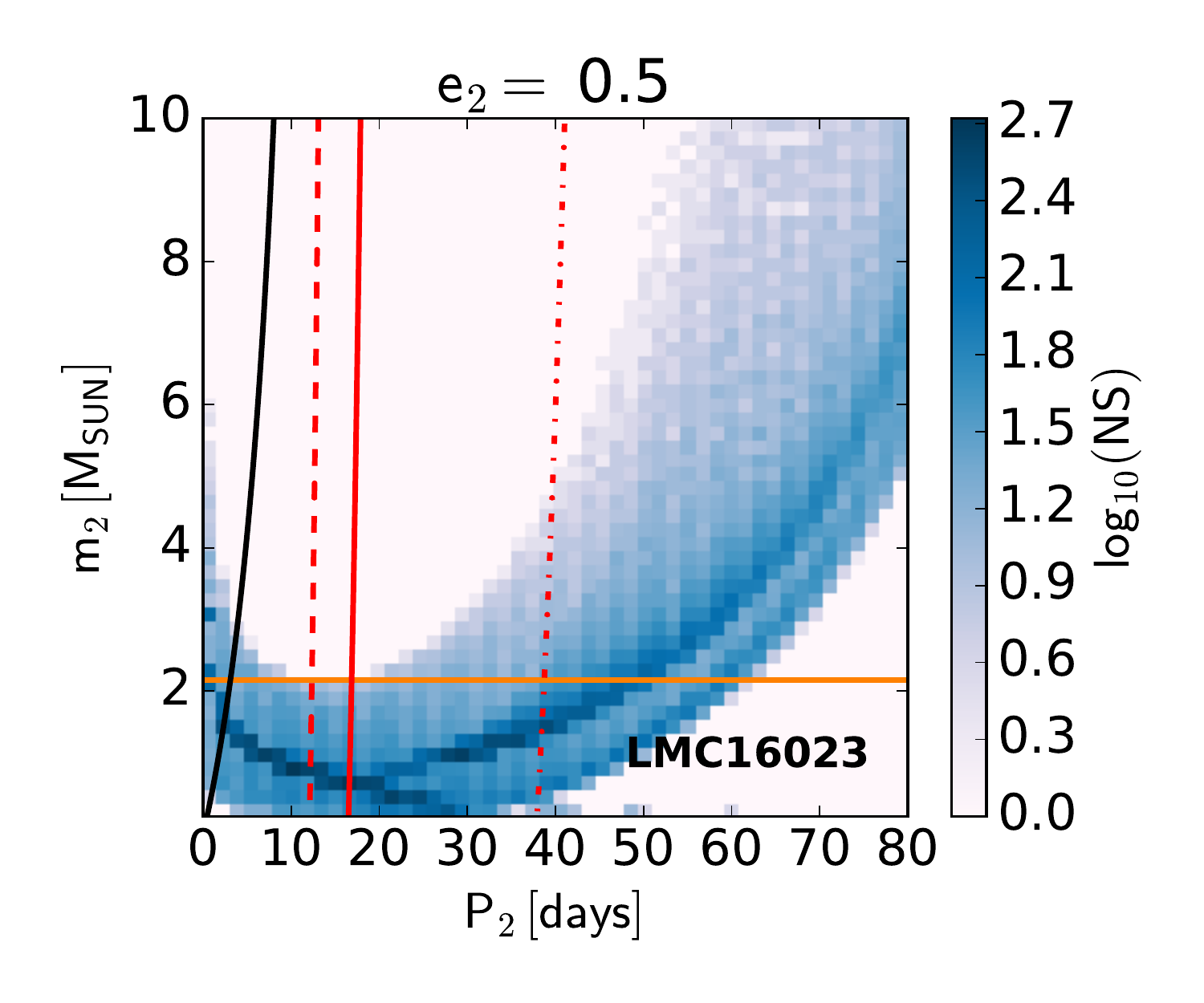} \\
        \end{tabular}
  \caption{Possible masses and periods of third bodies. All solutions from $68.3 \, \%$ area in the
  Fig.~\ref{fig.incl} for possible orientations $i_1 \rightarrow \pi - i_1$ and $I \rightarrow \pi - I$ are shown. 
  Logarithmic scale of a number of solutions (NS) is shownin blue. The orange line indicates the maximal third body mass according 
  to the limit of a third light from the LC solution. \textit{(Continued on next page.)}}
   \label{fig.m2p2}
\end{figure*}

\begin{figure*}
\ContinuedFloat 
\centering
        \begin{tabular}{@{}cc@{}}
                \includegraphics[width=87mm]{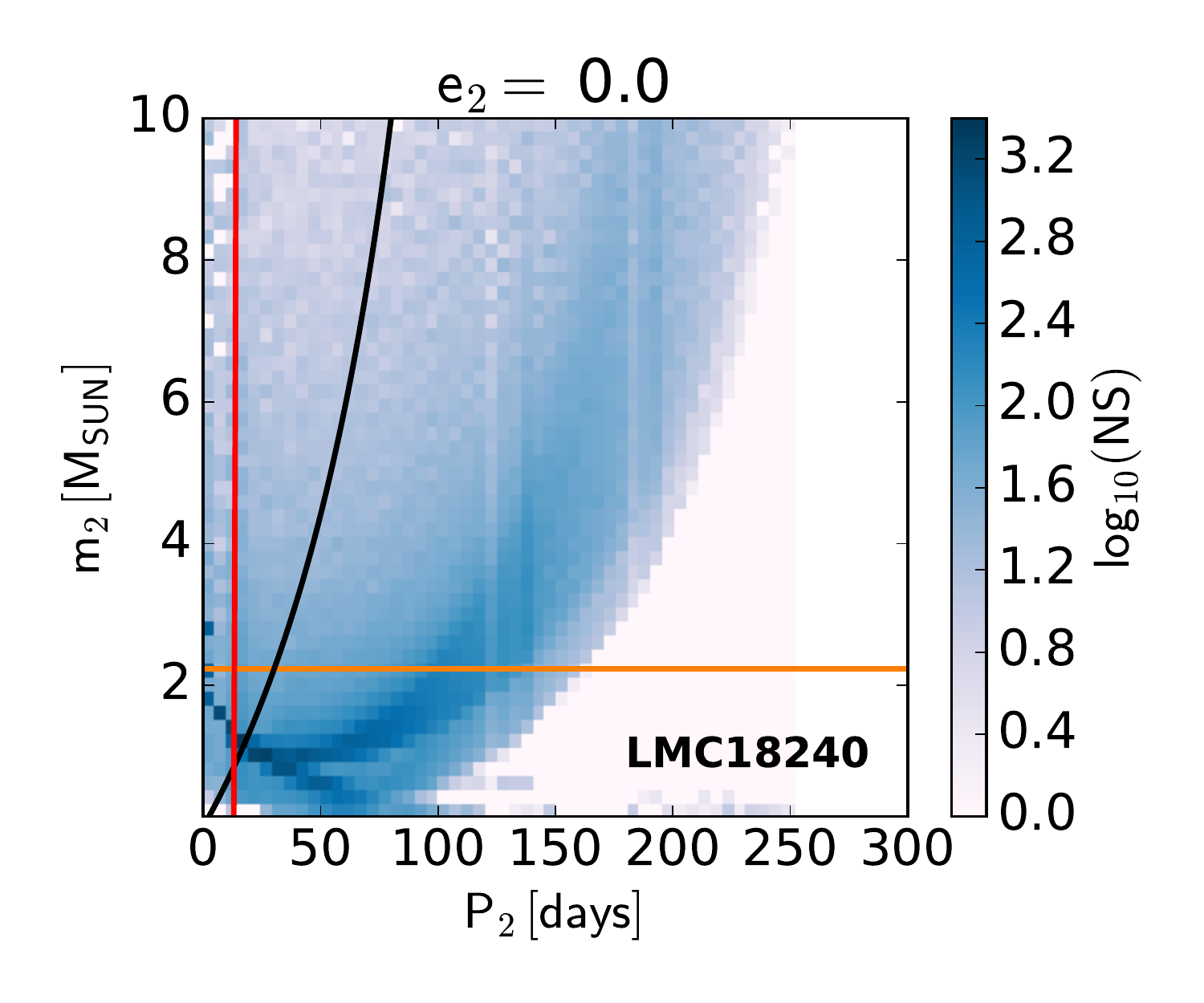} &
                \includegraphics[width=87mm]{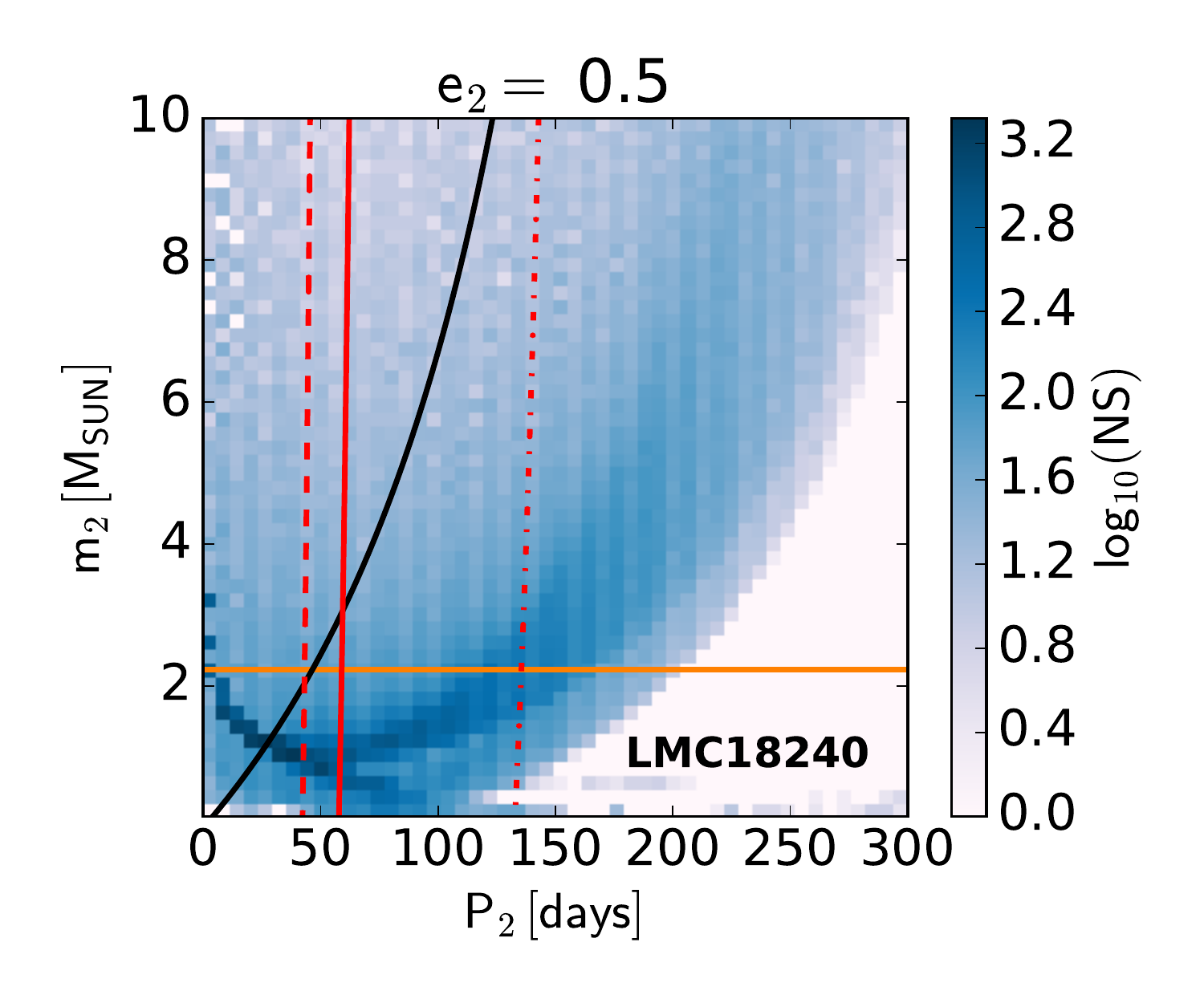} \\
                \includegraphics[width=87mm]{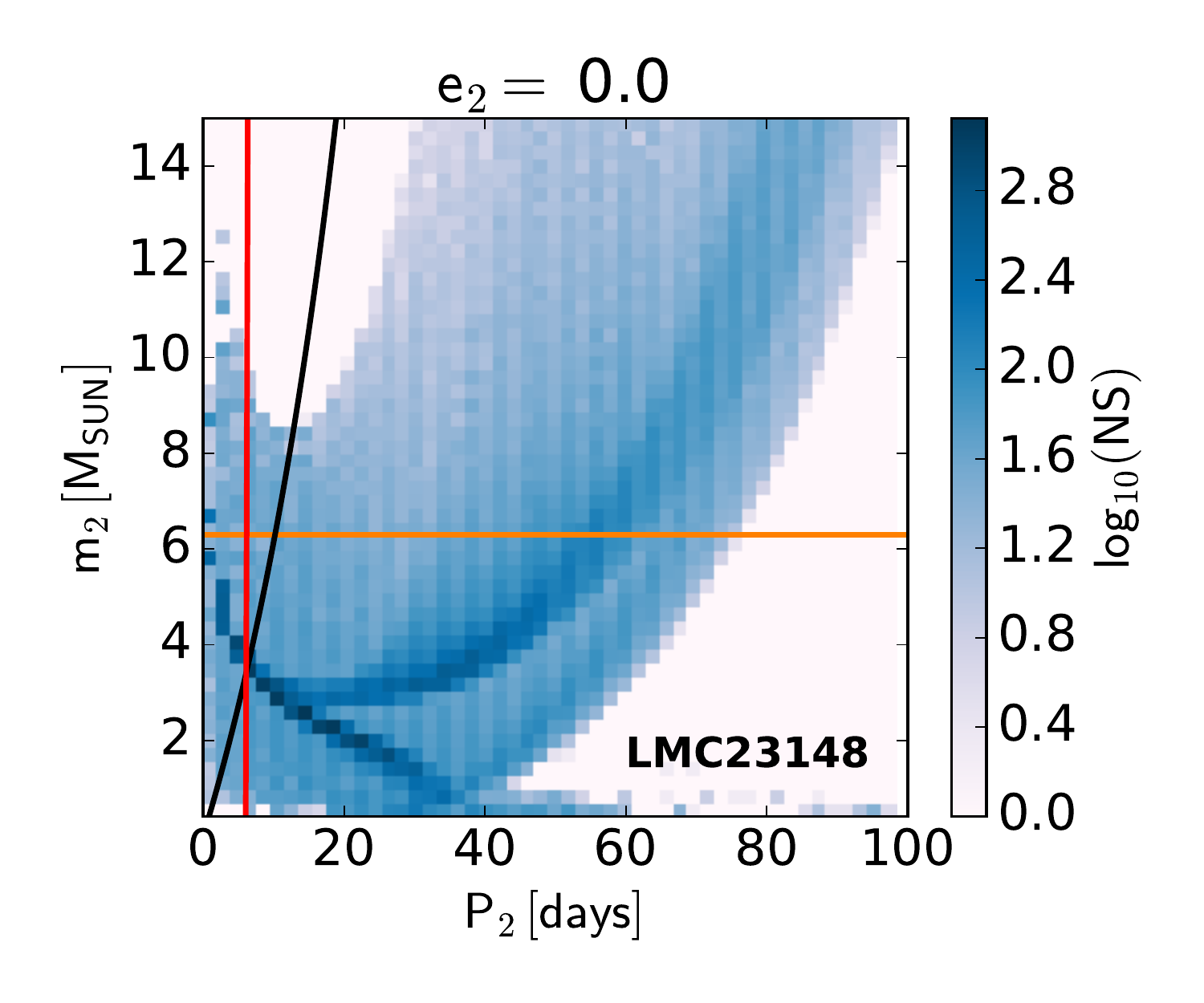} &
                \includegraphics[width=87mm]{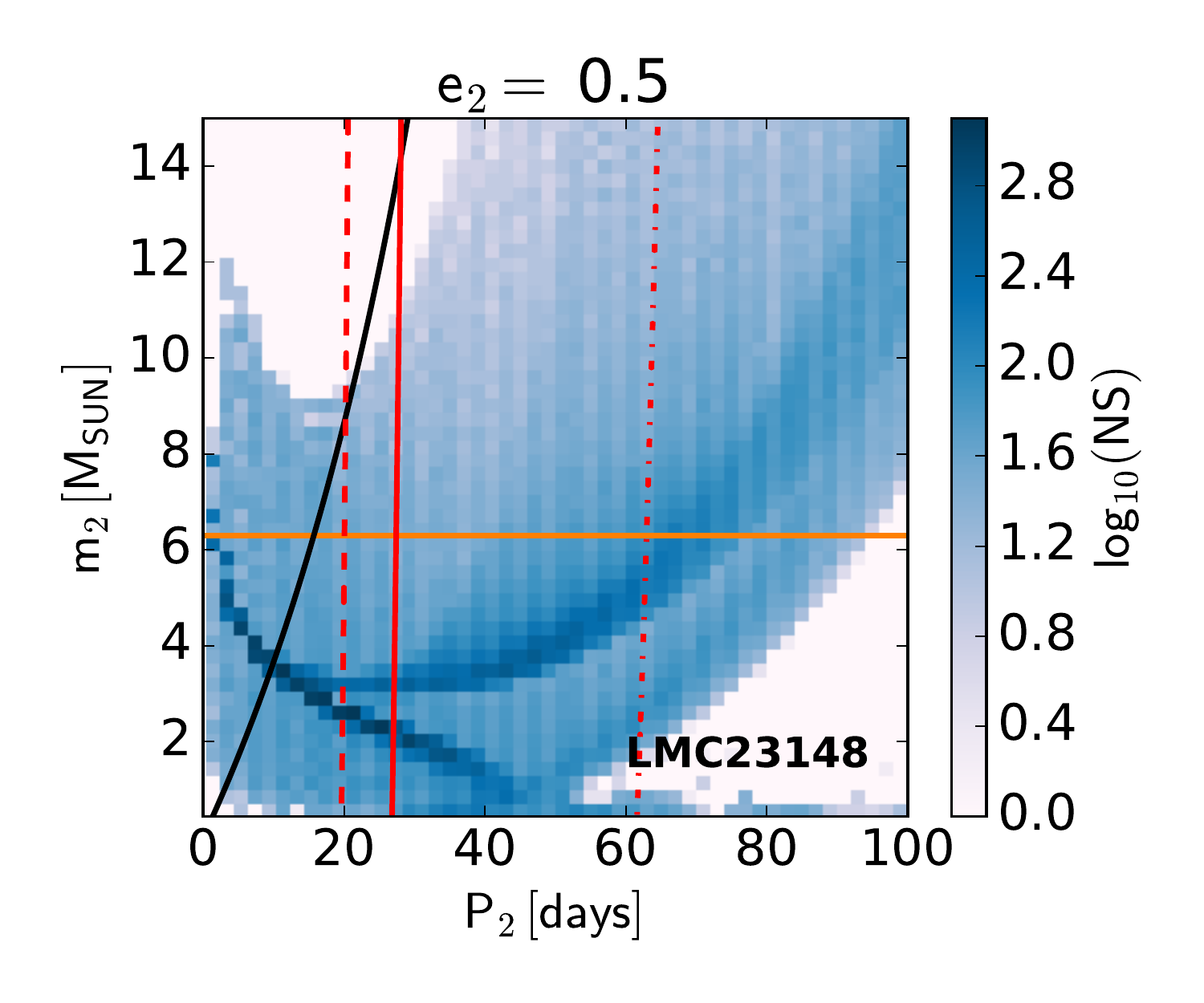} \\
                \includegraphics[width=87mm]{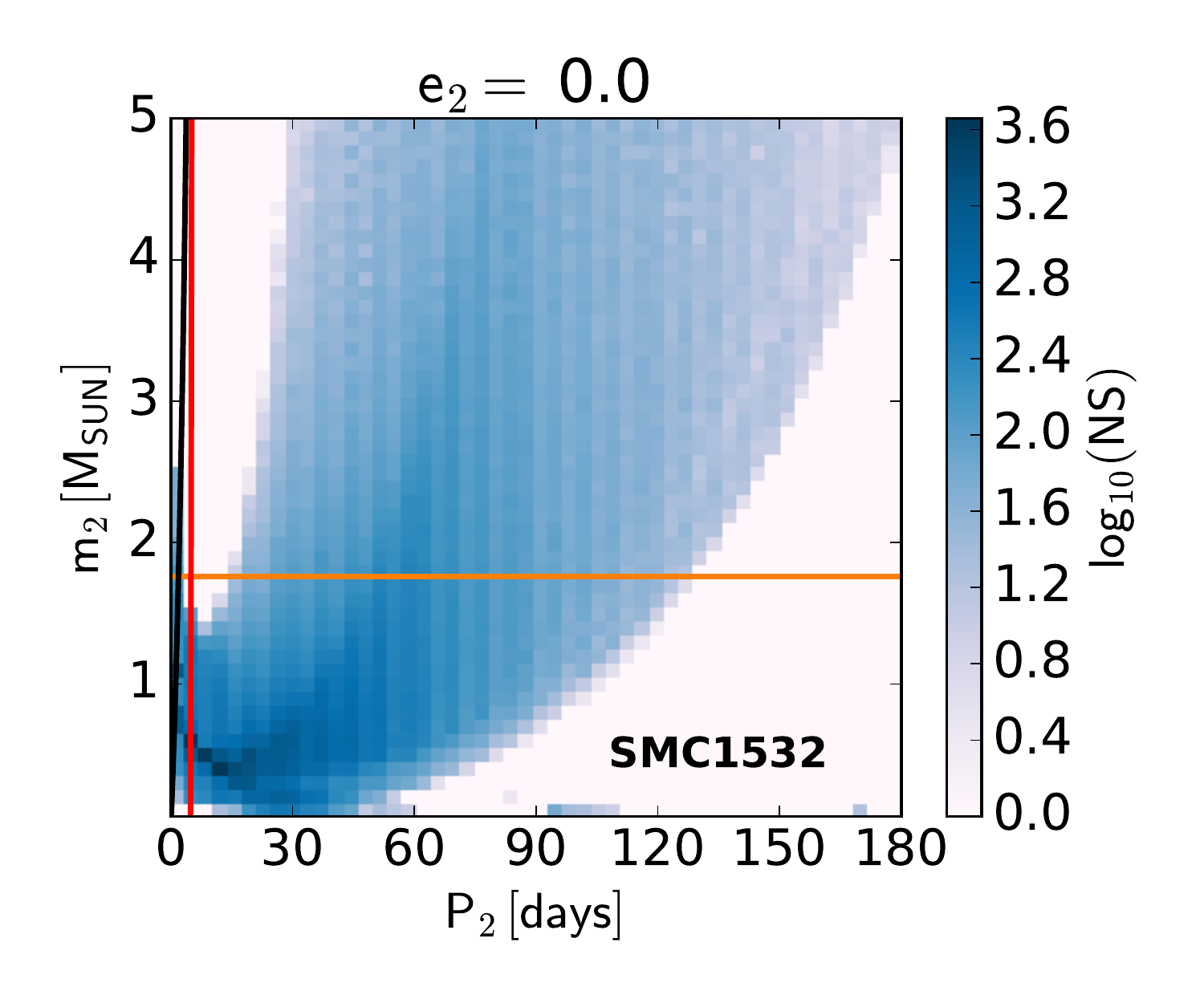} &
                \includegraphics[width=87mm]{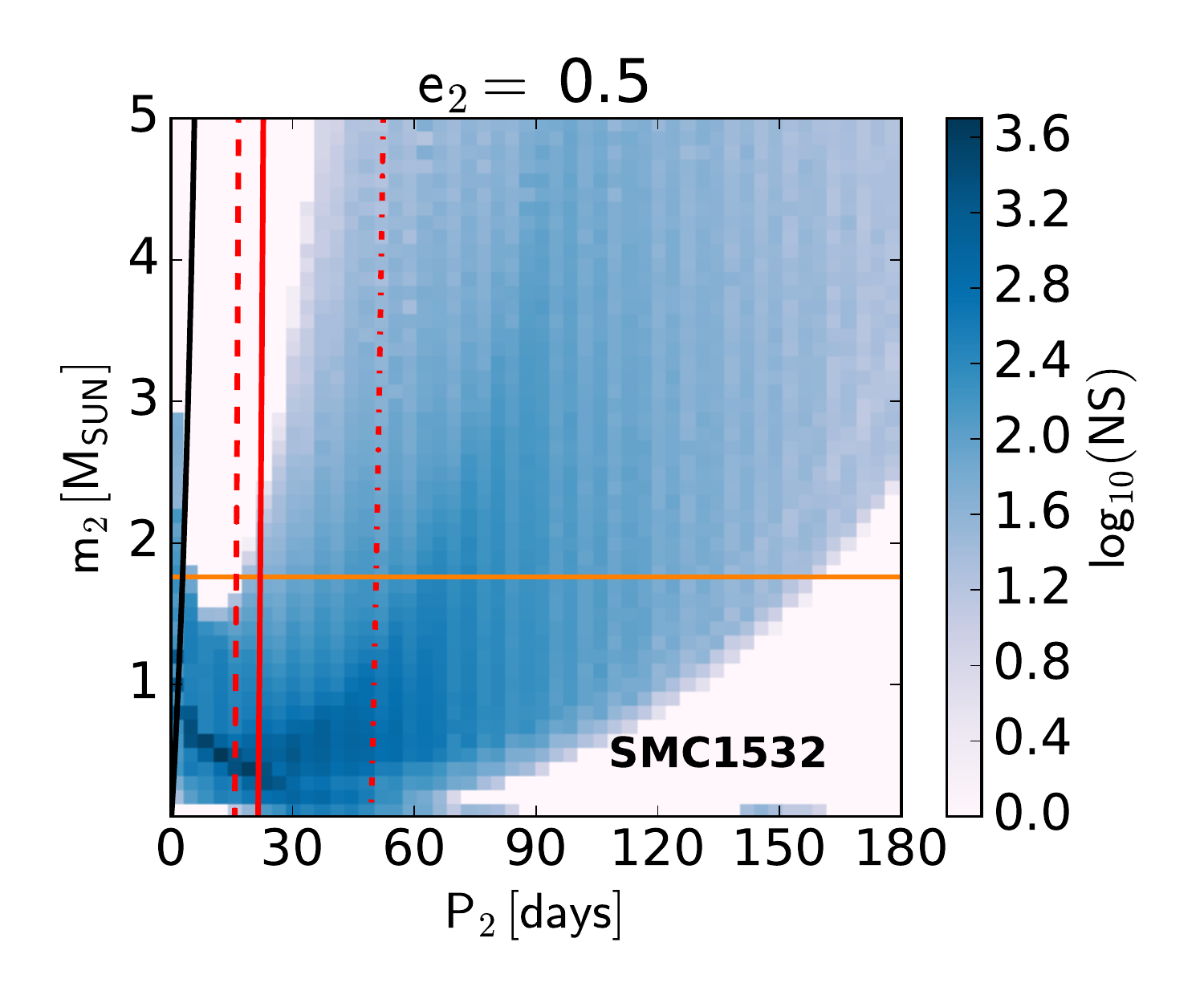} \\
        \end{tabular}
  \caption{\textit{(Continued from the previous page.)} Red solid line indicates the stability limit according to \citep{mardling_aarseth},
  dashed line indicates the stability limit according to \citep{sterzik_tokovinin2002} and dot-dashed line according to \citep{tokovinin2004}. 
  The blue line shows restrictions on ETVs. All figures are plotted for the extremal third body eccentricities $e_2 = 0$ and 
  $e_2 = 0.5$.}
\end{figure*}

\begin{figure*}
\ContinuedFloat 
\centering
        \begin{tabular}{@{}cc@{}}
                \includegraphics[width=87mm]{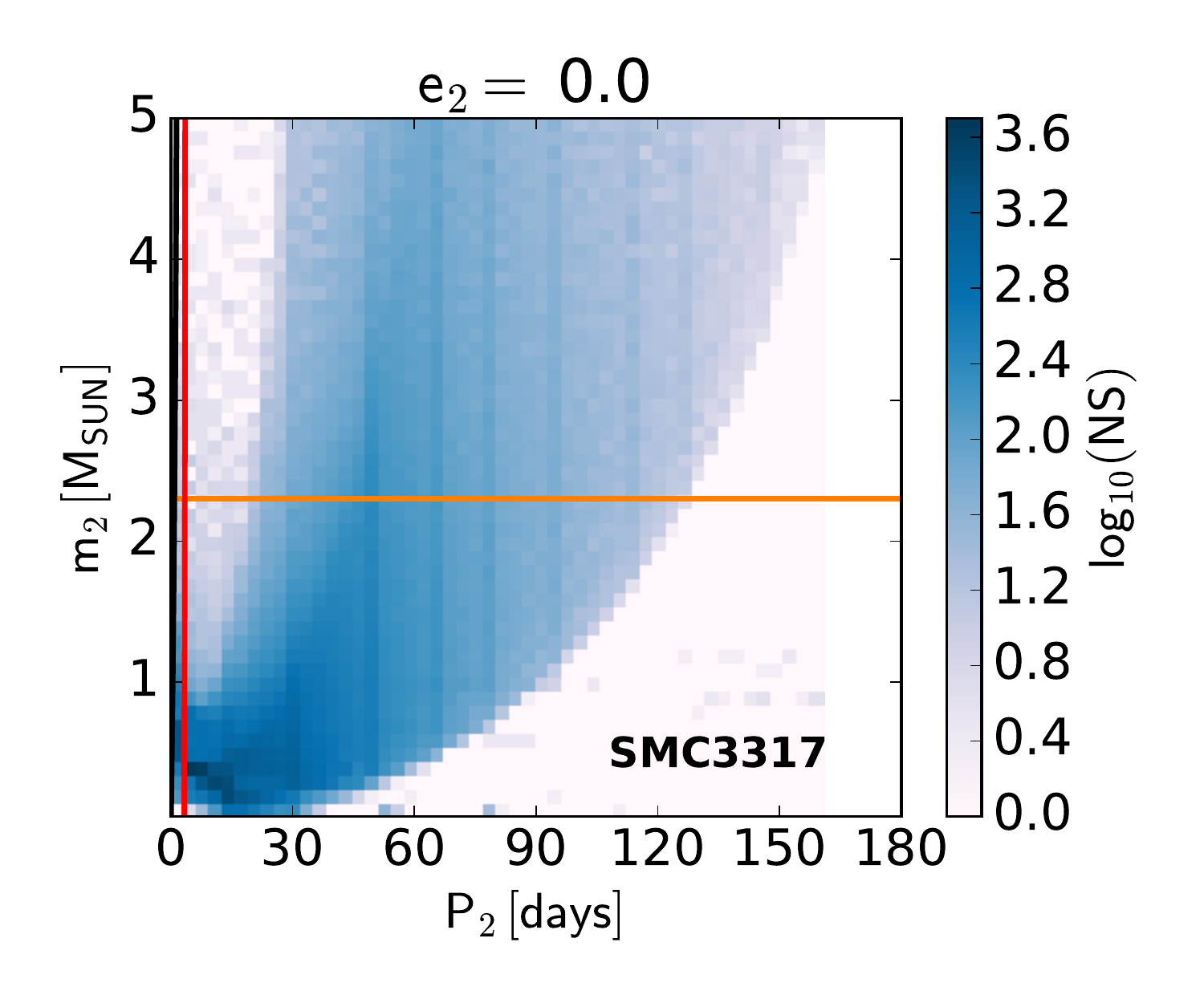} &
                \includegraphics[width=87mm]{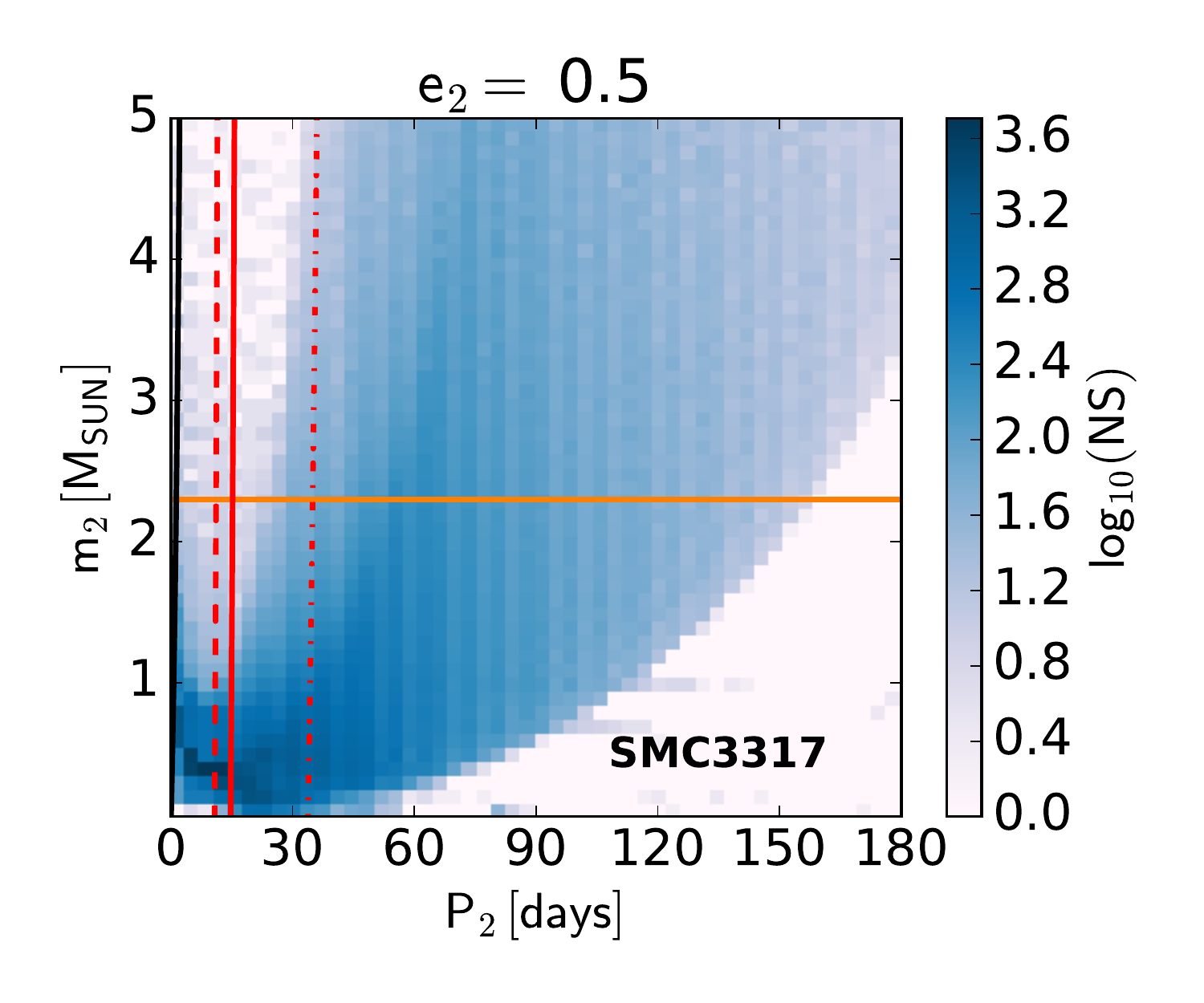} \\
                \includegraphics[width=87mm]{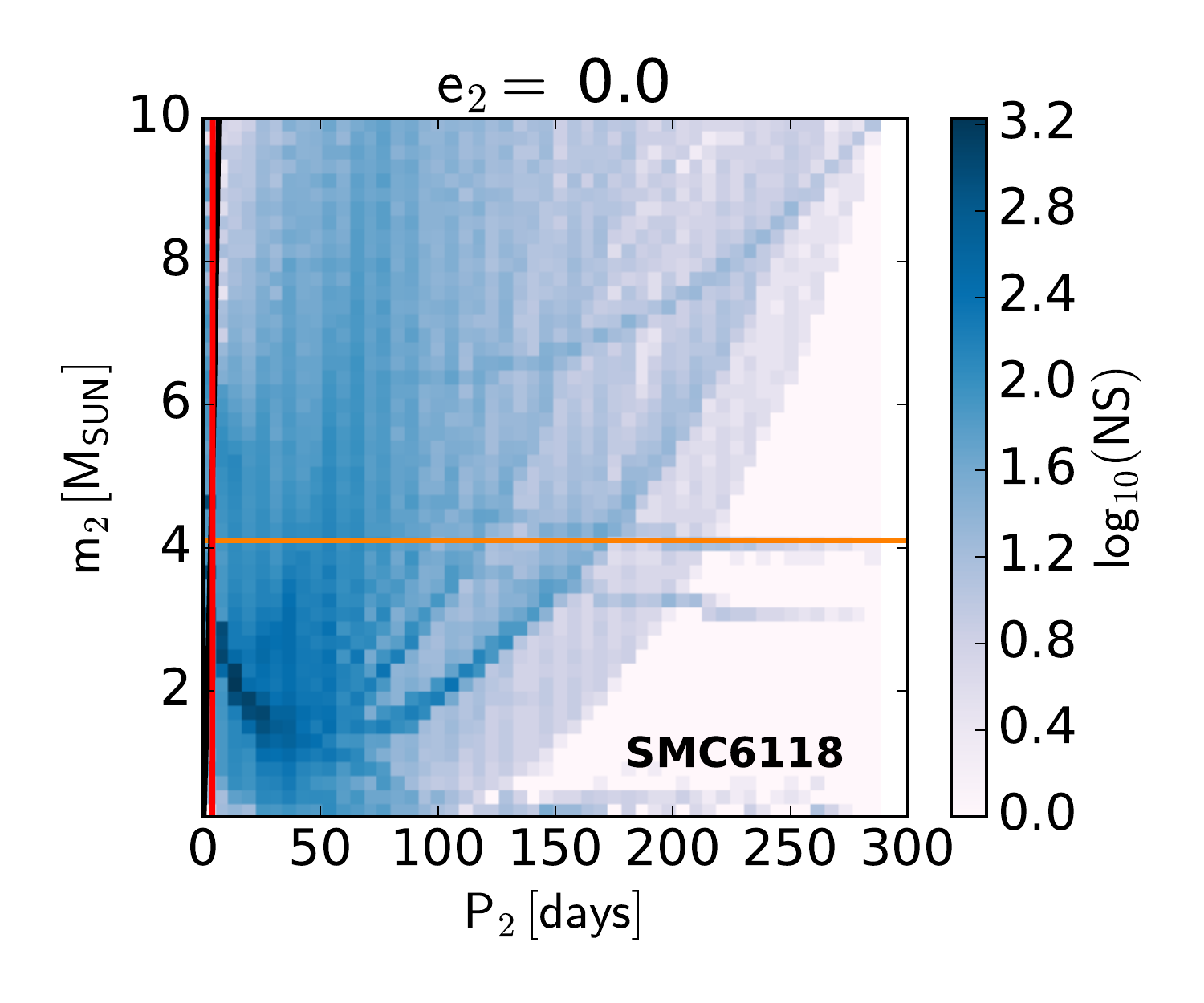} &
                \includegraphics[width=87mm]{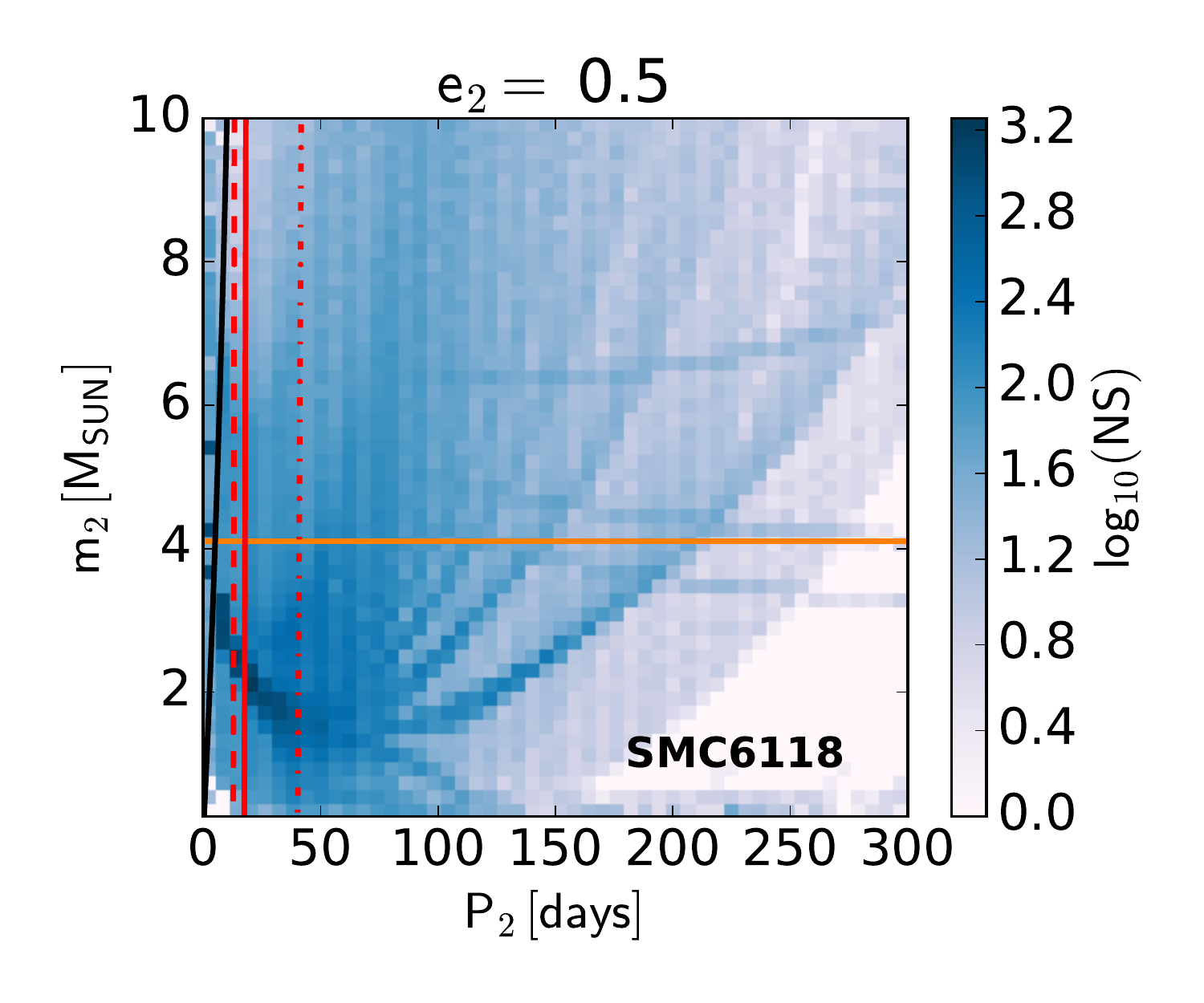} \\
        \end{tabular}
  \caption{\textit{continue}}
\end{figure*}

\FloatBarrier
\section{Eclipse timing residual diagrams}

\begin{figure*}[ht!]
\centering
        \begin{tabular}{@{}cc@{}}
                \includegraphics[width=87mm]{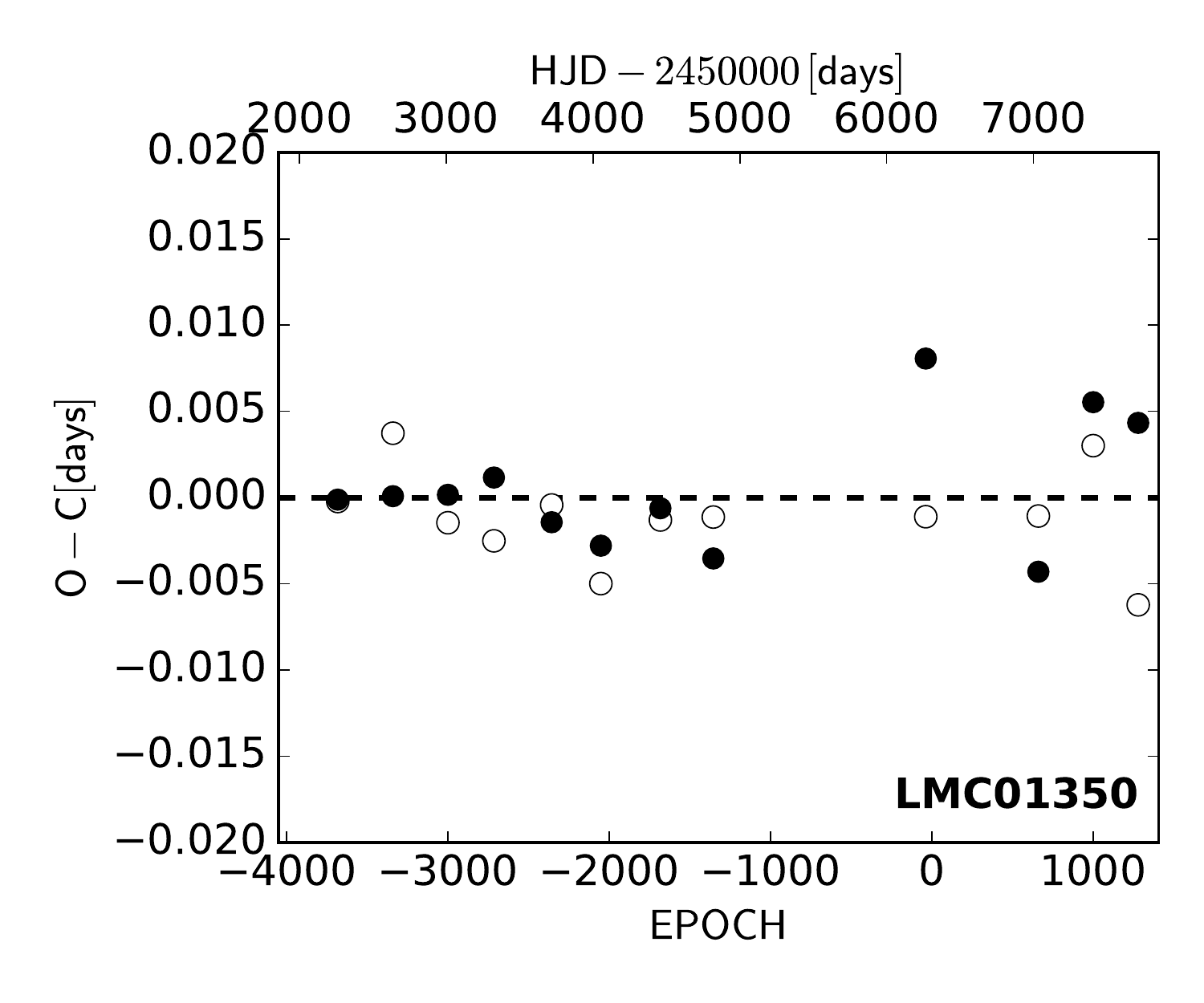} &
                \includegraphics[width=87mm]{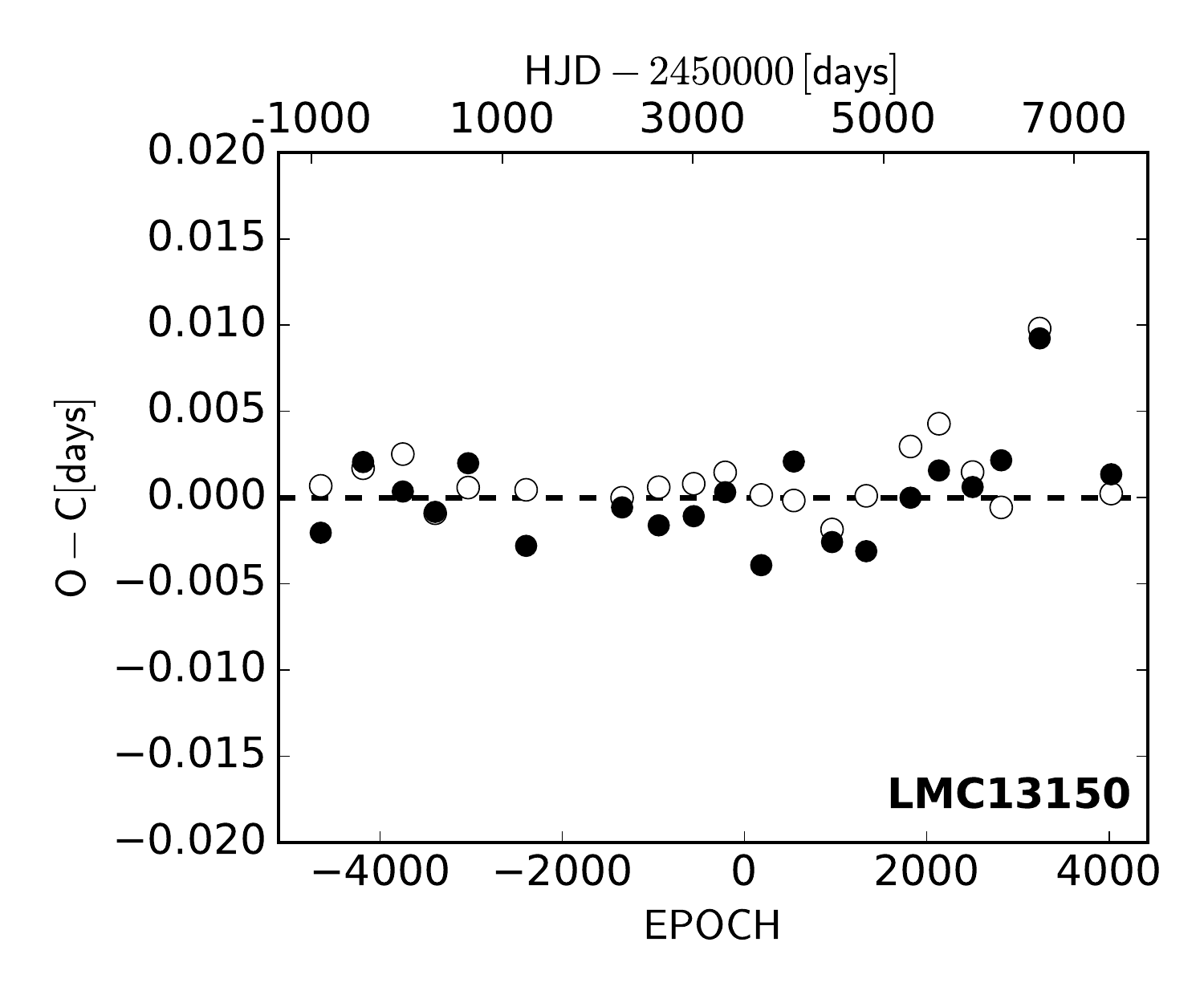} \\
                \includegraphics[width=87mm]{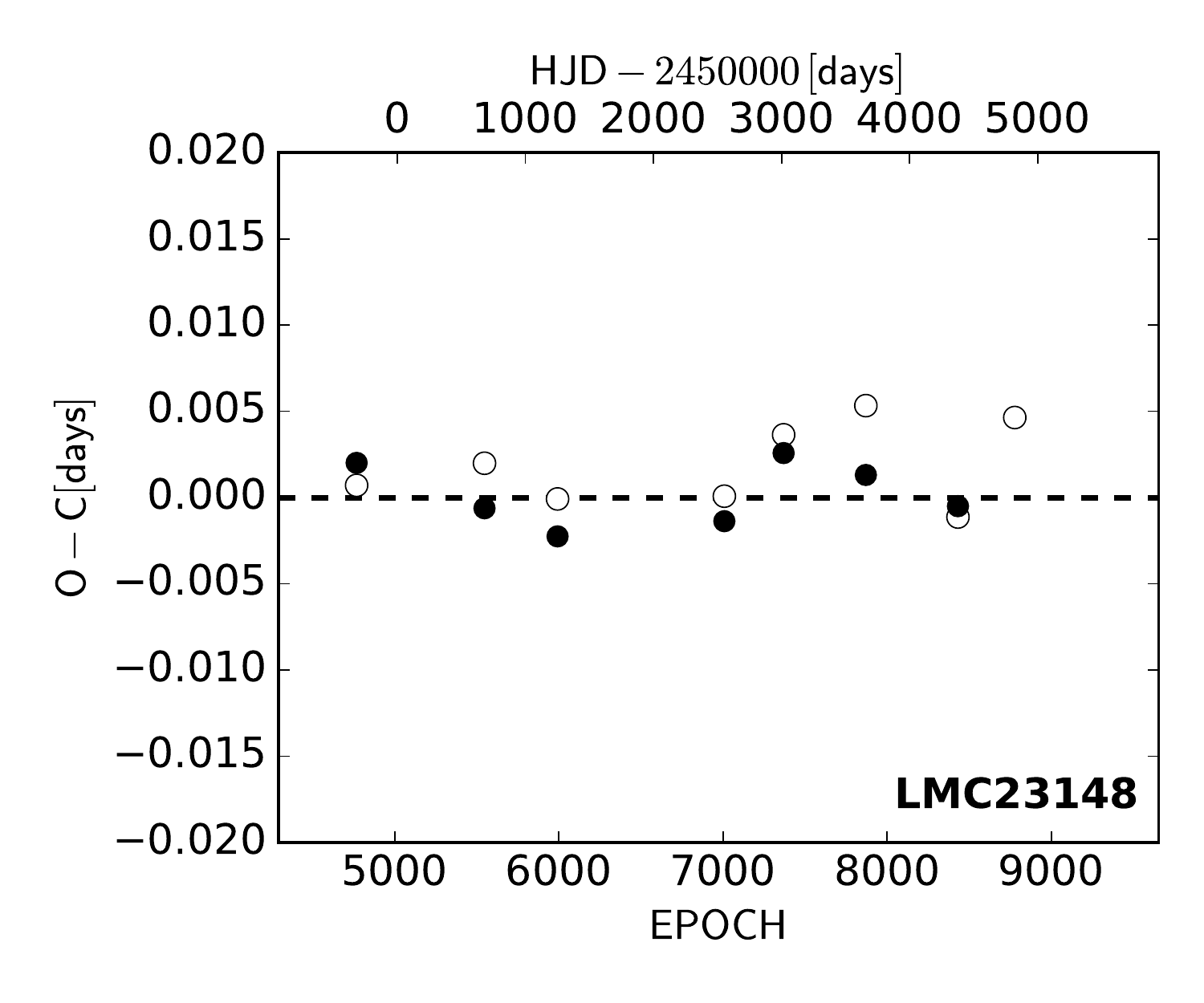} &
                \includegraphics[width=87mm]{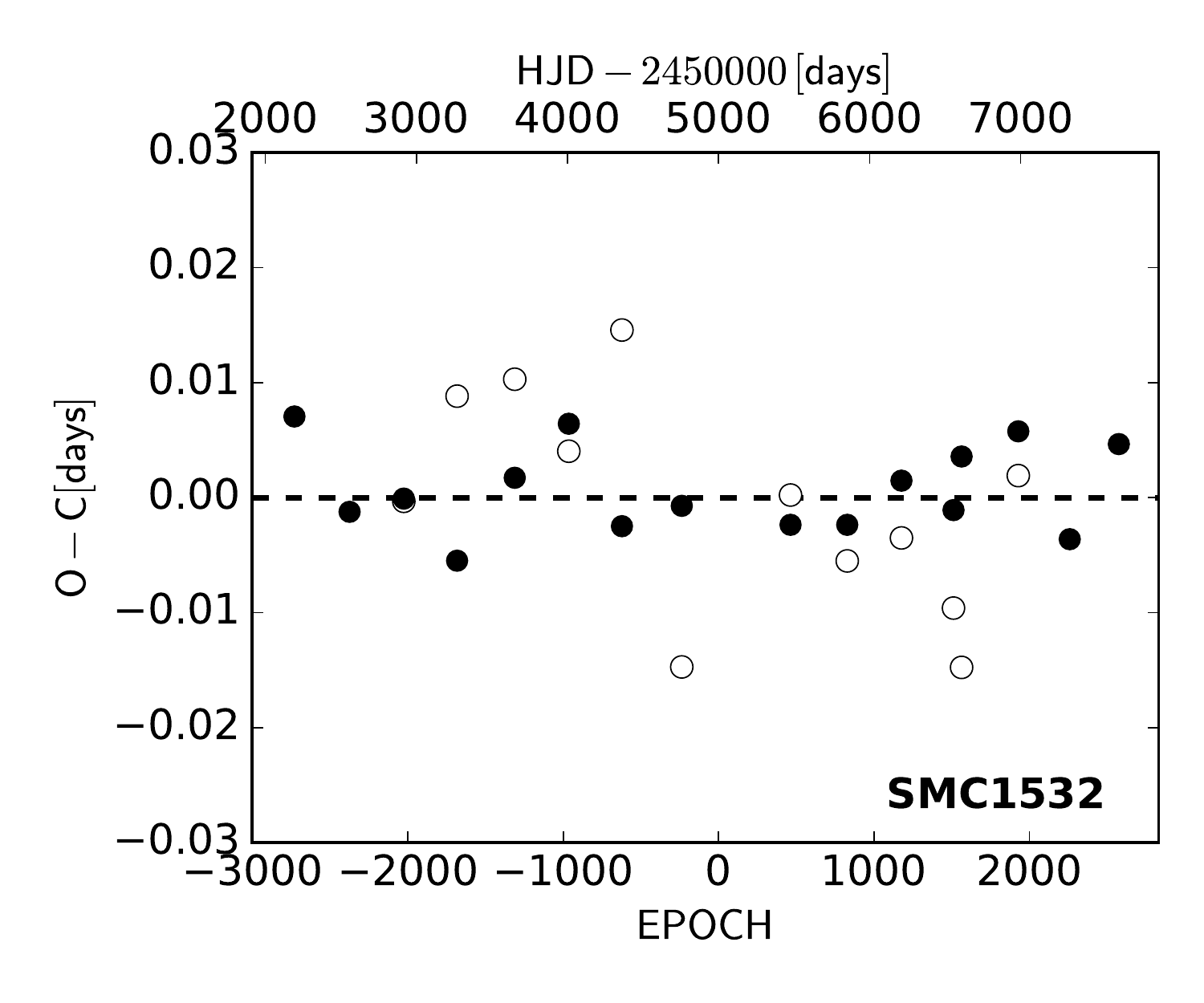} \\
                \includegraphics[width=87mm]{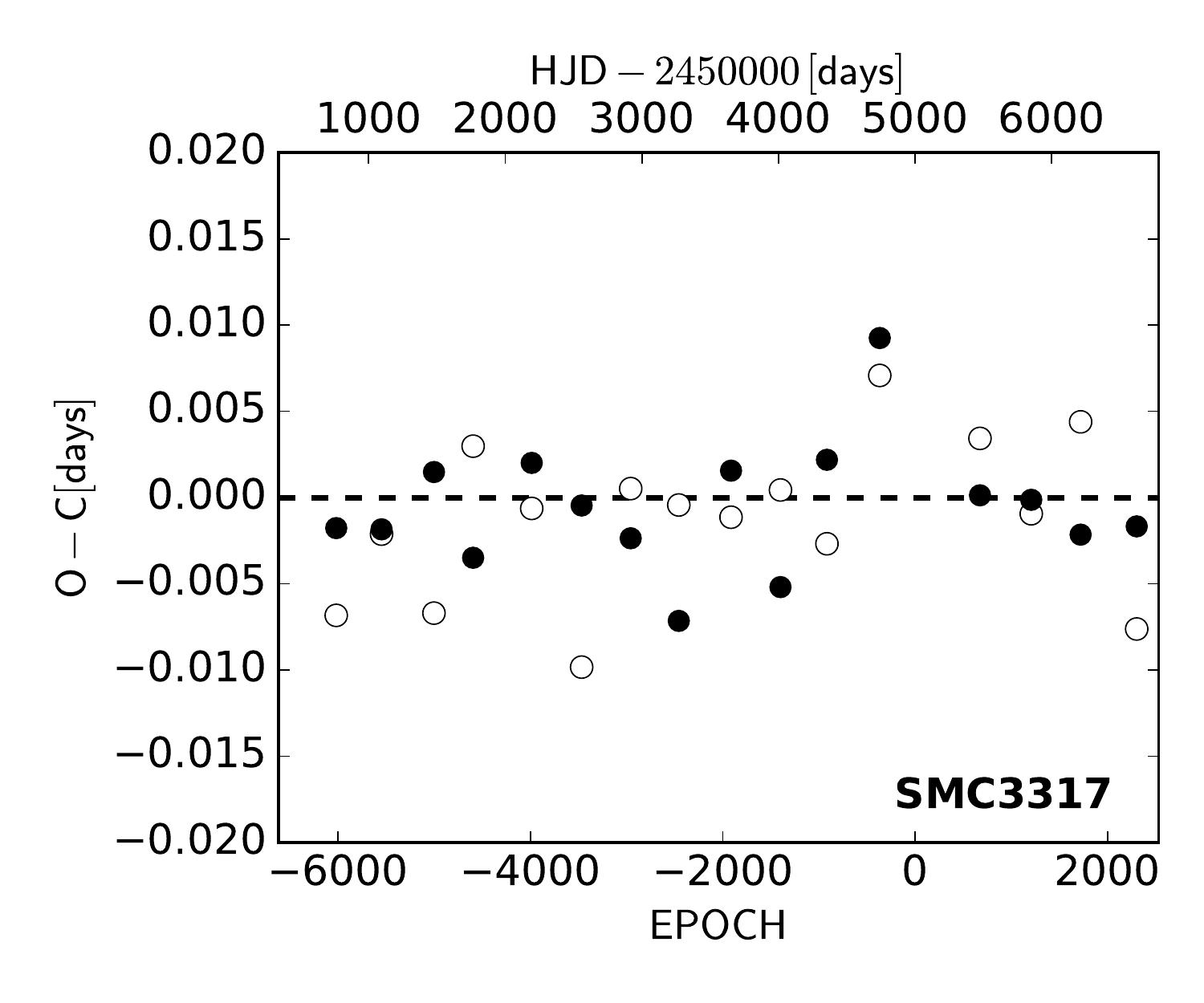} &
                \includegraphics[width=87mm]{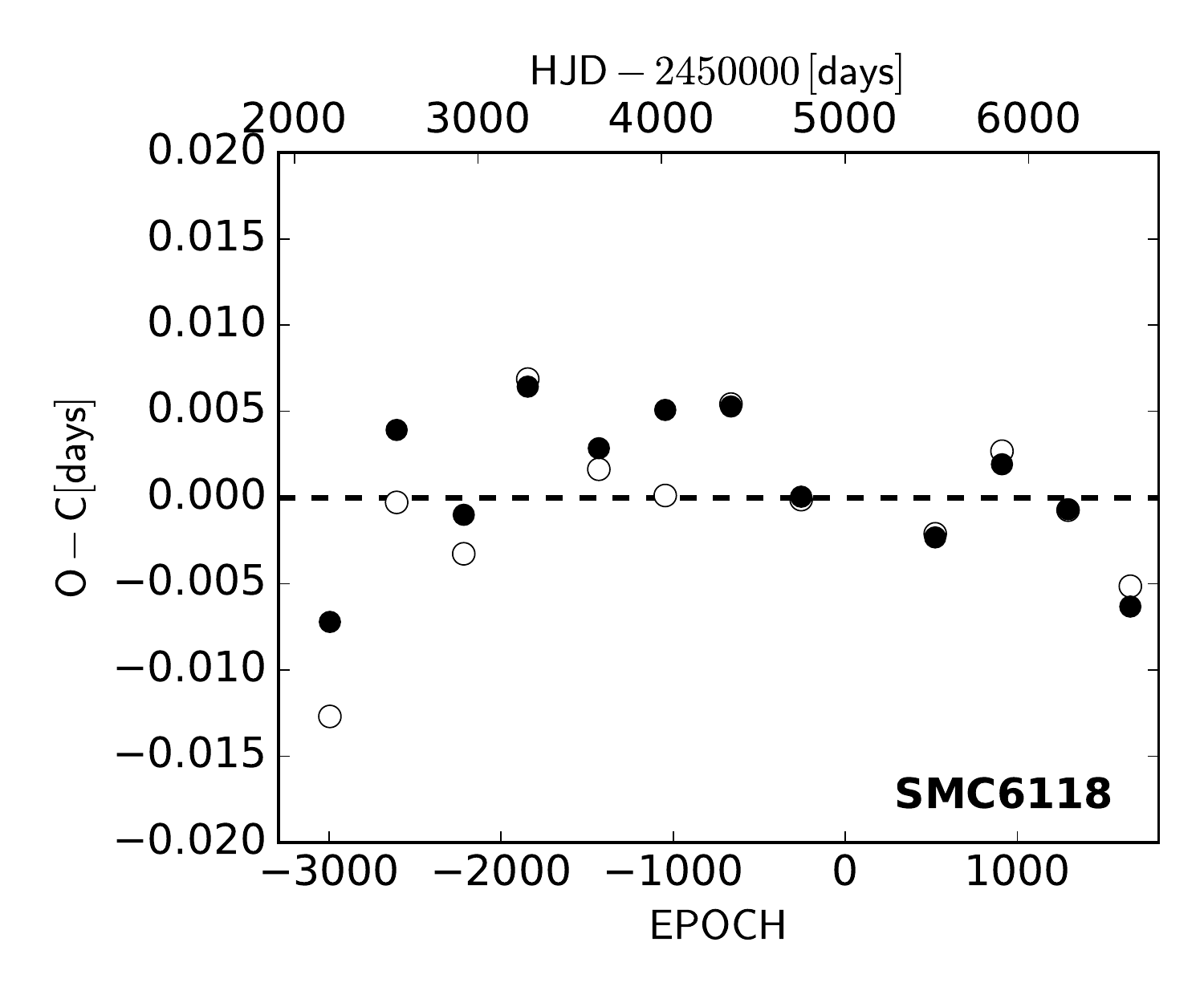} \\
        \end{tabular}
  \caption{Eclipse timing residual (observed $-$ computed) diagrams for selected systems with respect to mean linear ephemerides listed in 
  tables \ref{tab.ephem_lmc} and \ref{tab.ephem_smc}. Black and white points represent primary
   and secondary minima, respectively.}
   \label{fig.oc}
\end{figure*}

\newpage
\FloatBarrier
\section{Light curves of selected systems}

\begin{figure*}[ht!]
\centering
        \begin{tabular}{@{}cc@{}}
                \includegraphics[height=110mm]{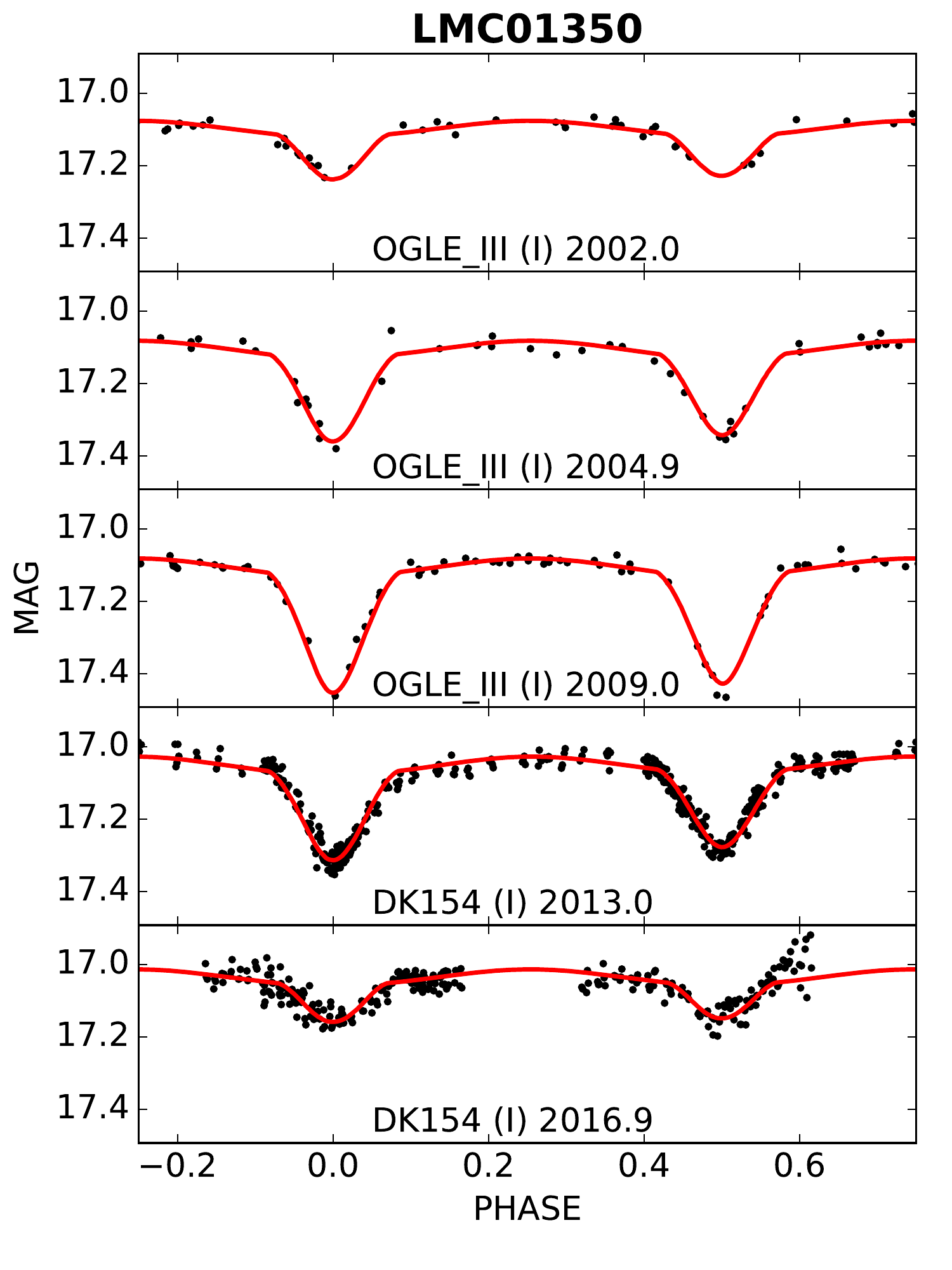} &
                \includegraphics[height=110mm]{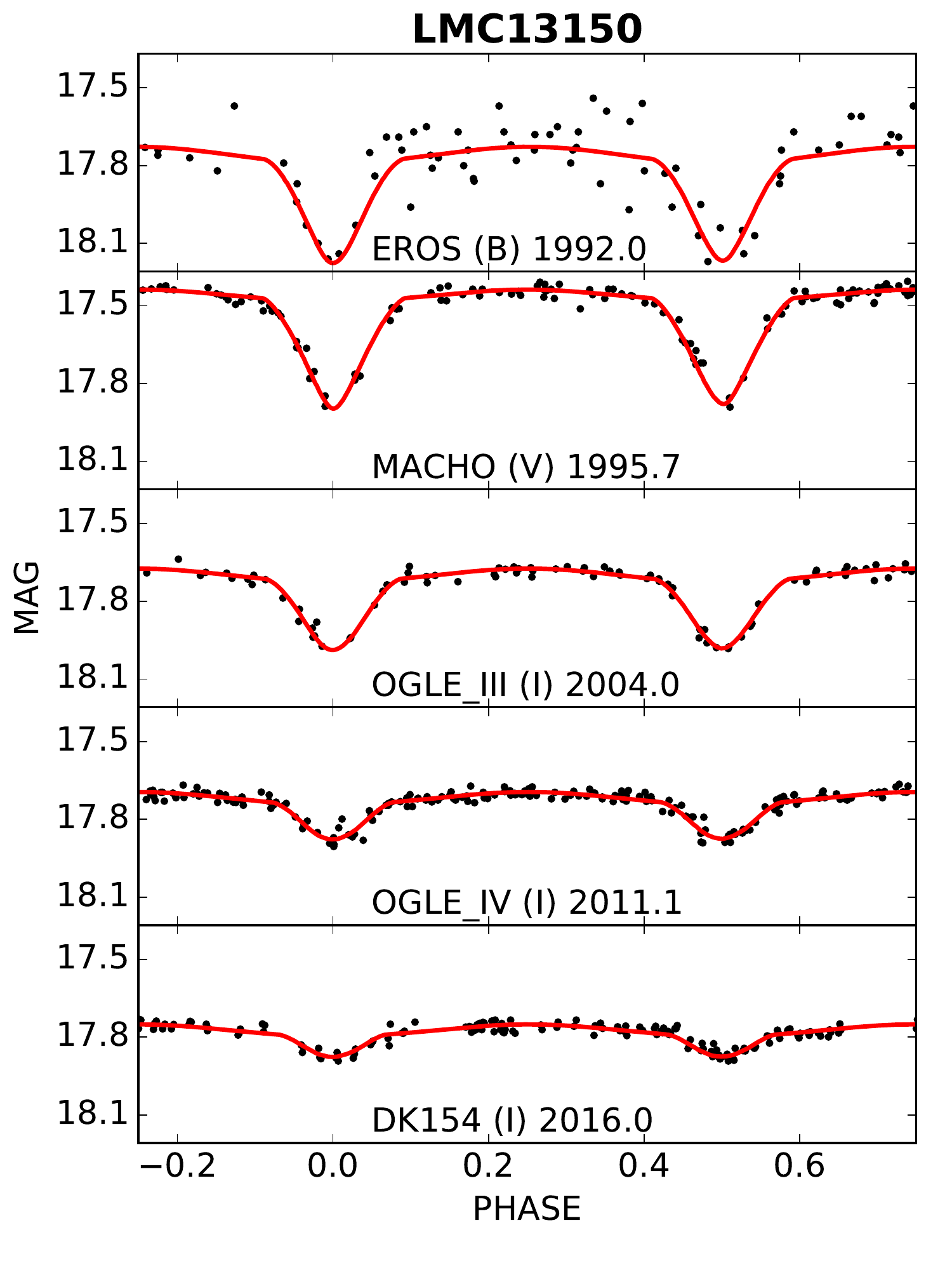} \\
                \includegraphics[height=110mm]{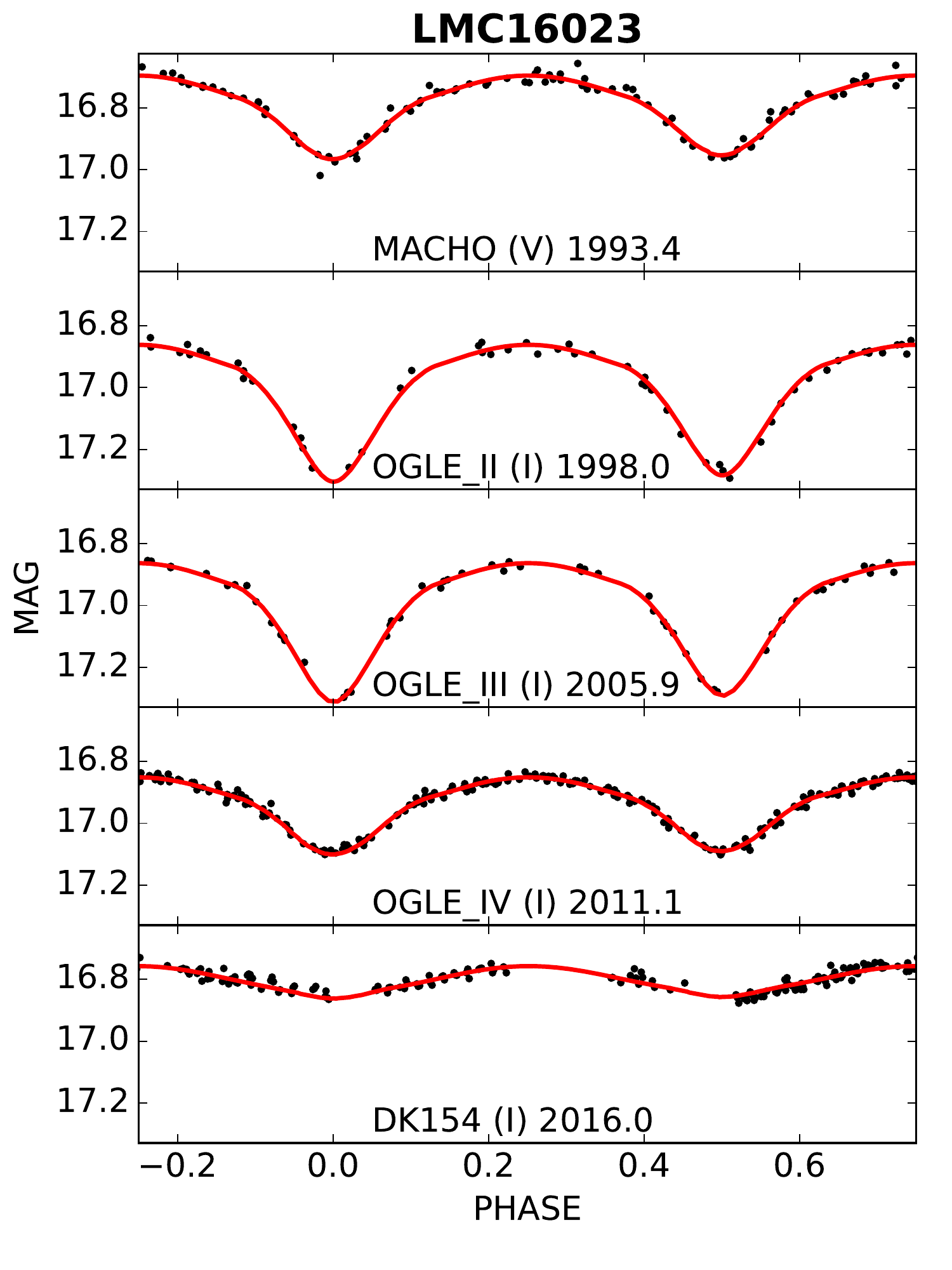} &
                \includegraphics[height=110mm]{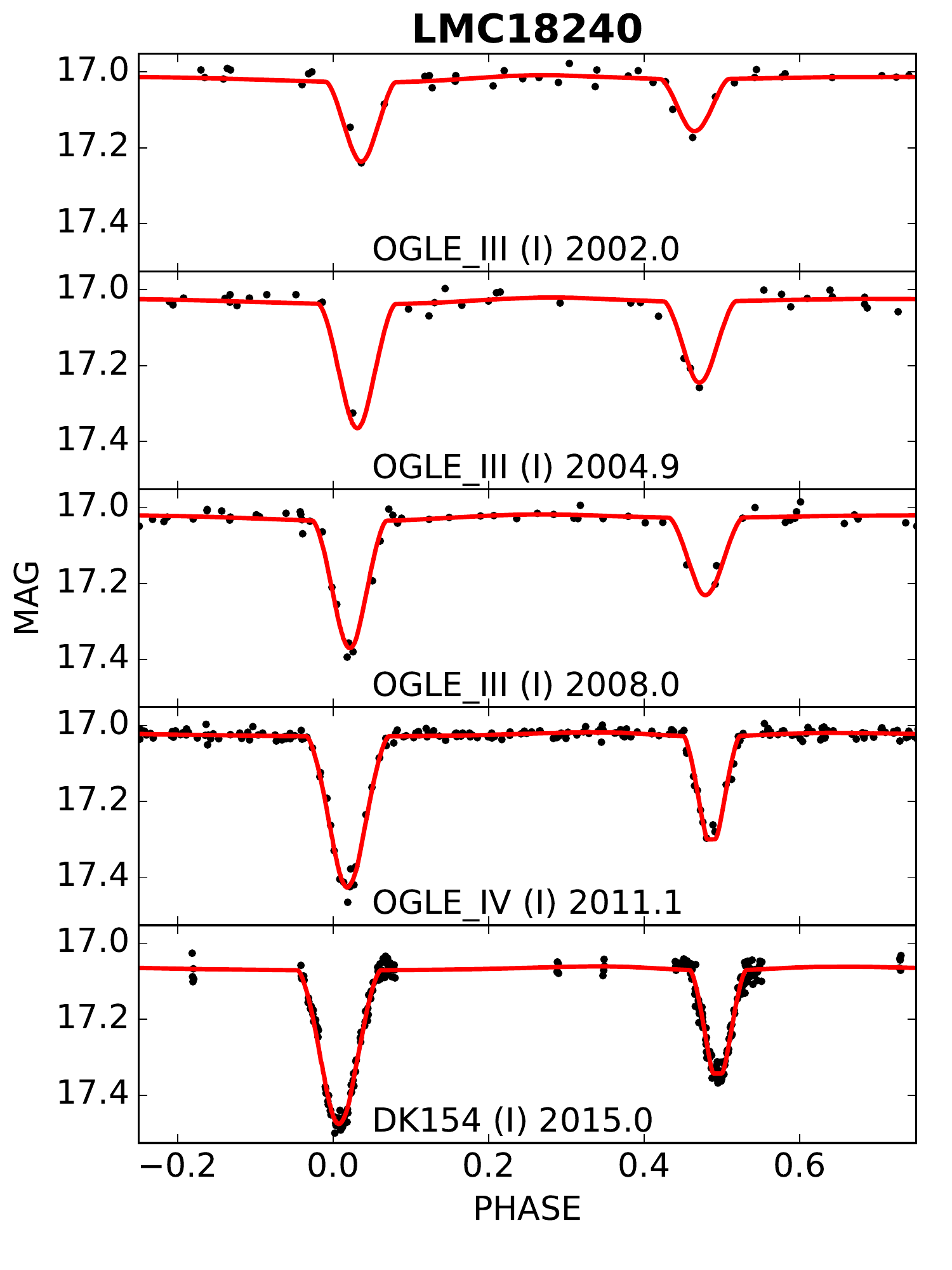} \\
        \end{tabular}
  \caption{Light curves of selected eclipsing binaries. Only several modeled LCs for each system 
  are shown. For each LC, survey designation, photometric band, and central time of given interval are listed.}
   \label{fig.lc}
\end{figure*}

\begin{figure*}
\ContinuedFloat 
\centering
        \begin{tabular}{@{}cc@{}}
                \includegraphics[height=110mm]{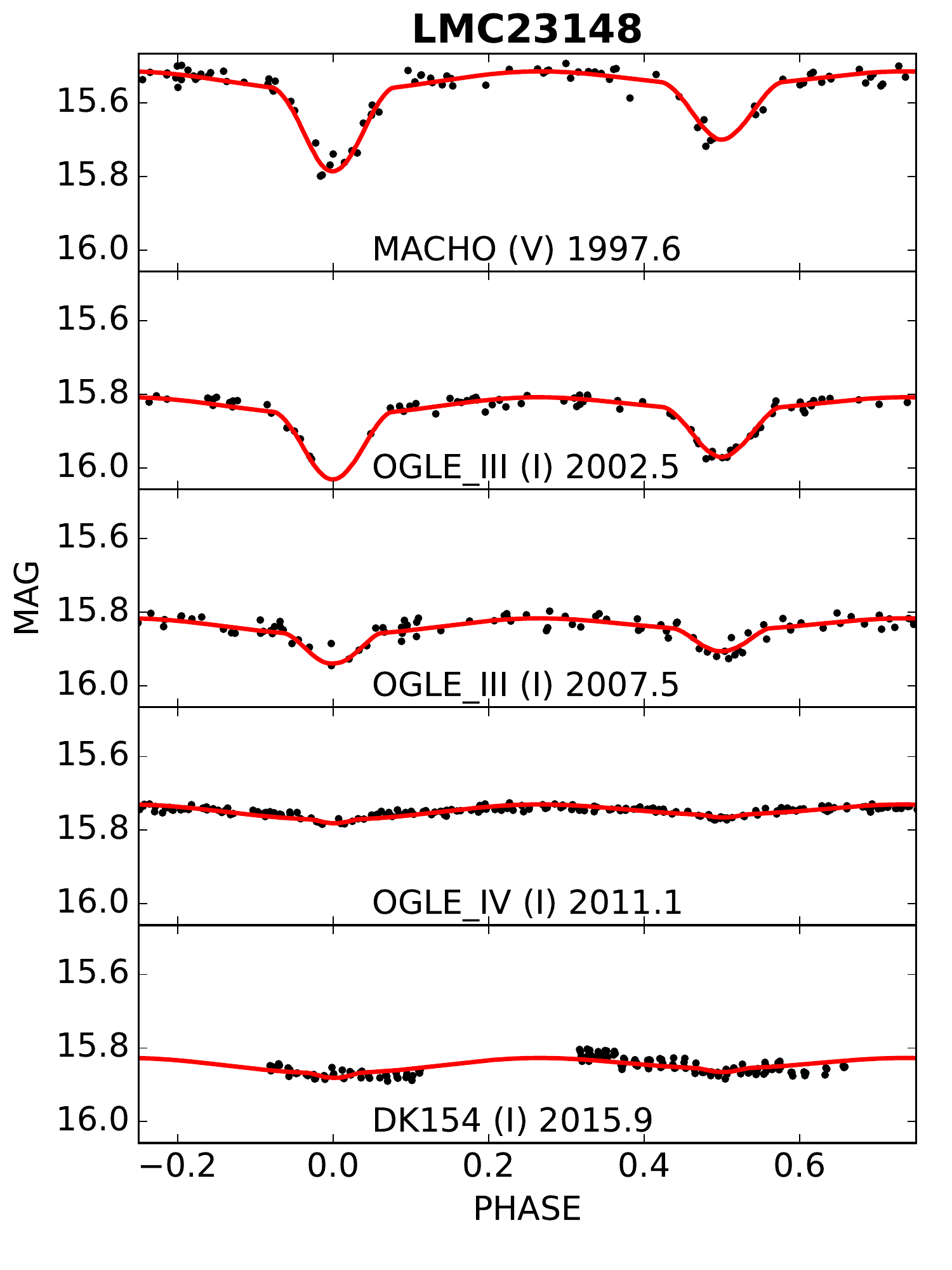} &
                \includegraphics[height=110mm]{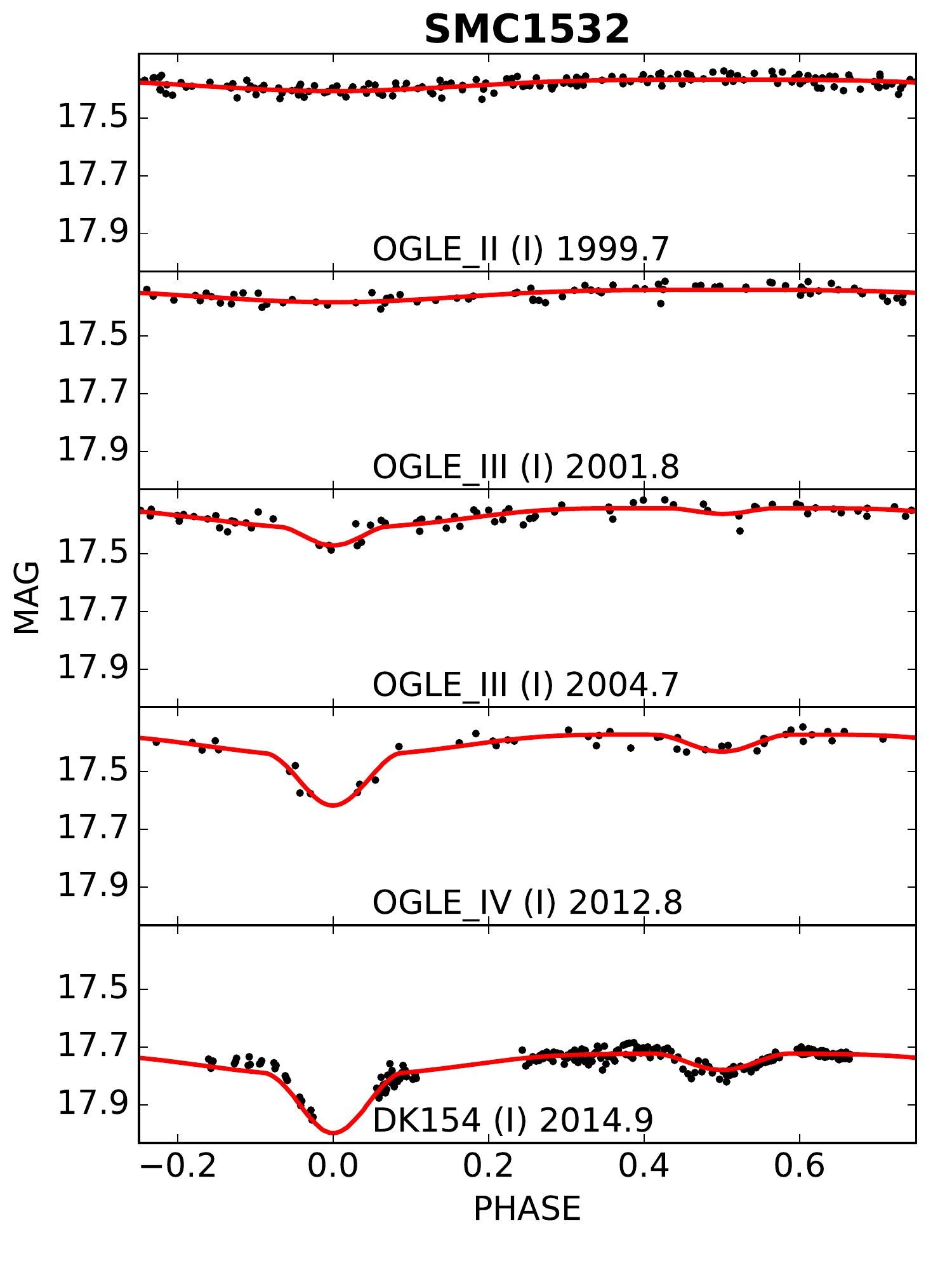} \\
                \includegraphics[height=110mm]{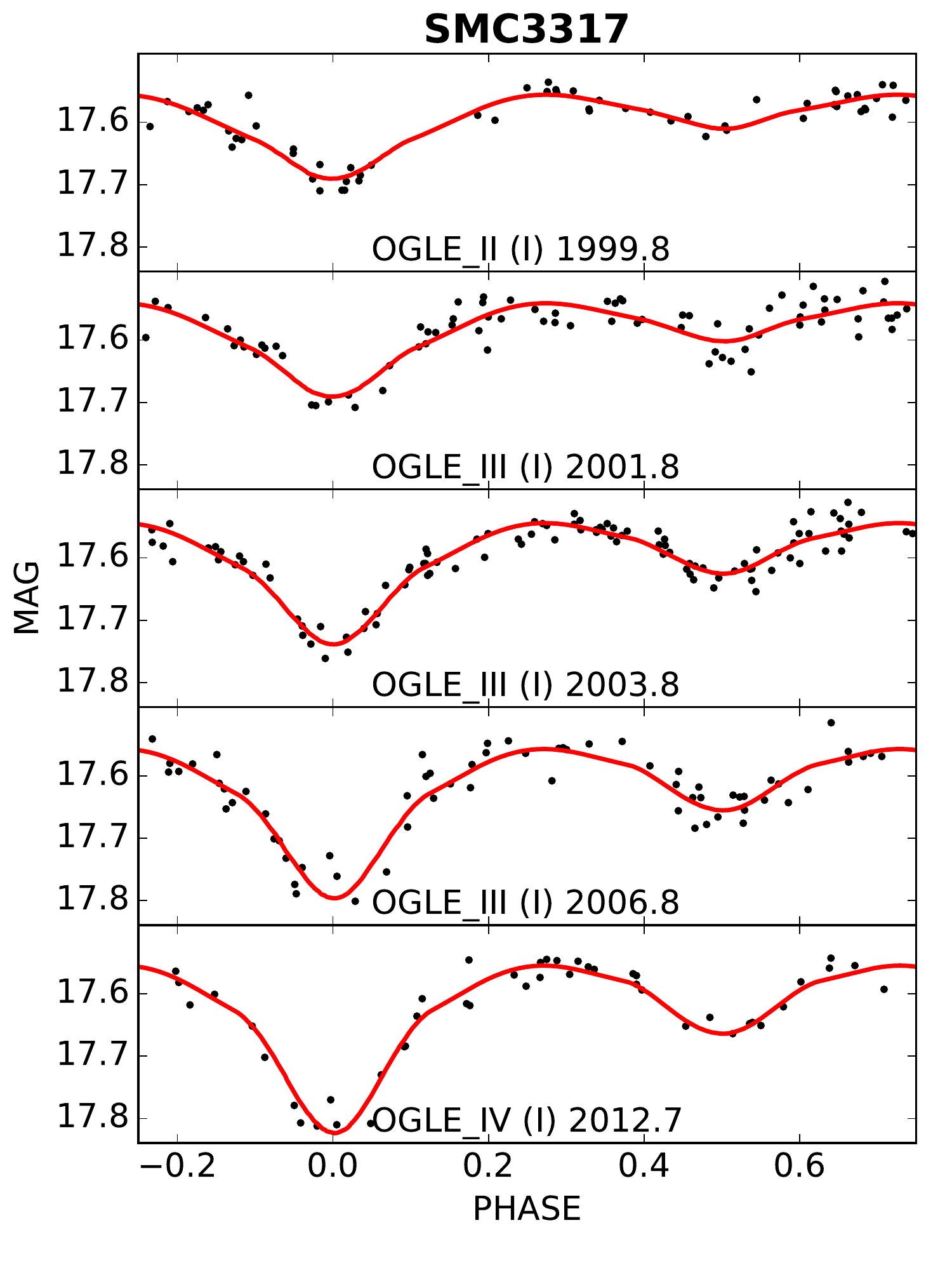} &
                \includegraphics[height=110mm]{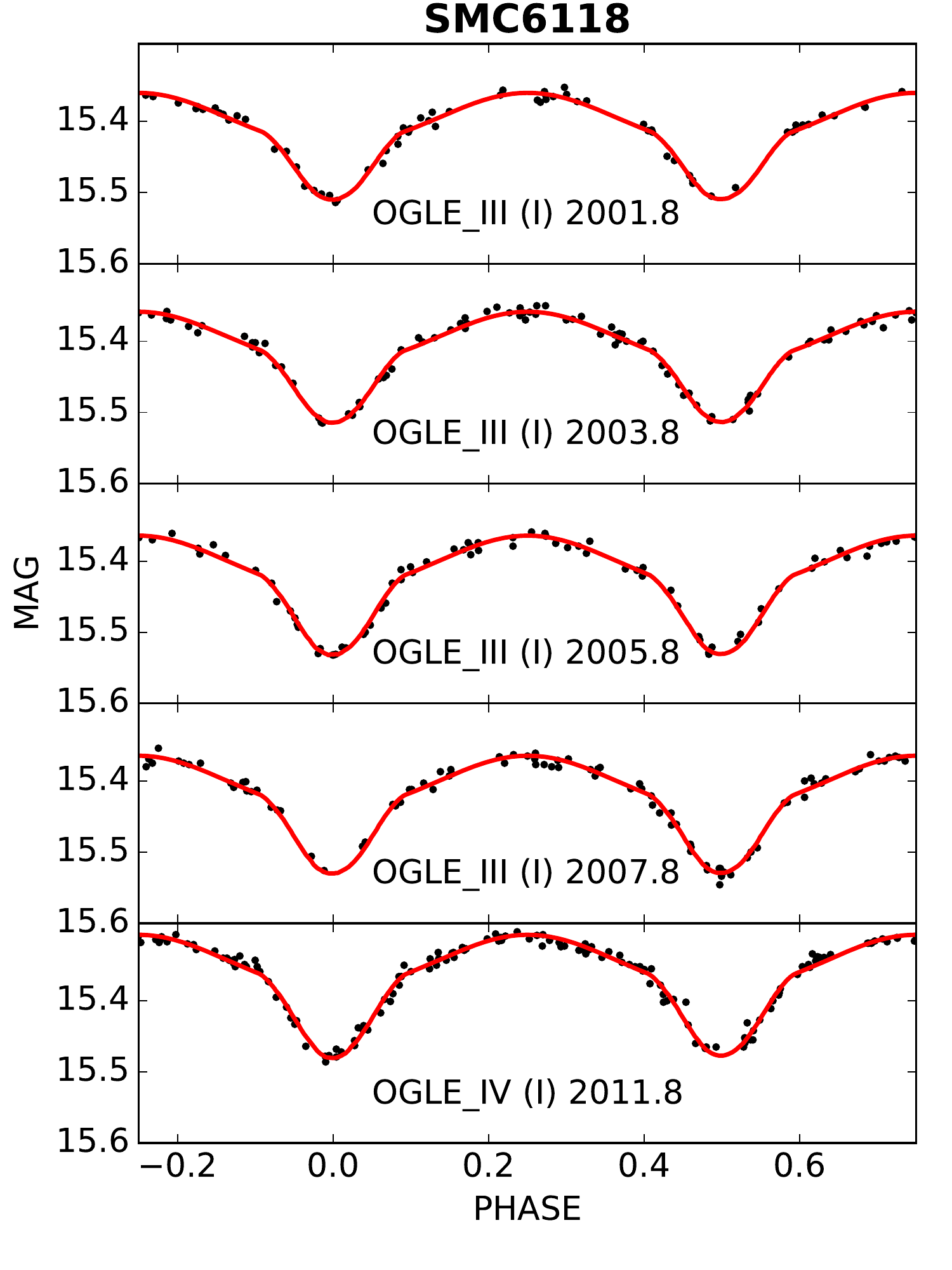} \\
        \end{tabular}
  \caption{\textit{continue}}
\end{figure*}

\newpage
\FloatBarrier
\section{Definition of symbols}
\begin{table*}[ht!]
\caption{Definition of symbols.}             
\label{tab.defsymb}      
\centering          
        \begin{tabular}{ c l c}   
        \hline\hline   
        
        & Parameter & Symbol \\
        \hline
        Mass & & \\
        & Inner binary components & $m_0$, $m_1$ \\
        & Third body & $m_2$ \\
        & Total mass of the inner binary & $M_1$ \\ 
        & Total mass of the system & $M_2$ \\
        & Mass ratio of inner binary components & $q$ \\
        \hline
        Eccentricity & Inner and outer orbit & $e_1$, $e_2$ \\
        \hline
        Seminajor axis & Inner and outer orbit & $a_1$, $a_2$ \\
        \hline
        Orbital angular momentum & Inner and outer orbit & $l_1$, $l_2$ \\
        \hline
        Argument of periastron & Inner and outer orbit & $\omega_1$, $\omega_2$ \\
        \hline
        Angular velocity & Nodal line precession & $\dot\Omega$ \\                             
        \hline
        Period & & \\
        & Orbital period of the inner binary & $P_1$ \\
        & Orbital period of the third component & $P_2$ \\
        & Sidereal period of the inner binary & $P_\mathrm{s}$ \\
        & Nodal line precession & $P_\mathrm{nodal}$ \\
        \hline
        Inclination & & \\
        & Observable inclination of the inner binary & $i_0$ \\
        & Angles between the invariant plane and the orbits & $i_1$, $i_2$  \\
        & Observable inclination of the third body & $i_3$, $i_\mathrm{3,inv}$ \\
        & Angle between invariant plane and plane tangent to the celestial sphere & $I$ \\
        & Mutual inclination of inner and outer orbit & $j$ \\
        \hline
        Reference epochs & & \\
        & Reference minimum of eclipsing binary & HJD0 \\
        & Periastron passage & $T_0$ \\
        & Passage nodal line through the plane tangent to the celestial sphere & $t_0$ \\
        
        \hline
        Photometric solution & & \\
        & Relative radii of EB components with respect to the semimajor axis & $R_0/a_1$, $R_1/a_1$ \\
        & Temperatures of primary and secondary component & $T_1$, $T_2$ \\
        & Relative luminosities of primary and secondary component & $L_\mathrm{X0}$, $L_\mathrm{X1}$ \\
        & Bolometric magnitudes of primary and secondary component & $M_\mathrm{bol0}$, $M_\mathrm{bol1}$ \\
        \hline
        Amplitude of ETVs & & \\
        & Dynamical delay & $A_\mathrm{phys}$ \\
        & R{\o}emer's delay & $A_\mathrm{LTE}$ \\
        \hline
        \end{tabular}
\end{table*}

\end{appendix}

\end{document}